\documentclass{pasa}

\usepackage{titlesec}

\usepackage{graphicx}
\usepackage{hyperref}
\usepackage{aas_macros}
\usepackage{fourier}
\usepackage[dvipsnames]{xcolor}
\usepackage{bm}

\makeatletter
\newcommand\level[1]{%
  \ifcase#1\relax\expandafter\chapter\or
    \expandafter\section\or
    \expandafter\subsection\or
    \expandafter\subsubsection\else
    \def\next{\@level{#1}}\expandafter\next
  \fi}
\newcommand{\@level}[1]{%
  \@startsection{level#1}
    {#1}
    {\z@}%
    {-3.25ex\@plus -1ex \@minus -.2ex}%
    {1.5ex \@plus .2ex}%
    {\normalfont\normalsize\bfseries}}

\DeclareMathAlphabet{\mathcal}{OMS}{cm}{m}{n}

\newcommand\be{\begin{equation}}
\newcommand\ee{\end{equation}}
\newcommand\bea{\begin{eqnarray}}
\newcommand\eea{\end{eqnarray}}
\newcommand{\de}{{\rm d}}
\newcommand{\g}{{\rm g}}
\newcommand{\vv}{{\rm v}}
\newcommand{\HH}{{\cal H}}
\newcommand{\bV}{\bm{V}} 
\newcommand{\bn}{\bm{n}}

\def\VEV#1{\left\langle #1 \right\rangle}

\usepackage{graphicx}
\usepackage{hyperref}
\usepackage{aas_macros}
\usepackage{fourier}
\usepackage[utf8]{inputenc}

\def\AU{\rm ~AU}
\let\ga\gtrsim
\let\la\lesssim
\usepackage{graphicx}
\usepackage{float}
\usepackage{subfiles}
\usepackage{color}
\usepackage{titlesec}
\usepackage{amssymb}
\usepackage{ulem}
\usepackage{epsfig}
\usepackage{subfigure}

\usepackage{aas_macros}
\usepackage{hyperref} 
\hypersetup{colorlinks,citecolor=blue,linkcolor=blue,urlcolor=blue}
\usepackage{xcolor}
\newcommand{\srs}[1]

\usepackage{graphicx}
\usepackage{epsfig}
\usepackage{subfigure}
\usepackage{aas_macros,braket}
\usepackage{kantlipsum}
\usepackage{verbatim}

\newcommand{\Ts}{T_{\rm S}}
\newcommand{\Tk}{T_{\rm K}}

\newcommand{\mx}{m_{\text{X}}}

\def\lsim{\,\lower2truept\hbox{${< \atop\hbox{\raise4truept\hbox{$\sim$}}}$}\,}
\def\gsim{\,\lower2truept\hbox{${> \atop\hbox{\raise4truept\hbox{$\sim$}}}$}\,}
\def\mnras{Monthly Notices of the Royal Astronomical Society}
\def\apj{The Astrophysical Journal}
\def\apjl{Astrophysical Journal Letters}
\def\aap{Astronomy and Astrophysics}
\def\physrep{Physics Reports}
\def\jcap{Journal of Cosmology and Astroparticle Physics}
\def\prd{Physical Review D}

\def\nat{Nature}
\def\apss{Astrophysics and Space Science}
\def\apjs{The Astrophysical Journal Supplement Series}
\def\aj{The Astronomical Journal}
\newcommand{\ltsim}{\mathrel{\raise.3ex\hbox{$<$\kern-.75em\lower1ex\hbox{$\sim$}}}}

\def\degree{\nobreak\ifmmode{^\circ}\else{$^\circ$}\fi}

\title[Fundamental Physics with the Square Kilometre Array]{Fundamental Physics with the Square Kilometre Array}

\author[Weltman et al.]
{A. Weltman,$^\#$
P. Bull,$^*$
S. Camera,$^*$
K. Kelley,$^*$
H. Padmanabhan,$^*$
J. Pritchard,$^*$
A. Racanelli,$^*$
S. Riemer-S\o{}rensen,$^*$
L. Shao,$^*$
S. Andrianomena,
E. Athanassoula,
D. Bacon,
R. Barkana,
G. Bertone,
C. Bonvin,
A. Bosma,
M.~Br\"uggen, 
C.~Burigana, 
C. B{\oe}hm, 
F. Calore, 
J. A. R. Cembranos, 
C. Clarkson, 
R.~M.~T.~Connors, 
\'A. de la Cruz-Dombriz, 
P.~K.~S.~Dunsby,
J. Fonseca,
N. Fornengo,
D. Gaggero,
I. Harrison,
J. Larena,
Y.-Z. Ma,
R. Maartens,
M. M\'endez-Isla,
S.~D.~Mohanty,
S.~G.~Murray, 
D. Parkinson,
A. Pourtsidou,
P. J. Quinn,
M. Regis,
P. Saha,
M. Sahl\'en,
M. Sakellariadou,
J. Silk,
T.~Trombetti,
F. Vazza,
T. Venumadhav, 
F. Vidotto,
F. Villaescusa-Navarro,
Y. Wang,
C. Weniger,
L.~Wolz,
F. Zhang,
B. M. Gaensler$^\dagger$ \\

(Author affiliations are listed in \S\ref{sec_affil} at the end of the paper.) \\

\affil{$^\#$Editor}
\affil{$^*$Section lead}
\affil{$^\dagger$Convenor} 
}

\jid{PASA}
\doi{10.1017/pas.\the\year.xxx}
\jyear{\the\year}

\usepackage{aas_macros}
\usepackage{hyperref} 
\hypersetup{colorlinks,citecolor=blue,linkcolor=blue,urlcolor=blue}

\hypersetup{draft}

\begin{document}

\begin{frontmatter}
\maketitle

\begin{abstract}
The Square Kilometre Array (SKA) is a planned large radio interferometer designed to operate over a wide range of frequencies, and with an order of magnitude greater sensitivity and survey speed than any current radio telescope. The SKA will address many important topics in astronomy, ranging from planet formation to distant galaxies. However, in this work, we consider the perspective of the SKA as a facility for studying physics. We review four areas in which the SKA is expected to make major contributions to our understanding of fundamental physics: cosmic dawn and reionisation; gravity and gravitational radiation; cosmology and dark energy; and dark matter and astroparticle physics. These discussions demonstrate that the SKA will be a spectacular physics machine, which will provide many new breakthroughs and novel insights on matter, energy and spacetime.
\end{abstract}

\end{frontmatter}

\setcounter{tocdepth}{5}
\tableofcontents

\section{Introduction}
\label{sec:intro}

The Square Kilometre Array (SKA) is a large international collaboration, with the goal of building the world's largest and most powerful radio telescope. The first phase of the SKA (``SKA1'') will begin operations in the early 2020s, and will comprise two separate arrays: SKA1-Low, which will consist of around 130\,000 low-frequency dipoles in Western Australia, and SKA1-Mid, which will be composed of $\sim$200 dishes in the Karoo region of South Africa \citep{dewdney2016,braun2017}. The second phase, SKA2 will be an order of magnitude larger in collecting area than SKA1, and will take shape in the late 2020s.

The science case for the SKA is extensive and diverse: the SKA will deliver spectacular new data sets that are expected to transform our understanding of astronomy, ranging from planet formation to the high-redshift Universe \citep{aaska2015}. However, the SKA will also be a powerful machine for probing the frontiers of fundamental physics. To fully understand the SKA's potential in this area, a focused workshop on ``Fundamental Physics with the Square Kilometre Array''\footnote{See {http://skatelescope.ca/fundamental-physics-ska/}.} was held in Mauritius in May 2017, in which radio astronomers and theoretical physicists came together to jointly consider ways in which the SKA can test and explore fundamental physics.

This paper is not a proceedings from this workshop, but rather is a white paper that fully develops the themes explored. The goal is to set out four broad directions for pursuing new physics with the SKA, and to serve as a  bridging document accessible for both the physics and astronomy communities. 
In \S\ref{sec_eor} we consider cosmic dawn and reionisation, in
\S\ref{sec_grav} discuss strong gravity and pulsars, in
\S\ref{sec_cosmology} we examine cosmology and dark energy, and in
\S\ref{sec_astropart} we review dark matter and astroparticle physics. In each of these sections, we introduce the topic, set out the key science questions, and describe the proposed experiments with the SKA.

\section{Cosmic Dawn and Reionisation}
\label{sec_eor}

Cosmic Dawn represents the epoch of formation of the first stars and  
galaxies that eventually contributed to the reionisation of the  
universe. This period is potentially observable through the 21-cm  
spin-flip transition of neutral hydrogen, redshifted to radio  
frequencies. In this section, we provide an overview of the ways in  
which we can use upcoming SKA
observations of cosmic dawn and of the epoch of reionisation (EoR) to place  
constraints on fundamental physics. These include the possible effects  
of warm dark matter on the 21-cm power spectrum during cosmic dawn,  
variations of fundamental constants such as the fine structure  
constant, measurements of the lensing convergence power spectrum,  
constraints on inflationary models, and cosmic microwave background (CMB) 
spectral distortions and  
dissipation processes. We describe foreseeable challenges  
in the detection and isolation of the fundamental physics parameters  
from the observations of cosmic dawn and reionisation, possible ways  
towards overcoming them through effective isolation of the astrophysics,  
synergies with other probes, and foreground removal techniques.

\subsection{Introduction}
Cosmologists seek to use the Universe as an experiment from which to learn
about new physics. There has already been considerable success in extracting
fundamental physics from the CMB and from large-scale structure (LSS) 
measurements
from large galaxy surveys. These CMB and LSS observations cover only a small 
fraction of the total observable Universe, both in terms of cosmic history and 
observable volume. A promising new technique for providing observations over 
the redshift range $z=3-27$ is by measurements of the 21-cm hyperfine line of 
neutral hydrogen, which can be observed redshifted to radio frequencies 
detectable by the SKA \citep{2015aska.confE...1K}. 

Since hydrogen is ubiquitous in intergalactic space, 21-cm observations offer a 
route to mapping out fluctuations in density, which contain information about 
cosmological parameters. As the 21-cm line is affected by various types of 
radiation, observing it gives a way to detect and study some of the first 
astrophysical objects, including stars and black holes. Once detected, the 
21-cm signal might also provide information about the high-redshift Universe 
that can constrain other physics, such as the effects of warm dark matter, 
annihilation or scattering of dark matter, the variation of fundamental 
constants, and possibly also tests of inflationary models 
\citep{2015aska.confE..12P}.

These are exciting times for cosmic dawn and reionisation, as the pathfinder 
experiments LOFAR \citep{Patil2017}, MWA \citep{Dillon2015}, PAPER 
\citep{ali2015}, and HERA \citep{DeBoer2017} have begun to collect data and set 
upper limits on the 21-cm power spectrum, while EDGES has reported a tentative 
detection \citep{bowman2018}. It is likely that in the next few years the 
cosmological 21-cm signal will open a new window into a previously unobserved 
period of cosmic history.

The rest of this section is organised as follows. In \S\ref{sec_eor_overview} 
we present a brief overview of the theory and observations related to cosmic 
dawn and the epoch of reionisation, and the various physical processes that 
influence the magnitude of the signal from these epochs. We summarise the status 
of observations in the field, including the upper limits to date from various 
experiments. We also provide a brief overview of the upcoming observations and 
modelling of the reionisation epoch. In \S\ref{sec:physics} we review aspects 
of fundamental physics that can be probed with the SKA, and in 
\S\ref{sec:challenges} we discuss some of the challenges to doing this. We 
provide a summary in \S\ref{sec:eor_conclusions}.

\subsection{Cosmic Dawn and Reionisation : theory and observations} 
\label{sec_eor_overview}

\subsubsection{Overview of the 21 cm signal}
\label{sec:overview}

The 21-cm line of neutral hydrogen corresponds to the transition between the 
singlet and triplet hyperfine levels of its electronic ground state, resulting 
from the interaction of proton and electron spins. The resulting transition has 
a rest frame frequency of 1.4 GHz, i.e., a wavelength of 21~cm. The electric 
dipole transition between the ground and excited hyperfine levels is forbidden 
due to parity; the lowest order transition occurs via a magnetic dipole, owing 
to which the triplet level has a vacuum lifetime of $\simeq 11 \ {\rm Myr}$. 
Due to this long lifetime, the dominant channels for the decay of the excited 
levels are either non-radiative \citep[atomic 
collisions;][]{1969ApJ...158..423A, Zygelman05}, or depend on the existing 
radiation field \citep[stimulated emission by CMB photons, or optical pumping by 
UV photons;][]{1952AJ.....57R..31W, 1958PIRE...46..240F}. This makes the 
relative population of the hyperfine levels a sensitive probe of the thermal 
state and density of the high-redshift intergalactic medium (IGM) and of early 
sources of ultraviolet radiation \citep{1975MNRAS.171..375S, 
1979MNRAS.188..791H, 1997ApJ...475..429M}.

Radio observations of this line are frequently used to map the velocity of 
neutral hydrogen (HI) gas in the Milky Way or in nearby galaxies, but currently 
it has not been detected in emission at redshifts $z>1$. When considering the 
21-cm line as a cosmological probe, it is standard to describe the measured 
intensity in terms of a brightness temperature\footnote{Brightness temperature 
is defined as the temperature that a blackbody would need to have to produce 
the observed surface brightness at a given observing wavelength.} and to 
consider the observed brightness temperature relative to some background 
source, typically either the CMB or a radio-bright point source. For cosmology, 
it is most useful to consider the case of the CMB backlight, for which the 
21-cm signal will then take the form of a spectral distortion over the whole 
sky.

\begin{figure}[tbp]
\begin{center}
\includegraphics[width = \columnwidth]{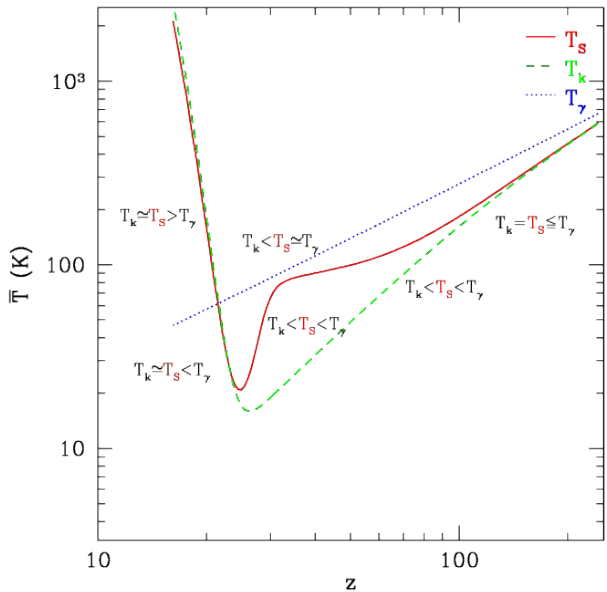}
\end{center}
\caption{Evolution of spin temperature $T_s$, gas temperature
$T_K$ and CMB temperature $T_{\gamma}$. This figure is taken from
\cite{Mesinger11}. \label{golbal}}
\end{figure}

The observable quantity is the brightness temperature, $\delta T_{\rm b}$, of 
the $21$-cm line against the CMB, which is set by radiative transfer through HI 
regions. 
The brightness temperature of 21 cm radiation can be expressed as
\begin{eqnarray}
\label{eq:delT} \delta T_{b}(\nu)  = && \frac{T_{s} -
T_{\gamma}}{1+z} (1 - e^{-\tau_{\nu_0}}) \nonumber \\
 \approx && 27 x_{\rm HI} (1+\delta_{b}) \left(\frac{H}{dv_r/dr + H}\right) 
\left(1 - \frac{T_{\rm cmb}}{T_{s}} \right) \nonumber \\
 \times & & \left( \frac{1+z}{10} \frac{0.15}{\Omega_{\rm m}
 h^2}\right)^{1/2} \left( \frac{\Omega_b h^2}{0.023} \right) {\rm
 mK},
\end{eqnarray}
where $T_s$ is the gas spin temperature, $\tau_{\nu_0}$ is the
optical depth at the 21-cm frequency $\nu_0$, $x_{\rm HI}$ is the neutral 
hydrogen fraction of the intergalactic medium, $\delta_{b}({\bf x},
z) \equiv \rho/\bar{\rho} - 1$ is the evolved (Eulerian) density
contrast of baryons, $H(z)$ is the Hubble parameter, $dv_r/dr$ is
the comoving gradient of the line of sight component of the
peculiar velocity, and all quantities are evaluated at redshift
$z=\nu_0/\nu - 1$; $\Omega_b$ is the present-day baryon density and $h$ is the 
present-day Hubble factor. Therefore, the brightness temperature of the 21-cm 
line is
very sensitive to the spin temperature of the gas and to the CMB temperature
\citep{Mesinger11}.

The 21-cm line is a unique window into cosmological epochs at which the 
universe is dominantly composed of neutral hydrogen atoms. These encompass the 
period from cosmological recombination (a redshift of $z = 1100$, or a proper 
time of $0.38 \ {\rm Myr}$ after the Big Bang) to the end of the reionisation 
era (a redshift of $z \simeq 6$, or a proper time of $\simeq 1.2 \ {\rm Gyr}$ 
after the Big Bang). Except for the last epoch, the rest of this period is 
unconstrained by current observations, and is fertile ground for exploration 
with new observations. There are several processes that contribute to the 
evolution of the brightness temperature of the 21-cm radiation. 
Observations of the brightness temperature, either through direct imaging or 
statistical measures of its fluctuations, can then inform us about the physical 
state of the neutral gas and the nature of its perturbations 
\citep{2015aska.confE...1K}.

\begin{enumerate}
 \item During the period from $z \simeq 1100$ to $z \simeq 200$, the gas 
temperature is kept close to that of the CMB by Thomson scattering of residual 
free electrons \citep{2012MNRAS.419.1294C}. Atomic collisions, and optical 
pumping by Lyman-$\alpha$ photons from the epoch of cosmological recombination, 
can lead to a small but non-negligible brightness temperature in the $21$-cm 
line \citep{2013JCAP...11..066F,2018PhRvD..98d3520B}.

 \item The epoch from $z \simeq 200$ to $z\simeq30$ is known as the Dark Ages; 
through this period, the CMB temperature and the gas temperature differ 
substantially, and atomic collisions are sufficiently fast to set the spin 
temperature to the latter and lead to a $21$-cm signal at a detectable level. 
The amplitude of the signal is set by the linear evolution of fluctuations on 
large scales \citep{2004PhRvL..92u1301L, 2007PhRvD..76h3005L}, and the bulk 
flows that set the baryonic Jeans scale \citep{2010PhRvD..82h3520T, 
2014PhRvD..89h3506A}. If detected, the $21$-cm signal from this epoch would be 
the ultimate probe of primordial cosmological fluctuations. Assuming cosmic 
variance limits, the $21$-cm signal could probe extremely faint inflationary 
gravitational wave backgrounds \cite[down to tensor-to-scalar ratios of $r \sim 
10^{-9}$;][]{2010PhRvL.105p1302M, 2012PhRvL.108u1301B} and low levels of 
primordial non-gaussianities \citep[down to parameters $f_{\rm NL} \simeq 
0.03$;][]{2006PhRvL..97z1301C, 2007ApJ...662....1P, joudaki2011, 
2015PhRvD..92h3508M}. Due to the low frequencies of the signal from this epoch, 
the observational prospects are not promising in the short to medium term.

 \item The period covering redshifts $z \simeq 30-15$ is called the Cosmic Dawn 
epoch, owing to the birth of the first stars (in sufficient numbers to affect 
21 cm observations). The radiation emitted by these first sources significantly 
changes the nature of the mean and fluctuating $21$-cm signal due to two main 
reasons: (i) optical pumping of the hyperfine levels due to Lyman-$\alpha$ 
photons, known as  the Wouthuysen-Field effect, which serves to couple the spin 
temperature of the gas to the ambient Lyman-alpha radiation \citep{Hirata06}, and 
(ii) heating of the gas by X-rays \citep{2006MNRAS.371..867F, 
2007MNRAS.376.1680P, 2014Natur.506..197F}. In addition, non-linear structure 
formation \citep{2006NewAR..50..179A, 2006ApJ...637L...1K} and baryonic bulk 
flows \citep{2012Natur.487...70V, 2012ApJ...760....3M, 2013MNRAS.432.2909F} 
imprint their effects on the signal. Primordial magnetic fields can also lead to 
features in the cosmological 21 cm signal during these epochs 
\citep{shiraishi2014}.
 
 \item Finally, during the epochs covered by $z \simeq 15-6$, the ionising 
photons from the radiation sources lead to the permeation of HII regions, and 
the mean signal drops, reaching close to zero as reionisation is completed.

\end{enumerate}

\begin{figure}
	\centering
	\includegraphics[width=3.2in]{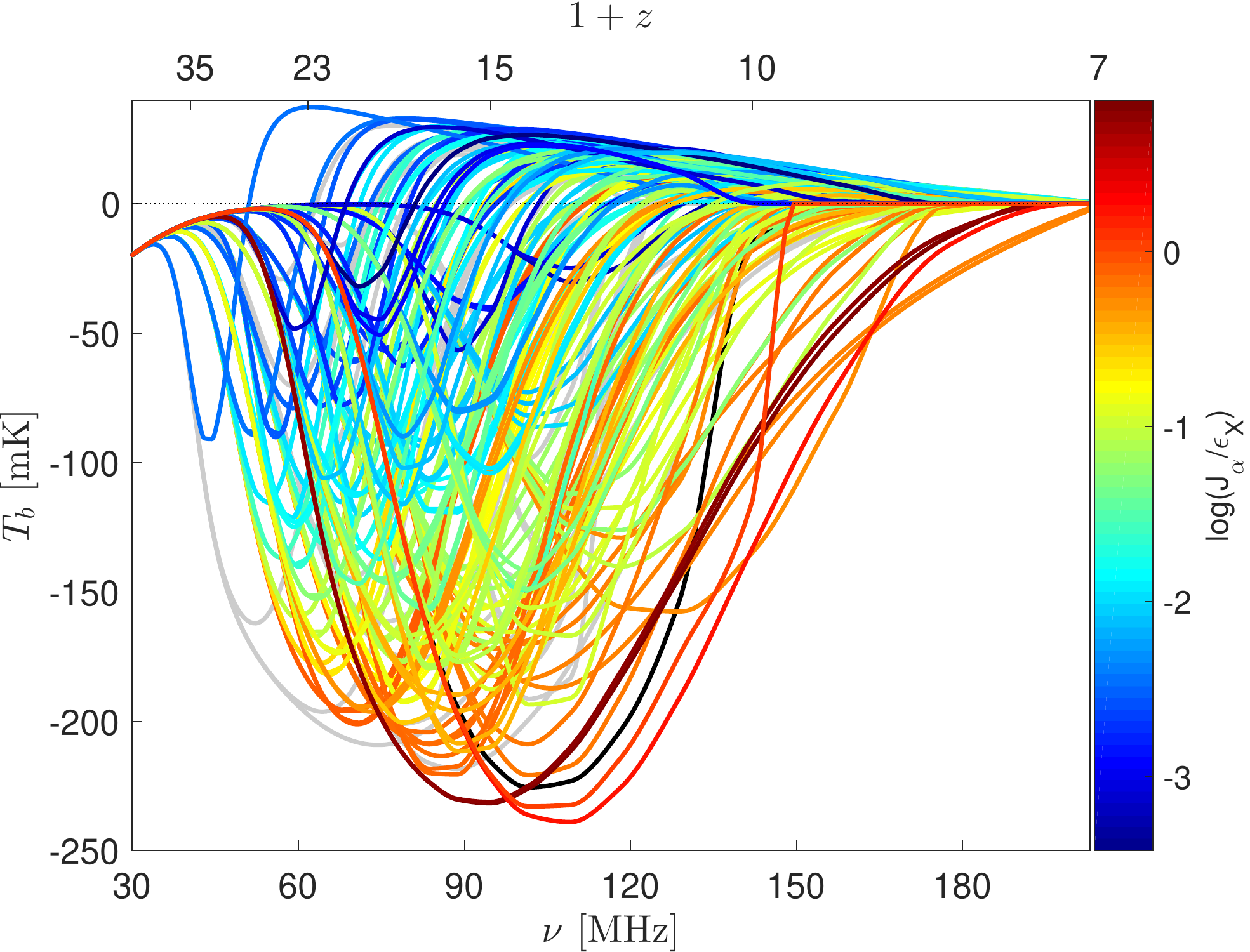}
	\caption{The 21 cm global signal as a function of redshift, for the 193 
different astrophysical models discussed in~\cite{2016arXiv160902312C}. The 
colour (see the colour bar on the right) indicates the ratio between the 
Ly$\alpha$ intensity (in units of erg s$^{-1}$ cm$^{-2}$ Hz$^{-1}$sr$^{-1}$) 
and the X-ray heating rate (in units of eV s$^{-1}$ baryon$^{-1}$) at the 
minimum point. Grey curves indicate cases with $\tau>0.09$, and a non-excluded 
case with the X-ray efficiency of X-ray sources set to zero; these cases are all 
excluded from the colour bar range. Figure taken 
from~\cite{2016arXiv160902312C}.} \label{fig:Tb2}
\end{figure}

Significant progress has been made in condensing these rich astrophysical 
effects into simple semi-analytical prescriptions that capture the large-scale 
features of the $21$-cm signal during this period \citep{2004ApJ...613....1F, 
2004ApJ...613...16F, 2007ApJ...669..663M,Mesinger11,2012Natur.487...70V}. For a 
fiducial model described by \citet{Mesinger11} and developed with the publicly 
available code 21CMFAST, the various evolutionary stages of the signal are 
illustrated in Figure~\ref{golbal}. The terms in the figure denote the spin 
temperature of the gas $T_s$, the CMB temperature $T_{\gamma}$, and the gas 
kinetic temperature $T_K$; the Figure illustrates  astrophysical effects on the 
signal that include decoupling from the CMB, the Wouthuysen-Field coupling, and 
X-ray heating. Figure~\ref{fig:Tb2} shows the wide range of possibilities for 
the sky-averaged signal (``the 21 cm global signal''). Its
characteristic structure of peaks and troughs encodes information
about global cosmic events. 
\cite{2016arXiv160902312C} discussed 193 different combinations of 
astrophysical parameters, illustrating the great current uncertainty in the 
predicted 21 cm signal. 

The most robust way of probing cosmology with the brightness temperature may be 
redshift-space distortions \citep[RSDs;][]{2005ApJ...624L..65B, 
2009astro2010S..82F}; see however \citet{2013PhRvL.110o1301S} and 
\citet{2015PhRvL.114j1303F}. Alternatively, a discussion of the bispectrum is 
provided by \citet{2006MNRAS.366..213S}. More futuristic possibilities include 
probing extremely weak primordial magnetic fields ($\sim 10^{-21} \ {\rm G}$ 
scaled to $z=0$) using their breaking of the line-of-sight symmetry of the 
$21$-cm power spectrum \citep{PhysRevD.95.083010}, and inflationary 
gravitational waves through the circular polarisation of the $21$-cm line 
\citep{2017arXiv170703513H, 2017arXiv170703514M}.

\subsubsection{Status of 21 cm experiments}
Observational attempts to detect the cosmological 21 cm signal have made 
significant progress in the last few years, with upper limits from 
interferometers beginning to make contact with the space of plausible models. 
Broadly speaking, there are two classes of 21 cm experiments: those attempting 
to measure the sky averaged ``global'' 21 cm signal, and those attempting to 
measure the 21 cm brightness temperature fluctuations. A natural comparison is 
to the CMB, where some experiments target either spectral distortions to the 
CMB black body, while others measure CMB anisotropies.

Experiments targeting the global signal include EDGES \citep{Bowman2008}, SARAS 
\citep{Patra2013}, LEDA\footnote{http://www.tauceti.caltech.edu/leda/}, SCI-HI 
\citep{voytek2014}, and a proposed lunar experiment DARE \citep{burns2012}. To 
detect the 21 cm global signal, in principle, only a single radio dipole is 
necessary, as its large beam will average over fluctuations to probe the 
averaged all sky signal. For these experiments, raw sensitivity is typically 
not the limiting factor; the main challenges are two-fold --- ensuring absolute 
calibration of the dipole and removing foregrounds. 

In \citet{Bowman2008}, EDGES reported a first lower limit on the duration of 
reionisation by searching for a sharp step in the 21 cm global signal, which 
is, in principle, distinguishable from the smooth foregrounds 
\citep{2010PhRvD..82b3006P}. More sophisticated techniques have been developed 
based upon forward modelling the signal, foregrounds, and instrument response 
in a Bayesian framework and prospects appear to be good \citep{harker}. 

Recently, EDGES reported a detection of the global 21 cm signal in absorption 
at a frequency of 78 MHz, corresponding to the redshift $z \sim 17$ 
\citep{bowman2018}. The absorption profile was flattened, with an amplitude 
about twice that predicted by several current models. The signal amplitude 
could possibly be evidence of interactions between (a subcomponent of) dark 
matter and baryons \citep[e.g.,][]{barkana2018a,barkana2018b,munoz2018}, which 
may have led to cooling of the IGM prior to reionisation.
Further investigation, as well as independent confirmation from other 
facilities, would lead to exciting prospects for constraining fundamental 
physics. 

In parallel, several new radio interferometers --- LOFAR, PAPER, MWA, HERA --- 
are targeting the spatial fluctuations of the 21 cm signal, due to the ionised 
bubbles during cosmic reionisation as well as Lyman-$\alpha$ fluctuations 
\citep{2005ApJ...626....1B} and X-ray heating fluctuations 
\citep{2007MNRAS.376.1680P} during cosmic dawn. These telescopes take different 
approaches to their design, which gives each different pros and cons. LOFAR in 
the Netherlands is a general purpose observatory with a moderately dense core 
and long baselines (in the case of the international stations, extending as far 
as Ireland). The MWA in Western Australia is composed of 256 tiles of 16 
antennas distributed within about 1-km baselines. PAPER (now complete) was 
composed of 128 dipoles mounted in a small dish and focussed on technological 
development and testing of redundant calibration. HERA in South Africa will be 
a hexagonal array of 330 x 14m dishes and, like PAPER, aims to exploit 
redundant calibration.

\begin{figure}
	\begin{center}
	\includegraphics[width=\linewidth]{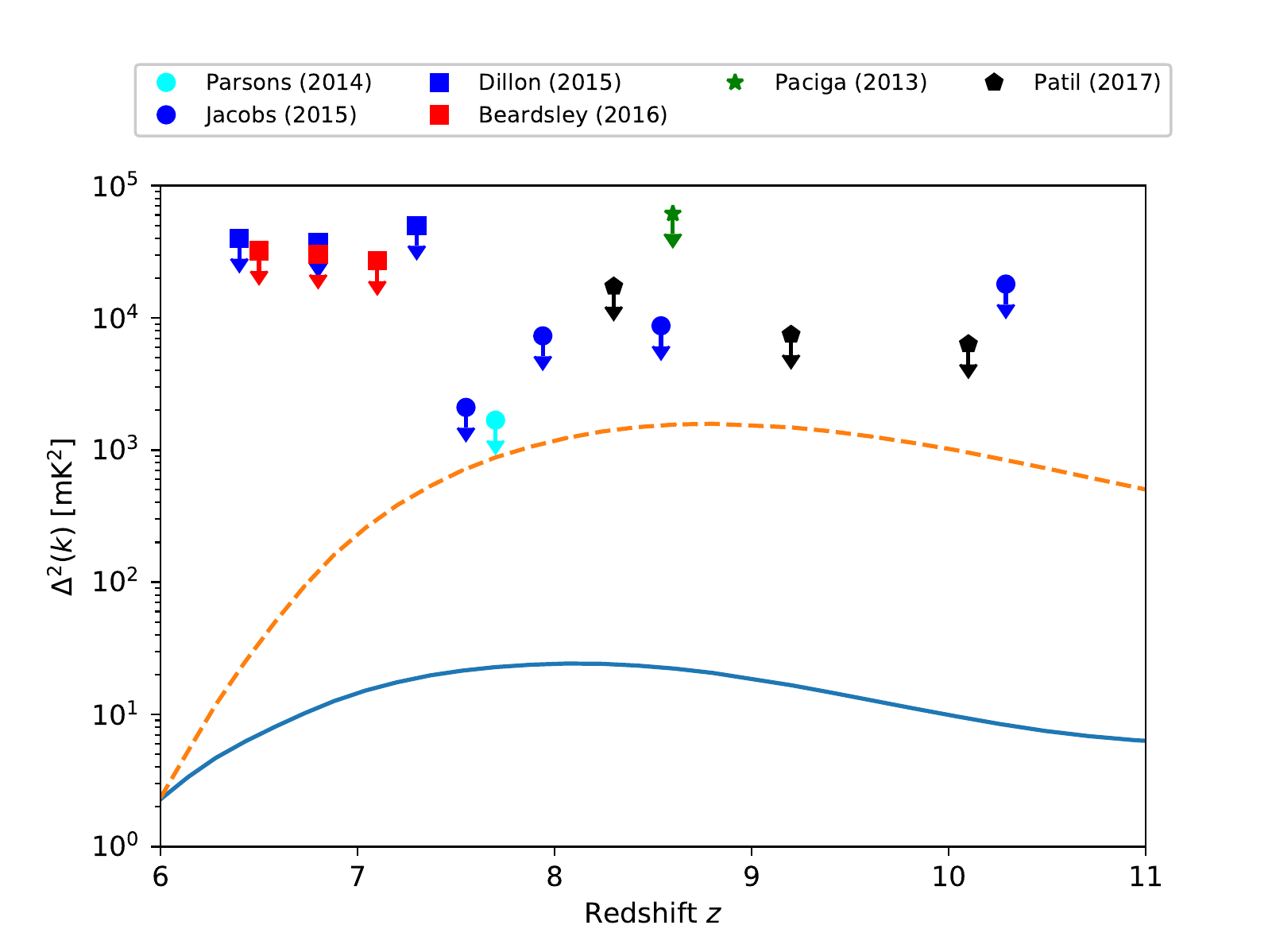}
	\caption{Summary of current constraints on the 21 cm power spectrum as 
a function of redshift. Since constraints are actually a function of both 
redshift and wavenumber $k$, only the best constraint for each experiment has 
been plotted. Here are plotted results for GMRT \citep{paciga2013}, PAPER32 
\citep{parsons2014, jacobs2015}, MWA128 \citep{Dillon2015, Beardsley2016} and 
LOFAR \citep{Patil2017}. Two comparison 21 cm signals calculated using 21CMFAST 
are shown to give a sense of the target range - one with fiducial values (solid 
blue curve) and a second with negligible heating (dashed orange curve).} 
\label{fig:ps_constraints}
    \end{center}
\end{figure}
These experiments have begun setting upper limits on the 21 cm power spectrum 
that are summarised in Figure~\ref{fig:ps_constraints}. At present, the best 
constraints are about two orders of magnitude above the expected 21 cm power 
spectrum. However, as noted earlier,  there is considerable uncertainty in 
these predictions, and in the case of an unheated IGM a much larger signal can 
be produced. \citet{pober2015} interpreted now-retracted upper limits from 
\citet{ali2015} as a constraint on the IGM temperature, ruling out an entirely 
unheated universe at $z=8.4$. The current upper limits typically represent only 
a few tens of hours of integration time, compared to the $\sim1000$ hours 
needed for the desired sensitivity. Systematic effects, especially instrumental 
calibration, currently limit the amount of integration time that can be 
usefully reduced. Overcoming these limitations is the major goal of all these 
experiments and steady progress is being made.

\subsection{Fundamental Physics from the Epoch of Reionisation}
\label{sec:physics}

In the previous section, we listed the main astrophysical and cosmological 
processes that contribute to the brightness temperature evolution of the 21 cm 
signal, and the status of the EoR 21 cm experiments. In this section, we 
provide glimpses into the details of some of the important constraints on 
fundamental physics that may be garnered from the epoch of reionisation and 
cosmic dawn.

\subsubsection{Cosmology from the EoR}

A key advantage of 21 cm observations is that they open up a new epoch of 
cosmological volume containing many linear modes of the density field, which 
can greatly increase the precision of cosmological parameter constraints. 
Typically, cosmology enters into the 21 cm signal through its dependence on the 
density field, so that the 21 cm signal can be viewed as a biased tracer in a 
similar way to low redshift galaxy surveys. The challenge is that obtaining 
fundamental physics from the 21 cm signal requires disentangling the 
``gastrophysics'' -- the effect of galaxies and other astrophysical sources on the hydrogen gas -- from the signature of physics. This is not an easy challenge, 
since the effect of astrophysics is typically dominant over that of fundamental 
physics effects, which are often subtle and desired to be measured at high 
precision. At this moment in time, our understanding of the nuances of both the 
21 cm signal and the observations is still relatively limited, but there are 
reasons for some optimism.

Broadly speaking, there are several routes to fundamental physics from the 21 
cm signal:
\begin{enumerate}
\item Treat the 21 cm signal as a biased tracer of the density field, and via 
joint analysis, constrain cosmological parameters.
\item Look for the direct signature of energy injection by exotic processes in 
the 21 cm signal, which is sensitive to the cosmic thermal history.
\item The clustering of ionised regions or heating will reflect the underlying 
clustering of galaxies, and so will contain information about the density 
field, e.g., non-gaussianity signatures or the lack of small scale structure due 
to warm dark matter.
\item Line of sight effects, such as weak lensing or the integrated Sachs-Wolfe 
(ISW) effect, where the 21 cm signal is primarily just a diffuse background 
source.
\item Look for unique signatures of fundamental physics, e.g., the variation of 
the fine structure constant, which do not depend in detail upon fluctuations in 
the 21 cm brightness.
\end{enumerate}

21 cm observations may also be useful in breaking degeneracies present in other 
data sets \citep{2017arXiv170504688K}. For example, measurements of the 
reionisation history may allow the inference of the optical depth to the CMB, 
breaking a degeneracy with neutrino mass \citep{2016PhRvD..93d3013L}.

\subsubsection{Exotic energy injection}

As discussed in Section \ref{sec:overview}, the 21 cm signal is sensitive to 
the underlying gas temperature through the 21 cm spin temperature. This makes 
the 21 cm line a rather unique probe of the thermal history of the Universe 
during the epoch of reionisation and the cosmic dawn. Provided that the IGM 
temperature is not too much larger than the CMB temperature (so that the $1 - 
T_{\rm CMB}/T_s$ term retains its dependence on $T_s$), we can use the Universe 
as a calorimeter to search for energy injection from a wide range of processes. 
Distinguishing different sources of heat will depend upon them having unique 
signatures in how that energy is deposited spatially or temporally.

After thermal decoupling at $z\sim150$, the gas temperature is expected to cool 
adiabatically, with a phase of X-ray heating from galaxies warming the gas, 
before the photoionisation heating during reionisation raises the temperature 
to $\sim10^4{\,\rm K}$ \citep[e.g.][]{2006MNRAS.371..867F,2012ApJ...760....3M}. 
There is considerable uncertainty in these latter stages, which depend upon 
poorly known properties of the galaxies and the cosmic star formation history. 

Many authors have put forward possible sources of more exotic heating including 
annihilating dark matter \citep[e.g.,][]{furlanetto2006dm, 
2013MNRAS.429.1705V}, evaporating primordial black holes \citep{clark2017, 
2008arXiv0805.1531M}, cosmic string wakes \citep{2010JCAP...12..028B}, and many 
more. In many cases, these might be distinguished from X-ray heating by (a) 
occurring before significant galaxy formation has occurred or (b) by depositing 
energy more uniformly than would be expected from galaxy clustering. 
Incorporation of dark matter annihilation models into simulations of the 21 cm 
signal suggests that plausible dark matter candidates might be ruled out by 
future 21 cm observations \citep{2013MNRAS.429.1705V}. Ultimately the physics of 
how dark matter annihilation produces and deposits energy as heating or 
ionisation is complex and requires consideration of the decay products and their 
propagation from the decay site into the IGM \citep{2015MNRAS.451.2840S}. 

Note that dark matter candidates may modify the thermal history through their 
effect on the distribution of galaxies too, as discussed in the next section.

\subsubsection{Warm dark matter effects}

Warm dark matter (WDM) is an important alternative to the standard cold
dark matter candidate. Although there have been a series of
studies on the constraints on the mass of the warm dark matter, a
large parameter space is still unexplored and is possible in
principle. These existing constraints include the lower limit on
the mass of a thermal WDM particle ($m_{X}\geq 2.3 \rm{keV}$) from Milky Way
satellites \citep{Polisensky11} and from Lyman alpha forest
data \citep{Narayanan01,Viel05,Viel08}.

A possible effect of warm dark matter during the reionisation and cosmic dawn 
epochs is distinguishable from both the mean brightness temperature  and the 
power spectra.
The key processes that are altered in the WDM model are the Wouthysen-Field 
coupling, the X-ray heating and the reionisation effects described in 
Sec.~\ref{sec:overview}. This is because the WDM can delay the first
object formation, so the absorption features in the $\delta T_{b}$
evolution could be strongly delayed or suppressed. In addition,
the X-ray heating process, which relies on the X-rays from the
first generation of sources, as well as the Lyman-alpha
emissivity, can be also affected due to the delayed first objects
\citep[see also Fig.~7 of][]{Pritchard12}. Of course, the magnitude of
the effects depends on the scale of interest. Finally,
reionisation is also affected because the WDM can delay the
reionisation process and therefore affect the ionisation fraction
of the Universe at redshift $\sim 10$ \citep[Figs. 8 and 9 of][]{Barkarna01}.

Examples of the effects of WDM models on the spin temperature of the gas will 
be discussed in \S\ref{wdm_hi}. For the case of WDM, the spin temperature, 
$\Ts$, stays near the CMB temperature, $T_{\gamma}$, at redshift $z>100$. The 
absorption trough occurs due to the fact that at a later stage, the X-ray 
heating rate surpasses the adiabatic cooling. Initially the mean collapse 
fraction in WDM models is lower than in CDM models, but it grows more rapidly 
in the heating of gas.

The mean brightness temperatures as a function of redshift (frequency) for such 
WDM models with $\mx=2,\,3,\,4\,\text{keV}$, respectively, is explored by 
\citet{Sitwell14} and elaborated on in \S\ref{wdm_hi}. It is shown that if the 
WDM mass is below the limit $\mx<10\,\text{keV}$, it can substantially change 
the mean evolution of $\overline{T}_{\rm b}(z)$. 

In addition to the mean temperature evolution, \citet{Sitwell14} also explored  
the power spectrum of WDM models, and showed a three-peak structure in  $k=0.08 
\, \text{Mpc}^{-1}$ and $k=0.18 \, \text{Mpc}^{-1}$ modes, which are associated 
with inhomogeneities in the Wouthuysen-Field coupling coefficient $x_{\alpha}$, 
the kinetic temperature $\Tk$, and the ionisation fraction $x_{\text{HI}}$, 
from high to low redshifts. As discussed in detail in \S\ref{wdm_hi}, the power 
at these specific modes can be boosted, depending upon the mass of the WDM 
particle. 

These variations in the mean temperature and fluctuations can be measured and 
tested using current interferometric radio telescopes. \citet{mesinger2013a} 
and \citet{Sitwell14} showed forecasts for the 1-$\sigma$ thermal noise levels 
for 2000 hours of observation time for the Murchison Widefield Array 
(MWA)\footnote{http://www.mwatelescope.org/}, the Hydrogen Epoch of 
Reionisation Array (HERA)\footnote{http://reionization.org}, and for SKA1-Low. 
On the other hand, there are major uncertainties in the evolution of 
high-redshift star-formation (in low-mass halos in particular), with a 
potentially complex history due to various astrophysical feedback mechanisms 
[including photo-heating, Lyman-Werner radiation (photons capable of dissociating molecular hydrogen), and supernova feedback; the 
latter includes hydrodynamic and radiative feedback as well as metal 
enrichment]. The estimates do indicate that next-generation radio observations 
may be able to test the excess power in the power spectra of brightness 
temperature for $\mx<10 \, \text{keV}$ models over a wide range of redshifts. 
The SKA, in particular, will provide a unique prospect of measuring the mean 
brightness temperature and the 21-cm power spectrum out to $z\simeq 20$. 
However, distinguishing WDM from CDM will require a clear separation from 
possible astrophysical effects.

\subsubsection{Measuring the fine structure constant with the SKA using the
21\,cm line}
The standard model of particle physics fails to explain the values of some 
fundamental ``constants'', like the mass
ratio of the electron to the proton, the fine structure constant, etc. 
\citep[see, e.g.,][]{Uzan:2010pm}. \citet{Dirac:1937ti} hypothesised that these
constants might change in space as well as in time.  Studies using the optical
spectra of distant quasars indicated, controversially, the existence of
temporal~\citep[e.g.,][]{Webb:2000mn} and spatial~\citep[e.g.,][]{Webb:2010hc} 
variations in the
fine structure constant, $\alpha$ (but see also Srianand et al.\ 2004\nocite{srianand2004} and the more recent results of Murphy et al.\ 2016\nocite{murphy2016} suggesting no significant cosmological variations). 
However, these results may be in tension with
terrestrial experiments using optical atomic clocks, which set a very stringent
limit on the temporal variation of $\alpha$~\citep{Rosenband:2008}.
Investigation along these lines has great significance to our understanding of
gravitation through the underlying equivalence principle~\citep{Shao:2016ezh},
as well as fundamental (scalar) fields and cosmology~\citep{Damour:2002mi}. It
could also provide an intriguing clue to the outstanding `cosmological
constant problem'~\citep{Parkinson:2003kf}.

In the case that $\alpha$ varies as a function of time (for example as a
cosmologically evolving scalar field), the evolution could be non-monotonic in
general.  Therefore, it would be greatly beneficial if we could measure 
$\alpha$ at various redshifts.
The  quasar spectra and optical atomic clocks mentioned previously only probe 
$\alpha$
at moderate redshifts, $0.5 \lesssim z \lesssim 3.5$ and $z \simeq 0$,
respectively. Hence,  reionisation and cosmic dawn provide an interesting 
avenue to probe the
possibility of a varying $\alpha$ at large $z$.  Because of its high
resolution in radio spectral lines, SKA1-Low has good prospects to use them 
(e.g., lines from
H{\sc i} and the OH radical) to determine 
$\alpha$~\citep{Curran:2007,SKA:2011}.  The covered redshifts for SKA1-Low will 
be, e.g., $z
\lesssim 13$ for the H{\sc i} 21\,cm absorption and $z \lesssim 16$ for the
ground-state 18\,cm OH absorption~\citep{Curran:2004mg}.  \citet{Khatri:2007yv}
proposed another method to measure $\alpha$, through the 21\,cm absorption of
CMB photons. They found that the 21\,cm
signal is very sensitive to variations in $\alpha$, such that a change of $1\%$
in $\alpha$ modifies the mean brightness temperature decrement of the CMB due
to 21\,cm absorption by $\gtrsim 5\%$ over the redshift range $30 \lesssim z
\lesssim 50$. It also affects, as a characteristic function of the redshift
$z$, the angular power spectrum of fluctuations in the 21\,cm absorption;
however, the measurement of the angular power spectrum at these redshifts 
(corresponding to the Dark Ages) would require
lower-frequency observations than those from the SKA. In summary, constraints 
on the variation
of $\alpha$ at various redshifts will significantly advance our basic
understanding of nature, and might provide clues to new physics beyond the
standard model~\citep{Uzan:2010pm}.

\subsubsection{Cosmic Shear and the EoR}
\label{sec:weaklensing}

Important information on the distribution of matter is encoded
by weak lensing of the 21-cm signal along the line of sight to the EoR 
\citep{2015aska.confE..12P}. \citet{Zahn:2005ap} and \citet{Metcalf:2009} 
showed that a large area survey at SKA sensitivity might have the potential to 
determine the lensing convergence power spectrum via the non-gaussianity of 21 
cm maps. It remains to be seen over what area SKA-Low surveys might have the 
sensitivity to measure cosmic shear, but the proposed deep EoR survey over 
100~deg$^2$ should be sufficient. This would measure how dark matter is 
distributed in a representative patch of sky, something feasible only with 
galaxy lensing toward unusually large galaxy clusters. This might offer the 
chance to match luminous matter with overall mass, thereby constraining the 
dark matter paradigm.

The convergence power spectrum can be estimated using the
Fourier-space quadratic estimator technique of
\cite{Hu:2001tn}, originally developed for lensing data on the CMB and expanded 
to 3D
observables, i.e., the $21$ cm intensity field $I(\theta,z)$ discussed by 
\citet{Zahn:2005ap} and \citet{Metcalf:2009}.

The convergence estimator and the corresponding lensing reconstruction noise 
are derived under the assumption that there is a gaussian distribution in 
temperature. This will not completely hold for the EoR, since reionisation 
introduces considerable non-gaussianity, but acts as a reasonable approximation.

The benefit of 21-cm lensing is that one can combine
data from multiple redshift slices. In Fourier space, fluctuations in 
temperature (brightness) are separated into wave vectors normal to the 
sightline $\mathbf{k_\perp}=\mathbf{l}/r$, with $r$ the angular diameter 
distance to the source redshift, and a
discretised parallel wave vector $k_\parallel =
2\pi j/\cal L$, where ${\cal L}$ is the depth of the volume observed. 
Considering modes with different values of $j$ to be orthogonal, an optimal 
estimator results from combining the estimators from separate $j$ modes without 
any mixing. The reconstruction noise of 3D lensing is then \citep{Zahn:2005ap}:
\begin{equation}
\resizebox{.9\hsize}{!}{$N(L,\nu) =\left[\sum_{j=1}^{j_{\rm max}} 
\frac{1}{L^2}\int \frac{d^2\ell}{(2\pi)^2}  \frac{[\mathbf{l} \cdot \mathbf{L} 
C_{\ell,j}+\mathbf{L} \cdot (\mathbf{L}-\mathbf{l})
C_{|\ell-L|,j}]^2}{2 C^{\rm tot}_{\ell,j}C^{\rm 
tot}_{|\mathbf{l}-\mathbf{L}|,j}}\right]^{-1}.$}
\end{equation}
Here, $C^{\rm tot}_{\ell,j}=C_{\ell,j}+C^{\rm N}_\ell$, where 
$C_{\ell,j}=[\bar{T}(z)]^2P_{\ell,j}$ with $\bar{T}(z)$ the mean observed 
brightness temperature at redshift $z$ due to the mean density of HI, and 
$P_{\ell,j}$ is the associated power spectrum of dark matter 
\citep{Zahn:2005ap}. 

\begin{figure}[h]
\centerline{
\includegraphics[width=\linewidth]{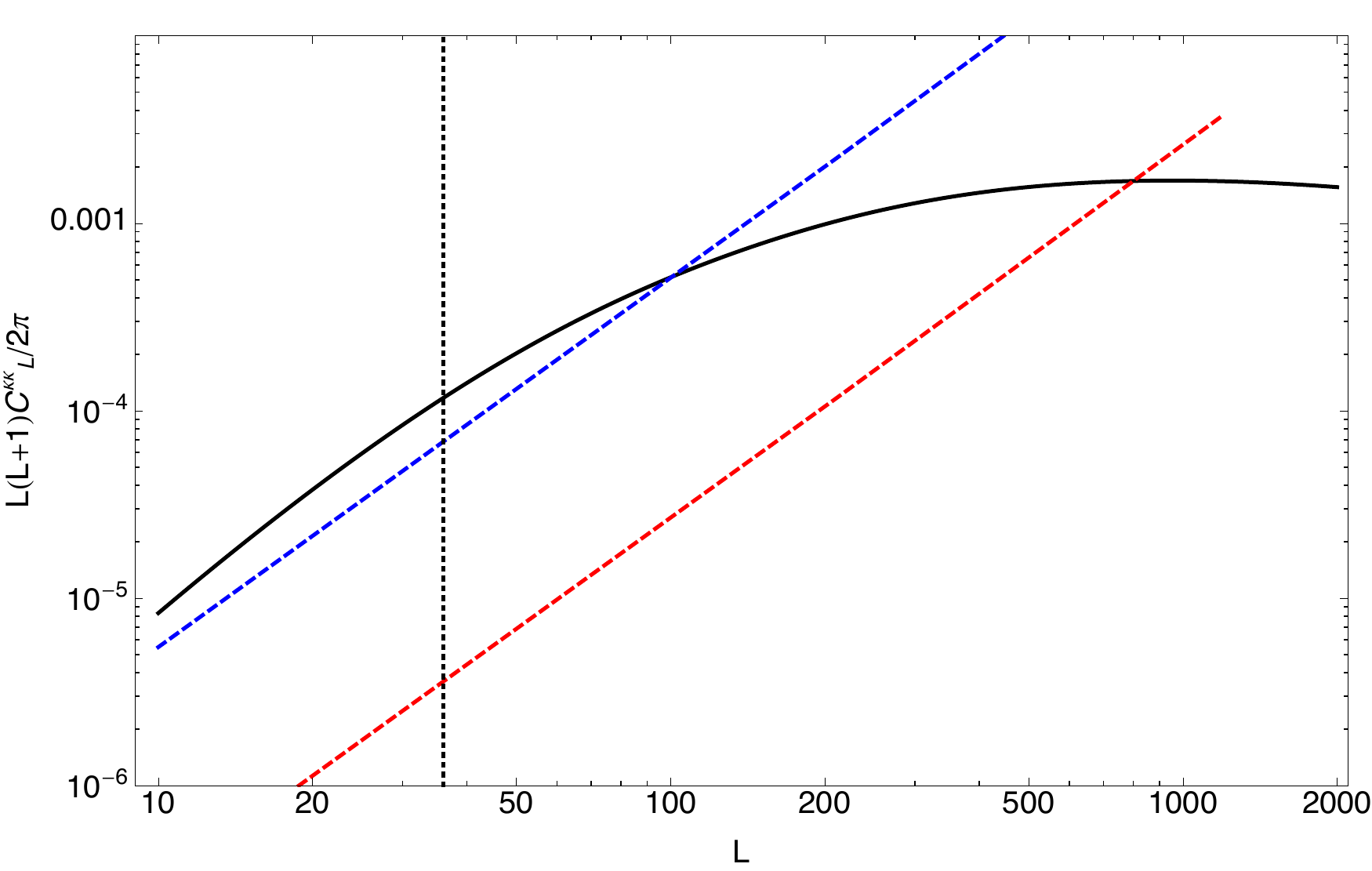}
}
\caption{The solid black line shows the power spectrum of the lensing 
convergence field, $C^{\kappa \kappa}_L$, for sources at $z=8$; dashed lines 
indicate the noise associated with lensing reconstruction, $N_L$.  The blue 
dashed line is for SKA1-Low with 10 8~MHz frequency bins around $z=8$, covering 
redshifts from $z \simeq 6.5$ to $z \simeq 11$.  The red dashed line is the same 
but for SKA2-Low. The vertical line represents an estimate of the lowest 
possible value of $L$ accessible in a 5-by-5 degree field.  Regions where noise 
curves fall below $C^{\kappa \kappa}_L$ indicate cases for which the typical 
fluctuations in the lensing deflection should be recoverable in a map. Figure 
taken from \citet{2015aska.confE..12P}.}
\label{fig:CLNL}
\end{figure}

Figure \ref{fig:CLNL} gives a sense of the sensitivity to the convergence power 
spectrum that might be achieved with SKA-Low after 1000 hours integration on a 
20-deg$^2$ field. It should be feasible to measure the signal associated with 
lensing over a range of angular scales.  Increasing the survey area would allow 
access to large angular scales, where the signal-to-noise is greatest. This 
measurement would be significantly improved with the larger sensitivity of SKA2 
\citep{2017arXiv170801235R}.  For redshifts after reionisation, the power 
spectrum of weak lensing should be better measured using SKA-Mid and the 21-cm 
intensity mapping approach discussed above, but covering a much wider sky area 
\citep{PourtsidouMetcalf:2014}.

\subsubsection{Integrated Sachs-Wolfe effect}

In \S\ref{sec:overview}, we provided an overview of the 21 cm brightness  
temperature fluctuation and its dependence on cosmological and astrophysical 
parameters. While we have thus far focused on high redshifts, it will be 
possible to use 21 cm measurements from {\it after}\/ reionisation in order to 
obtain constraints on various cosmological models. In this case, the 21 cm 
emission comes from hydrogen atoms within galaxies. The intensity (or 
equivalently temperature) fluctuations can be mapped on large scales, without 
resolving individual galaxies; this measurement is known as 21 cm intensity 
mapping (IM).
In this section, we consider using the post-reionisation power spectrum of the 
temperature brightness measured by the SKA, and the cross-correlation of SKA IM 
measurements with SKA galaxy number counts, in order to detect the integrated 
Sachs-Wolfe effect.
As examples of measurements that can be obtained with this observable, we look 
at the IM constraining power to test statistical anisotropy and inflationary 
models.

\cite{Raccanelli:21cmISW} presented a study on using the cross-correlation of 
21-cm surveys at high redshifts with galaxy number counts; the formalism and 
methodology is described in that paper.
The use of 21-cm radiation instead of the (standard) CMB can provide a 
confirmation of the detection of the integrated Sachs-Wolfe effect, which will 
be detected by several instruments at different frequencies at the time, and 
hence influenced by different systematics.

The ISW effect~\citep{sachs67, crittenden96, nishizawa14} is a gravitational 
redshift due to the time-evolution of the gravitational potential when photons 
traverse underdensities and overdensities in their journey from the last 
scattering surface to the observer. This effect produces temperature 
fluctuations that are proportional to the derivative of gravitational 
potentials.

The ISW effect has been detected~\citep{nolta04, pietrobon06, ho08, 
giannantonio08a, raccanelli08, giannantonio12, planckisw, planckisw2015} 
through cross-correlation of CMB maps at ${\rm GHz}$-frequencies with galaxy 
surveys. It has also been used to constrain cosmological 
parameters~\citep{giannantonio08dgp, massardi10, bertacca11, 
Raccanelli:2014ISW}.

\begin{figure}[b!]
\includegraphics[width=0.99\columnwidth]{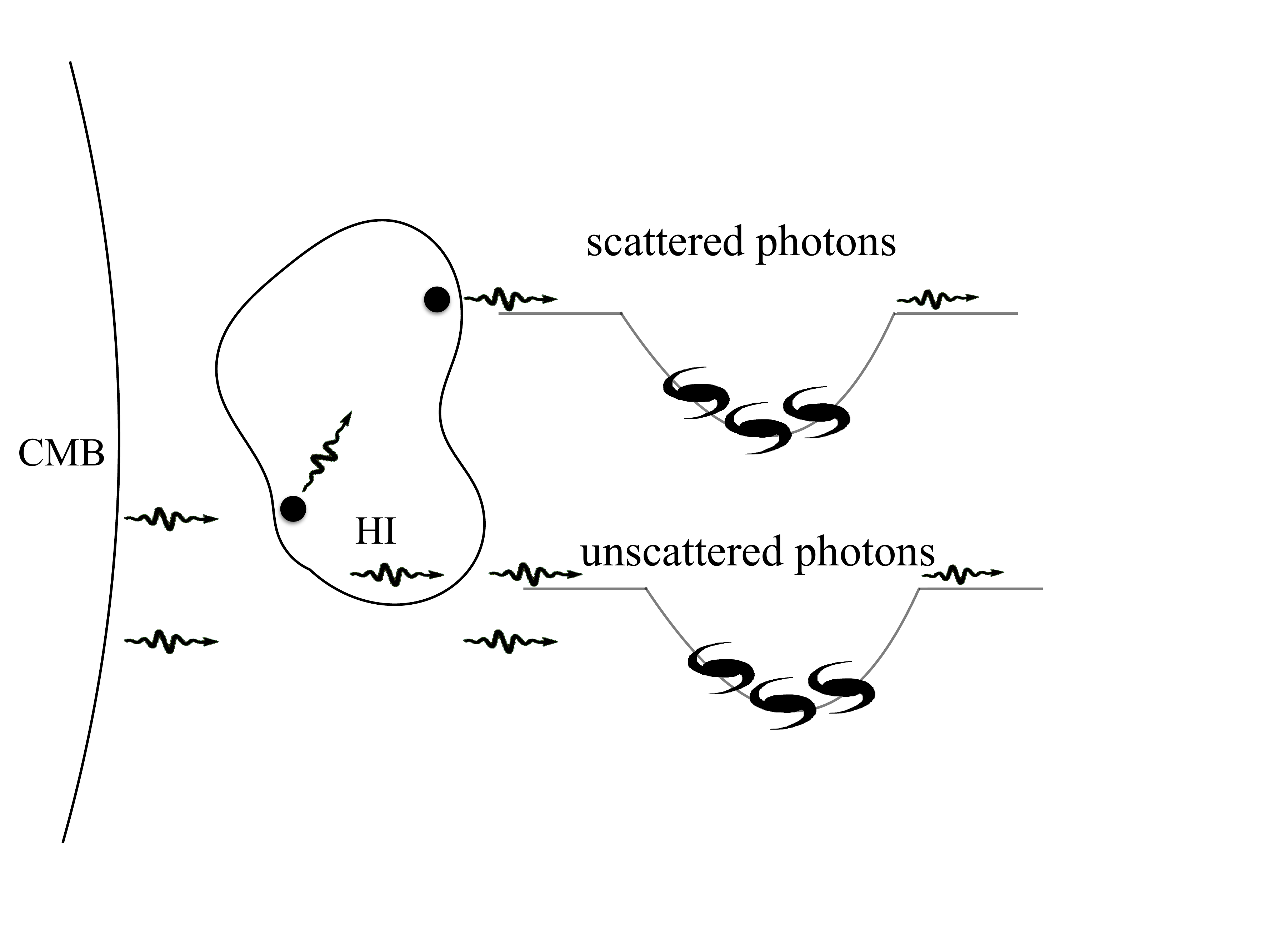}
\caption{
{\it Illustration:} Radiative transfer of CMB photons through neutral hydrogen 
gas clouds induces fluctuations at 21 cm frequencies (due to absorption or 
emission, depending on the relative temperatures of the intergalactic medium 
and the CMB). The majority of the signal is comprised of unscattered CMB photons 
at the Rayleigh-Jeans tail of its blackbody spectrum. These photons later 
undergo line-of-sight blue- or red-shifting as they travel through the evolving 
gravitational potential wells. Figure taken from \citet{Raccanelli:21cmISW}.}
\label{fig:illustration}
\end{figure}

Similar to the CMB, the 21 cm background at high redshifts, described by the 
brightness temperature fluctuation in \S\ref{sec:overview}, will also 
experience an ISW effect from the evolution of gravitational potential wells 
(see Fig.~\ref{fig:illustration}). The dominant signal present is that of 
unscattered CMB photons, and therefore its late-time ISW signature is highly 
correlated with the signature at the peak CMB frequencies. A complementary 
measurement at 21 cm frequencies is promising as it represents an independent 
detection of the ISW effect, measured with different instruments and 
contaminated by different foregrounds. As the 21 cm background is set to be 
observed across a vast redshift range by upcoming experiments, there should be 
ample signal-to-noise for this detection.
The ISW effect on those CMB photons that \textit{do} interact with the neutral 
hydrogen clouds at high redshifts provide a source of observable signal. 
Assuming the CMB fluctuations can be efficiently subtracted from the 21 cm 
maps, this signal can potentially be detected in the data as well.

To detect the ISW effect, one would cross-correlate the brightness temperature 
maps with galaxy catalogues. In the case when the photons are unscattered, the 
detection is more difficult to obtain. The detection depends on a series of 
parameters of the 21 cm detecting instrument, such as the observing time, the 
frequency bandwidth, the fractional area coverage and the length of the 
baseline.
The results weakly depend on the details of the galaxy survey used. Different 
surveys give slightly different results, but do not lead to a dramatic change 
in the overall signal-to-noise ratio. Targeting specific redshift ranges and 
objects could help. 
The main advantage for detecting the ISW effect is due to the large area of the 
sky covered. If we assume the standard general relativity (GR) and $\Lambda$ 
cold dark matter ($\Lambda$CDM) cosmology, the ISW effect is mostly important 
during the late-time accelerated phase, so low-redshift galaxies are to be 
targeted.
The use of a tomographic analysis in the galaxy catalogue and the combination 
of different surveys \citep[see, e.g.,][]{giannantonio08a, bertacca11} can  
improve the detection of the signal in the case of the LSS-CMB correlation.

\subsubsection{Statistical Anisotropy}
Most inflationary models predict the primordial cosmological perturbations to 
be statistically  homogeneous and isotropic. CMB observations, however, indicate 
a possible departure from statistical isotropy in the form of a dipolar power 
modulation at large angular scales. A 3$\sigma$ detection of the dipolar power 
asymmetry, i.e., a different power spectrum in two opposite poles of the sky, 
was reported based on analysis of the off-diagonal components of angular 
correlations of CMB anisotropies with WMAP and Planck data on large scales 
\citep{Hansen:2004vq, Gordon:2005ai, Eriksen:2007pc, Gordon:2006ag, 
Ade:2013nlj,Akrami:2014eta,1502.02114,Ade:2015sjc,Aiola:2015rqa}. The 
distribution of quasars at later times was, however, studied 
by~\cite{Hirata:2009ar}, and showed an agreement with statistical isotropy on 
much smaller angular scales. 

A significant detection of deviation from statistical isotropy or homogeneity 
would be inconsistent with some of the simplest models of inflation, making it 
necessary to postulate new physics, such as non-scalar degrees of freedom. It 
would, moreover, open a window into the physics of the early Universe, thus 
shedding light upon the primordial degrees of freedom responsible for inflation.

The off-diagonal components of the angular power spectrum of the 21 cm
intensity fluctuations can be used to test this power asymmetry, as discussed 
in detail by \cite{Shiraishi:2016}. One can also constrain the rotational 
invariance of the universe using the power spectrum of 21 cm fluctuations at 
the end of the Dark Ages. The potential ability to access small angular scales 
gives one the opportunity to distinguish the dipolar asymmetry generated by a 
variable spectral index, below the intermediate scales at which this vanishes. 
One can compute the angular power spectrum of 21 cm fluctuations sourced by the 
dipolar and quadrupolar asymmetries, including several non-trivial scale 
dependencies motivated by theories and observations.
By the simple application of an estimator for CMB rotational asymmetry 
\citep{Pullen:2007tu,Hanson:2009gu}, we can forecast how well 21 cm surveys can 
constrain departures from rotational invariance.
Results for dipolar and quadrupolar asymmetries, for different models and 
surveys, are discussed by~\cite{Shiraishi:2016}, who show that the planned SKA 
may not reach the same precision as future CMB experiments in this regard; 
however, an enhanced SKA instrument could provide the best measurements of 
statistical anisotropy, for both the dipolar and quadrupolar asymmetry.

The SKA could, though, provide some constraining power for asymmetry parameters 
since 21 cm measurements have different systematics and come from an entirely 
different observable compared to the CMB.
Moreover, 21 cm surveys provide an independent probe of broken rotational 
invariance, and as such, would help in disentangling potential biases present 
in previous CMB experiments.

\subsubsection{Tests of inflation}
Measurements of IM from SKA can be used to constrain inflationary models via 
limits on the matter power spectrum, in particular the spectral index and its 
`running'.

Single-field slow-roll inflation models predict a nearly scale-invariant power 
spectrum of perturbations,
as observed at the scales accessible to current cosmological experiments.
This spectrum is slightly red, showing a non-zero tilt.
A direct consequence of this tilt are
nonvanishing runnings of the spectral indices, $\alpha_s=\mathrm d n_s/\mathrm 
d\log k$, 
and $\beta_s=\mathrm d\alpha_s/\mathrm d\log k$, which in the minimal 
inflationary scenario
should reach absolute values of $10^{-3}$ and $10^{-5}$, respectively. This is 
of particular importance for primordial-black-hole (PBH) production in the 
early universe, 
where a significant increase in power is required at the scale corresponding to 
the 
PBH mass, which is of order $k \sim 10^5$ Mpc$^{-1}$ for solar-mass PBHs 
\citep{astro-ph/9901268,astro-ph/0511743}.
It has been argued that a value of the second running $\beta_s = 0.03$,
within 1$\sigma$ of the Planck results,
can generate fluctuations leading to the formation of $30\, M_{\odot}$ 
primordial black holes if extrapolated to the smallest scales 
\citep{1607.06077}.

The measurements of 21 cm intensity mapping can be used to measure these 
runnings.
A fully-covered 1-kilometre-baseline interferometer, observing the epoch of 
reionisation, will be able to measure the running $\alpha_s$ with $10^{-3}$ 
precision, enough to test the inflationary prediction. However, to reach the 
sensitivity required for a measurement of $\beta_s\sim 10^{-5}$, a dark-ages 
interferometer, with a baseline of $\sim 300$ km, will be required. Detailed 
analyses of 21~cm intensity mapping experiments forecasts for this (including 
comparisons with CMB and galaxy surveys) measurements have been made recently 
\citep{Munoz:2017, Pourtsidou:2016dzn, Sekiguchi:2017}.

\subsubsection{Free-free emission from cosmological reionisation}

As we know, the CMB emerges from the thermalisation epoch, at $z \sim 
10^{6}-10^{7}$, with a blackbody (BB) spectrum 
thanks to the combined effect of Compton scattering and photon 
emission/absorption processes (double Compton and bremsstrahlung) in the cosmic 
plasma, which, at early times, are able to re-establish full thermal 
equilibrium in the presence of arbitrary levels of perturbing processes.
Subsequently, the efficiency of the scattering and above radiative processes 
decrease because of the expansion of the Universe and the consequent combined 
reduction of particle number densities and temperatures, and it was no longer 
possible to achieve the thermodynamical equilibrium. 

The CMB spectrum measurements at frequencies between 30~GHz and 600~GHz from 
the FIRAS instrument on board the NASA 
COBE\footnote{\url{http://lambda.gsfc.nasa.gov/product/cobe/}}
satellite confirm the hot big bang model, at the same time providing the main 
constraints about the deviations from a BB 
possibly caused by energy dissipation mechanisms in the cosmic plasma
\citep{1996ApJ...473..576F,2002MNRAS.336..592S}. 
Recent observations at long wavelengths have been carried out 
with the TRIS experiment \citep{2008ApJ...688...24G} and the ARCADE-2 balloon 
\citep{2011ApJ...730..138S,2011ApJ...734....6S}.
High accuracy CMB spectrum observations at long wavelengths ($0.5 \lsim \lambda 
\sim15$~cm) have been proposed for the DIMES \citep{1996astro.ph..7100K} space 
mission, 
with the aim of probing (i) dissipation processes at high redshifts ($z \gsim 
10^5$), 
resulting in Bose-Einstein like distortions \citep{1970Ap&SS...7...20S},
and (ii) low redshifts mechanisms ($z \lsim 10^4$) before or after the 
photon-matter decoupling, generating Comptonisation and free-free (FF) 
distortions \citep{1991ApJ...371....8B} that, for positive (negative) 
distortion parameters, are characterised, respectively, 
by a decrement (an excess) at intermediate wavelengths and an excess (a 
decrement) at long wavelengths. 
The distorted spectrum is mainly determined by the energy fractional exchange 
involved in the interaction,
the time and kind of the dissipation mechanism, and the density of baryonic 
matter.

\begin{figure}[t]
\minipage{0.49\textwidth}
\includegraphics[width=\linewidth]{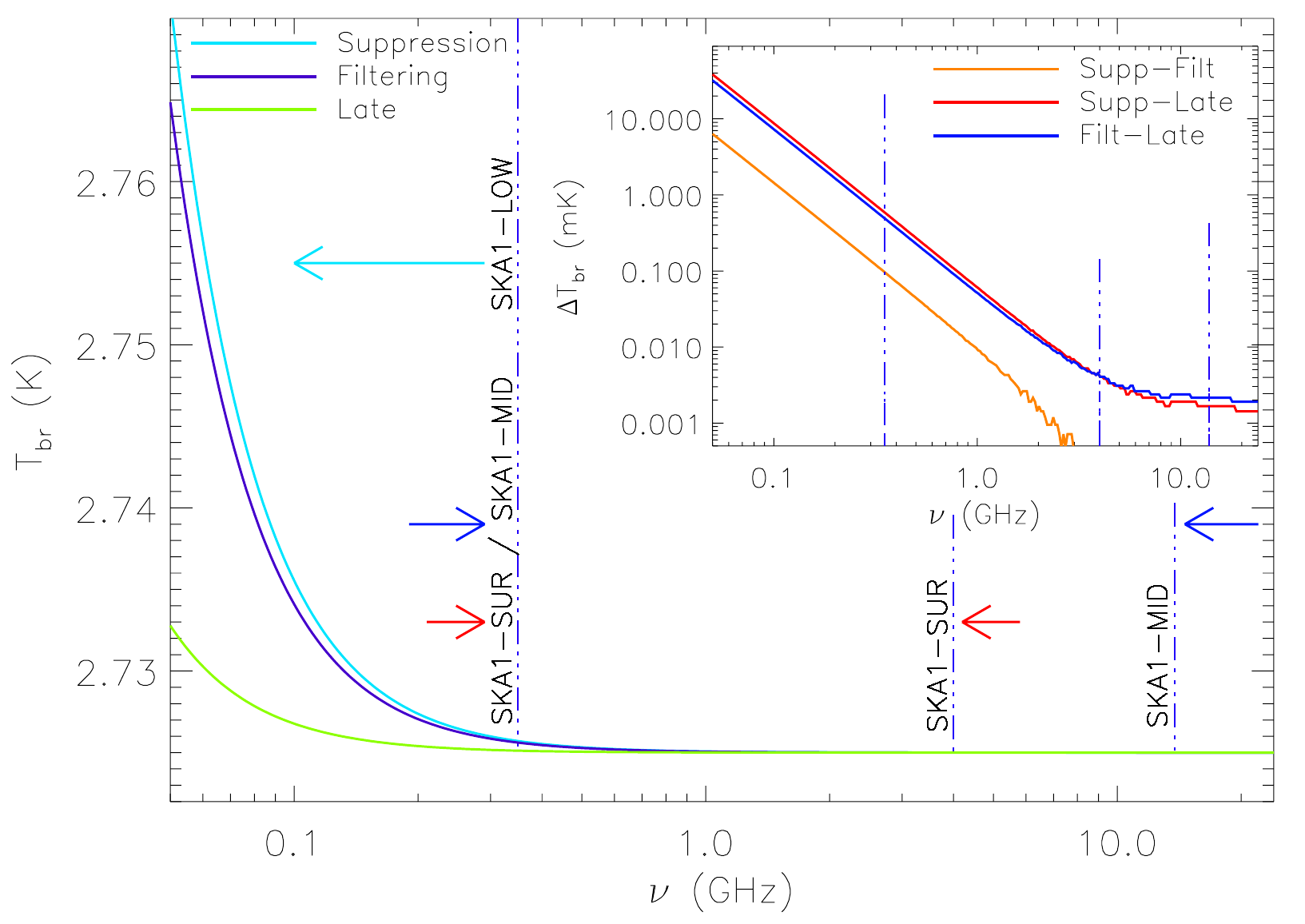}
\endminipage\hfill
\caption{Free-free diffuse signal 
in the interval of frequencies covered by SKA2  
computed for two astrophysical reionisation models 
(a {\it late} phenomenological prescription
is also shown).
The inset displays the absolute differences between the three models. The 
vertical lines specify the frequency coverage of SKA1 configurations. Taken 
from \cite{2015aska.confE.149B}. These curves define the minimal FF signal 
theoretically expected. For extreme models, like those considered by 
\cite{1999ApJ...527...16O}, the FF excess could be even $\sim 70$ times larger.}
\label{fig:cmb_dist}
\end{figure}

Cosmological reionisation, one of the three main mechanisms predicted to 
generate departures from a perfect BB \citep{sunyaevkhatri2013}, produces 
electron heating
which causes coupled Comptonisation and free-free distortions. The amplitude of 
Comptonisation distortion is proportional to the
energy fractional exchange occurred in the process. The Comptonisation 
parameter that characterises this energy exchange, denoted by $u$, is expected 
to have a typical minimum value of $10^{-7}$ from reionisation (and maximum 
values up to $\sim {\rm few} \times 10^{-6}$, achieved by including various 
types of sources).
For example, assuming the radiative feedback mechanisms proposed in the {\it 
filtering} and the {\it suppression} prescriptions
\cite{2008MNRAS.385..404B} obtained values of
the Comptonisation parameter produced by astrophysical reionisation of
$u \simeq (0.965 - 1.69) \times 10^{-7}$ (see Fig.~\ref{fig:cmb_dist}).

The SKA's high sensitivity and resolution can provide us relevant information 
to improve the current knowledge of the CMB spectrum and of the energy exchanges 
in cosmic plasma.
Furthermore, the SKA will help the modeling of Galactic emissions and 
extragalactic foregrounds, a substantial advancement being necessary to detect 
and possibly characterise the expected tiny spectral distortions. 
The extragalactic radio foreground is weaker than the Galactic 
radio emission but, in contrast to the Galactic foregrounds that represent the 
main limitation in CMB spectrum observations, 
it is difficult to separate it from the cosmological background 
by analysing its angular distribution properties in the sky because of the 
limited resolution of experiments devoted to CMB monopole temperature, 
particularly at low frequencies.
The accurate determination of the extragalactic source number counts from the 
deep SKA surveys allow to compute the source background,
improving the quality of its separation in CMB spectrum studies.

SKA will trace the neutral hydrogen distribution and the transition from 
the essentially neutral to the highly ionised state of the IGM during the dawn 
age and the 
 the reionisation epoch using the 21 cm redshifted line \citep[see, 
e.g.,][]{2008MNRAS.384.1525S}.
At the same time, it could directly reconstruct the evolution of ionised 
material by observing the FF emission produced by ionised halos.  
Reionisation models based on both semi-analytical
approaches \citep{2004MNRAS.347..795N} and numerical computations 
\citep{2011MNRAS.410.2353P} allow to estimate the expected signal.
Dedicated, high resolution observations may allow one to distinguish  the 
free-free spectral distortions by ionised halos from those by diffuse ionised 
IGM.
With SKA2-Low we could discover up to $\sim 10^{4}$ individual FF emission 
sources per squared degree at $z>5$, 
understanding the different contributions from ionised halos and from the 
diffuse ionised IGM to the global FF cosmological signal \citep[more details 
are provided by][]{2015aska.confE.149B}.

In conclusion, SKA precise number counts, particularly at frequencies from 
$\sim 1$ to a few GHz, will be crucial for a precise analysis of dedicated CMB 
spectrum measurements.
The precise mapping of large and dedicated regions of the sky with the SKA 
extremely good capability of producing interferometric images represents an 
interesting opportunity
to observe diffuse free-free emission anisotropies from large to small angular 
scales and individual halos. Moreover, implementing SKA with 
very compact configurations and ultra-accurate calibrators could be, in 
principle, a way to detect the absolute level of diffuse free-free emission.

\subsection{Detection Prospects and Challenges with the SKA}
\label{sec:challenges}

Having provided an overview of various fundamental physics constraints which 
may be achievable with the SKA observations of Cosmic Dawn and Reionisation, we 
present here a brief summary of the detection prospects, synergies with other 
probes at these epochs, and the foreground mitigation challenges, which  are 
relevant to recover fundamental physics constraints from these epochs.

\subsubsection{Challenges From EoR Astrophysics}

The astrophysics of the 21 cm line necessarily presents a `systematic' in the 
study of fundamental physics and cosmology. This is especially true at the 
epoch of reionisation and cosmic dawn, in which the various astrophysical 
processes described in \S\ref{sec:overview} lead to effects which need to be 
isolated effectively for the measurement of cosmological parameters. Modelling 
the astrophysics accurately is crucial to be able to distinguish the 
fundamental physics, and the power spectrum may need to be convolved with 
astrophysical models \citep[e.g., by using codes similar to 21CMFAST][described 
in \S\ref{sec:overview}]{Mesinger11}, in order to place competitive constraints 
on cosmology. 

Bayesian inference may be used to interpret the brightness temperature power 
spectrum in the context of a model, and to place constraints on cosmological 
parameters. In order to do this, analytic or semi-analytic techniques 
\citep[e.g.,][]{2004ApJ...613....1F, pritchard2008} are essential, since fast 
and accurate model parameter evaluation is required. It can be shown that this 
`astrophysical separation' can be effectively achieved in the post-reionisation 
universe using a halo model formalism to describe HI and obtain the 
uncertainties in the parameters from all the astrophysical constraints  
\citep{hpar2017,hparaa2017}. The combination of astrophysical constraints at 
these epochs can be shown to lead to 60\%--100\% uncertainty levels in the 
measurement of the HI power spectrum \citep{hptrcar2015}, which provides a 
measure of the `astrophysical degradation' relevant for forecasting 
cosmological and fundamental physics parameters. Similar modelling techniques 
applied to the high-redshift observations, though expected to be significantly 
harder, may be used to isolate the astrophysical effects for accurate 
constraints on the fundamental physics and cosmological parameters as described 
in the previous sections.

\subsubsection{Synergies between 21 cm and galaxy surveys}
\label{sec:gal21cmsyn}

Cross-correlating different astrophysical probes can eliminate the systematic 
effects in the measurements, and thus enable tighter constraints on the 
fundamental physics from the epoch of reionisation. Several large-area surveys 
of galaxies in the epoch of reionisation that overlap SKA1 and SKA2 are 
planned, using, e.g., the Hyper-SuprimeCam on Subaru (Lyman-$\alpha$ emitters, 
LAEs), Euclid (Lyman-break galaxies, LBGs), the Large Synoptic Survey Telescope 
(LSST, LBGs), and the Wide-Field Infrared Survey Telescope (WFIRST, LAEs and 
LBGs). 

Galaxy samples from such surveys will provide important calibrations of 
galaxy-population properties during the EoR, such as their clustering strength 
and star-formation-rate density. Towards later phases of reionisation, 
fluctuations in the neutral hydrogen fraction govern the brightness 
temperature. These fluctuations in turn depend on the properties of the sources, 
including their clustering \citep{2013ExA....36..235M}. Combining data on the 
population of source galaxies with the global brightness temperature signal 
measured with SKA at these epochs, the fraction of reionisation that is caused 
by the galaxies can be constrained \citep{2016arXiv160902312C}. 
Cross-correlation of the SKA brightness temperatures with the LAE/LBG samples 
from galaxy surveys provide additional constraints on the reionisation history, 
e.g., to what extent different galaxy populations contribute to reionisation, 
the evolution of the ionisation fraction, and the topology of reionisation 
\citep{0004-637X-836-2-176}. The brightness temperature can further be 
correlated with the properties of galaxies directly, e.g., from the Euclid or 
LSST wide+deep surveys \citep{2015aska.confE.145B}.

Using targeted observations with near/mid-infrared instruments, e.g. the James 
Webb Space Telescope and the European Extremely Large Telescope, currently 
uncertain source properties such as the net ionising flux and escape fraction 
can be constrained spectroscopically \citep[e.g.,][]{2016ApJ...827....5J}. The 
galaxy luminosity function within ionised bubbles identified in the 21 cm 
brightness temperature maps can also be constrained using these 
cross-correlations.

In the post-reionisation universe, cross-correlations can be used to understand 
the general life cycle of galaxies, which is determined by their star-formation 
activity in relation to the available gas reservoirs. 
The star-formation rate has been observed to peak at redshift 2 
\citep{2014ARA&A..52..415M} whereas observations of the HI energy density, 
$\Omega_{\rm HI}$, with redshift suggest very subtle to non-existing evolution 
of the gas densities \citep{2009ApJ...696.1543P}. This could imply that the 
molecular phase of hydrogen is the dominant ingredient in galaxy evolution 
processes \citep[e.g.,][]{Lagos:2015gpa,2016MNRAS.462.1749S}, though it is also 
tightly connected to the atomic as well as the ionised fractions of the 
hydrogen.

Mapping the intensity fluctuations of the 21 cm brightness temperature has been 
attempted in the post-reionisation universe with the Green Bank Telescope at 
$z\approx 0.8$ \citep{2013MNRAS.434L..46S}. Cross-correlating the data with 
complementary optical galaxy surveys \citep{2013ApJ...763L..20M} increases the 
detectability of the signal as well as giving a constraint on the average HI 
contents of the optical objects \citep{2017arXiv170308268W}. 

The SKA provides ample opportunities to extend existing observations to bigger 
volumes and higher redshifts \citep{Santos:2015vi}. In particular, SKA-Low can 
supply novel information via its proposed intensity mapping experiment in the 
higher frequencies of the aperture array at $3<z<6$. These observations will be 
crucial to understand the transitioning process of the cold gas after the epoch 
of reionisation as well as the distribution of HI gas in relation to the 
underlying halo mass and host galaxy properties. Additionally, the 
cross-correlations of the high-redshift HI datasets with either galaxy surveys 
or intensity maps of other spectral lines will reveal universal scaling 
relations of galaxy formation and evolution processes.

\subsubsection{Foreground Modelling}
One of the most significant challenges for an EoR detection is that of the 
overwhelming foregrounds.  
The problem is typically broken into three independent components -- Galactic 
synchrotron (GS), which contributes around 70\% of the total foreground 
emission \citep{Shaver1999}; extragalactic (EG) sources (predominantly compact) 
which contribute about 27\% \citep{2013ExA....36..235M}; and finally Galactic 
free-free emission which constitutes the remaining $\sim 1\%$. 
Altogether, these foregrounds are expected to dominate the EoR signal 
brightness by up to 5 orders of magnitude, though this figure reduces to 2--3 
when considering the interferometric observable: angular brightness 
fluctuations \citep{Bernardi2009}.
Furthermore, each source is expected to predominantly occupy a different region 
of angular-spectral space \citep{Chapman2016}.

All foreground mitigation techniques rely on first subtracting measured 
components, such as bright compact EG sources in the field-of-view 
\citep{Pindor2011}, and a diffuse sky model. 
While significant advances have been made in deep targeted observations of the 
foregrounds by various instruments 
\citep{Bernardi2009,Bernardi2010,Ghosh2011,Yatawatta2013,Jelic2014,Asad2015,
Remazeilles2015,Offringa2016,Procopio2017,Line2017}, due to their overwhelming 
dominance, even the residuals (from faint unmodelled sources and 
mis-subtraction) necessitate a robust mitigation approach.

The key to signal extraction lies in its statistical differentiation from the 
foregrounds, and it is well known that such a separation occurs naturally in 
the frequency (line-of-sight) dimension. 
While the signal is expected to exhibit structure on scales of $\sim$ MHz, the 
foregrounds are predominantly broadband emission, creating a smooth spectral 
signature. 
Leveraging this insight, several techniques for foreground residual mitigation 
have arisen in the past decade. 
Broadly, they may be split into two categories: (i) foreground subtraction, in 
which a smooth spectral model is fit and subtracted, and (ii) foreground 
avoidance, in which Fourier modes which are known to be foreground-dominated 
are eschewed.

\paragraph{Foreground Subtraction}
Foreground subtraction utilises the smoothness of the spectral dependence of 
the foregrounds in order to fit a smooth model to each angular pixel along the 
frequency axis. 
The best-fit model is subtracted, in the hope that the residuals are primarily 
the EoR signal. 

Specific methods in this technique have been further categorised by whether 
they are `blind': that is, whether they specify a parametric form to be fit, or 
whether the form is blindly identified by a statistical method. \\

\noindent
{\em Parametric Methods.} The earliest example of foreground modelling was the 
fitting of smooth polynomials of varying order 
\citep[e.g.,][]{McQuinn2006,Bowman2006}.
A more statistical approach is that of `correlated component analysis' (CCA) 
\citep{Ricciardi2010}, which invokes an empirical parametric form for each of 
the foreground components along with a linear mixing algorithm. 
For an application of CCA to simulated data, see \cite{Bonaldi2015}.
These methods have the inherent advantage of simplicity and the ability to 
impose any physical knowledge of the foreground structure directly.
Conversely, they suffer from the potential to over-fit and destroy the signal, 
as well as from ambiguity in the specification of a parametrisation. \\

\noindent
{\em Non-Parametric Methods.} One may alternatively propose a set of arbitrary 
bases to assume the role of a mixing matrix in the process of blind source 
separation. 
This alleviates the potential for over-fitting, and removes the ambiguity of 
form specification, to the detriment of simplicity and ability to directly 
input prior knowledge.
The most well-known implementations of this approach are fast independent 
component analysis \citep[FastICA;][]{Chapman2012} and generalised 
morphological component analysis \citep[GMCA;][]{Chapman2013}.
The latter appears to be the most robust approach in the foreground subtraction 
category \citep{Chapman2015}, and has been used as part of the LOFAR EoR 
pipeline \citep{Patil2017}.

\paragraph{Foreground Avoidance}
An inherent danger with foreground subtraction methods is the fact that, even 
post-subtraction, residuals may dominate over the signal due to overfitting or 
mis-subtraction. 
A more conservative route lies in first representing the data as a cylindrical 
power spectrum, i.e., separating line-of-sight modes, $k_{||}$, from 
perpendicular modes, $k_\perp$.
In this space, the foreground contributions are seen to occupy a low-$k_{||}$ 
region known as the `wedge'. 
This region has a reasonably sharp demarcation, and its complement is 
designated the EoR `window' \citep{Liu2014,Liu2014a}.
In principle, a final averaging purely over window modes yields a pristine 
power spectrum of the signal, and this has been employed by the PAPER project 
\citep{ali2015} and can inform instrument design \citep[e.g.,][]{DeBoer2017}.

This approach has the major drawback that a wide range of high-signal modes are 
unused \citep{Chapman2016, Liu2014a}.
A more optimal general approach was developed by \cite{Liu2014,Liu2014a}, based 
on the minimum variance estimator formalism of \cite{Liu2011}.
This method hinges upon defining the data covariance of the `junk' (i.e., the 
instrumentally distorted foregrounds and other systematics), either empirically 
\citep{Dillon2015} or parametrically \citep{Trott2016a}, and consistently 
suppresses modes which are foreground-dominated, optimally using all 
information.

A difficulty with parametric covariance is the suitable specification of the 
complex foreground models in the presence of instrumental effects. 
Accordingly, the Cosmological HI Power Spectrum Estimator 
\citep[CHIPS;][]{Trott2016a}, for example, employs simplistic prescriptions, 
with EG sources obeying empirical power-law source counts and uniform spatial 
distributions and GS emission obeying an isotropic power-law angular spectrum.
Recent studies have begun to relax these simplifications, for example, 
\cite{Murray2017} define the EG point-source covariance in the presence of 
angular clustering.

\paragraph{Summary and Outlook}
A number of systematic comparisons of foreground mitigation methods have been 
performed. 
\cite{Chapman2015} compared foreground subtraction methods and found that GMCA 
proves the most robust to realistic foreground spectra. 
\cite{Alonso2015} unified a number of subtraction methods under a common 
mathematical framework and showed that for a large suite of fast simulations 
the methods perform comparably.
\cite{Chapman2016} compared subtraction with avoidance, finding that they are 
complementary: avoidance recovers small scales well, while subtraction recovers 
large scales well. 
More specifically, \cite{Jacobs2016} compared the entire data pipelines used 
for the MWA analysis, including a basic avoidance technique 
\citep[$\epsilon$ppsilon;]{barry2019}, empirical covariance \citep{Dillon2015} and parametric 
covariance \citep{Trott2016a}. 
For the MWA data, each was shown to perform comparably.

Looking to the future, several challenges have been identified. 
One such challenge is the potential for polarisation leakage, which may induce 
a higher amplitude of small-scale structure on the foregrounds, obscuring the 
signal \citep{Moore2015,Asad2015,Asad2016,Asad2017}.
Another challenge is to improve the fidelity of EG source covariances. 
In particular, to date a distribution of source sizes has not been considered, 
and neither is the faint-source population constrained to any significant 
degree at EoR-pertinent frequencies.
More theoretically, attempts to consistently unify the avoidance and 
subtraction approaches must be furthered in order to extract maximal information 
from the data \citep[see, e.g.,][for examples of Bayesian 
frameworks]{Ghosh2015,Sims2016a,Lentati2017}.
Finally, an assortment of instrumental effects such as baseline mode-mixing 
\citep{Hazelton2013} must be overcome.

Despite these challenges, the increasing depth of low-frequency targeted 
foreground observations along with theoretical advancement of foreground 
techniques ensures that the EoR cannot hide forever.

\subsection{Summary}
\label{sec:eor_conclusions}

We have identified some of the key areas where SKA observations of the 21 cm 
signal are likely to impact fundamental physics as:
\begin{enumerate}
\item Cosmological parameters, especially neutrino mass and constraints on warm 
dark matter models (and other possible properties of dark matter).
\item Variations in fundamental constants (e.g., the fine structure constant).
\item Detecting the integrated Sachs-Wolfe effect in cross-correlation with 
galaxy catalogues.
\item Constraints on inflationary models and measurement of the runnings of the 
spectral index.
\item Tests of statistical anisotropy.
\item CMB spectral distortions and dissipation processes.
\end{enumerate}

We have indicated the challenges in the detection of the EoR signal with 
upcoming experiments, including the systematic imposed by the uncertainties in 
the astrophysics during these epochs, and ways to effectively isolate this to 
recover the underlying fundamental physics. We also briefly described synergies 
with other surveys during the same epochs, which allow cross-correlations that 
eliminate systematic effects to a large extent. Finally, we commented on the 
challenges from foregrounds at these frequencies, and the techniques for the 
foreground mitigation by both subtraction and avoidance methods.

Overall, the combination of (i) accurate astrophysical modelling of 
reionisation and the first stars, (ii) advances in detection techniques and 
foreground mitigation, and (iii) synergies with various other cosmological 
probes promises an optimistic outlook for observing the epochs of cosmic dawn 
and reionisation with the SKA, and for deriving exciting fundamental physics 
constraints from these as yet unobserved phases of the universe.

\section{Gravity and Gravitational Radiation}
\label{sec_grav}

Gravity plays a crucial role in astrophysics on all scales. While Einstein's General Theory of Relativity is our best theory, meeting all observational tests to date, there remain a number of open problems in astrophysics and cosmology that have, at their heart, the question of whether GR is the correct theory of gravity. In this section, we consider the ways in which the SKA will bring new opportunities for tests of theories of gravity at various length scales.

\subsection{Introduction}

\subsubsection{General Relativity and Modified Gravity}

To date, GR has passed every test with flying colours. The
most stringent of these have been carried out in the solar system and with binary
pulsars~\citep{Will:2014kxa, Stairs:2003eg, Wex:2014nva, Shao:2016ezh,
Kramer:2016kwa}, where a wide range of deviations from GR have been essentially
ruled out with extremely high precision.  The recent direct measurement of
gravitational waves by Advanced LIGO/Virgo has produced a new opportunity
to validate GR in a very different physical situation, i.e., a highly dynamical,
strong-field spacetime~\citep{TheLIGOScientific:2016src, Abbott:2017vtc}, and a
growing variety of cosmological tests of gravity are beginning to be carried
out with ever-increasing precision \citep{Joyce:2014kja,Bull2015}. These
are just a few of the regimes in which new gravitational phenomena could be
hiding, however \citep{Baker:2014zba}, and most have not yet been tested with
the high precision that is characteristic of solar system tests. Furthermore,
intriguing clues of possible deviations from GR have been emerging (e.g., in
recent studies of dark matter and dark energy) but are far from
decisive, and remain open to interpretation. Finally, GR may turn out to be the
low-energy limit of a more fundamental quantum gravity theory, with hints of
the true high-energy theory only arising in relatively extreme physical
situations that we have yet to probe. As such, testing GR across a broader
range of physical regimes, with increasing precision, stands out as one of the
most important tasks in contemporary fundamental physics. The SKA 
will be a remarkably versatile instrument for such tests, as we will discuss
throughout this section.

An important tool in extending tests of GR into new regimes has been the
development of a variety of alternative gravity theories
\citep{Clifton:2011jh}. These give some ideas of what possible deviations from
GR could look like, and help to structure and combine observational tests in a
coherent way. While there are many so-called {\it modified gravity} theories in
existence, it is possible to categorise them in a relatively simple way,
according to how they break Lovelock's theorem \citep{Lovelock:1971yv}. This is
a uniqueness theorem for GR; according to Lovelock's theorem, GR is the only
theory that is derived from a local, four-dimensional action that is at most
second order in derivatives only of the spacetime metric. Any deviation from
these conditions {\it breaks} the theorem, giving rise to an alternative non-GR
theory that may or may not have a coherent structure. For example, one can add
additional gravitational interactions that depend on new scalar or tensor
degrees of freedom (e.g. Horndeski or bigravity models respectively), add extra
dimensions (e.g., Randall-Sundrum models), introduce non-local operators (e.g.
non-local gravity), higher-order derivative operators (e.g., $f(R)$ theory), or
even depart from an action-based formulation altogether (e.g., emergent
spacetimes). Each of these theories tends to have a complex structure of its
own, which is often necessary to avoid pathologies such as {\it ghost}\ degrees
of freedom, derivative instabilities and so forth. Viable theories are also
saddled with the need to reduce to a theory very close to GR in the solar
system, due to the extremely restrictive constraints on possible deviations in
that regime. The result is that most viable modified gravity theories predict
interesting new phenomena --- for example screening mechanisms that shield non-GR
interactions on small scales as in Chameleon gravity \citep{Khoury:2003aq,Khoury:2003rn}, --- which in turn inform the
development of new observational tests. Unsuccessful searches for these new
phenomena can constrain and even rule out specific subsets of these theories,
and test GR in the process.

\paragraph{Testing relativistic gravity with radio pulsar timing}
\label{sec:pulsar:timing}

Pulsar timing involves the use of large-area radio telescopes or arrays to record the so-called times of arrival (TOAs) of pulsations from rotating radio pulsars.
Millisecond pulsars
(MSPs) are especially stable celestial clocks that allow timing precision at
the nanosecond level~\citep{Taylor:1992kea, Stairs:2003eg}. Such precision
enables unprecedented studies of neutron star astronomy and fundamental
physics, notably precision tests of gravity theories~\citep{Wex:2014nva,
Manchester:2015mda, Kramer:2016kwa}.

The TOAs from pulsar timing depend on the physical parameters that describe the 
pulsar system. These include
the astrometric and rotational parameters
of the pulsar, velocity dispersion in the intervening interstellar medium, and the motion of the telescope in the solar system (including the
movement and the rotation of the Earth). If the pulsar is in a binary system,
the TOAs are also affected by the orbital motion of the binary, which in turn depend on the underlying gravity theory~\citep{Damour:1991rd,
Edwards:2006zg}. Deviations from GR --- if any --- will manifest in TOAs, and
different kinds of deviations predict different {\it residuals} from the GR
template.

The double pulsar J0737--3039~\citep{Kramer:2006nb} represents the
state-of-the-art in the field. Five independent tests have already been made possible with
this system. GR passes all of them.
When the SKA is operating, the double pulsar will provide completely new tests,
for example measuring the Lense-Thirring effect~\citep{Kehl:2016mgp}, which
probe a different aspect of gravitation related to the spin.

What makes the field of testing gravity with pulsar timing interesting is that,
although the double pulsar represents the state-of-the-art, other pulsars can
outperform it in probing different aspects of gravity~\citep{Wex:2014nva}. For
example, the recently discovered triple pulsar system (with one neutron star
and two white dwarfs) is the best system to constrain the universality of free
fall (UFF) for strongly self-gravitating bodies~\citep{Ransom:2014xla, Shao:2016ubu,2018arXiv180702059A}.
UFF is one of the most important ingredients of the strong equivalence
principle~\citep[SEP;][]{Will:2014kxa}. When UFF is violated, objects with
different self-gravitating energies could follow different
geodesics~\citep{Damour:1991rq}. When the SEP is violated, for a binary
composed of two objects with different self-gravitating energies, it is very
likely that a new channel to radiate away orbital energy will open. If dipole radiation exists (in addition to the quadrupole radiation in GR), a
binary will shrink faster, resulting in a new contribution to the time
derivative of the orbital period~\citep{Damour:1991rd}. For example, this
happens in a class of scalar-tensor theories~\citep{Damour:1996ke}, and in
these theories, the dipole radiation might also be enhanced due to the strong
field of neutron star interiors. Binary pulsars have provided the best
constraints for this phenomenon~\citep{Freire:2012mg, Shao:2017gwu}.

Pulsars can be used to test the validity of theories~\citep{deCesare:2016axk,deCesare:2016dnp}  that lead to time variation of Newton's gravitational constant. A time varying Newton's constant will contribute to the decay
of the binary orbit as~\citep{Damour:1988zz, Nordtvedt:1990zz}
\begin{equation}
{\dot P\over P}=-2{\dot G\over G}\left[ 1-\left( 1+{m_c\over 2M} \right) s
\right] ~,
\end{equation}
where $P$, $m_c$, $M$ stand for the orbital period, the companion mass, and the sum of
the masses of the pulsar and its companion, respectively, and $s$ denotes a {\it sensitivity}\
parameter. Currently the strongest constraint on the temporal variation of the
gravitational constant results from lunar laser ranging (LLR) analysis, which
sets~\citep{Williams:2004qba}
\begin{equation}\label{eq:Gdot:LLR}
{\dot G\over G}=(4\pm 9)\times 10^{-13} \, {\rm yr}^{-1}~.
\end{equation}
Pulsar timing of PSRs~J1012+5307~\citep{Lazaridis:2009kq},
J1738+0333~\citep{Freire:2012mg} and J1713+0747~\citep{Zhu:2018etc} has achieved
limits comparable to Equation~(\ref{eq:Gdot:LLR}).

Binary pulsars can also be used to test cosmological models that lead to local
Lorentz invariance (LLI) violation. In particular, some modified gravity
models, such as the TeVeS~\citep{Bekenstein:2004ne} or the D-material
universe \citep[a cosmological model motivated from string
theory that includes a vector field;][]{Elghozi:2015jka} imply  violation of LLI.  Possible
violation of LLI results in modifications of the orbital motion of binary
pulsars~\citep{Damour:1992ah, Shao:2012eg, Shao:2014oha}, as well as to
characteristic changes in the spin evolution of solitary
pulsars~\citep{Nordtvedt:1987, Shao:2013wga};
for the latter, LLI also leads to
spin precession with respect to a fixed direction~\citep{Shao:2012eg}.
Hence, LLI violation implies changes in the
time-derivative of the orbit eccentricity, of the projected
semi-major axis, and of the longitude of the periastron, while it
changes the time-behaviour of the pulse profile. The strongest current constraints on LLI
violation are set from pulsar experiments, using the timing of binary pulsars.

There is also the potential for the SKA to search for the predicted effects of quantum gravity. Specifically, in a pulsar-black hole (BH) binary system, the disruption effect due to
  quantum correction can lead to a different gravitational time delay and
  interferometry of BH lensing. Recently, the discovery of PSR~J1745--2900~\citep{2013Natur.501..391E, 2013ApJ...775L..34R, Shannon:2013hla}
  orbiting the Galactic centre black hole Sgr A* opens up the possibility for
  precision tests of gravity~\citep{Pen:2013qva}. The radio pulses emitted from
  the pulsar can be lensed by an intervening black hole that is in between the
  pulsar and observer.  Therefore, the gravitational time-delay effect and
  interferometry between the two light rays can be used to investigate the
  possible quantum deviations from standard Einstein
  gravity~\citep{Pen:2013qva}.  According to~\citet{Pen:2013qva}, the fractal
  structure of the BH surface due to quantum corrections can destroy any
  interference between the two light rays from the pulsars. In the future, the 
  SKA will find a large number of pulsar-black hole binary
  systems, with which we will be able to perform stringent tests of gravity.

Finally, binary pulsars haven been used to constrain a free parameter of a
higher-derivative cosmological model, obtained as the gravitational sector of a
microscopic model that offers a purely geometric
interpretation for the Standard Model~\citep{Chamseddine:2006ep}.  By studying the propagation of
gravitons~\citep{Nelson:2010rt}, constraints were placed on the parameter that
relates coupling constants at unification, using either the quadrupole formula
for gravitational waves emitted from binary pulsars~\citep{Nelson:2010ru}  or
geodesic precession and frame-dragging
effects~\citep{Lambiase:2013dai}.  These constraints will be improved once more
rapidly rotating pulsars close to the Earth are observed. Clearly such an approach
can be used for several other extended gravity
models~\citep{Capozziello:2014mea, Lambiase:2015yia}.

Since the SKA will provide better timing precision and discover more pulsars, all
the above tests will be improved significantly~\citep{Shao:2014wja}.

\paragraph{Black-hole physics and Sgr A*}
Testing black hole physics is an intriguing and challenging task for modern
astronomy. Relativity predicts that any astrophysical black
hole is described by the Kerr metric and depends solely on its mass and angular momentum (or equivalently spin).
Sagittarius~A* (Sgr~A*), which is the closest example of a supermassive black hole (SMBH), is an ideal laboratory with which the SKA can test gravity theories and the no-hair theorem \citep{Kramer:2004hd}.

\begin{figure}
\center
\includegraphics[scale=0.5]{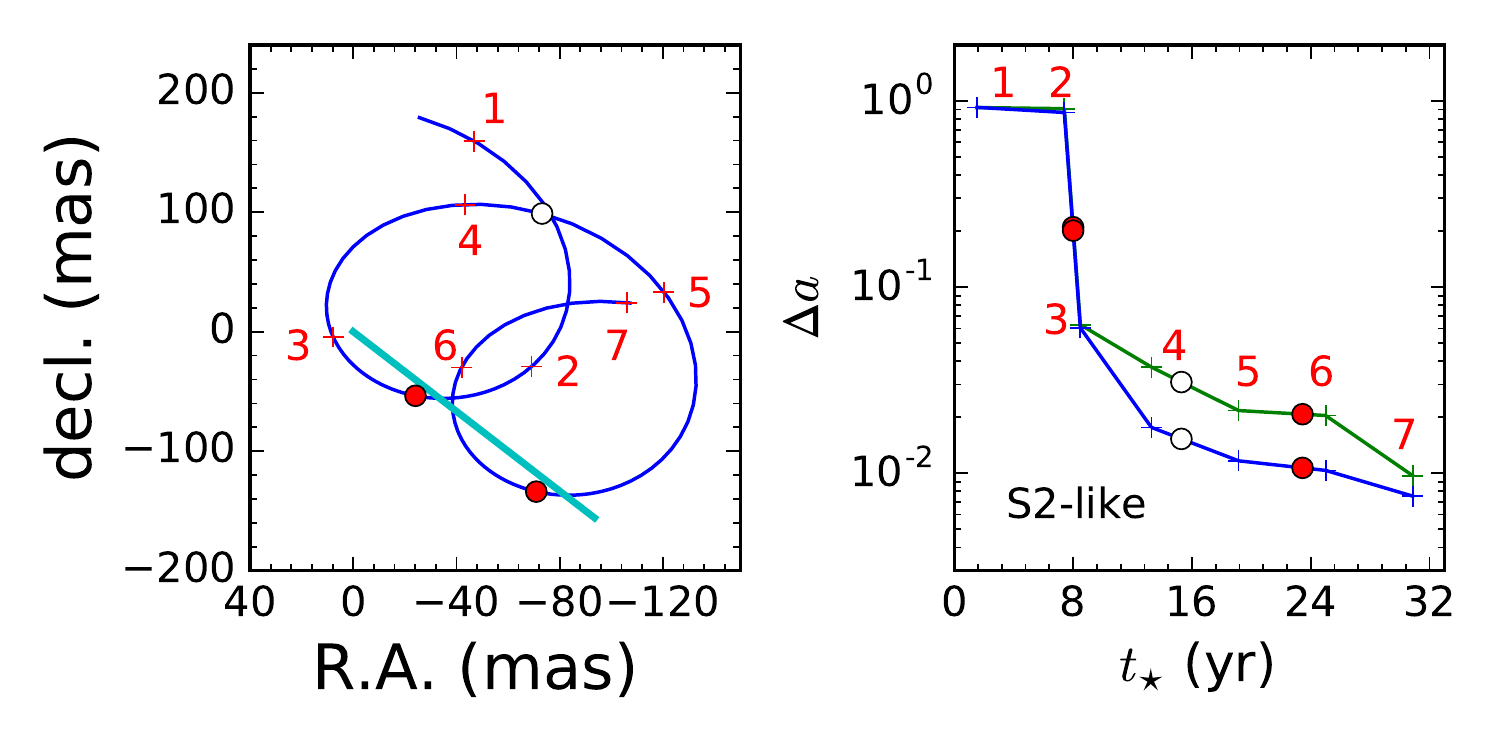}
\caption{{\em Left:} Apparent trajectories on the sky: blue for the
  pulsar and cyan for the SMBH. {\em Right:} Accuracy on the recovered
  spin magnitude, with green showing results when TOAs on their own are used,
  and blue showing results from combining both timing and proper motion
  information.~\citep{2017ApJ...849...33Z}.  The filled red and empty
  white circles mark the pericentre and apocentre, respectively,
  of the pulsar orbit.  The curves are interpolated from the computed
  accuracies at the epochs labelled 1--7.}
\label{fig:mcmc_fits_S2_like}
\end{figure}

Pulsars are extremely precise natural clocks due to their tremendous rotational stability.  Thus, a relativistic
binary of a pulsar and Sgr~A* would be a robust tool for testing relativity in stronger gravitational fields than is available from
pulsar binaries with stellar mass companions. Such a test will be important since strong field predictions can be fundamentally different between
GR and a number of alternative gravity theories \citep[see][for a review]{2016CQGra..33k3001J}.

The Galactic centre (GC) hosts a large number of young and massive stars within the inner parsec, which can be the progenitors of
pulsars~\citep[e.g.,][]{2006ApJ...643.1011P, 2013ApJ...764..155L}. The
population of normal pulsars can be hundreds within distance of $<4000AU$ from
Sgr A*~\citep[e.g.,][]{2014ApJ...784..106Z, 2004ApJ...615..253P,
2014MNRAS.440L..86C}.  The orbits of the innermost ones could be as tight as $\sim100$--$500\,AU$ from the SMBH~\citep{2014ApJ...784..106Z}.
Furthermore, a magnetar recently discovered in this region~\citep{2013ApJ...775L..34R, 2013Natur.501..391E} also suggests that a population of normal pulsars is likely to be present near the GC, since magnetars are rare pulsars.

To reveal pulsars in the GC region, a high frequency (usually $>9$\,GHz) radio survey is needed as there is severe radio scattering by the interstellar medium at low frequencies. Radio surveys so far have not found any normal pulsars
in the innermost parsec of the GC~\citep[e.g.,][]{2009ApJ...702L.177D,
2010ApJ...715..939M, 2011MNRAS.416.2455B}.  SKA1-Mid would be capable
of revealing pulsars down to $2.4$\,GHz with spin periods $\sim 0.5$s in this region~\citep{2015aska.confE..45E}.
The timing accuracy of pulsars for SKA after $\sim1$ hour integration can
reach $\sigma_T\simeq100\,\mu$s~\citep{2012ApJ...747....1L} at a
frequency of $\ga15\,$GHz, and $\sigma_T\simeq0.1$--$10\,$ms if the
frequency is between $\ga 5\,$GHz and $\la 15\,$GHz.  Besides the timing
measurements, proper motions would be measurable for these pulsars.
Finally, the baselines of the SKA are expected to be up to $\sim 3000\,$km, and
thus it can provide image resolution up to $2\,$mas at $10\,$GHz~\citep{2012PASA...29...42G}
and astrometric precision reaching $\sim10\,\mu$as~\citep{2004NewAR..48.1473F}.

The relativistic effects cause orbital precession of the pulsars orbiting Sgr~A*, in both the argument of pericentre and the orbital plane. A number of previous studies have focussed on the relativistic effects according to the orbital averaged precession over multiple orbits \citep[e.g.,][]{Wex:1998wt,
2004ApJ...615..253P, 2012ApJ...747....1L, 2016ApJ...818..121P} or the resolved
orbital precession within a few orbits~\citep{2010ApJ...711..157A,
2010ApJ...720.1303A}.  These studies
implement post-Newtonian techniques based
on~\citet{1976ApJ...205..580B}, \citet{1986AIHS...44..263D}, and
\citet{2006MNRAS.369..655H}, or a mixed perturbative and numerical
approach~\citep{2010ApJ...720.1303A}. For a pulsar orbiting an SMBH, it is
also feasible to implement full relativistic
treatments~\citep{2015ApJ...809..127Z, 2017ApJ...849...33Z}.

The TOAs of pulsars rotating around Sgr A* are affected by a
number of relativistic effects, e.g., Einstein delay and Shapiro
delay~\citep{1986AIHS...44..263D, Taylor:1992kea}. The orbital precession
caused by frame-dragging and quadrupole moment effects also impact the
TOAs. Recent studies have found that the frame-dragging effect in TOAs for
a pulsar-Sgr A* binary are quite strong compared to the timing accuracies of the
pulsar~\citep{2012ApJ...747....1L,2016ApJ...818..121P}, i.e., orders of
$10$--$100$\,s per orbit while the timing accuracies are typically
$\sim0.1$\,ms~\citep{2017ApJ...849...33Z}. Current TOA modelling assumes that the orbital precession
increases linearly with time. However, it is found to be inaccurate compared to
the TOA accuracy; thus, more sophisticated modelling of TOAs are needed, e.g.,
explicitly solving the geodesic equation of the pulsars and the propagation
trajectories of the photons~\citep{2017ApJ...849...33Z}.

Frame-dragging and quadrupole momentum effects can be tightly constrained by observing relativistic pulsar-Sgr A* binaries. If the
orbital period of a pulsar is $\sim0.3$\,yr, the frame-dragging and the quadrupole
moment effect of the SMBH can be constrained down to $\sim10^{-2}$--$10^{-3}$ and
$\sim10^{-2}$, respectively, within a decade, providing timing accuracies of $\sigma_{\rm
T}\sim100\mu$s~\citep{2012ApJ...747....1L}. By monitoring a normal pulsar with
an orbital period of $\sim2.6$\,yr and an eccentricity of $0.3$--$0.9$, and assuming a
timing accuracy of $1$--$5$\,ms, the magnitude, the line-of-sight
inclination, the position angle of the SMBH spin can be constrained with
$2\sigma$ errors of $10^{-3}$--$10^{-2}$, $0.1^\circ$--$5^\circ$ and
$0.1^\circ$--$10^\circ$, respectively, after $\sim$8~years~\citep{2017ApJ...849...33Z}.
Even for pulsars in orbits similar to the currently detected stars S2/S0-2 or S0-102, the spin of the
SMBH can still be constrained within $4$--$8\,$yr \citep{2017ApJ...849...33Z}; see Figure~\ref{fig:mcmc_fits_S2_like}. Thus, any pulsar
located closer than $\sim 1000$\,AU from the SMBH is plausible for GR spin measurements
and tests of relativity.

Combining timing and astrometric measurements of GC pulsars, the mass and
distance of Sgr~A* can be constrained with extremely high accuracy. If the
proper motion of pulsars can be determined with an accuracy of $10\,\mu$as
along with timing measurements, the mass and the distance of the SMBH,
 can be constrained to about $\sim 1\,M_\odot$ and $\sim1\,$pc, respectively~\citep{2017ApJ...849...33Z}.

It is important to note, however, that GC pulsars would experience gravitational perturbations from other masses, such as stars or other stellar
remnants. These (non-relativistic) perturbations may obscure the spin-induced signals
outside $\ga100$--$400\,\AU$~\citep{2010PhRvD..81f2002M,
2017ApJ...834..198Z}. Outside this region, how to remove this Newtonian
``foreground'' remains an unsolved problem. One possible filtering strategy may be to use wavelets~\citep{2014MNRAS.444.3780A}.

\paragraph{Cosmological tests of gravity}

While GR has proven robust against all observational and experimental tests that have been carried out so far, most of these have been restricted to the solar system or binary pulsar systems -- i.e. firmly in the small-scale, weak field regime. The recent LIGO gravitational wave detection has added a valuable strong field test of GR to the roster, but it is the relatively poorly-constrained cosmological regime that has perhaps the greatest chance of offering a serious challenge to Einstein's theory. The application of GR to cosmology represents an extrapolation by many orders of magnitude from where the theory has been most stringently tested, out to distance scales where unexpected new gravitational phenomena -- specifically, dark matter and dark energy -- have been discovered to dominate the Universe's evolution. While it may yet be found that these have `conventional' explanations, perhaps in terms of extensions to the Standard Model of particle physics, the fact remains that they have so far {\it only} been detected through their gravitational influence. As such, it is of utmost importance to examine whether the extrapolation of GR out to cosmological distances could be to blame for the appearance of these effects -- perhaps we are interpreting our observations in the context of the wrong gravitational theory?

Cosmological tests of GR are still in their infancy, however. While most `background' cosmological parameters are now known to better than 1\% precision, additional parameters that describe possible deviations from GR are considerably less well constrained. Recent measurements of the growth rate of large scale structure have been made at the 10\% level, for example, while many alternative theories of gravity have never even been subjected to tests beyond a comparison with background parameter constraints from (e.g.) the CMB. It is clear, then, that there is some way to go before constraints on GR in the cosmological regime approach the accuracy that has been achieved in the small-scale, weak field limit.

The SKA is expected to play a central role in a multitude of high-precision tests of GR in cosmological settings, often in synergy with other survey experiments in different wavebands. In this section, we consider several examples of how SKA1 and SKA2 will contribute to precision cosmological tests of GR, including: growth rate and slip relation measurements with galaxy clustering and weak lensing observations; tests of gravity and dark energy using the 21-cm intensity mapping technique; detecting relativistic effects on ultra-large scales; peculiar velocity surveys; and void statistics.

On linear sub-horizon scales, there are two main ways in which deviations from GR can affect cosmological observables: by modifying how light propagates, and by modifying how structures collapse under gravity \citep{Amendola:2012ky}. Both effects can be probed using large statistical samples of galaxies, for example by measuring the weak lensing shear and redshift-space distortion signals. At optical wavelengths, these observations are the preserve of photometric (imaging) and spectroscopic redshift surveys respectively, but radio observations offer several alternative possibilities for getting at this information.

\paragraph{Radio weak lensing:} Effective weak lensing surveys can be performed using radio continuum observations \citep{Brown:2015ucq}, where the total emission from each galaxy is integrated over the entire waveband to increase signal-to-noise. SKA1-Mid has excellent $u-v$ plane coverage, making it possible to image large numbers of galaxies and measure their shapes. It will perform a large continuum galaxy survey over an area of several thousand square degrees \citep{2015aska.confE..18J},  achieving a sky density of suitable lensed sources of 2.7 arcmin$^{-2}$ at a mean redshift of $\sim 1.1$ \citep{2016MNRAS.463.3674H}. 
This is a substantially lower number density than contemporary optical
surveys; for example, the Dark Energy Survey (DES) will yield $\sim 12$ arcmin$^{-2}$ at a mean redshift of
0.6. However, forecasts suggest that the two surveys should constrain cosmological
parameters with a similar level of accuracy --- for example, both SKA1 and
DES lensing surveys should produce $\mathcal{O}(10\%)$ constraints on the
parameter $\Sigma_0$, which parametrises deviations of the lensing potential
from its GR behaviour \citep{2016MNRAS.463.3674H}. This is mainly due to the
stronger lensing signal from a significant high-redshift tail of continuum
sources that compensates for the lower source number density. Corresponding
forecasts for SKA2 suggest that a number density of 10 arcmin$^{-2}$ will be
achievable at a mean redshift of 1.3, for a survey covering 30\,000 deg$^2$,
yielding $\sim 4\%$ constraints on $\Sigma_0$ \citep{2016MNRAS.463.3674H},
surpassing what will be possible with Euclid. While SKA alone will produce
strong constraints on modified gravity lensing parameters, the combination
of SKA with optical lensing surveys should be the ultimate goal, as the two
different methods have very different systematics that should mostly drop
out in cross-correlation, producing much `cleaner' lensing signals with
enhanced signal-to-noise \citep{2016MNRAS.463.3686B,2017MNRAS.464.4747C}.

\begin{figure}[t]
\includegraphics[width=0.5\textwidth]{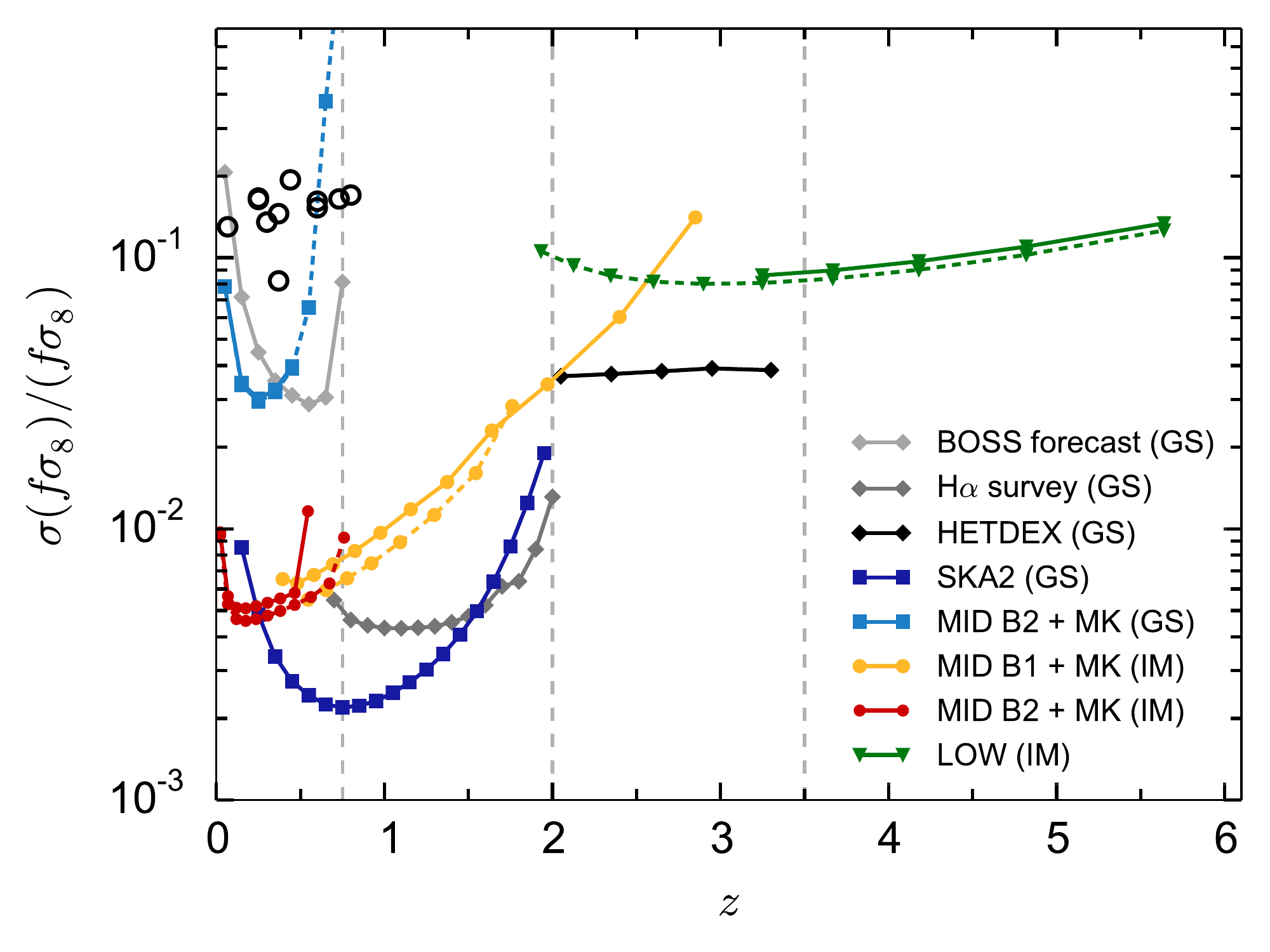}
\caption{Comparison of predicted constraints on the growth rate,
$f\sigma_8$, from RSD measurements with various SKA and contemporary
optical/near-infrared surveys. `GS' denotes a spectroscopic galaxy survey, while `IM' denotes an intensity mapping survey. The open circles show a compilation of recent RSD measurements. Taken from \cite{Bull:2015lja}.}
\label{fig:rsd}
\end{figure}

\paragraph{Redshift-space distortions and peculiar velocities from HI
galaxies} 
\label{sec_rsd}
SKA1 will have the sensitivity and spectral resolution to perform
several different types of spectroscopic galaxy surveys, using the 21cm
emission line from HI. The simplest is a redshift survey, where the 21cm
line is detected for as many galaxies as possible, with a signal-to-noise
ratio sufficient only to get a fix on each redshift. Both SKA1 and SKA2 will
be able to perform very large redshift surveys; the SKA1 version will be
restricted to quite low redshifts, due to the steepness of the sensitivity
curve for HI \citep{Yahya:2014yva, Harrison:2017pcu}, while the SKA2 version
will be essentially cosmic variance limited from redshift 0 to $\sim 1.4$
for a survey covering 30,000 deg$^2$ \citep{Yahya:2014yva, Bull:2015lja}.
Precise spectroscopic redshifts allow the galaxy distribution to be
reconstructed in 3D down to very small scales, where density fluctuations
become non-linear, and galaxies have substantial peculiar velocities due to
their infall into larger structures. These velocities distort the 3D
clustering pattern of the galaxies into an anisotropic pattern, as seen in
redshift-space. The shape of the anisotropy can then be used to infer the
velocity distribution, and thus the rate of growth of large-scale structure.
HI redshift surveys with SKA1 and SKA2 will both be capable of precision
measurements of these RSDs, with SKA1 yielding $\sim 10\%$ measurements of
$f\sigma_8$ (the linear growth rate multiplied by the normalisation of the
matter power spectrum) in several redshift bins out to $z \approx 0.5$, and
SKA2 yielding $\lesssim 1\%$ measurements out to $z \approx 1.7$
\citep{Bull:2015lja}. See Figure~\ref{fig:rsd} for a comparison with other
surveys.

Note that redshift surveys are not the only possibility -- one can also try
to spectrally resolve the 21cm lines of galaxies with high signal-to-noise
ratios, and then measure the width of the line profile to obtain their rotation velocities. This can then be used in conjunction with the Tully-Fisher (TF) relation that connects rotation velocity to intrinsic luminosity to directly measure the distances to the galaxies, making it possible to separate the cosmological redshift from the Doppler shift due to the peculiar velocity of the galaxy. Direct measurements of the peculiar velocity are highly complementary to RSDs, as they measure the growth rate in combination with a different set of cosmological parameters (i.e., they are sensitive to $\alpha = f[z] H[z]$). The recovered velocity field can also be cross-correlated with the density field (traced by the galaxy positions), resulting in a significant enhancement in the achievable growth rate constraints if the source number density is high enough \citep{Koda:2013eya}. SKA1 will be able to perform a wide, highly over-sampled TF peculiar velocity measurement at low redshift \citep[cf., the sensitivity curves of][]{Yahya:2014yva}, potentially resulting in better constraints on the growth rate than achievable with RSDs. The peculiar velocity data would also be suitable for testing (environment-dependent) signatures of modified gravity due to screening, as discussed by \citet{Hellwing:2014nma} and \citet{Ivarsen:2016xre}.

\paragraph{21-cm intensity mapping} \label{sec_grav_im}
21-cm intensity mapping \citep{Battye:2004re,Chang:2007xk} is an innovative technique that uses HI to map the three-dimensional large-scale structure of the Universe. Instead of detecting individual galaxies like traditional optical or radio galaxy surveys, HI intensity mapping surveys measure the intensity of the redshifted 21-cm emission line in three dimensions (across the sky and along redshift).  

The possibility of testing dark energy and gravity with the SKA using 21-cm intensity mapping has been studied extensively \citep{Santos:2015vi}. 
More specifically, it has been shown that an intensity mapping survey with SKA1-Mid can measure cosmological quantities like the Hubble rate $H(z)$, the angular diameter distance $D_{\rm A}(z)$, and the growth rate of structure $f\sigma_8(z)$ across a wide range of redshifts \citep{Bull:2014rha}, at a level competitive with the expected results from Stage IV optical galaxy surveys like Euclid \citep{Amendola:2016saw}. For example, a very large area SKA1-Mid intensity mapping survey can achieve sub-$1\%$ measurements of $f\sigma_8$ at $z<1$ \citep{Bull:2015lja}.

However, the intensity mapping method is still in its infancy, with the major issue being foreground contamination (which is orders of magnitude larger than the cosmological signal) and systematic effects.
These problems become much more tractable in cross-correlation with optical galaxy surveys, since systematics and noise that are relevant for one type of survey but not the other are expected to drop out \citep{Masui:2012zc,Pourtsidou:2015mia,Wolz:2015lwa}. Therefore, cross-correlating the 21-cm data with optical galaxies is expected to alleviate various systematics and lead to more robust cosmological measurements. 

As an example, we can consider cross-correlating an HI intensity mapping survey with SKA1-Mid with a Euclid-like optical galaxy clustering survey, as discussed by \citet{Pourtsidou:2016dzn}. Assuming an overlap $A_{\rm sky} = 7000 \, {\rm deg}^2$, it was found that very good constraints can be achieved in ($f\sigma_8, D_{\rm A}, H$) across a wide redshift range $0.7 \leq z \leq 1.4$, where dark energy or modified gravity effects are important (see Table \ref{tab:SKAIMcross}). Furthermore, it was found that combining such a survey with CMB temperature maps can achieve an ISW detection with a signal-to-noise ratio $\sim 5$, which is similar to the results expected from future Stage IV galaxy surveys. Detecting the ISW effect in a flat Universe provides direct evidence for dark energy or modified gravity.

\begin{table}
\begin{center}
\caption{\label{tab:SKAIMcross} Forecasted fractional uncertainties on $\{f\sigma_8, D_{\rm A}, H\}$ assuming the SKA1-Mid intensity mapping and Euclid-like spectroscopic surveys.}  
\begin{tabular}{cccc}
\hline
$z$ & $\delta(f\sigma_8)/(f\sigma_8)$ & $\delta D_{\rm A}/D_{\rm A}$ & $\delta H/H$\\
\hline
0.7&0.04&0.03&0.02\\
   0.8&0.05&0.03&0.02\\
0.9&0.05&0.03&0.03\\
1.0&0.06&0.04&0.03\\
1.1&0.07&0.04&0.03\\
1.2&0.08&0.05&0.03\\
1.3&0.10&0.06&0.03\\
1.4&0.11&0.06&0.04\\
\hline
\end{tabular}
\end{center}
\end{table}

\paragraph{Relativistic effects on ultra-large scales} Thanks to the unmatched depth of continuum radio galaxy surveys, the large sky coverage, and the novel possibilities available with HI intensity mapping, the SKA will probe huge volumes of the Universe, thus allowing us to access the largest cosmic scales. Scales close to the cosmic horizon and beyond carry valuable information on both the primeval phases of the Universe's evolution and on the law of gravity. 

On the one hand, peculiar inflationary features such as primordial non-gaussian imprints are strongest on the ultra-large scales. On the other hand, if we study cosmological perturbations with a fully relativistic approach, a plethora of terms appears in the power spectrum of number counts besides those due to Newtonian density fluctuations and RSDs \citep{2011PhRvD..84d3516C,Bonvin:2011bg,2012PhRvD..86f3514Y,Jeong:2011as,Alonso:2015uua}. For instance, lensing is known to affect number counts through the so-called magnification bias; but other, yet-undetected effects like time delay, gravitational redshift and Sachs-Wolfe and integrated Sachs-Wolfe-like terms also contribute on the largest cosmic scales. To measure such relativistic corrections would mean to further thoroughly confirm Einstein's gravity, in a regime far from where we have accurate tests of it. Otherwise, if we found departures from the well known and robust relativistic predictions, this would strongly hint at possible solutions of the dark matter/energy problems in terms of a modified gravity scenario \citep{Lombriser2013,Baker:2014zva,Baker:2015bva}.

Alas, measurements on horizon scales are plagued by cosmic variance. For
instance, forecasts for next-generation surveys show that relativistic
effects will not be detectable using a single tracer
\citep{Camera:2015fsa,Alonso:2015sfa} and primordial non-gaussianity
detection is limited to $\sigma(f_{\rm NL})\gtrsim 1$
\citep{Camera:2014bwa,Raccanelli:2014ISW}. This calls for the multi-tracer
technique (MT), developed for biased tracer of the large-scale cosmic
structure and able to mitigate the effect of cosmic variance
\citep{Seljak:2008xr,Abramo:2013awa,Ferramacho:2014pua}.
\citet{Fonseca:2015laa} showed that the combination of two contemporaneous
surveys, a large HI intensity mapping survey with SKA1 and a Euclid-like
optical/near-infrared (NIR) photometric galaxy survey, will provide
detection of relativistic effects, with a signal-to-noise of about 14.
Forecasts for the detection of relativistic effects for other combinations
of radio/optical surveys are discussed by \citet{Alonso:2015sfa}.

\paragraph{Void statistics}

As a particular case for the SKA, we consider number counts of voids, and
forecast cosmological parameter constraints from future SKA surveys in
combination with Euclid, using the Fisher-matrix method (see also \S\ref{sec:lssneutrinos}). Considering that
additional cosmological information is also available in, e.g., shapes/profiles, accessible with the SKA, voids are a very promising new cosmological probe. 

\begin{figure*}[t]
\includegraphics[trim={4cm 0.9cm 4cm 1cm},clip,width=\textwidth]{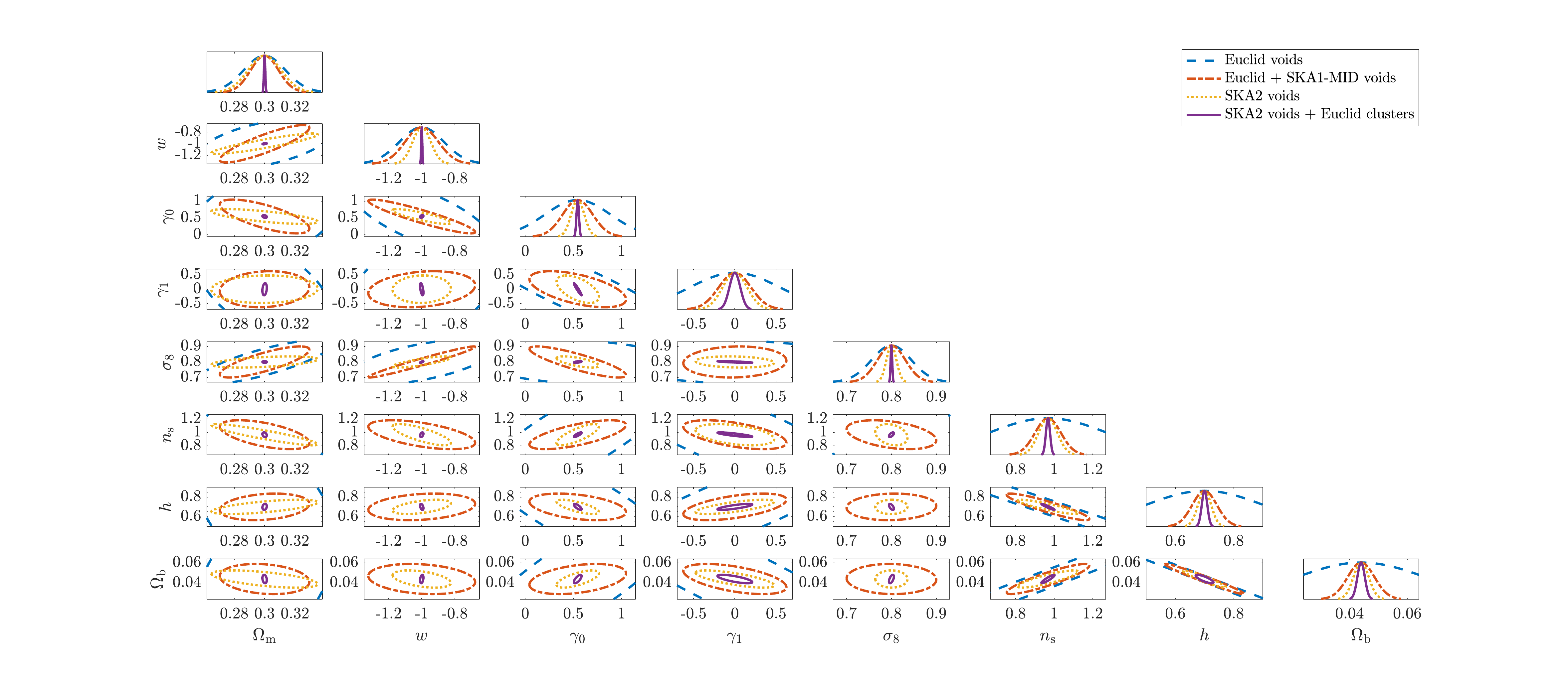}
\caption{\label{fig:voidmgforecast} Forecast 68\% parameter confidence constraints for a flat wCDM model with time-dependent growth index of matter perturbations. Note the considerable degeneracy breaking between the Euclid and SKA1 void samples, and between the SKA2 void and Euclid cluster samples. SKA1-Mid covers 5000~deg$^2$, $z = 0-0.43$. SKA2 covers 30,000 deg$^2$, $z=0.1-2$. Euclid voids covers 15,000 deg$^2$, $z=0.7-2$. Euclid clusters covers 15,000~deg$^2$, $z=0.2-2$. The fiducial cosmological model is given by $\{\Omega_{\rm m} = 0.3, w = -1, \gamma_0 = 0.545, \gamma_1 = 0, \sigma_8 = 0.8, n_{\rm s} = 0.96, h = 0.7, \Omega_{\rm b} = 0.044\}$. We have also marginalised over uncertainty in void radius and cluster mass \citep{Sahlen:2016kzx}, and in the theoretical void distribution function \citep{Pisani:2015jha}.}
\end{figure*}

We consider a flat wCDM cosmology (i.e., a CDM cosmology with a constant equation
of state, $w$) with a modified-gravity model described by a
growth index $\gamma(a) = \gamma_0 + \gamma_1(1-a)$ \citep{DiPorto:2012ey}.
The void distribution is modelled following \citet{Sahlen:2015wpc} and
\citet{Sahlen:2016kzx}, here also taking into account the galaxy density and
bias for each survey \citep{Yahya:2014yva,Raccanelli:2015vla}. The results
are shown in Figure~\ref{fig:voidmgforecast}. 
The combined SKA1-Mid and Euclid void number counts could achieve a
precision $\sigma(\gamma_0) = 0.16$ and $\sigma(\gamma_1) = 0.19$,
marginalised over all other parameters. The SKA2 void number counts could improve on this, down to $\sigma(\gamma_0) = 0.07$, $\sigma(\gamma_1) = 0.15$. By using the powerful degeneracy-breaking complementarity between clusters of galaxies and voids \citep{Sahlen:2015wpc,Sahlen:2016kzx,2019PhRvD..99f3525S}, SKA2 voids + Euclid clusters number counts could reach $\sigma(\gamma_0) = 0.01$, $\sigma(\gamma_1) = 0.07$. 

\subsection{Gravitational-Wave Astronomy}

\subsubsection{Understanding gravitational-wave sources}

Gravitational waves (GWs) may be sourced by an astrophysical object (compact objects
such as neutron stars and black holes) or they can be of a cosmological origin.
Binaries of coalescing compact objects constitute the main goal of ground-based
interferometers. Processes operating in the early universe may lead to a stochastic gravitational wave background, offering a unique opportunity to understand the laws that operated at such high energies, as gravitational waves are out of thermal equilibrium since the Planck scale. Possible sources of gravitational waves of cosmological origin are inflation, particle production, preheating, topological defects like cosmic strings, and first order phase transitions.

\subsubsection{Detection of gravitational waves with SKA galaxy surveys}

Galaxy catalogues can be used to detect GWs; the idea of looking at the
angular motion of sources both in the Milky Way~\citep{Jaffe:2004,
Book:2011} and on extragalactic scales dates back to the 1980s \citep[see,
e.g.,][]{Linder:1986, Linder:1988, Braginsky:1990, Kaiser:1997}.

The possibility of detecting GWs using SKA galaxy surveys has been
investigated recently by~\cite{Raccanelli:2016GW}, by looking at what has
been defined `cosmometry', i.e., the high redshift equivalent of astrometry:
the passage of a stochastic gravitational wave background (SGWB) will cause
the angular position of distant sources to oscillate. The oscillations have
a zero average, but the RMS is proportional to the strain of the passing GWs. Therefore, by means of a statistical analysis of galaxy correlations, it could be possible to detect GWs from the early Universe.

Another possibility comes from using galaxy catalogues obtained with the SKA
and their statistics to detect the presence of an SGWB from the effect of
tensor perturbations; gravitational waves are tensor perturbations, and so a
background of them will have effects on galaxy clustering and gravitational
lensing statistics \citep[see also][]{CosmicRulers1, CosmicRulers2}.

\subsubsection{Pulsar timing arrays}

Pulsar timing arrays (PTAs) use the `quadrupole correlation' (the
Hellings-Downs curve) in the timing residuals from an array of pulsars, aiming
to detect low-frequency GWs in the frequency range
$10^{-9}-10^{-6}$\,Hz~\citep{Hellings:1983fr, Foster:1990, Hobbs:2017oam}.
The Parkes PTA collaboration~\citep[PPTA;][]{Manchester:2012za} was established
in 2004, followed in 2007 by the European PTA~\citep[EPTA;][]{Kramer:2013kea},
and the North American Nanohertz Observatory for Gravitational
Waves~\citep[NANOGrav;][]{McLaughlin:2013ira}. PPTA, EPTA, and NANOGrav form
the International PTA collaboration~\citep[IPTA;][]{Verbiest:2016vem} to share
data and algorithm among different PTAs. When the SKA is online, it will boost
the PTA efforts to detect low-frequency GWs~\citep{Kramer:2004hd,
Janssen:2014dka}.

There are various GW sources for PTAs~\citep{Janssen:2014dka}. For example,
cosmic strings, one-dimensional topological defects, arise naturally in many
field theories as a particular class of false vacuum
remnants~\citep{Jeannerot:2003qv}.  A loop of invariant string length $\ell$,
has a period $T=\ell/2$ and oscillates at a fundamental frequency
$\omega=4\pi/\ell$. Hence, it radiates gravitational waves with frequencies
that are multiples of $\omega$ and decays in a lifetime $\ell/(100 G\mu)$,
where $G$ is Newton's constant and $\mu$ is the mass per unit length for cosmic strings. The
loop contribution to the stochastic gravitational wave background is expressed
in terms of the frequency $f$ as,
\begin{equation}
\Omega_{\rm GW}={f\over \rho_{\rm c}} {d\rho_{\rm GW}\over df}~,
\end{equation}
where $\rho_{\rm c}$ denotes the critical energy density of the universe, and
$\rho_{\rm GW}$ depends on the string linear density and therefore on the
temperature of the phase transition followed by spontaneous symmetry breaking
leading to the cosmic string production.  Pulsar timing experiments are able to
test the spectrum of gravitational waves at nanohertz frequencies, while
LIGO/Virgo detectors are sensitive in the 10--1000\,Hz band.

For understanding the impact of SKA on pulsar timing based GW detection, it is
important to estimate the total number of MSPs that can
be discovered with the SKA and the typical root mean square (RMS) noise level of pulsar TOAs that can be attained. 

In one survey scenario~\citep{Smits:2011technote}, 
SKA1-Mid is expected to detect 1200 
MSPs in 53 days of telescope time, and this number will climb up to 6000 MSPs
with SKA2-Mid \citep{2009A&A...493.1161S}. It is predicted that one timing
observation for 250 MSPs at a signal-to-noise ratio of $\sim 100$ each---the
level at which GW detection becomes feasible for anticipated sources---can be
obtained with 6 to 20 hours of telescope time on SKA2-Mid. 

The timing precision is determined by the noise budget of the measured TOAs. 
The RMS of the pulse phase jitter noise and the radiometer noise, 
the most important noise sources at 100 ns timing precision level, 
can be estimated by \citep{Cordes:2010fh, Wang:2015bsa}
\begin{equation}\label{jitter}
\sigma_{\rm j} \approx 0.28 W\sqrt{\frac{P}{t}} \;,
\end{equation}
\begin{equation}\label{rad}
\sigma_{\rm r} \approx \frac{W S}{F\sqrt{2\Delta f t}}\sqrt{\frac{W}{P-W}} \;. 
\end{equation}
Here $P$ is the pulsar period, $t$ is the integration time, 
$W$ is the effective pulse width, $F$ is the flux density, 
$\Delta f$ is the bandwidth, and $S=\frac{2\eta k}{A_{\rm e}}T_{\rm sys}$ 
is the system equivalent flux density \citep{2013tra..book.....W}, 
where $\eta$ is the 
system efficiency factor ($\sim 1.0$), $T_{\rm sys}$ is the system temperature, 
$A_{\rm e}$ is the effective collecting area, and $k$ is 
Boltzmann's constant. Using the design parameters for SKA2 
and the relevant physical parameters for individual pulsars 
obtained from simulations \citep{2009A&A...493.1161S}, 
one finds that for SKA2 the pulse phase jitter noise will 
be the dominant noise source, comparable to the radiometer 
noise for most of MSPs. 
The RMS of total noise $\sigma_t$ for measured TOA is the 
quadratic summation of jitter noise and radiometer noise, i.e., 
$\sigma_{t}^{2} = \sigma_{j}^{2} + \sigma_{r}^{2}$. 

\begin{figure}
\includegraphics[scale=0.38]{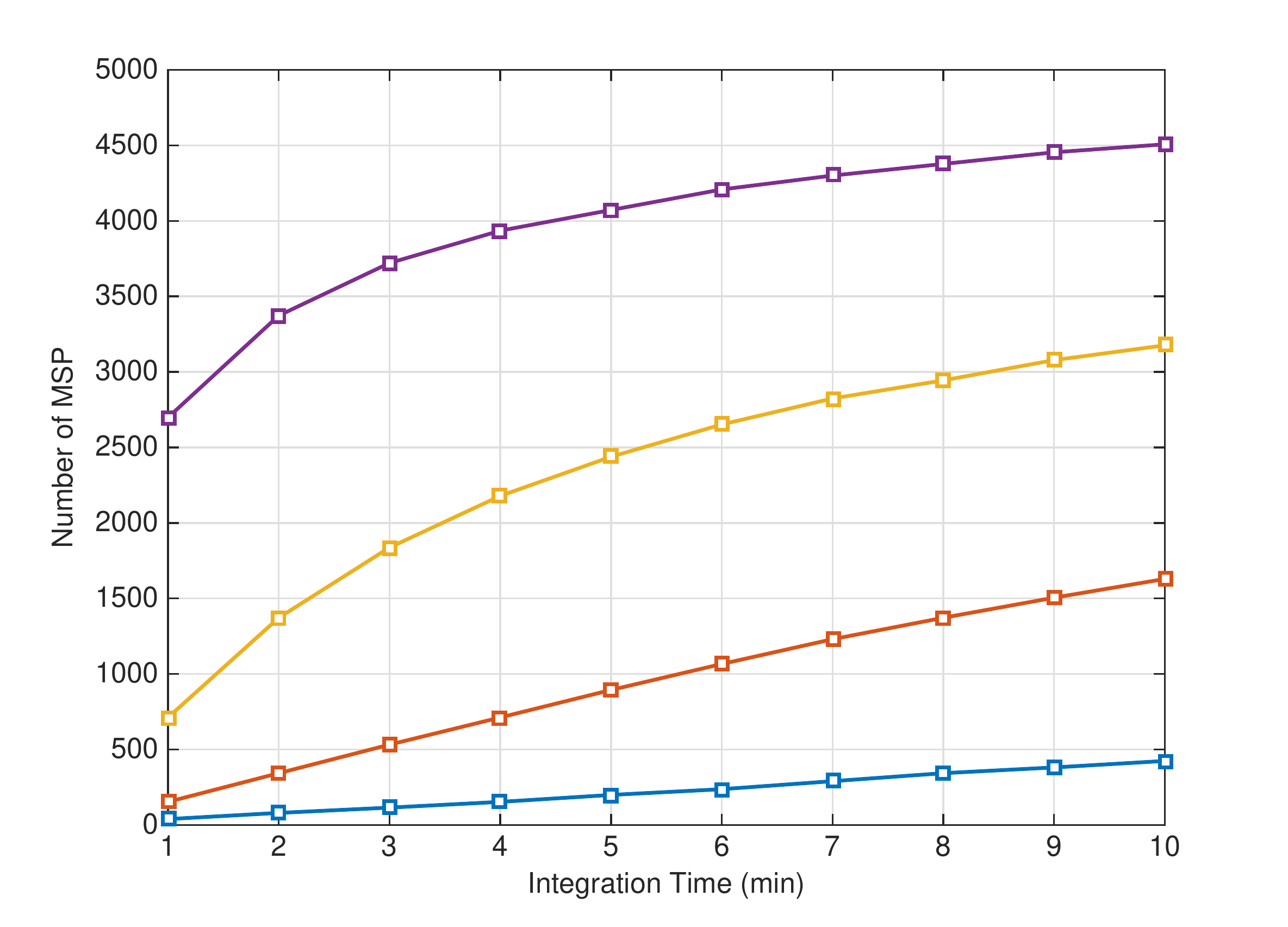}
\caption{Numbers of MSPs that can archive a certain RMS noise level (or better) with varying integration time. Colour lines indicate different RMS noise levels (from bottom to top): 50 ns (blue), 100 ns (red), 200 ns (yellow), and 500 ns (purple).}\label{fig:skamsp}
\end{figure}

Figure~\ref{fig:skamsp} shows the number of MSPs that can achieve 
50~ns, 100~ns, 200~ns, and 500~ns timing precisions, respectively,
with varying 
integration time, $t$. It turns out that if we choose $t=5$ min for SKA2-Mid, 
then there can be about 900 MSPs \citep[out of 6000 MSPs considered
by][]{2009A&A...493.1161S} timed to an RMS level of 
100~ns or better. One caveat of our calculation is that we have 
not considered red timing noise, which is usually less than 
100 ns for MSPs~\citep{Shannon:2010bv}. Assessing the timing noise 
in terms of amplitude and spectral index of individual 
MSPs is one of the most crucial tasks in the data analysis 
for detecting GWs with PTAs \citep[e.g.,][]{Arzoumanian:2015liz}.

Based on these estimates, it appears that a SKA-era PTA with $\sim 1000$
MSPs timed to $\lesssim 100$~ns at a cadence of one timing observation every
two weeks may be feasible.  Such a PTA will reach a sensitivity that will
allow, for example, a $10^{10}$~$M_\odot$ redshifted chirp mass supermassive
black hole binary (SMBHB) to be detected out to $z\simeq 28$ and a
$10^{9}$~$M_\odot$ redshifted chirp mass SMBHB to be detected out to
$z\simeq 1-2$. This will enable high confidence detection of GWs from some
of the existing optically identified SMBHB
candidates~\citep{PhysRevLett.118.151104}.

Besides the stochastic GW signal from the unresolved SMBHB population that may
be detected with SKA1 itself~\citep{Janssen:2014dka}, it is likely that some
individual SMBHBs will stand out above the SGWB and
become resolvable.  The data analysis challenge of resolving multiple sources
from a background population is likely to be  a significant one  given the
large  number of SMBHBs that will be uncovered by  an SKA-era PTA. The PTA data
analysis methods for resolvable sources 
\citep[e.g.,][]{Ellis:2012zv,Wang:2014ava,Zhu:2015tua,Wang:2015nqa,Wang:2017jxw} 
must be able to handle multiple sources while
taking into account (i) the SGWB from unresolved sources
that acts as an unmodelled noise source, and  (ii) instrumental and timing noise
characterisation across $\sim 10^3$ MSPs.  Previous
studies~\citep[e.g.,][]{Babak:2011mr,Petiteau:2012zq} of multiple source detection have
assumed a simplified model of the GW signal in which the so-called {\it pulsar
term}
is dropped and the signal is embedded in white noise with no SGWB. Further
development of data analysis methods  that can work without these simplifying
assumptions is required. 

\subsection{Primordial gravitational waves (B-modes): polarised foregrounds with SKA}

\def\lsim{\,\lower2truept\hbox{${< \atop\hbox{\raise4truept\hbox{$\sim$}}}$}\,}
\def\gsim{\,\lower2truept\hbox{${> \atop\hbox{\raise4truept\hbox{$\sim$}}}$}\,}

The angular power spectrum of polarised anisotropies in the CMB can be
decomposed into E-modes, mainly generated by perturbations of scalar type in the
primordial Universe, and B-modes that could be mainly contributed at low multipoles, $\ell$, (i.e., large
angular scales) by primordial tensor metric
perturbations.\footnote{Vector perturbations generate both E- and B-modes,
but their contributions are predicted to be typically much less relevant, 
except for specific models.} Detecting and characterising primordial B-modes likely represents the
unique way to firmly investigate the stochastic field of primordial
gravitational waves through the analysis of tensor perturbations they
produce.  Although other mechanisms can produce tensor perturbations, the
multipole dependence of primordial B-modes generated by cosmic inflation is
relatively well predicted while their overall amplitude, related to the
ratio, $r = T/S$, of tensor to scalar primordial perturbations depends on
the inflation energy scale. For this reason, the detection of primordial B-modes
received a special attention in current and future CMB polarisation
experiments \citep[see, e.g.,][and references
therein]{Andre:2013nfa,2016SPIE.9904E..0XI,2016arXiv161208270C}.

The foreground signal from extragalactic radio sources \citep[see, e.g.,][and references
therein]{2016arXiv160907263D} generates the most relevant source of contamination
for CMB analyses in total intensity and in polarisation at sub-degree angular scales
up to a frequency of $\sim 100$ GHz. The precise modeling of
radio source contribution to polarisation anisotropies at small scales 
is crucial for the accurate treatment and subtraction of the B-mode signal
generated by the lensing effect produced on CMB photons by cosmic structures, intervening 
between the last scattering surface and the current time. Improving lensing subtraction
implies a better understanding of the primordial B-modes at intermediate and low multipoles.
Thus, the precise assessment of radio source foreground is fundamental for
CMB angular power spectrum analyses, and especially for the discovery
of primordial B-modes, particularly in the case of low values of $r$, and,
in general, to accurately characterise them. This step needs precise measurements of the contribution of radio
sources down to several factors below the source detection threshold of the CMB
experiment, related to the noise, background and foreground amplitudes
depending on the considered sky area. Indeed, for microwave surveys with
$\sim$ arcmin resolution and sensitivity from tens to few hundreds of mJy, 
sources below the detection limit largely contribute to polarisation fluctuations. 
However, modern radio interferometers with sensitivity and resolution much better than 
those of CMB experiments can reveal these source populations. 
At the same time, complementary high-sensitivity polarisation observations 
at the frequencies where CMB experiments are carried out will provide 
useful data for both generating adequate masks and
statistically characterising source populations to subtract their
statistical contribution below the detection
thresholds in angular power spectrum analyses. 
The SKA will allow researchers to perform deep observations of polarised sources only
up to $\sim 20$ GHz, but probing the extremely faint tail of their flux density distribution.  
The estimation of the source polarisation fluctuations from SKA data requires one to extrapolate them to higher frequencies 
where CMB anisotropies are better determined, and the related errors decrease with the decreasing of flux density detection
limits. Thus, the combination of ultra-deep SKA surveys with millimetric observations 
will be very fruitful to characterise source
contribution to polarisation fluctuations at small scales, and then
to improve lensing and delensing treatment from high to low multipoles. 

\begin{figure}
\hskip -2mm
\includegraphics[width=9cm]{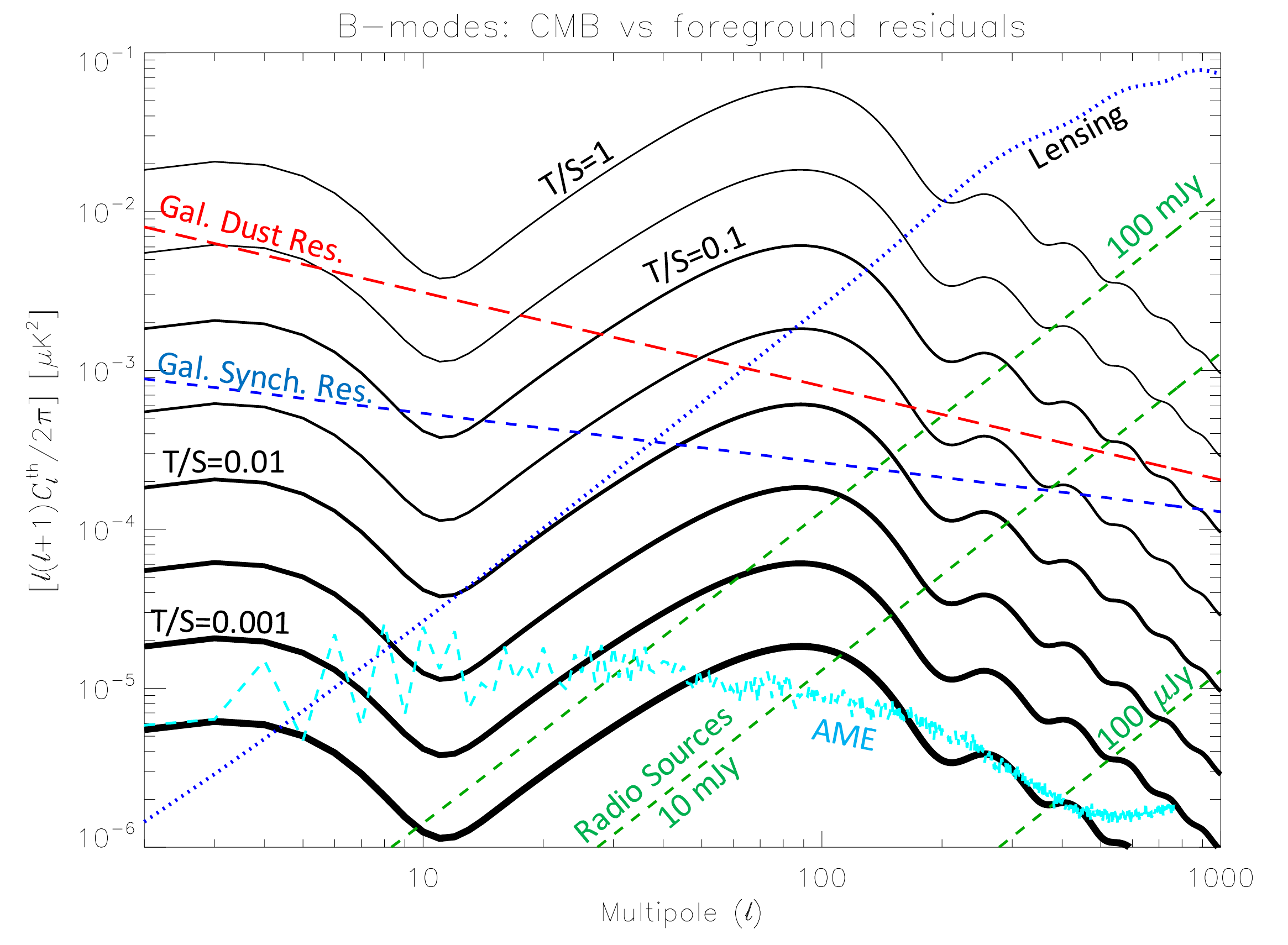}
\caption{The CMB primordial B-mode polarisation angular power spectrum for different tensor-to-scalar perturbation ratios (from $1$ to $3 \times 10^{-4}$; solid black lines) and, separately, the lensing contribution (blue dots). They are compared with estimates of potential residuals from Galactic foregrounds (at 70 GHz) and angular power spectrum from polarised radio sources (at 100 GHz) below different detection thresholds (green dashes; from top to bottom, 100 mJy and 10 mJy, representative of thresholds achievable, respectively, in current and future CMB experiments, and 100 $\mu$Jy representative of potential improvement discussed here). Red long dashes show typical potential residuals from Galactic polarised dust emission extrapolated from 353 GHz assuming an error of 0.01 in the dust grain spectral index. Blue dashes show typical potential residuals from Galactic polarised synchrotron emission extrapolated from 30 GHz assuming an error of 0.02 in the synchrotron emission spectral index. Azure dashes show an estimate of  Galactic AME angular power spectrum scaled from total intensity to polarisation assuming a polarisation degree of 2\%  with, conservatively, all the power in the B-mode (a power two times smaller is expected assuming equal power in E- and B-modes).}
\label{fig:cmb_vs_fore_4SKA}
\end{figure}

An accurate description of SKA continuum surveys can be found in \cite{2015aska.confE..67P}. 
The very faint tail of source counts can be firmly assessed by deep and
ultra-deep surveys. Furthermore, it is possible to extend the analysis to even weaker flux densities,
below the sensitivities of the considered surveys, 
using methods as described by \citet{2012ApJ...758...23C}. This is particularly promising at low frequencies 
in the case of the continuum surveys dedicated to non-thermal emission in clusters and filaments. 
The ultra-deep SKA survey aimed at studying the star-formation history of the Universe is planned to have a
RMS sensitivity of some tens of nJy per beam with a resolution at arcsec level or better,\footnote{According to \cite{2012ApJ...758...23C}, considering, as example, frequencies around 1.4 GHz, 
a source confusion limit of about 10 nJy is derived, thus indicating
that, for surveys with RMS sensitivity of tens of nJy, 
source confusion will not be a crucial limitation over a wide set of frequencies \citep{2015aska.confE.149B}.}
to be compared, for example, with the sensitivity levels of tens of $\mu$Jy 
of present determination of radio source counts at GHz frequencies \citep[see,
e.g.,][]{2001A&A...365..392P,2012ApJ...758...23C},
thus representing a substantial improvement for the measurement of number counts and fluctuations of very faint sources.
Figure~\ref{fig:cmb_vs_fore_4SKA} compares the CMB primordial B-mode angular power spectrum for different values of $r$ with the
lensing signal and potential residual foregrounds. The signal of the B-mode angular power spectrum for radio sources
is displayed for various detection limits, adopting the radio source
fluctuation conservative model of \cite{TucciToffolatti2012}. 
Even considering errors from frequency extrapolation (as generously accounted in the Figure assuming a very prudential threshold for the signal),
this analysis shows that SKA characterisation of radio source
polarisation properties will allow one to reduce their potential residual impact on polarisation fluctuations at essentially negligible levels.
 
Moreover, with the SKA it will be possible to improve our understanding of Galactic diffuse foregrounds at low frequencies,
where polarised synchrotron emission peaks, a key point for B-mode searches, considering that CMB experiments are
typically carried out at higher frequencies.
This will have particular implications for (i) Galactic synchrotron emission models, (ii) the tridimensional treatment of the Galaxy, as well as
(iii) the component of its magnetic field coherent on large scale \citep{2008A&A...477..573S,2009A&A...507.1087S,2010RAA....10.1287S,2011A&A...526A.145F,2012APh....36...57F} that rely on modern numerical methods
\citep{1998ApJ...509..212S,2009A&A...495..697W} and turbulence \citep{2002ApJ...575L..63C}.
For cosmological applications it may be critical to better characterise also 
the anomalous microwave emission (AME) that is found to be correlated with dust emission in the far-infrared and is
generated by rapidly
spinning small dust grains having an electric dipole moment. While its spectrum likely peaks  
at $\sim 15$--50 GHz, its polarisation degree on very wide sky areas, likely at the percent level, is still almost unknown.
SKA2 will allow to derive accurate maps of AME at low frequencies.

In Figure~\ref{fig:cmb_vs_fore_4SKA} the CMB B-mode angular power spectrum is compared with potential contaminations  
from Galactic emissions (evaluated in a sky region excluding the $\sim 27$\% sky fraction mostly affected by Galactic emission) estimated on the basis of Planck 2015
results \citep{2016A&A...594A..10P,2016A&A...594A..25P}.
As is well known, the polarised emission from Galactic dust mostly impacts CMB B-mode analyses \citep{2015PhRvL.114j1301B,2016A&A...586A.133P}, but for detecting and characterising B-modes for 
$r \lsim$ some $\times 10^{-2}$, the accurate understanding of all types of polarised foreground emissions, including synchrotron and AME, is also crucial.    

Many cosmological studies are based on analyses carried out on very wide sky areas, thus calling for accurate, large sky coverage
Galactic radio emission mapping.
The SKA1 continuum surveys \citep{2015aska.confE..67P} at 1.4~GHz and at 120~MHz, to be performed integrating for about 1--2 years, 
will have a $\sim$ 75\% sky coverage. 
A comparison in terms of sensitivity per resolution element with the radio surveys 
currently exploited in CMB projects data analysis \citep{2008A&A...479..641L}
allows one to appreciate the significant improvement represented by future SKA surveys.
The SKA 1.4-GHz survey has in fact a sensitivity target $\sim 20$ times better than the
best currently available all-sky 1.4 GHz radio surveys, while the SKA 120-MHz
survey is planned to improve in sensitivity of about a factor of 4 with respect to the 408 MHz Haslam map \citep{1982A&AS...47....1H}.

\subsubsection{Galaxy-GW cross-correlation}
SKA galaxy maps can be cross-correlated with GW maps from, e.g., laser
interferometers to obtain novel measurements potentially able to probe
gravity in new ways. One such possibility involves the correlation of GW
maps with galaxy catalogues in order to determine the nature of the
progenitors of binary black holes. This can be also used to obtain
ultra-high precision estimation of cosmological parameters \citep[e.g.,][]{Cutler:2009qv}, to test cosmological models~\citep{Camera:2013xfa}, 
or to set constraints on the relation between distance and redshift~\citep{Oguri:2016}.

Using the same approach,~\citet{Raccanelli:2016PBH} suggested that the
cross-correlation between star-forming galaxies and GW maps could constrain 
the cosmological model in which primordial black holes are the dark matter (see \S\ref{sec:PBHs} for a discussion of this possibility). Here we follow the same approach, focusing on stellar-mass PBHs, in the mass window probed by the LIGO instrument.

Using galaxy number counts, one can observe the
correlation between galaxies and the hosts of binary black hole mergers.
We can compute what constraints on the abundance of PBHs as dark matter
(DM) \citep[for the same model of][]{1603.00464} can be set by
cross-correlating SKA surveys with LIGO and Einstein Telescope (ET) GW maps.
Considering angular power spectra $C_\ell$, that can be computed as \citep[see, e.g.,][]{raccanelli08, Pullen:2012}:
\begin{align}
\label{eq:ClXY}
C_{\ell}^{XY}(z,z') &= \left< a^X_{\ell m}(z) a^{Y\,^*}_{\ell m}(z') \right> \nonumber\\
&= r \int \frac{4\pi dk}{k} \Delta^2(k) W_{\ell}^X(k, z) W_{\ell}^Y(k, z') \, ,
\end{align}
with $W_{\ell}^{[X,Y]}$ the window functions of the distribution of sources 
of different observables, $\Delta^2(k)$ the dimensionless matter power spectrum, and $r$ the 
coefficient of cross correlation.


We consider radio survey maps from the S-cubed
simulation\footnote{http://s-cubed.physics.ox.ac.uk}, using the SEX and SAX catalogs,
for continuum and HI, respectively, applying a cut to the simulated data appropriate for the assumed flux limit for the considered cases.
For all surveys we take $f_{\rm sky} =0.75$.

The distribution of GW events can be estimated by:
\begin{equation}
\label{eq:ngw}
\frac{d N_{GW}(z)}{dz} \approx \mathcal{R}^m(z) \tau_{\rm obs} \frac{4\pi\chi^2(z)}{(1+z)H(z)} \, ,
\end{equation}
where $\mathcal{R}^m(z)$ is the merger rate, 
$\tau_{\rm obs}$ the observation time and $H(z)$ the Hubble parameter.

An important factor to understand the nature of the progenitors of binary black hole mergers is given by the halo bias of the hosts of the mergers. 
We assume that mergers of objects at the endpoint of stellar evolution
would be in galaxies hosting large numbers of stars, hence
being in halos of $\sim 10^{11-12} M_\odot$, the great majority of mergers of primordial
binaries will happen within haloes of $< 10^{6} M_\odot$, as demonstrated by~\cite{1603.00464}.
The discrimination can happen then because these 2 types of haloes are connected to
different amplitudes of the gaalxy bias. Specifically, we take galaxies hosting stellar
GW binaries to have properties related to SFG galaxy samples. Therefore, we can
use $b^{\rm Stellar}_{GW} = b_{\rm SFG}$, assuming its redshift
dependent values as in \cite{Ferramacho:2014pua}.
Conversely, small haloes hosting the majority of primordial black hole mergers are expected to have $b\lesssim 0.6$, approximately constant in our redshift range~\citep{Mo:1995}.
Thus, assuming SFGs bias as $b(z) > 1.4$, we take, as the threshold for a model discrimination, $\Delta_b = b_{\rm SFG} - b_{\rm GW} \gtrsim 1$; 
given that this threshold should in fact be larger at higher redshifts, our assumption can be seen as conservative.

Taking the specifications of planned future surveys, we forecast the measurement uncertainties with the Fisher matrix formalism~\citep{Tegmark:1997}:
\begin{equation}
\label{eq:Fisher}
F_{\alpha\beta} = \sum_{\ell} \frac{\partial C_\ell}{\partial 
\vartheta_\alpha}
\frac{\partial C_\ell}{\partial
\vartheta_\beta} {\sigma_{C_\ell}^{-2}} \, , 
\end{equation}
with $\vartheta_{\alpha, \beta}$ being the parameters to be measured, and the power spectra derivatives are computed at fiducial values $\bar \vartheta_{\alpha}$ and the measurement errors are $\sigma_{C_\ell}$.

We compute the amplitude of the cross-correlation marginalising over the
galaxy bias factor, assuming a prior of 1\% precision on galaxy bias from external measurements.

We imagine multiple GW detector configurations, observing the whole sky; naturally, the precision of the localization of gravitational wave events has a fundamental role in determining the constraining power on the cross-correlation galaxies-GW, and also the maximum redshift observable has.
We choose the following specifications for GW interferometers:
\begin{itemize}
	\item aLIGO: $\ell_{\rm max} = 20$, $z_{\rm max} = 0.75$;
	\item LIGO-net: $\ell_{\rm max} = 50$, $z_{\rm max} = 1.0$;
	\item Einstein Telescope: $\ell_{\rm max} = 100$, $z_{\rm max} = 1.5$;
\end{itemize}
where with LIGO-net we assume a network of interferometers including the current LIGO detectors, VIRGO, and the planned Indian IndIGO and the Japanese KAGRA instruments, in order to achieve a few square degrees of angular resolution.
For assigning statistical redshifts to radio continuum catalogues we follow the technique of~\citet{Kovetz:2016}.

\begin{figure}
\centering
\includegraphics[width=\columnwidth]{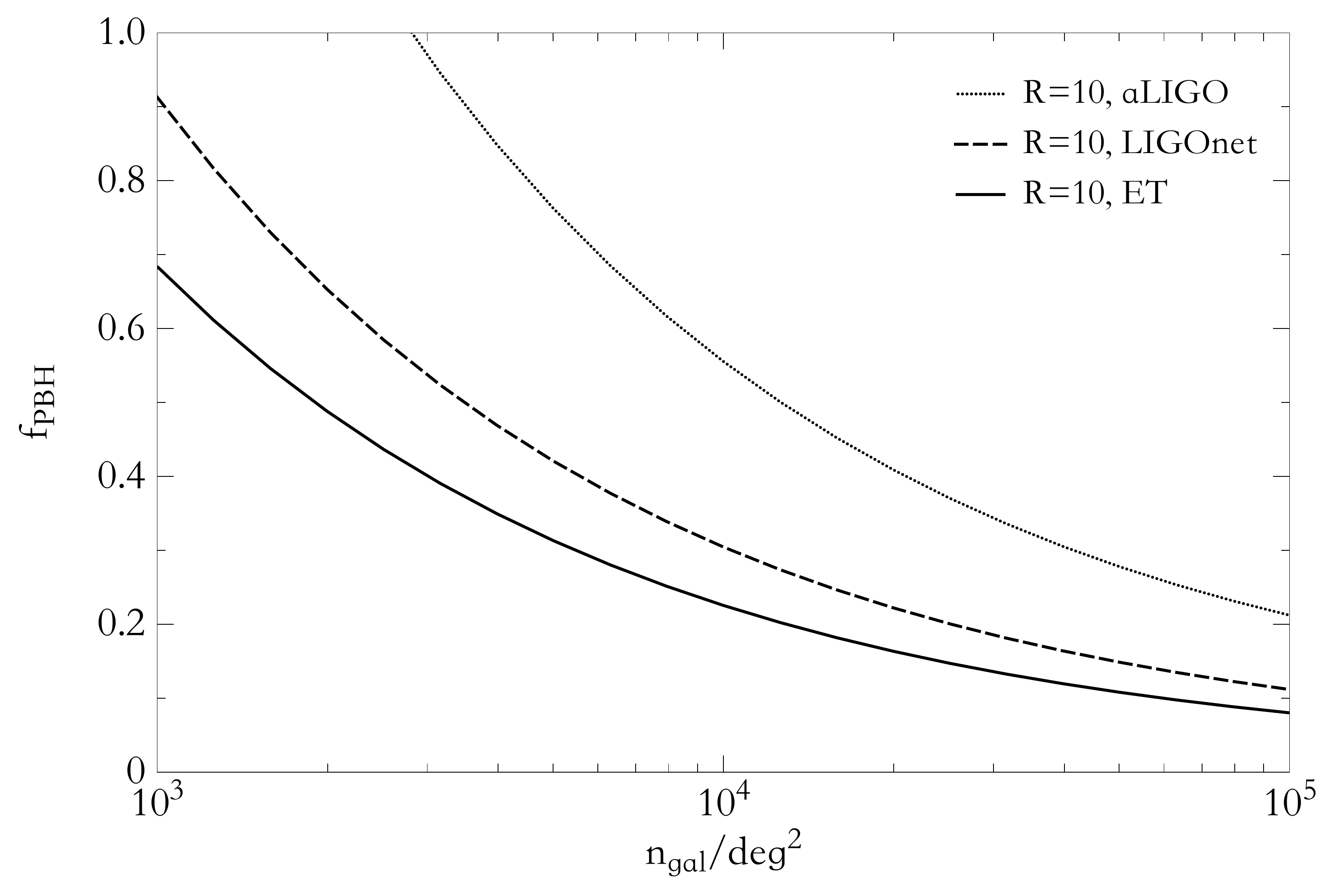}
\caption{Forecast fraction of DM in PBH, for different fiducial experiment sets. 
For details on GW experiments, see text.
}
\label{fig:pbh_exp}
\end{figure}

In Figure~\ref{fig:pbh_exp} we plot the forecasts (at 1-$\sigma$ level) for three different GW interferometer configurations, after five years of data collection, assuming a merger rate $\mathcal{R}=10$~Gpc$^3$~yr$^{-1}$, correlated with SKA radio surveys.
The correlation of HI and continuum surveys will not be different if correlating with LIGO and a future LIGO-NET. On the other hand, in the ET case, there will be an effect due to the maximum redshift probed.

As one can see, aLIGO will be able to derive only weak constraints on the effects of PBHs as DM when correlating GWs with a survey detecting a few thousand sources per deg$^2$ (or observing for a longer time).
However, future correlations of LIGO-net or the Einstein Telescope with radio surveys in continuum with some redshift information can deliver a clear detection for $f_{\rm PBH}=1$, or the hints of PBHs that comprise a small part of the dark matter. Other current and future constraints on PBHs as dark matter are discussed in \S\ref{sec:PBHs}.

\paragraph{Pulsar timing array}

Multiple resolvable systems may be detectable in the pulsar timing data due to 
the fact that an SKA-era PTA can detect SMBHBs residing at high redshifts. 
Correlating the GW signal with optical variability in follow up observations can 
teach us much about the physical process in accreting SMBHBs. 
The distance reach of the SKA-era PTA also implies that high redshift SMBHB 
candidates pinpointed in time-domain optical observations, such as PG~1302--102~\citep{Graham:2015gma},  
can be followed up in the GW window. The direct detection of the GW signal 
from an optically identified candidate SMBHB will confirm its true nature.

Accretion onto an SMBHB may produce periodic variability in the light curve of a quasar~\citep{Macfadyen:2006jx,Graham:2015gma,Hayasaki:2015qoa}. Since the 
study of quasar optical variability is a key scientific goal for LSST~\citep{2009arXiv0912.0201L}, numerous SMBHB candidates may be discovered during the survey lifetime. 
The LSST cadence per object will yield $\sim 1$~measurement of flux per week, which implies that source variability frequency up to $10^{-6}$~Hz can be detected.  This yields a good overlap with SMBHB orbital frequencies in the $[5\times 10^{-10}, 10^{-6}]$~Hz interval that are observable with PTAs. Similarly, the LSST coverage of active galactic nuclei (AGN) will reach a redshift of $~7.5$, overlapping the SKA era PTA distance reach for SMBHBs.

LSST full science operation is scheduled to begin around 2023, which coincides with the start date of SKA1. Thus, LSST and SKA1 will have a  substantial overlap in observation of sources over a common period of time. There will also be some overlap with SKA2 when it starts around $2030$. In the absence of a significant overlap, SKA2-era PTA-based  GW searches can be correlated against optically identified candidates in archival LSST data.
LSST observations can help narrow down the parameter space to be searched in the PTA data analysis. This is especially important if the source is strongly evolving. The sky location of the source can tell us which MSPs to time with higher precision and faster cadence.

There will be significant data analysis challenges involved in linking PTA-based 
GW searches in the SKA era with LSST. 
The sky localisation accuracy of a PTA-based search for SMBHBs 
depends on the source brightness and sky location. 
Given the typical localization error on the sky of approximately 
$100~{\rm deg}^2$~\citep{PhysRevLett.118.151104} for bright sources, 
a source detected by a SKA-era PTA is likely to be associated with a large 
number of variable objects. However, the frequency of optical variability 
and that of the GW signal, the latter being quite accurately measurable, 
are likely to be strongly related and this can help in significantly narrowing 
down the set of candidates to follow up.

\subsection{Summary}

Gravity and gravitational radiation are central topics in modern
astrophysics. The SKA will have a major impact in this field, via:
\begin{enumerate}
\item better timing of binary pulsar systems, in order to probe new aspects of the
gravitational interaction, for example, measurement of the Lense-Thirring effect;

\item discovery of pulsar binaries orbiting stellar-mass black holes or Sgr~A*, which will enable novel tests such as the no-hair theorem and
even some quantum-gravity scenarios;

\item galaxy clustering, weak lensing, 21-cm intensity mapping, peculiar velocity
surveys, and void statistics, with which we can study gravitational interactions at
cosmological scales with great precision;

\item cross-correlation of radio weak lensing surveys and HI intensity mapping with
optical lensing surveys and optical galaxy clustering surveys, respectively, in order to
reduce associated systematics and achieve better sensitivities;

\item synergies with other gravitational-wave observations (e.g., the B-modes in
the primordial gravitational waves), using SKA galaxy surveys and polarisation
foreground observations;

\item direct detection of gravitational waves at nanohertz frequencies with pulsar
timing arrays. 
\end{enumerate}

Studies of gravitation with the SKA will not be limited to the above
items. Various possible synergies with other large surveys at optical (e.g., with LSST) and other wavelengths in the SKA era are expected to be highly productive.

\section{Cosmology and Dark Energy}
\label{sec_cosmology}

As large optical and NIR galaxy surveys like DES, Euclid, and LSST begin to
deliver new insights into various fundamental problems in cosmology, it will
become increasingly important to seek out novel observables and independent
methods to validate and extend their findings. A particularly rich source of
new observational possibilities lies within the radio band, where gigantic
new telescope arrays like the SKA will soon perform
large, cosmology-focussed surveys for the first time, often using innovative methods that will strongly complement, and even surpass, what is possible in the optical. We discuss a number of such possibilities that have the chance to significantly impact problems such as understanding the nature of dark energy and dark matter, testing the validity of general relativity and foundational assumptions such as the Copernican Principle, and providing new lines of evidence for inflation. These include radio weak lensing, 21cm intensity mapping, Doppler magnification, TF peculiar velocity surveys, multi-tracer searches for primordial non-Gaussianity, full-sky tests of the isotropy of the matter distribution, and constraints on the abundance of primordial black holes.

\subsection{Introduction}

Cosmology has blossomed into a mature, data-driven field, with a diverse set of precision observations now providing a concordant description of the large scale properties of the Universe. Through a variety of cosmological observables, we can examine the {\it expansion} history of the Universe, described by the evolution of the Friedmann-Lema\^itre-Robertson-Walker (FLRW) scale factor $a$ as a function of time; its {\it geometry}, given by its spatial curvature; and the {\it growth} history of structures in the Universe, describing the degree to which overdensities have grown in amplitude over time due to gravitational collapse. Different types of observations constrain these aspects of the Universe's evolution to a greater or lesser degree; for instance, RSDs constrain the growth history only, while distance measurements with Type Ia supernovae constrain the expansion history only. The overall picture is highly encouraging, with broad agreement found across a range of very different observables that probe a number of different eras across cosmic time. 

The successes of the precision cosmology programme have led us to something
of a crisis however. Our extremely successful descriptive model of the
Universe --- $\Lambda$CDM --- fits the vast majority of observations with
great precision, but is mostly constructed out of entities that have so far
defied any proper fundamental physical understanding. The shakiest
theoretical pillars of $\Lambda$CDM are dark energy, dark matter, and
inflation. The first two make up around 26\% and 69\% of the cosmic energy
density today respectively, and yet lack any detailed understanding in terms
of high energy/particle theory or conventional gravitational physics. The
latter is responsible for setting the geometry of the Universe and for
sowing seed inhomogeneities that grew into the large-scale structure we see
today, but also lacks a specific high-energy theory description. What is
more, these components are all tied together by General Relativity, a
tremendously well-tested theory on solar system scales that we essentially
use unchanged in cosmology --- an extrapolation of some nine orders of
magnitude in distance. The concordance model is therefore built on a
foundation of several phenomenological frameworks --- each of them compelling
and well-evidenced, but lacking in the fundamental physical understanding
that, say, the Standard Model of particle physics provides --- and tied
together by an extrapolation of a theory that has only really been proven on
much smaller scales.

Cosmology, then, has its work cut out for the foreseeable future. Measuring parameters of the $\Lambda$CDM model to ever-increasing precision is not enough if we aspire to an in-depth physical understanding of the cosmos --- we must develop and test new, alternative theories; seek out novel observables that can stress $\Lambda$CDM in new, potentially disruptive ways; and discover and carefully analyse apparent anomalies and discordant observations that could expose the flaws in $\Lambda$CDM that might lead us to a deeper theory.

This work is well under way, with a series of large optical and NIR galaxy
surveys leading the charge. Experiments such as DES, Euclid, and LSST will
measure multiple galaxy clustering and lensing observables with sufficiently
great precision to test a number of key properties of dark energy, dark
matter, inflation, and GR on cosmological scales. Their analyses rely on
detailed modelling of the large-scale structure of the Universe, plus painstakingly-developed analysis tools to recover small signals from these enormously complex datasets. Over time, they will likely discover a good many anomalies, some of which may even be hints of beyond-$\Lambda$CDM physics. This is exciting and profound work, but will probably not be enough to settle cosmology's biggest questions on its own. Instead, we will need to independently confirm and characterise the `anomalies', so that we can ultimately build a coherent picture of whatever new physics is behind them. This will require alternative methods beyond what is provided by optical/NIR surveys, including different observables, and different analysis techniques.

This is where the SKA arguably has the most to offer cosmology. While the
SKA will be able to measure many of the same things as contemporary
optical/NIR surveys --- galaxy clustering and lensing, for example --- it
will do so using markedly different observing and analysis techniques. This
is extremely valuable from the perspective of identifying and removing
systematic effects, which could give rise to false signals and anomalies, or
otherwise compromise the accuracy of the measurements. Weak lensing
observations in the radio will have very different systematics compared with
optical surveys, for example, as atmospheric fluctuations and other point
spread function uncertainties should be much reduced, while shape
measurement uncertainties might be quite different for an interferometer.
The SKA will provide a large cosmological survey dataset in the radio to be compared and cross-correlated with the optical data, allowing the sort of joint analysis that will be able to confirm anomalies or flag up subtle systematic effects that a single survey would not be able to on its own.

\begin{figure}
\hspace{-1em}
\includegraphics[width=0.5\textwidth]{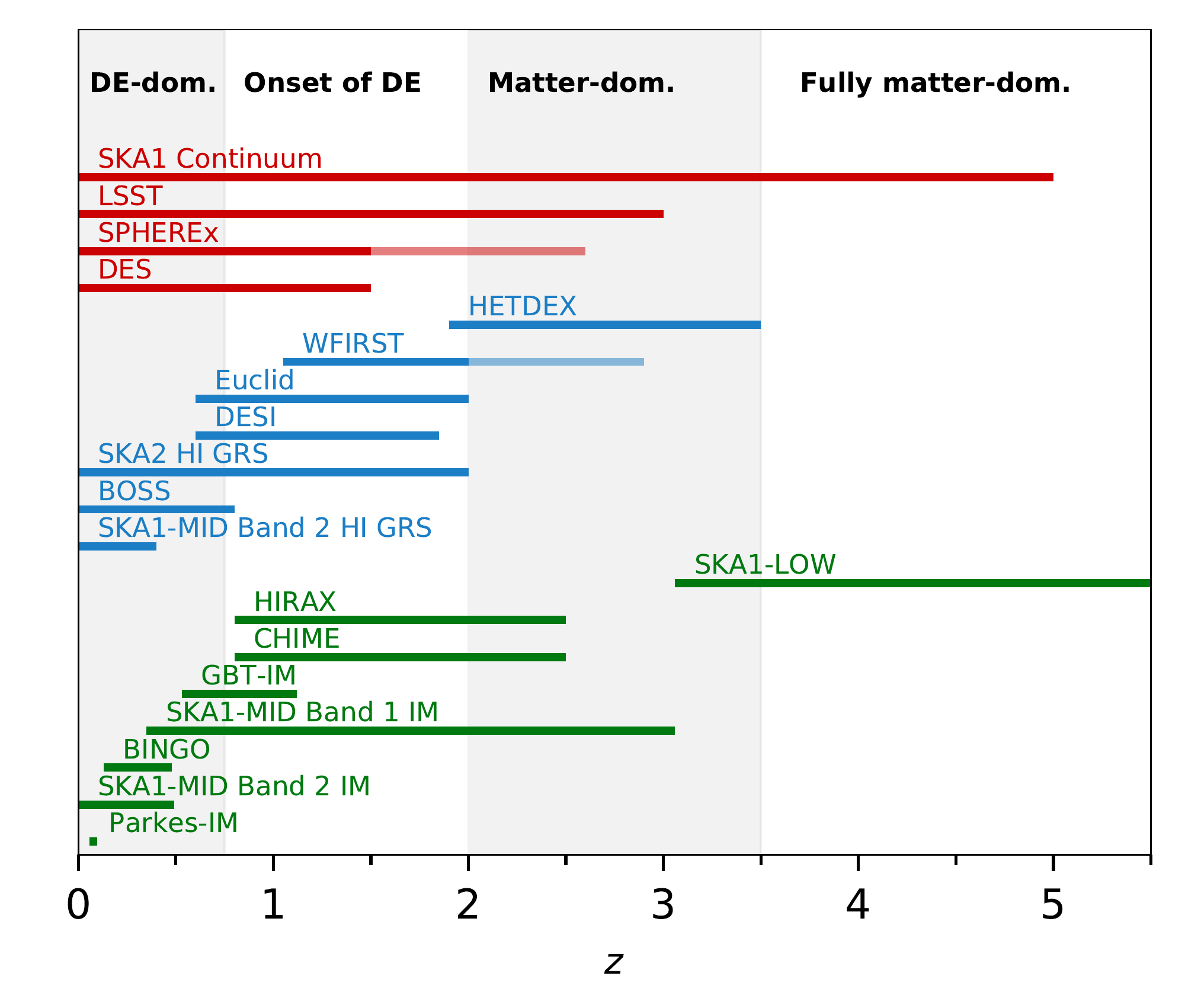}
\vspace{-2em}
\caption{The approximate redshift ranges of various current and future
large-scale structure surveys. 21cm intensity mapping surveys are shown in
green (bottom), spectroscopic galaxy redshift surveys in blue (middle), and
photometric/continuum surveys in red (top). WFIRST and SPHEREx both have
secondary samples (with lower number density or photometric precision),
which are shown as paler colours. Taken together, the SKA surveys offer full
coverage of the redshift range from 0 to $\gtrsim 6$, using multiple survey
methods. The grey bands show an approximate division of the full redshift
range into different eras, corresponding to the dark-energy-dominated
regime, the onset of dark energy, the matter-dominated regime, and the fully matter-dominated regime.}
\label{fig:surveyz}
\end{figure}

The fact that radio telescopes work so differently from their optical
counterparts also opens up the possibility of making novel measurements that
would otherwise be difficult and/or time consuming at higher frequencies.
The intrinsically spectroscopic nature of radiometers makes it possible to
perform efficient intensity mapping surveys, making it easier to access
large-scale structure at higher redshifts, for example. The flexible angular
resolution of radio interferometers (one can re-weight the baselines on the
fly to achieve different effective resolutions) could also be useful for,
e.g., hybrid lensing studies involving both shape measurement and galaxy kinematics. While exploitation of the novel capabilities of radio instrumentation is only just beginning in cosmological contexts, there is a great deal of promise in some of the new observables that have been proposed. Taken together with the precision background and growth constraints from the SKA and other sources, perhaps one of these new observables will provide the vital hint that collapses some of cosmology's great problems into a new understanding of fundamental physics.

In this section, we therefore focus on the novel contributions that radio
telescopes, and in particular the SKA, will bring to observational
cosmology. For completeness, we will briefly mention more conventional
observations that are possible with the SKA, such as spectroscopic BAO measurements, but defer to previous works for detailed discussions of these.

\subsection{Tests of cosmic acceleration}
\label{sec:DE-MG}
The cause of the accelerating expansion of the Universe is one of the greatest open questions in fundamental physics. Possible attempts at explanation include Einstein's cosmological constant, often associated with the energy density of the QFT vacuum; additional very light particle fields such as quintessence; or some modification to the theory of gravity. Any one of these explanations requires either the introduction of exciting new physics beyond the Standard Model, or a much deeper understanding of the relationship between quantum field theory and GR.

In order to learn about the phenomenology of this new energy component, it
is useful to try to measure at least two quantities: the energy density of
the dark energy today, quantified by the parameter $\Omega_{{\rm DE},0}$,
and its equation of state (pressure to density ratio) as a function of
redshift, $w(z)$. The former has been measured with good precision by CMB,
supernova, and large-scale structure experiments over the past 15--20 years, which have established extremely strong evidence that dark energy is the dominant component of the cosmic energy density in the late Universe. The task now is to pin down the latter, as this offers some hope of being able to differentiate between some of the different scenarios.

Unfortunately, the space of possible dark energy models is very large and diverse, and many models can be tuned to reproduce almost any $w(z)$ that could be observed. Determining the equation of state to high precision remains an important task however, as one can still draw a number of useful conclusions from how it evolves. The most important thing to check is whether the equation of state at all deviates from the cosmological constant value, $w=-1$. If dark energy truly is a cosmological constant, then understanding how the QFT vacuum gravitates, and solving various severe fine-tuning issues, becomes the key to understanding cosmic acceleration. If the equation of state is {\it not} constant, however, this points to the presence of new matter fields or modifications of GR as the culprit.

Beyond this, it is also useful to know whether $w$ ever dips below $-1$. An equation of state below this is said to be in the `phantom' regime \citep{Caldwell:1999ew}, which would violate several energy conditions for a single, minimally-coupled scalar field. A field that has additional interaction terms (e.g. with the matter sector) {\it can} support a phantom effective equation of state however \citep{Raveri:2017qvt}, and so finding $w < -1$ would be a strong hint that there are additional interactions to look for.

Finally, the actual time evolution of the equation of state can also provide some useful clues about the physics of dark energy. Many models exhibit a `tracking' behaviour for example, where $w(z)$ scales like the equation of state of the dominant component of the cosmic energy density at any given time (e.g. $w_m = 0$ during matter domination and $w_r = 1/3$ during radiation domination). Oscillating equations of state, or those that make dark energy non-negligible at early times (`early dark energy'), correspond to more exotic models.

In this section, we briefly discuss two methods for constraining the redshift evolution of dark energy with the SKA: measuring the distance-redshift relation with 21cm intensity mapping experiments, and measuring the expansion directly using the redshift drift technique. For more in-depth forecasts and discussion of distance and expansion rate measurements that will be possible with SKA, see \cite{Bull:2015lja}. See \S\ref{sec_grav} for predictions of typical $w(z)$ functions for a variety of dark energy and modified gravity models.

\subsubsection{BAO measurements with 21cm intensity maps}

The baryon acoustic oscillation (BAO) scale provides a statistical `standard ruler' that can be used to constrain the distance-redshift relation, and therefore the abundance and equations of state of the various components of the Universe. The BAO feature is most commonly accessed through the 2-point correlation function of galaxies from large spectroscopic galaxy surveys like BOSS and WiggleZ, and presents as a `bump' in the correlation function at separations of $\sim 100 h^{-1}$ Mpc. It has been found to be extremely robust to systematic effects, and can in principle be measured out to extremely high redshift. Current constraints are mostly limited to $z \lesssim 1$ however, except for a handful of datapoints at $z \sim 2.4$ from Lyman-$\alpha$ forest observations.

The SKA will add to this picture by providing another route to BAO
measurements --- through the 21cm intensity mapping method. IM
uses fluctuations in the aggregate brightness temperature of the
spectral line emission from many unresolved galaxies to reconstruct a
(biased) 3D map of the cosmic matter distribution. This has the advantage of
dramatically improving survey speed, since all the flux from all of the sources
(even very faint ones) contributes to the signal. Galaxy surveys, on the
other hand, must apply some detection threshold in order to reject noise
fluctuations from their catalogue, and so most of the available flux is
therefore thrown away (except for around sufficiently bright sources).

\begin{figure}
\centering
\includegraphics[width=0.48\textwidth]{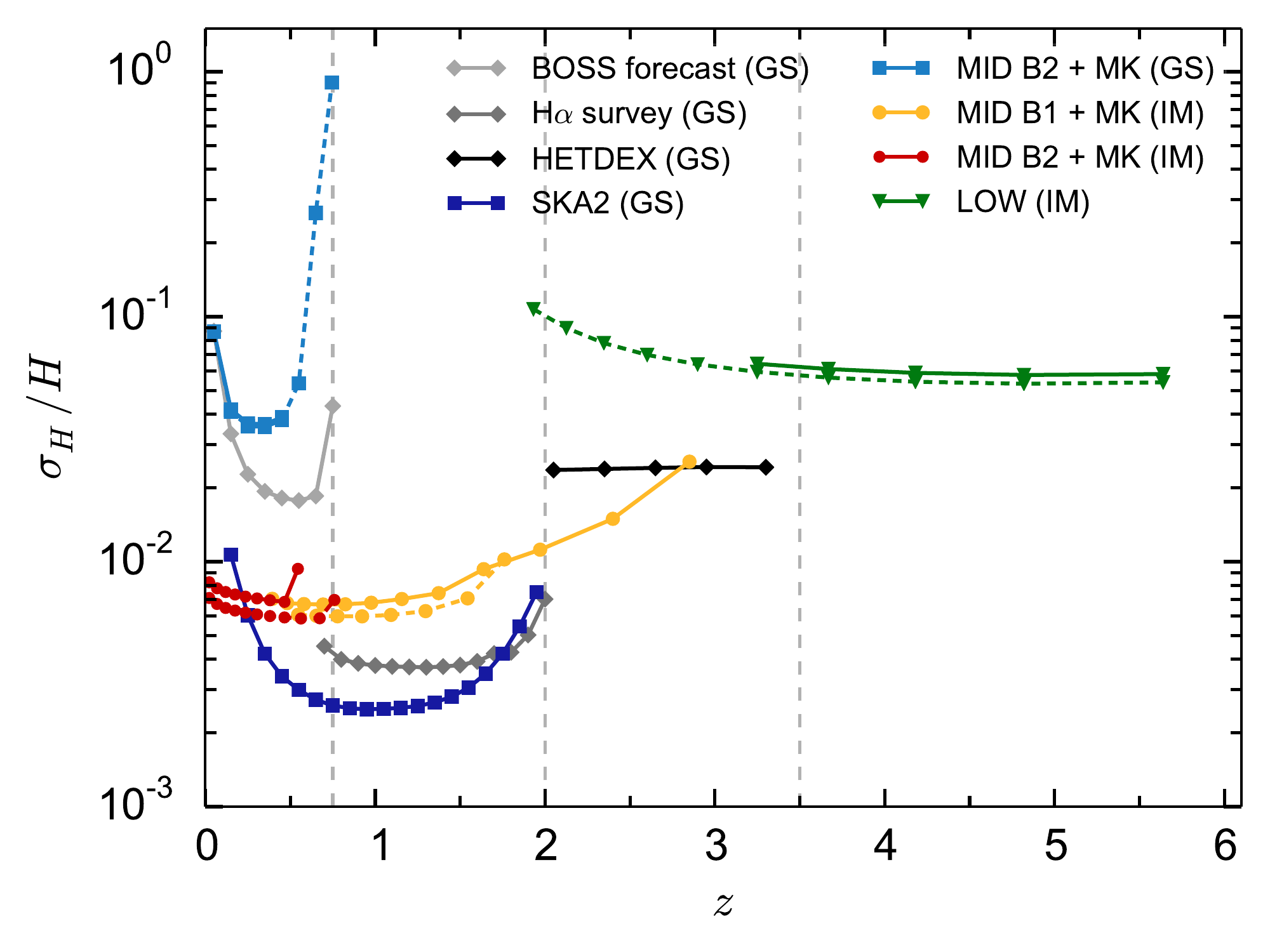}
\caption{Forecasts for the fractional error on the expansion rate, $H(z)$, expected to be achieved with various galaxy surveys (GS) and intensity mapping surveys (IM), from \cite{Bull:2015lja}. SKA surveys will be able to effectively survey volumes at higher redshifts than optical/NIR experiments, and with SKA2 will ultimately achieve better precision in the $0 \lesssim z \lesssim 2$ regime as well. {\it Figure reproduced with permission, from \cite{Bull:2015lja}.}}
\label{fig:imbao}
\end{figure}

The SKA will significantly improve upon existing BAO measurements in two
main ways. First, it will be able to access the BAO signal over
significantly larger volumes of the Universe than current or even future
surveys. Existing BAO measurements are limited in accuracy mostly due to
sample variance, and so can only be improved by increasing the survey area
or extending the redshift range. Future spectroscopic galaxy surveys like
DESI and Euclid will also extend measurements to higher redshifts, over
larger survey areas (see Fig.~\ref{fig:surveyz}), but 21cm intensity mapping
surveys with the SKA will surpass all of them in terms of raw volume. A
SKA1-Mid Band 1 IM survey will potentially be able to survey the redshift range
$0.4 \lesssim z \lesssim 3$ over $\sim 25,000$ deg$^2$, although resolution
considerations will result in slightly poorer constraints than a
spectroscopic galaxy survey with the same footprint. SKA2 will be able to
perform a spectroscopic HI galaxy survey over a similar area out to $z
\approx 2$ (sample variance limited out to $z \approx 1.5$), and so is
expected to essentially be the last word in BAO measurement in this regime.
Fisher forecasts for constraints on the expansion rate with various galaxy
and intensity mapping surveys are shown for comparison in Figure~\ref{fig:imbao}.

Secondly, SKA will be capable of detecting the BAO at significantly higher redshifts than most galaxy surveys, with SKA1-Mid Band 2. While dark energy dominates the cosmic energy density only at relatively low redshifts, $z < 1$, many dark energy models exhibit a tracking behaviour that means that their equation of state deviates most significantly from a cosmological constant at $z \gtrsim 2-3$. Precision determinations of $w(z)$ at $z > 2$ may therefore be more discriminating than those in the more obvious low redshift regime that is being targeted by most spectroscopic galaxy surveys.

\subsubsection{Redshift drift as a direct probe of expansion}
\label{sec_z_drift}

Most probes of acceleration rely on measuring distances or the expansion rate, using standard rulers or candles. An interesting alternative is to observe the so-called {\it redshift drift}, which is the time-variation of the cosmological redshift, $\de z/\de t$ \citep{1962ApJ...136..319S, Loeb:1998bu}. This allows a very direct measurement of the expansion rate, as
\be
\frac{\de z}{\de t} = (1+z)H_0 - H(z),
\ee
and has the advantage of giving a `smoking gun' signal for cosmic
acceleration --- the redshift drift can be positive only in accelerating
cosmological models. While the existence of an {\it apparent} cosmic
acceleration is well-established, much of the evidence from probes that are
interpreted in a model-dependent way, i.e., within the context of a
(perturbed) FLRW model. A number of non-FLRW cosmologies have been proposed
in the past that {\it appear} to be accelerating when distance/expansion
rate measurements are interpreted within an assumed FLRW model, but in which
the expansion of space is actually {\it decelerating} locally everywhere
\citep{Clarkson:2010uz, Andersson:2011za, Bull:2012zx}. This effect is
normally achieved though the introduction of large inhomogeneities, which
distort the past lightcone away from the FLRW behaviour, but which still
reproduce the isotropy of the Universe as seen from Earth. While this kind
of model has essentially been ruled out as a possible explanation for dark
energy by other observables \citep[see, e.g.,][]{Bull:2011wi, Zibin:2011ma},
the question of whether smaller inhomogeneities could cause non-negligible
biases in estimates of background cosmological parameters is still very much
open \citep[e.g.,][]{Clarkson:2011br, Bonvin:2015kea, Fleury:2016fda}. Redshift drift provides an independent and arguably more direct way of measuring cosmic acceleration, and so represents a promising observable for studying these effects and, eventually, definitively determining their size. The independence of redshift drift from other probes is also advantageous for breaking degeneracies in measurements of dark energy observables such as the equation of state \citep{2012PhRvD..86l3001M, Kim:2014uha, Geng:2014ypa}.

In principle, one can measure the redshift drift effect by tracking the
change in redshift of spectral line emission over some period of time. To
get an estimate of the magnitude of this effect, we note that $H_0 = 100 h\,
{\rm km}\,{\rm s}^{-1}\,{\rm Mpc}^{-1} \sim 10^{-10}$~yr$^{-1}$, so
observing the redshift drift over a time baseline of $\Delta t = 10$ years
would require a spectral precision of $\Delta z \sim 10^{-9}$, corresponding
to a frequency shift in, e.g., the 21cm line of $\Delta \nu \sim 1$ Hz.
Plugging in exact numbers for $\Lambda$CDM, the required spectroscopic
precision is actually more like 0.1 Hz if one wishes to measure the cosmic
acceleration directly at $z \approx 1$ \citep{Klockner:2015rqa}. Achieving
this sort of precision is challenging, as a number of systematic effects
must be controlled in a consistent manner for over a decade or more. From a
practical standpoint, the best way forward seems to be to perform
differential measurements of the time-dependence of line redshifts over many
thousands, if not millions, of galaxies. SKA2 will provide the requisite
sensitivity and spectral precision to perform this test for millions of HI
galaxies out to $z \sim 1.5$. More details, including an examination of
systematics such as peculiar accelerations, are given by \cite{Klockner:2015rqa}.

\subsection{Cosmological tests of General Relativity}
\label{sec:gravity}

General Relativity has been exquisitely tested for a wide range in
gravitational potential $\sim GM/rc^2$ and tidal field strength $\sim
GM/r^3c^2$ \citep{Psaltis:2008bb}; this includes tests in our solar system and extreme environments such as binary pulsars. Nevertheless, there is a dearth of direct tests of GR for tidal strengths $<10^{-50}$, which also happens to correspond to the domain in which we notice dark matter and dark energy. It is therefore of great interest to test General Relativity in a cosmological context.

Now we turn to the other explanation for accelerated expansion: that we are
mistaken about the law of gravity on very large scales. This can generate
acceleration by changing the geometric part of the Einstein equation by
modifying the Einstein-Hilbert action, and so requires no extra `dark
fluid'. Many such models have been proposed \citep[for a review, see][]{Bull2015}, but most of these produce a expansion history similar or identical to those predicted by dark energy models. Alternative observational tests are needed to distinguish between dark energy and modified gravity, through measurement of the growth of cosmic structures.

Cosmological observations are sensitive to the effects of gravity in diverse ways. In a Universe described by a perturbed FLRW metric, observations such as RSD are sensitive to the time-part of the metric, while gravitational lensing is affected by both the time-part and the space-part. These elements of the metric are themselves related to the density distribution of matter via the Einstein field equations (or classically, the Poisson equation). 

A simple test of gravity, then, is to examine whether the combination of different cosmological observations behaves as expected in General Relativity, or if a simple modification fits the observations better. If we model the Universe with a perturbed FLRW model,
\begin{equation}
\de s^2 = - (1+2\Psi)\de t^2 + (1-2\Phi) a^2 \de x_i\wedge\de x^i,
\label{eqn:pert-frw}
\end{equation}
where $\Psi$ and $\Phi$ are the two gauge-invariant Bardeen potentials and $a(t)$ is the cosmological scale factor. We can parametrise a range of modifications to gravity by the ratio $\eta=\Psi/\Phi$, and an additional factor $\mu$ in Poisson's equation relating $\Psi$ and overdensity. We can then calculate observables for various values of $\eta$ and $\mu$ and fit these to the cosmological probe data.

A more sophisticated approach is to write down a general action for linear cosmological perturbations of theories of gravity \citep[e.g.,][]{2016JCAP...08..007L} that contains parameters $\alpha_i$ characterising the theories. Again, observables can be calculated for particular values of $\alpha_i$, and so the permitted range of gravity theories fitting cosmological data can be assessed.

\subsubsection{Growth rate measurements with peculiar velocities}

Many dark energy and modified gravity models are capable of mimicking a
$\Lambda$CDM expansion history, and so could be indistinguishable from a
cosmological constant based on the equation of state of dark energy alone.
This is not the case for the growth history, however, which is typically substantially modified regardless of the background evolution. This is because modifications to GR tend to introduce new operators/couplings in the action, which lead to new terms in the evolution equations with distinct redshift- and scale-dependences.

A useful illustration can be found in the Horndeski class of general single scalar field modifications to GR. In the sub-horizon quasi-static limit (where spatial derivatives dominate over time derivatives), the linear growth equation for matter perturbations can be written as \citep{Baker:2013hia, Gleyzes:2017kpi}
\be
\ddot{\Delta}_M + \mathcal{H}\dot{\Delta}_M -\frac{3}{2}\Omega_M(a) \mathcal{H}^2 \xi \Delta_M = 0,
\ee
where overdots denote conformal time derivatives, $\mathcal{H} = a H$ is the conformal Hubble rate, $\Delta_M$ is the matter density perturbation, and $\xi(k, a) = 1$ in GR. The modification to the growth source term is restricted to have the form
\be
\xi(k, a) = \frac{f_1(a) + f_2(a) / k^2}{f_3(a) + f_4(a) / k^2},
\ee
where $\{f_n\}$ are arbitrary functions of scale factor that depend on the new terms added to the action. Within Horndeski models, all new terms in the action contribute to $\xi(k, a)$, and will cause deviations from GR growth at some scale and/or redshift. As such, we see that tests of growth can be more decisive in searching for deviations from GR than the equation of state.

The growth history can be constrained through a number of observables, for example the redshift-dependent normalisation of the matter power spectrum, $D(z)$, as probed by lensing or galaxy surveys; the ISW effect seen in the CMB and/or galaxy surveys, and the growth rate, $f(z) = d\log D / d\log a$, primarily measured through probes of the cosmic peculiar velocity field. In this section we will concentrate on the growth rate, as it exhibits fewer degeneracies with other cosmological parameters than the growth factor, $D(z)$, and can be measured with significantly higher signal-to-noise than the ISW effect.

The most precise growth rate constraints to date come from the RSD effect,
which makes the 3D correlation function of galaxies anisotropic as seen by
the observer. The effect is caused by the addition of a Doppler shift to the
observed redshift of the galaxies, due to the line-of-sight component of
their peculiar velocities. The growth rate can only be measured in
combination with either the galaxy bias, $b(z)$, or overall normalisation of
the power spectrum, $\sigma_8$, using the RSD technique, as these terms also
enter into the quadrupole (or ratio of quadrupole to monopole) of the galaxy
correlation function. As discussed in \S\ref{sec_rsd} and by \cite{Bull:2015lja}, HI galaxy redshift surveys and 21cm intensity mapping surveys with SKA are expected to yield sub-percent level constraints on the combination $f \sigma_8$ out to $z \sim 1.7$.

The SKA will also provide a more direct measurement of the peculiar velocity
field, through observations of galaxy rotation curves and the TF relation
\citep{Tully:1977fu}. The TF relation is an empirical relationship between
the intrinsic luminosity of a galaxy and its rotational velocity. Assuming
that it can be accurately calibrated, the TF relation can therefore be used
to convert 21cm line widths --- which depend on the rotation velocity ---
into distances (which can be inferred from the ratio of the intrinsic
luminosity and observed flux of the galaxy). Comparing the measured distance
with the one inferred from the redshift of the galaxy then gives the
peculiar velocity \citep[e.g.,][]{Springob:2007vb}.

This method has the disadvantage of being restricted to relatively low
redshifts --- the error on the velocity typically scales $\propto
(1+z)$ --- and relying on a scaling relation that must be calibrated
empirically. Nevertheless, direct observations of the peculiar velocity
field are sensitive to the combination $f H \sigma_8$ instead of $f
\sigma_8$, and so can provide complementary information to the RSD
measurements (and help break parameter degeneracies). The SKA and its
precursors will be able to perform suitable spectrally-resolved surveys of
many tens of thousands of HI galaxies out to $z\sim 0.3-0.4$ over most of
the sky --- essentially the widest and deepest TF velocity survey possible.
As well as providing a valuable independent probe of the velocity field,
these data can also be combined with the clustering information from a
traditional redshift survey extracted from the same survey dataset,
resulting in a significant improvement in the precision on $f\sigma_8$
compared with either probe individually \citep{Koda:2013eya}. A full-sky
survey with SKA precursor surveys WALLABY and WNSHS should be capable of
putting a joint RSD+TF constraint of $\sim 4\%$ on $f \sigma_8$ in a single
$z \approx 0$ redshift bin, for example \citep{Koda:2013eya}.

\subsubsection{Radio weak lensing}

\noindent Weak lensing maps the coherent distortions of galaxy shapes across
the sky \citep[see, e.g.,][for a review]{Bartelmann:1999yn}. With the path taken by light from distant galaxies determined by the matter distribution along the line of sight, and the response of curvature to that matter distribution, lensing represents an excellent probe of the theory of gravity. Dividing sources into tomographic redshift bins also allows us to track structure growth over cosmic time.

The SKA will be capable of detecting the high number densities of resolved,
high redshift star-forming galaxies over large areas necessary for weak
lensing surveys \citep{2016MNRAS.463.3686B}, with expected number densities
of $\sim 2-3$~arcmin$^{-2}$ over 5000~deg$^2$ for SKA1 and $\sim
12$~arcmin$^{-2}$ over 30,000~deg$^2$ for SKA2, giving comparable
raw source numbers to DES and Euclid, respectively. Doing weak lensing
in the radio band also has a number of distinct advantages, including the
expectation of a higher redshift source population
\citep[e.g.,][]{Brown:2015ucq,2016MNRAS.463.3674H} and information on
intrinsic alignments from polarisation \citep{2011MNRAS.410.2057B} and
rotational velocity \citep{2013arXiv1311.1489H} information. Foremost,
however, is the advantage of being able to combine weak lensing measurements
between SKA and optical surveys, forming cross power spectra $C_{\ell}^{XY}$
(where $X,Y$ label shear measurements for the two different experiments and
$i,j$ different redshift bins): \begin{equation} \label{eqn:limber_cross} C
^{X_{i}Y_{j}} _\ell = \frac{9H_0^4 \Omega_{\rm m}^2}{4c^4} \int_0^{\chi_{\rm
h}} {\rm d} \chi \, \frac{g^{X_i}(\chi) g^{Y_j}(\chi)}{a^2(\chi)} P_{\delta}
\left(\frac{\ell}{f_K(\chi)},\chi \right), \end{equation} where $a(\chi)$ is
the scale factor of the Universe at co-moving distance $\chi$, $f_K(\chi)$
is the angular diameter distance, $P_{\delta}(k, \chi)$ is the matter power
spectrum and $g^{i}(\chi)$ are the lensing kernels.

Using {\em only}\ these cross-experiment power spectra to form cosmological
constraints has been shown to retain almost all of the statistical power
available from the intra-experiment (i.e. $C_{\ell}^{XX}$) power spectra
\citep{2016MNRAS.463.3674H}, whilst removing wavelength-dependent
systematics that can otherwise cause large biases in the parameter estimation \citep[see][for a demonstration on real data]{2017MNRAS.464.4747C,2016MNRAS.456.3100D}.

\begin{figure*}
\centering
\includegraphics[width=0.48\textwidth]{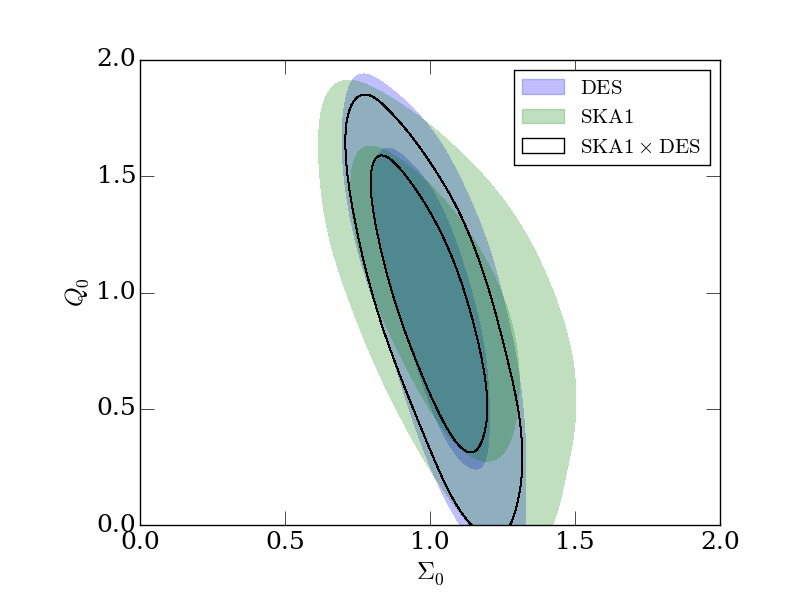}\includegraphics[width=0.48\textwidth]{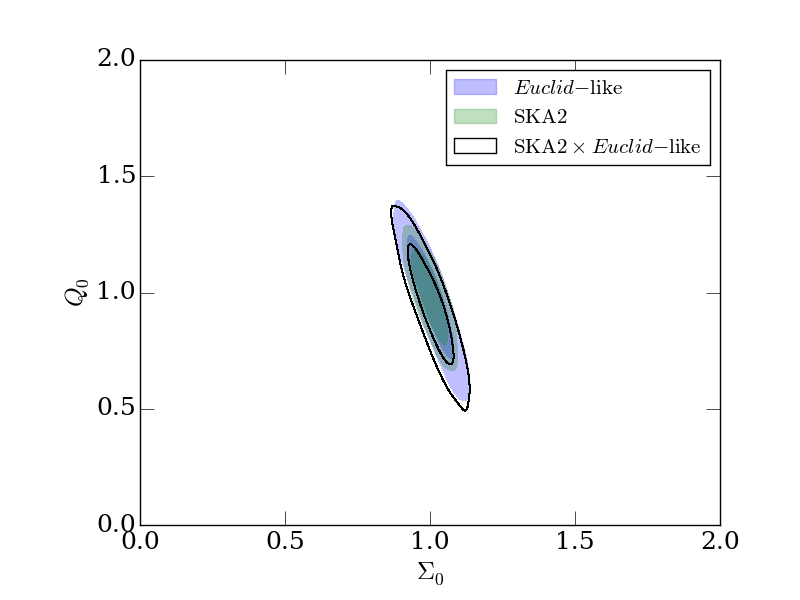}
\caption{SKA1 (left) and SKA2 (right) constraints on modified gravity parameters as described in the text, from optical-only (blue), radio-only (green) and radio$\times$optical cross-correlation-only (empty contours) cosmic shear power spectrum measurements. The forecasts were created using Markov chain Monte Carlo forecasts from the \textsc{CosmoSIS} toolkit \citep{2015A&C....12...45Z} and are marginalised over the base $\Lambda$CDM parameters. {\it Figure reproduced with permission, from \cite{2016MNRAS.463.3674H}.}}
\label{fig:wl_mg}
\end{figure*}

Figure \ref{fig:wl_mg} shows constraints on modified gravity parameters as
specified by \cite{2015PhRvD..92b3003D} (with $R=\eta$ and $\Sigma=\mu(1+\eta)/2$ in the notation specified here for Eq.~\ref{eqn:pert-frw}), showing the equivalent constraining power of both SKA-only and SKA$\times$optical to that expected from premier optical surveys. Similar constraints are available in the $w_0$-$w_a$ plane, with $\sim30\%$ constraints available from SKA1 and $\sim10\%$ constraints from SKA2 (both when combined with Planck CMB measurements). Note that the empty contours do refer to the cross-correlation alone, not to the combination of radio and optical. It is clear from this, as we mentioned above, that the cross-correlation contain as much constraining power as the auto-correlations.

\subsubsection{Doppler magnification}

\noindent Gravitational lensing consists of shear and convergence, $\kappa$.
While the shear is determined only by the matter distribution along the line
of sight, the convergence also has contributions from the Doppler,
Sachs-Wolfe, Shapiro time-delay, and ISW effects~\citep{Bonvin:2008ni, Bolejko:2012uj, Bacon:2014uja, Kaiser:2014jca, Bonvin:2016dze}. These contributions  modify the distance between the observer and the galaxies at a given redshift and consequently they change their observed size. The main contributions are gravitational lensing and a Doppler term: $\kappa=\kappa_\g+\kappa_\vv$, where
\bea\nonumber
\kappa_\g&=&\frac{1}{2r}\int_{0}^{r} dr'\frac{r-r'}{r'}\Delta_\Omega(\Phi+\Psi)\, ,\\ 
\kappa_\vv&=&\left(\frac{1}{r\HH} -1\right)\bV\cdot\bn\, , \nonumber\label{kappav}
\eea
where $r=r(z)$ is the comoving distance, $\Delta_\Omega$ is the 2-sphere
Laplacian which acts on the gravitational potentials $\Phi$ and $\Psi$,
$\HH$ is the conformal Hubble rate and $\bV\cdot\bn$ is the peculiar
velocity of a source projected along the line of sight $\bn$. Can we observe
$\kappa_\vv$? This contribution to the convergence has so far been neglected
in lensing studies, but it has been shown that it can be measured in
upcoming surveys~\citep{Bacon:2014uja, Bonvin:2016dze}, and can improve
parameter estimation as we now discuss.  Further details are provided by \cite{Bonvin:2016dze}.

For a given object, its peculiar velocity, $\bV$, is induced by nearby matter clustering, and so we expect the Doppler convergence to be strongly correlated with the observed galaxy number density, giving a  signal in the cross-correlation $\xi=\langle \Delta(z,\bn)\,\kappa(z',\bn') \rangle$. For an over-density, objects in front of it in redshift space will appear disproportionately larger than those behind, giving a clear dipole in $\xi$. In general the correlation between $\Delta=b\, \delta -\frac{1}{\HH} \partial_r(\bV\cdot\bn)$, which includes local bias $b(z)$ and an RSD term, and $\kappa_\vv$,   is given by
\begin{align}
&\xi_\vv(r,d,\beta)=\frac{\HH(z)}{\HH_0}f(z)\left(1-\frac{1}{\HH(z)r(z)} \right)\label{xiv}\\
&\times\left\{\left(b(z)+\frac{3f(z)}{5}
\right)\nu_1(d)P_1(\cos\beta)-\frac{2f(z)}{5}\nu_3(d)P_3(\cos\beta)
\right\}~, \nonumber 
\end{align}
where $f={d\ln D}/{d \ln a}$ is the growth rate ($D$ is the growth function), $P_\ell$ are the Legendre polynomials of order~$\ell$, and~$\nu_\ell$ is the power spectrum integrated against the~$\ell$'th spherical Bessel function. $\beta$ is the angle between the points where $\Delta$  and $\kappa$ are measured with respect to the line of sight. Here we have used the plane-parallel approximation, which makes the multipole expansion transparent~-- $P_1$ is a dipole and $P_3$ an octopole. The RSD contribution alters the coefficient of the second term in dipole. The correlation with RSD also induces an octopole in the $P_3$ term.

Multipole patterns in $\xi$ can be optimally extracted by integrating against the appropriate Legendre polynomial, $P_1(\cos\beta)$ in the case of the dipole. This implies we can optimally measure Doppler magnification in a survey of volume $V$ using the estimator
\be\nonumber
\xi_{\rm dip}(d)=\frac{3}{4\pi}\frac{\ell_p^5}{d^2V} \sum_{ij} \Delta_i \kappa_j \cos\beta_{ij}\delta_K(d_{ij}-d)\, 
\ee
where we associate to each pair of pixels $(i, j)$ of size~$\ell_p$ a separation $d_{ij}$ ($\delta_K(d_{ij}-d)$ selects pixels with separation~$d$) and an orientation with respect to the line-of-sight~$\beta_{ij}$. We measure the galaxy number count~$\Delta_i$ and convergence~$\kappa_j$ in each pixel respectively. A similar estimator can be constructed for the octopole. 
The dipole becomes, on average in the continuous limit,
\begin{align}\nonumber
&\langle\hat{\xi}_{\rm dip}\rangle(d)\simeq\frac{\HH(z)}{\HH_0}f(z)\left(1-\frac{1}{\HH(z)r(z)} \right)\left(b(z)+\frac{3f(z)}{5} \right)\nu_1(d)\, .\label{meandip}
\end{align}
In general this estimator also includes a dipole contribution from the normal lensing term since objects behind over-densities are magnified, but below~$z\sim1$ it is the Doppler term which dominates, so we neglect it here.

\begin{figure}
\centering
\includegraphics[width=0.48\textwidth]{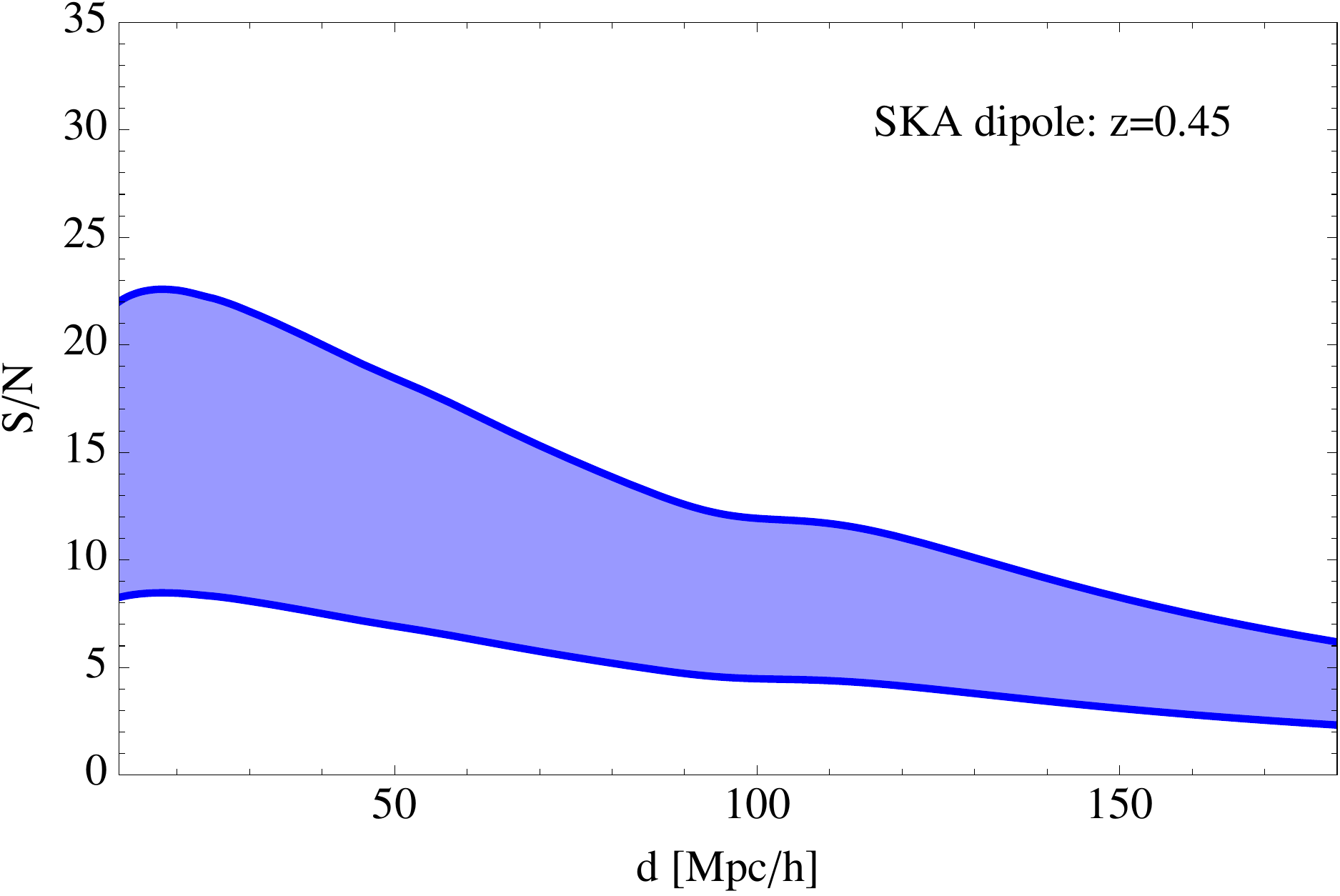}
\caption{\label{fig:SN_SKA} Signal-to-noise ratio for the Doppler magnification dipole as a function of separation, for a redshift bin $0.4<z<0.5$ in an SKA Phase 2 HI galaxy survey. The higher bound is for an intrinsic error on the size measurement of $\sigma_\kappa=0.3$, and the lower bound is for $\sigma_\kappa=0.8$. For the octopole the signal-to-noise is about an order-of-magnitude smaller. {\it Figure reproduced with permission, from \cite{Bonvin:2016dze}.}}
\end{figure}

We present an example forecast of the expected signal-to-noise for the SKA2
galaxy survey in Figure~\ref{fig:SN_SKA}. We present it for a broad range of the expected error on size measurements~$\sigma_\kappa$, and we assume that an intrinsic size correlation will have a negligible dipole. For a range of separations $12\leq d \leq 180$\,Mpc/$h$, combined over $0.1\leq z\leq 0.5$ (assuming that uncorrelated redshift bins), the cumulative signal-to-noise ratio is 35 (93) for the dipole and 5 (14) for the octopole, for $\sigma_\kappa = 0.3$ (0.8). The SKA should therefore allow a highly significant detection of the Doppler magnification dipole, and a firm detection of the octopole. 

\begin{figure}
\centering
\includegraphics[width=0.48\textwidth]{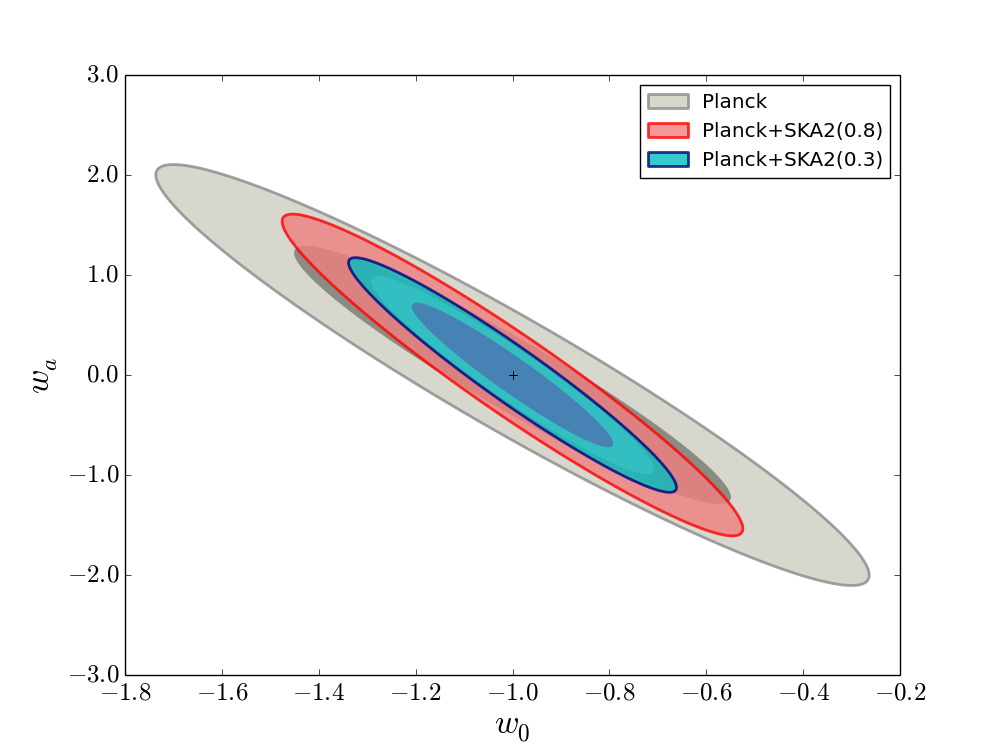}
\caption{\label{fig:waw0} Joint constraints on the $w_0$ and $w_a$ parameters, marginalised over all other parameters (except the bias, which is fixed), for Planck (T+P+lensing) alone, and Planck combined with an SKA2 HI galaxy survey. We use the dipole at separation 12\,Mpc/$h \leq d \leq$ 180\,Mpc/$h$. 
{\it Figure reproduced with permission, from \cite{Bonvin:2016dze}.}
}
\end{figure}

As an example of the improvement to parameter estimation that the Doppler
dipole will give, in Figure~\ref{fig:waw0} we show the constraints on
$w_0-w_a$ (marginalised over all other parameters, but fixing the bias model)
from Planck (temperature, polarisation, and CMB lensing) alone, and Planck combined with an SKA2 HI galaxy survey. Comparing with 
constraints from RSDs~\citep[see, e.g., Fig. 10 of][]{Grieb:2016uuo} we find that slightly better constraints are expected to be provided the Doppler magnification dipole, while similar constraints are expected for the SKA shear measurements. This is also the case for constraints on modifications to gravity.

In summary, extracting the dipole of the density-size cross correlation is a
novel new probe which is complementary to other lensing and RSD
measurements. This will help improve constraints from the SKA2 galaxy
survey. Furthermore, if we measure both the dipole and the RSD quadrupole,
we can test for the scale-independence of the growth rate, because the
quadrupole is sensitive to the gradient of the velocity whereas the dipole
is sensitive to the velocity itself. In addition, it should be possible to
reconstruct the peculiar velocity field directly from measurements of the
Doppler magnification dipole~\citep{Bacon:2014uja}.

\subsubsection{Cross-correlations with 21cm intensity maps}

A very promising way to test dark energy and gravity with the SKA is using
the HI intensity mapping technique \citep{Santos:2015vi}. A large sky HI IM
survey with SKA1-Mid can provide precise measurements of quantities like the
Hubble rate, $H(z)$, the angular diameter distance, $D_{\rm A}(z)$, and
$f\sigma_8(z)$, which depends on how dark energy and gravity behave on
large scales, across a wide range of redshifts \citep{Bull:2014rha}. A major challenge for intensity mapping experiments is foreground contamination and systematic effects. Controlling such effects becomes much easier in cross-correlation with optical galaxy surveys, since noise and systematics that are survey specific are expected to drop out \citep{Masui:2012zc, Wolz:2015lwa, Pourtsidou:2015mia}.

Hence, cross-correlating the intensity mapping maps with optical galaxy data
is expected to mitigate various systematic effects and to lead to more
robust cosmological constraints. As discussed earlier in
\S\ref{sec_grav_im}, we follow \citet{Pourtsidou:2016dzn} by
considering cross-correlation of an SKA1-Mid HI intensity mapping survey with a
Euclid-like optical galaxy survey, assuming an overlap $A_{\rm sky} =
7000$~deg$^2$. The results are shown in Table~\ref{tab:SKAIMcross}: we can
expect very good measurements of the growth of structure, the angular
diameter distance and the Hubble rate across a redshift range where the
effects of dark energy or modified gravity are becoming important. We note
again that an additional advantage of these forecasts is that they are
expected to be more robust than the ones assuming auto-correlation
measurements, due to the mitigation of various systematic effects.
\citet{Pourtsidou:2016dzn} also showed that a large sky intensity mapping
survey with the SKA, combined with the Planck CMB temperature maps, can detect the ISW effect with a signal-to-noise ratio $\sim 5$, a result competitive with Stage~IV optical galaxy surveys. The detection of the ISW effect provides independent and direct evidence for dark energy or modified gravity in a flat Universe.

Another way to test the laws of gravity on large scales is using the $E_{\rm G}$ statistic \citep{Zhang:2007nk,Reyes:2010tr,Pullen:2014fva,Pullen:2015vtb}. In Fourier space, this is defined as
\begin{equation}
E_{\rm G}(k,z) = \frac{c^2k^2(\Phi-\Psi)}{3H^2_0(1+z)\theta(k)} , 
\end{equation} 
where $\theta \equiv \mathbf{\nabla \cdot v}/H(z)$ is the peculiar velocity perturbation field. We can construct a Fourier space estimator for $E_{\rm G}$ as \citep{Pullen:2014fva}
\begin{equation}
\hat{E}_{\rm G}(\ell,\bar{z})=\frac{c^2\hat{C}_\ell^{\rm g\kappa}}{3H^2_0\hat{C}_\ell^{\rm g\theta}} \, ,
\end{equation}
and it can be further written as a combination of the galaxy-convergence
angular cross-power spectrum $C^{\rm g\kappa}_\ell$, the galaxy angular
auto-power spectrum $C^{\rm gg}_\ell$, and the RSD parameter $\beta =
f/b_g$. This estimator is useful because it is galaxy bias free in the
linear regime. Using HI instead of galaxies, we can use 21-cm IM clustering surveys with the SKA in combination with optical galaxy, CMB, or 21-cm lensing measurements to measure $\hat{E}_{\rm G}$. \citet{Pourtsidou:2015ksn} considered various survey combinations and found that very precise ($< 1\%$) measurements can be achieved.

\subsection{Tests of inflation}
\label{sec:inflation}

In the $\Lambda$CDM model, the Universe is flat, homogeneous, and has perturbations characterised by an almost-scale-invariant power spectrum of Gaussian perturbations, generated by a period of accelerated expansion in the early Universe known as inflation \citep{Bardeen:1983qw,Mukhanov:1985rz,astro-ph/0504097}.
This primordial power spectrum creates overdensities that we can observe through temperature anisotropies in the CMB \citep{astro-ph/9609105}, through brightness fluctuations in the 21-cm hydrogen line \citep{astro-ph/0010468,2004PhRvL..92u1301L}, and with the cosmological large-scale structure, once these perturbations grow nonlinear \citep{astro-ph/9506072}.

The simplest model of inflation is slow-roll inflation, in which the
expansion is driven by a single minimally-coupled potential-dominated scalar
field with a nearly flat-potential. Any deviations from such a simple model,
for example if there are multiple fields contributing to the generation of
fluctuations, or some change in the couplings, will lead to modified
spectrum of density perturbations that can be detected by large-scale structure surveys. We consider two such modifications: the presence of primordial non-Gaussianity and the production of primordial black holes.

\subsubsection{Primordial non-Gaussianity}

Non-Gaussian distributed fluctuations in the primordial gravitational potential represent one of the so-called `four smoking guns of inflations'. In particular, non-standard inflationary scenarios are expected to generate a large level of non-Gaussianity \citep[see, e.g.,][]{astro-ph/0406398, 1003.6097, 1004.0818}. If we restrict ourselves to local-type non-Gaussianity, Bardeen's gauge-invariant potential can be written as a perturbative correction to a Gaussian random field $\phi$, whose amplitude is parameterised by the parameter $f_{\rm NL}$, i.e.\
\begin{equation}
\label{eq:fnl}
\Phi=\phi+f_{\rm NL}\left(\phi^2-\langle\phi^2\rangle\right) .
\end{equation}
The current tightest bounds on $f_{\rm NL}$ come from measurement of the local bispectrum of the CMB \citep{planckfnl}, and amount to
\be\label{pfnl}
f_{\rm NL}=0.8 \pm 5.0~(1\sigma).
\ee
Albeit effectively ruling out models of inflation that generate a
large amount of local-type primordial non-Gaussianity, Planck constraints and even future CMB experiments are not expected to improve significantly the current bounds. This calls for new data. The CMB is localised at recombination, giving only 2-dimensional
information about the bispectrum and higher-order. Galaxy surveys can access
the distribution of matter in three dimensions, thus having access to a larger number of modes than those accessible to CMB experiments, thus delivering the next level of precision.

In the linear regime, local primordial non-Gaussianity generates a
scale-dependence of the clustering of biased tracers of the cosmic large-scale structure \citep[see, e.g.,][]{0710.4560}, reading
\begin{equation}
b(z,k) = \bar{b}(z)+\Delta b(z, k),
\end{equation}
where the non-Gaussian modification to $\bar{b}(z)$, the scale-independent Gaussian bias, is
\begin{equation}
\label{eq:ng-bias}
= \bar{b}(z)+ [\bar{b}(z)-1] f_{\rm NL}\delta_{\rm ec} \frac{3 \Omega_{m}H_0^2}{c^2k^2T(k)D(z,k)}.
\end{equation}
Here, $T(k)$ is the transfer function (normalised such that $T(k)\to1$ when $k\to0$), and $\delta_{\rm ec}\approx 1.45$ is the critical value of the matter over-density for ellipsoidal collapse. Because of the $1/k^2$ dependence, such a signal for non-Gaussianity is strongest on the largest cosmic scales, which are accessible by a large-area galaxy clustering survey with the SKA, using either the H{\sc i}-21cm emission \citep{Camera2013} or the radio continuum emission \citep{Raccanelli:2014ISW,Raccanelli17} of galaxies. With the SKA2 HI galaxy redshift survey, it should be possible to reach $\sigma_{f_{\rm NL}} $ close to 1 \citep{Camera:2014bwa}. Even the best next-generation galaxy surveys will not be able to bring $\sigma (f_{\rm NL})$ below 1, using single tracers of the matter distribution \citep{Alonso:2015uua}; this represents a cosmic variance floor to the capacity of galaxy surveys with a single tracer.

\begin{figure}[ht]
\centering
\includegraphics[width=0.5\textwidth]{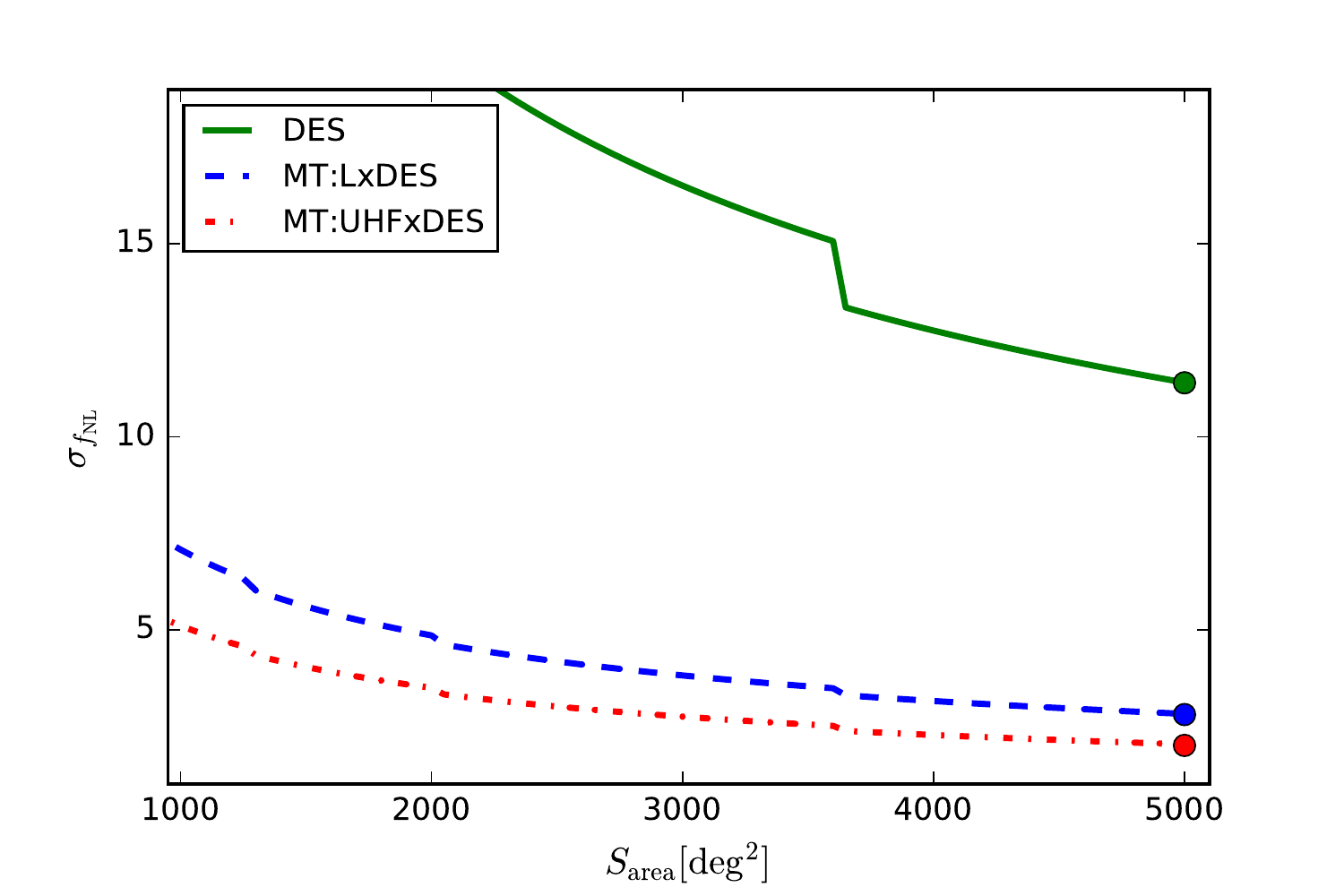}
\caption{Constraints on $\sigma (f_{\rm NL})$ against sky area for DES on its own (solid, green) and for multi-tracer of DES and MeerKAT (dashed, blue: low redshift band, dot-dashed, red: high redshift band). This calculation considers estimates for the full photometric sample of DES, i.e., ``red'' early type galaxies with  ``blue'' galaxies full of young stars. {\it Figure reproduced with permission, from \cite{Fonseca:2016xvi}.}}
\label{fig:fnlMT}
\end{figure}

The multi-tracer technique, which combines the auto- and cross-correlations of two or more tracers of the underlying cosmic structure, is able to overcome the problem of cosmic variance, thus allowing us to the measure the ratio of the power spectra without cosmic variance \citep{Seljak:2008xr}.
The MT technique is more effective when the bias and other
features of the tracers are as different as possible.
\citet{Ferramacho:2014pua} have shown that the identification of radio
populations in continuum galaxy catalogues allows us to push the limit on
primordial non-Gaussianity below $f_{\rm NL}=1$, in particular when redshift
information for radio continuum galaxies is recovered by
cross-identification with optical surveys \citep{Camera:2012ez}. 
\citet{Alonso:2015sfa} and \citet{Fonseca:2015laa} have subsequently shown that an SKA1
IM survey combined with LSST can achieve $\sigma (f_{\rm
NL})<1$. \cite{Fonseca:2016xvi} have also shown that even the precursor
MeerKAT (intensity mapping) and DES (clustering of the red and blue
photometric galaxy samples, combined) can improve on the
Planck constraint of Equation \eqref{pfnl} (see Fig.~\ref{fig:fnlMT}).
\cite{Fonseca:2015laa} also illustrated how detection of primordial
non-Gaussianity is tightly related to other relativistic effects important
on the scale of the horizon. Failure in properly accounting for all these
ultra-large scale corrections may lead to biased results in future
cosmological analyses \citep{Camera:2014sba,Camera:2014dia}.

\subsubsection{Primordial black holes}
It is customary to parametrise deviations from perfect scale invariance by a few variables, which capture the change in the shape of the power spectrum at some pivot scale $k_*$. 
The first of these numbers is the scalar tilt $(1-n_s)$, which expresses a constant offset in the power-law index. Higher derivatives, or runnings, of the power spectrum, are the scalar $\alpha_s=\mathrm d n_s/\mathrm d \log k$, and the second running $\beta_s\equiv\mathrm d \alpha_s/\mathrm d \log k $.

The scalar perturbations, $\zeta_{\bf k}$, have a two-point function 
given by
\begin{equation}
\VEV{\zeta_{\bf k}^{}\zeta_{\bf k'}^*} = P_{\zeta}(k) (2\pi)^3 \delta_D(\bf k+k'),
\end{equation}
where $P_\zeta(k)$ is the scalar power spectrum, for which 
we can define an amplitude as
\begin{align}
\log \Delta^2_s(k) &\equiv \log \left [\dfrac{k^3}{2\pi^2} P_\zeta(k) \right] = \log A_s + (n_s-1) \log\left(\dfrac k {k_*}\right) \nonumber \\ &+ \frac 1 2 \alpha_s \log^2\left(\dfrac k {k_*}\right)+ \frac 1 6 \beta_s \log^3\left(\dfrac k {k_*}\right),
\label{eq:Delta2}
\end{align}
where $A_s$ is the scalar amplitude. At the pivot scale of $k_*=0.05$
Mpc$^{-1}$, Planck has measured a scalar amplitude $A_s = 2.092 \times
10^{-9}$, with tilt $n_s=0.9656$ \citep{2018arXiv180706209P}.

The primordial perturbations, $\zeta$, generated during inflation, create matter overdensities $\delta\equiv \rho/\bar \rho -1$, where $\rho$ is the energy density and $\bar{\rho}$ its spatial average.
These matter perturbations source the temperature fluctuations in the CMB and later on grow to seed
the large-scale structure of the universe.
In linear theory, matter and primordial perturbations
are related to each other through a transfer function 
$\mathcal T(k)$,
so that the matter power spectrum is:
\begin{equation}
P_{\delta} (k) = \mathcal T^2(k) P_{\zeta}(k).
\end{equation}

In single-field slow-roll inflation, scale invariance is predicted to extend
over a vast range of scales \citep{0907.5424,1502.02114}.
However, we only have access to a small range of wavenumbers around the CMB pivot scale $k_*=0.05$ Mpc$^{-1}$.
The amplitude, $A_s$, of the (scalar) power spectrum and its tilt, $n_s$, give us information
about the first two derivatives of the inflaton potential 
when this scale, $k_*$, exited the horizon during inflation.
Higher-order derivatives of this potential produce non-zero runnings,
which for slow-roll inflation generically have values $\alpha_s\sim(1-n_s)^2$ and 
$\beta_s\sim(1-n_s)^3$, beyond the reach of present-day cosmological
experiments \citep{1007.3748}.
Next-generation cosmological experiments, including SKA galaxy surveys and 21-cm measurements, can measure these numbers. 

Slow-roll inflation models generally predict $|\alpha_{\rm s}| \sim 0.001$
and $|\beta_{\rm s}|\sim 10^{-5}$. Any large deviation from these values
would disfavour single-field inflation models. \citet{Pourtsidou} showed
that combining a Stage IV CMB experiment with a large sky 21-cm IM survey
with SKA2-Mid can yield $\sigma (\alpha_{\rm s}) \simeq 0.002$, while a high
redshift ($3<z<5$) intensity mapping survey with a compact SKA2-Low-like
instrument gives $\sigma (\alpha_{\rm s}) \simeq 0.0007$. Reaching the
required precision on the second running, $\beta_{\rm s}$, is difficult and can only be achieved with very futuristic interferometers probing the Dark Ages \citep{Munoz:2017}. 

A detection of $\alpha_s$, or $\beta_s$, would enable us to distinguish between inflationary models with otherwise equal predictions,
and would shed light onto the scalar power spectrum over a wider $k$ range. 

In the absence of any salient features in the power spectrum, such as small-scale non-Gaussianities, 
the power in the smallest scales will be determined by the runnings of the scalar amplitude.
This is of particular importance for primordial black hole  production in the early universe, 
where a significant increase in power is required at the scale corresponding to the 
PBH mass, which is of order $k \sim 10^5$ Mpc$^{-1}$ for solar-mass PBHs \citep{astro-ph/9901268,astro-ph/0511743}.
It has been argued that a value of the second running $\beta_s = 0.03$,
within 1$-\sigma$ of Planck results,
can generate fluctuations leading to the formation of $30\, M_{\odot}$ PBHs if extrapolated to the smallest scales \citep{1607.06077}.

Combining galaxy-clustering SKA measurements with future CMB experiments will enhance measurements of these parameters, so that we will be able to measure significant departures from single-field slow-roll inflation. 
Moreover, long baseline radio interferometers, observing the epoch of reionisation, will be able to measure the running $\alpha_s$ with enough precision to test the inflationary prediction. However, to reach the sensitivity required for a measurement of $\beta_s\sim 10^{-5}$, a Dark Ages interferometer, with a baseline of $\sim 300$ km, will be required.

A large positive value of the second running, $\beta_s$, has consequences for 
PBH formation.  
There has been interest in PBHs as a dark matter candidate \citep[see, e.g.,][]{CarrHawking,Meszaros,1607.06077},
since they could explain some of the gravitational-wave events 
observed by the LIGO collaboration \citep{2016PhRvL.116f1102A,1603.00464}.

If they are to be the dark matter (see \S\ref{sec:PBHs} for a full discussion of this possibility), PBHs could have formed in the primordial universe from very
dense pockets
of plasma that collapsed under their own gravitational pull.
The scales in which stellar-mass PBHs were formed are orders of magnitude
beyond the reach of any cosmological observable. 
However, if the inflationary dynamics were fully determined by a single field, 
one could extract information about the potential $V(\phi)$ at the smallest
scales from $V(\phi_*)$ at the pivot scale (and its derivatives) by extrapolation.

The formation process of PBHs is poorly understood \citep{astro-ph/9901268},
so one can as a first approximation assume that PBHs form at the scale at which $\Delta_s^2(k)$ becomes of order unity. 
It is clear that any positive running, if not compensated by a negative running of
higher order, will create enough power in some small-enough scale to have $\Delta_s^2(k)=1$.
Nonetheless, the mass of the formed PBHs is required to be larger than $\sim 10^{15}$~g (i.e., $\sim10^{-18} \, M_{\odot}$), 
to prevent PBH evaporation before $z=0$, 
which sets a limit on the smallest scale where PBHs can form of
of $\sim 10^4$~km.

In order to produce PBHs of $\sim 30 \, M_{\odot}$, as suggested
by~\cite{1603.00464} to be the dark matter, the relevant scale is $\sim
10$~pc. This would force the second running to be as large as $\beta_s \approx 0.03$, which will be tested at high significance by SKA2 galaxy surveys and IM measurements.

Detailed investigations of constraints on inflationary parameters related to
PBH production and observational constraints have been performed recently,
by authors including
\citet{Byrnes1,Byrnes2,Byrnes3,Germani,Munoz,Pourtsidou,Sekiguchi:2017}. We
refer to those papers for accurate and thorough observational constraints
and predictions. 

\subsection{Tests of Fundamental Hypotheses}

\subsubsection{Tests of the Cosmological Principle}
\label{sec:cosmo-principle}
Testing the foundations of the standard cosmological model is an important part of strengthening the status of this model. One of the basic pillars of cosmology is the large-scale  FLRW geometry, in other words, the cosmological principle: on large enough scales the universe is {\it on average} spatially homogeneous and isotropic. This principle consists of two parts: 

{\bf Statistical isotropy of the Universe around us:} There is a large body
of separate evidence that the Universe is isotropic, on average, on our past
lightcone. The strongest such evidence comes from the observed level of
anisotropies of the CMB. The observed dipole in the CMB is consistent with
our proper motion with respect to the CMB rest frame
\citep[see][]{Kogut:1993ag,Aghanim:2013suk}. Thus, once corrected for this
proper motion, the CMB does indeed appear isotropic around us to one part in
$10^{5}$, a level perfectly consistent with the standard model of cosmology
supplemented by small fluctuations generated early during a phase of
inflation. In addition, a generic test of Bianchi models presented by \cite{Saadeh:2016sak}  with CMB strongly disfavours large-scale anisotropic expansion.

{\bf The Copernican Principle:} we are typical observers of the Universe; equivalently: we are not at a special spatial location in the Universe.
Relaxation of this principle has sometimes been invoked as a solution to the
dark energy problem \citep[see,
e.g.,][]{GarciaBellido:2008nz,February:2009pv}, but studies of kinetic
Sunyaev-Zeldovich
effects have strongly disfavoured such solutions; see \cite{Bull:2011wi}
and \cite{Clifton:2011sn}. However, the principle itself remains to be tested accurately, irrespective of the actual solution to the dark energy problem. 

It is clear that these two ingredients, which, when combined, imply the cosmological principle, have different scientific statuses. On the one hand, the observed statistical isotropy around us is easily constrained by direct observations down our past lightcone. On the other hand, the Copernican Principle provides a prescription about what happens off our past lightcone, both in our causal past and outside of our causal past. Assessing its validity is therefore much more difficult.

One can find detailed accounts of various ways one can constrain the
large-scale geometry of the Universe from observations in two recent reviews
\citep{Clarkson:2010uz,Clarkson:2012bg}. Some detailed discussions of the
prospects of the SKA for future tests of the cosmological principle are
presented by \cite{Schwarz:2015pqa}. In particular, the SKA will be ideal to
measure the cosmic radio dipole and to test if it aligns with the CMB
dipole, as it should be the case in standard cosmology. A recent analysis of
the WISE-2MASS optical catalogue by \cite{Bengaly:2016amk} has not found any
significant anisotropy in the large-scale structure distribution, but the
SKA will allow us to pinpoint the direction and amplitude of the dipole with
great accuracy \citep[e.g., with SKA2, one will be able to determine the
direction of the dipole to within 1~degree; see][]{Schwarz:2015pqa}, and to compare them directly with the CMB measurement, since the SKA will probe a super-horizon size volume.\\
Tests of the Copernican Principle, on the other hand, are much harder to
design, and are usually much less precise. However, two promising techniques
have emerged, which allow one to get some information on what happens off
our past lightcone. First, a direct comparison of the transverse and radial
scales of BAOs gives one access to a test of possible anisotropies in the
local expansion rate of the Universe away from us \citep[see][]{Maartens:2011yx,February:2012fp}.

\begin{figure}[ht]
\centering
\includegraphics[width=0.5\textwidth]{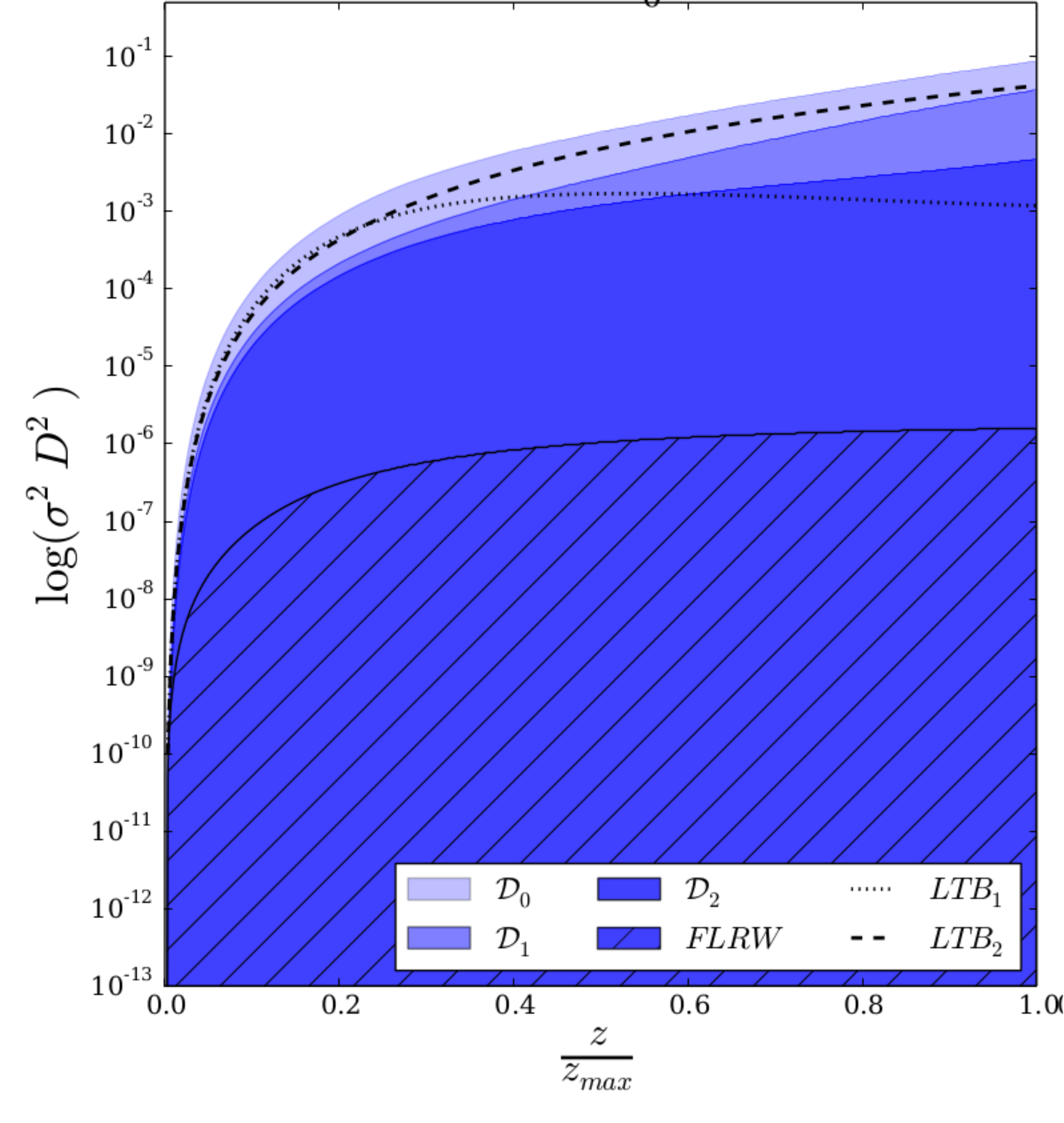}
\caption{Constraints on the matter shear normalised by the angular distance,
$D$, as a function of redshift on our current past lightcone. The blue
regions, from light to dark, correspond to the upper 2-$\sigma$ contours
reconstructed from currently available data (i.e., simulation ${\cal
D}_{0}$), forecast, $D(z)$ and redshift-drift data (i.e., simulation, ${\cal
D}_{1}$) and finally all of the above, including $H(z)$ data from
longitudinal BAO measurements (i.e., simulation, ${\cal D}_{2}$). The hatched region corresponds to the intrinsic shear present in a perturbed FLRW model with a uv-cutoff of 100~Mpc. For comparison we also show two spherically symmetric but inhomogeneous models, one with a homogeneous bang time $t_{B}(r)=0$ (labelled LTB$_1$) and one without (labelled LTB$_2$). }
\label{fig:CPtest}
\end{figure}

Second, direct measurements of the redshift-drift, while a remarkable probe
of the nature of dark energy (see \S\ref{sec_z_drift}), can also help constrain the Copernican
Principle, as presented by \cite{Bester:2015gla,Bester:2017lrt}.
\cite{Bester:2017lrt} use a fully relativistic way of reconstructing the
metric of the Universe from data on our past lightcone, with a minimal set
of a priori assumptions on the large scale geometry. Focusing on spherically
symmetric (isotropic) observations around a central observer (in the
$\Lambda$-Lema\^itre-Tolman-Bondi class), one can characterise any departure
from homogeneity by the scalar shear of the cosmological fluid,
$\sigma^{2}=\frac{1}{2}\sigma_{ij}\sigma^{ij}$. Figure~\ref{fig:CPtest}
presents constraints on this shear from current optical data (label ${\cal
D}_{0}$), and for a forecast with radio-astronomy data (labels ${\cal
D}_{1}$ and ${\cal D}_{2}$) generated around a fiducial $\Lambda$CDM model.
${\cal D}_{0}$ uses Type~Ia supernova data from \cite{suzuki2012} to
determine the angular distance $D(z)$, cosmic chronometre data  from
\cite{moresco2011,moresco2015} to determine the longitudinal expansion rate
$H_{\|} (z)$, and stellar ages from \cite{sneden1996} to put a lower bound
on the age of the Universe $t_{0}$. ${\cal D}_{1}$ uses only Type~Ia
supernova data from \cite{suzuki2012} and forecast for SKA2 BAO in intensity
mapping for $D(z)$, as well as forecast for a redshift drift experiment like
CHIME, from \cite{Chime2014}. Finally, ${\cal D}_{2}$ consists of all the
combined inputs of ${\cal D}_{0}$ and ${\cal D}_{1}$. Details of the methods
are presented by \cite{Bester:2017lrt} and references therein.  
One clearly sees that the redshift drifts are the best data to improve on current constraints.

\begin{figure}[ht]
\centering
\includegraphics[width=0.5\textwidth]{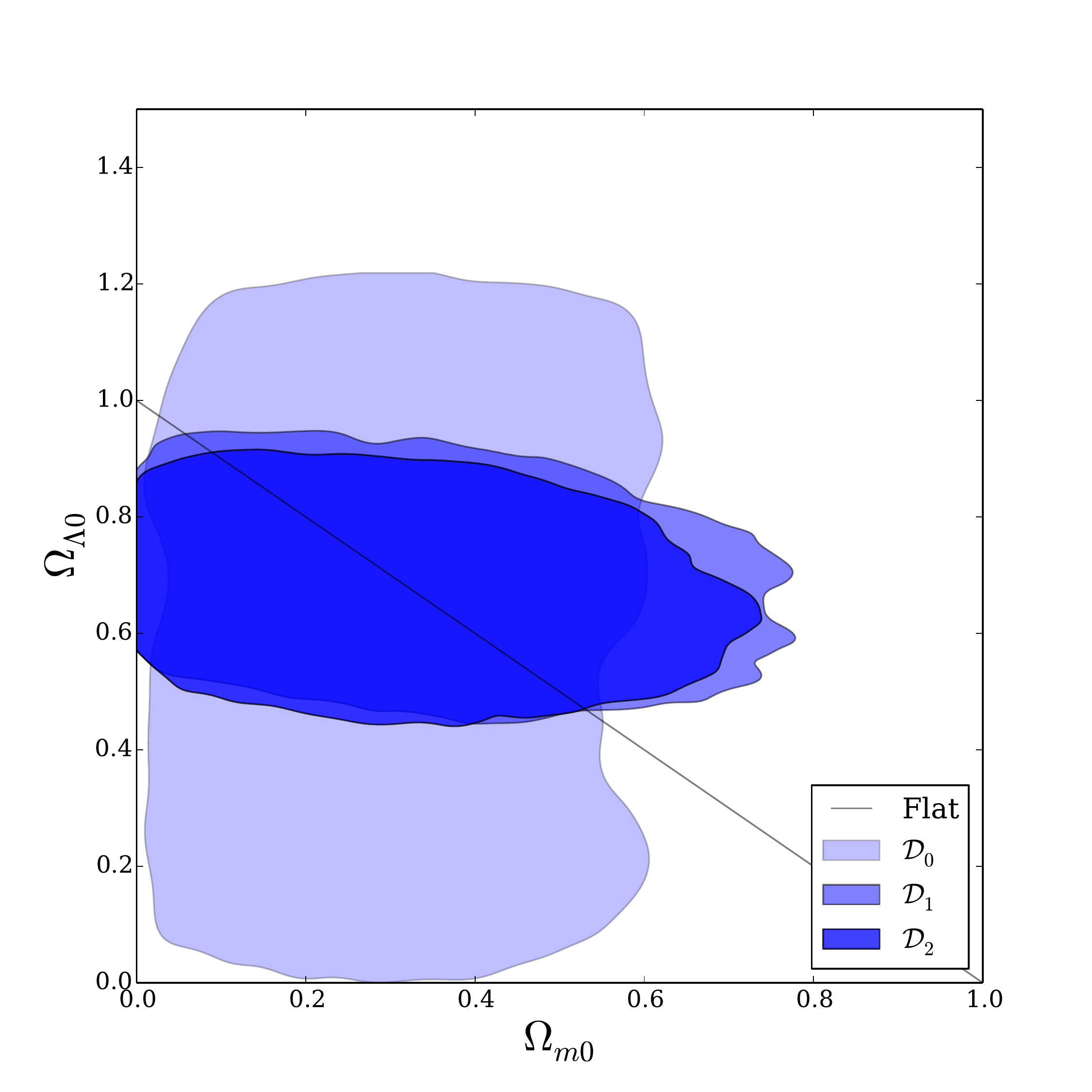}
\caption{2-$\sigma$ constraints on $\Omega_{\Lambda}$ and $\Omega_{m}$ on the worldline of the central observer today for the various combinations of data presented in the text.}
\label{fig:OmL}
\end{figure}

Besides, they also allow for a remarkable determination of the value of the
cosmological constant {\em without}\ assuming the Copernican Principle: as shown in Figure~\ref{fig:OmL}, in the class of spherically-symmetric, but inhomogeneous models, the inclusion of redshift drift data allows one to constrain $\Omega_{\Lambda}$ at less that $10$\%.

\subsubsection{Tests of local Lorentz invariance}
\label{sec:LLV}

Cosmological models inspired from fundamental theories may lead to violation
of LLI. The strongest constraints on such proposals
will be set by pulsar experiments, as discussed in
\S\ref{sec:pulsar:timing}. These constraints are proportional to the timing
precession of binary pulsars, and hence will be dramatically improved with
the SKA project.

A cosmological model that leads to LLI violation and has so far passed all
other tests is the D-material Universe~\citep{Elghozi:2015jka}, a model
which may appear as the low-energy limit of certain brane
theories~\citep{Mavromatos:2007sp} in the context of string theories with
large extra dimensions. This cosmological model aims at providing a
justification for the phenomenological $\Lambda$CDM model, which relies on
the existence of two unknown quantities, namely a positive cosmological
constant, $\Lambda$, and CDM component, both introduced in order to fit current astrophysical data.

According to string theory, matter consists of one-dimensional objects, the
strings. Different vibrations of a string represent different particle
types, while splitting and joining of elementary strings represent particle
interactions. String ends live on a surface that can be thought of as a
large massive object, a Dirichlet brane (D-brane), in spacetime. Branes of
different dimensionality, depending on the particular string theory, are
thought to be embedded within a higher dimensionality background, the bulk.
In this framework, let us consider a compactified (3+1)-dimensional brane
propagating in a higher dimensional bulk populated by zero-dimensionality
(point-like)  D-branes, called D-particles, since they have all their
spatial dimensions wrapped around compact space.  As the (3+1)-dimensional
brane moves in the bulk, D-particles cross it, resulting in foamy
structures. Since branes are by definition the collection of the end points
of open strings, particle excitations (described by open strings) propagate
in a medium of D-particles. Thus, brane-puncturing massive D-particles can
be captured by electrically neutral matter open strings, a process that is
described by the Dirac-Born-Infeld action. This scenario leads to a
bi-metric theory\footnote{A bi-metric theory has two metrics: (1) the `Einstein
frame' metric, $g_{\mu\nu}$, that satisfies the canonically Einstein-Hilbert
action, and (2) a modified physical metric that matter and radiation `feel'; it depends on $g_{\mu\nu}$ but also on scalar and vector fields.} \citep{Mavromatos:2007sp}, with a vector field appearing naturally as the result of the recoil velocity field of D-particles. The recoil results in a metric deformation of the neighbouring spacetime, and in Lorentz invariance being locally broken. The latter implies the emergence of vector-like excitations that can lead to an early era of accelerated expansion, in the absence of an inflaton field, and contribute to large scale structure (enhancing the dark matter component) and galaxy formation \citep{Ferreras:2007kw,Ferreras:2009rv,Mavromatos:2009xh}.
The D-material universe has been shown to be in agreement with gravitational
lensing phenomenology~\citep{Mavromatos:2012ha}. Moreover, the medium of
D-particles leads to recoil velocity field condensates that induce an effective mass for the graviton~\citep{Elghozi:2016wzb}, in agreement with the constraints imposed from the Advanced LIGO interferometric data \citep{2016PhRvD..93l2004A,2016PhRvL.116f1102A,Abbott:2017vtc}.
So far, the D-material universe is in agreement with observational constraints \citep{Mavromatos:2012ha,Elghozi:2016wzb}

This, so far successful, cosmological model with the advantage of being based on a microscopic theory, can be further tested with the SKA~\citep{Janssen:2014dka}. LLI violation leads to modifications of the orbital dynamics of binary pulsars, as well as to modifications of the spin evolution of solitary pulsars~\citep{Shao:2013wga}, while for the latter it also leads to a spin precession with respect to a fixed direction~\citep{Shao:2012eg}. Hence, LLI violation implies changes in the time-derivative of the orbital eccentricity, of the projected semi-major axis, and of the longitude of the periastron, while it changes the time-behaviour of the pulse profile.

Since the accuracy of timing precession of binary pulsars will be
significantly improved with the SKA project, one expects to further
constrain models leading to LLI violation, such as the cosmological model
mentioned above. For a given pulsar, the timing precision scales with the
signal-to-noise ratio of its pulse profile. As simulations have indicated,
if the SKA improves the signal-to-noise ratio of pulse profiles by a factor of
10, the Lorentz-violating coefficients will be constrained by the same
factor, within only a 10-year cycle of observations. Combining these SKA
observations with 20 years of pre-SKA data, one may be able to constrain the
Lorentz-violating coefficients up to a factor of 50 \citep{Shao:2014wja}.

\subsection{Summary}\label{sec:conclusions}
In this section, we have reviewed how data from the SKA will open a new era
for radio cosmology, allowing us to test the foundations of the concordance
cosmological model to unprecedented accuracy. Furthermore, we argue that the
SKA's commensality with other observational campaigns, aimed at scrutinising
the Universe's large-scale structure in the optical and NIR bands, will allow us to have independent checks of crucial cosmological observations, de facto reinforcing statistical analyses on long-standing problems such as the nature of dark energy or the validity of general relativity on cosmological scales.

Below, we list the main points considered in the section:
\begin{itemize}
\item \textit{Tests of cosmic acceleration (see \S\ref{sec:DE-MG}).}
The zeroth order test to understand whether the late-time cosmic expansion is truly due to a cosmological constant term or if it is a dark energy component that dominates the Universe's present-day evolution is to check the constancy of the equation of state of dark energy, $w(z)$. The SKA will be able to do this both at the level of background and cosmological perturbations. The latter will be achieved mostly via BAO measurements, for which 21cm intensity mapping will represent a unique added value of the SKA, compared to usual galaxy surveys. The former is envisaged through measurements of the redshift drift, which will allow us to probe the Hubble parameter directly and not as an integrated quantity, e.g., as for type Ia supernovae. 

\item \textit{Tests of gravity (\S\ref{sec:gravity}).}
Although general relativity has been tested
to exquisite precision in the solar system and in strong gravity regimes,
we still extrapolate it for orders of magnitude when we use it to interpret
cosmological data. For this reason, the possibility of deviations from
Einsteinian gravity are particularly interesting in the context of dark
energy, for which a modified gravity model may represent a viable
alternative. The main means by which the SKA will test this hypothesis is
the study of the growth of large-scale structure. On the one hand, the SKA
will complement optical/NIR surveys such as  those to be performed by Euclid or LSST in quantifying deviations from general relativity at the level of the matter power spectrum, employing 21cm intensity mapping, HI and continuum galaxy number counts, as well as radio weak lensing cosmic shear. On the other hand, the SKA depth and sky area will allow us to probe for the first time the largest cosmic scales, which see the peak of as-yet-undetected relativistic effects.

\item \textit{Tests of inflation (\S\ref{sec:inflation}).}
Those same extremely large scales where relativistic effects hide also retain pristine information about inflation. One of the most robust predictions of inflation is a certain amount of non-Gaussianity in the distribution of primordial density fluctuations. By probing the growth of structures on the scale of the horizon and, in particular, by cross-correlating multiple tracers of the underlying dark matter distribution, we will be able to push the limits on primordial non-Gaussianity, eventually reaching sub $f_{\rm NL}=1$ precision. Moreover, the study of the matter power spectrum over a wide range of scales will allow us to test the hypothesis of primordial black holes, for which a significant increase in power is required at the scale corresponding to the primordial black hole mass. 

\item \textit{Tests of the cosmological principle (\S\ref{sec:cosmo-principle}).}
By measuring the cosmic radio dipole and comparing it to the observed CMB dipole, the SKA is ideally suited to test the hypothesis of statistical isotropy of the Universe around us. Furthermore, redshift drift measurements can also help constrain the Copernican Principle, in particular by putting strong bounds on inhomogeneous cosmological models, such as $\Lambda$-Lema\^itre-Tolman-Bondi cosmologies. 
\end{itemize}

\section{Dark Matter and Astroparticle Physics}
\label{sec_astropart}

\subsection{Introduction}

\begin{figure}
\includegraphics[width=\columnwidth]{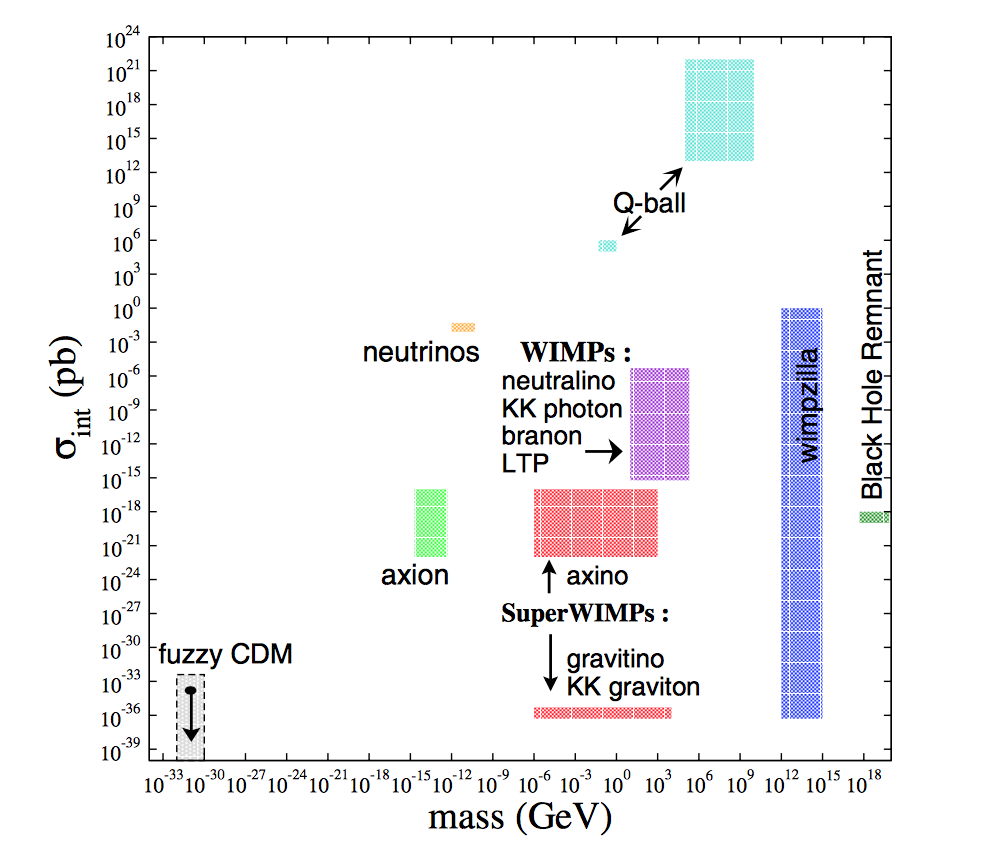}
\caption{The mass and cross-section (in picobarns, where 1~pb~$=10^{-40}$~m$^{-2}$) for various dark matter particle candidates.  Figure taken from~\cite{par_space}.}
\label{fig:space}
\end{figure}

The detection of dark matter remains a key goal of modern cosmology and
astrophysics.  After three decades of searching, the case for its existence
remains stronger than ever, with measurements from
Planck~\citep{2018arXiv180706209P} reinforcing the hypothesis that massive,
non-luminous matter comprises 26\% of the total energy density of the
Universe. Radio astronomy in particular has played a critical role in
constraining the properties and evolution of DM halos since their initial
prediction, particularly in the observation of HI rotation curves well
beyond the optical radius of galaxies~\citep[see,
e.g.,][]{1981AJ.....86.1791B, 1981AJ.....86.1825B, 1985ApJ...295..305V,
1989A&A...223...47B}. Future radio observations may also be crucial for identifying the DM among the many suggested candidates. Weakly interacting massive particles (WIMPs) have been a primary focus, a category of new, principally fermionic, particles predicted from extensions to the standard model of particle physics.  However, attention has also turned to a variety of other candidates: 
recent observations at LIGO~\citep{1603.00464} have re-invigorated the search for primordial black holes, and the search for axions has received significant support in recent years 
\citep[see Fig.~\ref{fig:space} and][for an overview of the dark matter parameter space]{par_space}.

The development of the SKA marks a significant advancement in radio
astronomy and offers the possibility of direct or indirect detection of dark
matter. One of the major challenges in doing this is to disentangle the DM
signal from astrophysical signals. With its huge improvement in sensitivity,
resolution, and versatility, the SKA will massively increase our
understanding of astrophysical backgrounds and facilitate disentanglement.
Our key goal in compiling this work is to bring together the areas in which
the SKA and its precursors can make its greatest contribution to both
cosmology and particle physics. \S\ref{sec:dmprop} looks at ways the SKA may
help to constrain general DM properties; \S\ref{sec:dmsearch} reviews the
search for DM candidates and details ways in which the SKA can support the
search for WIMPs, axions and PBHs in particular; and \S\ref{sec:astrop} investigates ways in which the SKA can constrain astroparticle properties.

\subsection{Dark Matter Properties}
\label{sec:dmprop}

The evidence for dark matter on galaxy scales comes from
21 cm line observations of rotation curves, which do not decline beyond the
optical image of gas rich galaxies. However, it is still debated how predominant
the dark matter is in the inner parts of such galaxies, since the mass models
are degenerate, so that additional dynamical criteria have to be brought to
bear. These are not straightforward, and a debate is ongoing on the validity
of using stellar velocity dispersions to settle this
issue~\citep[see for example] [and references therein]{2017ASSL..434..209B}.

Attempts to constrain the geometrical shape of the dark matter halo, using the
flaring of the HI layer beyond the optical radius, are also unexpectedly
difficult. For bright galaxies, the current picture is rapidly changing, as
there is more and more evidence for a complicated baryon cycle, with both
accretion and outflows related to star formation activity. This affects the 
kinematics of extraplanar HI gas, which rotates slower than the HI in the 
stellar disk~\citep[for example][for NGC 891]{Oost07}. Of course smaller 
galaxies could be more quiescent, but for those the thickness of the HI layer 
might play a role.

Through HI intensity mapping and observation of the HI power spectrum, SKA will be able to provide new insights into galaxy formation and evolution, thus providing greater clarity on the properties of DM. Most particularly, such observations will provide a window into DM distribution, DM halo abundance and clustering, and the thermal nature of DM. 

\subsubsection{Dark matter distribution}

\begin{figure}
\includegraphics[width=0.5\textwidth]{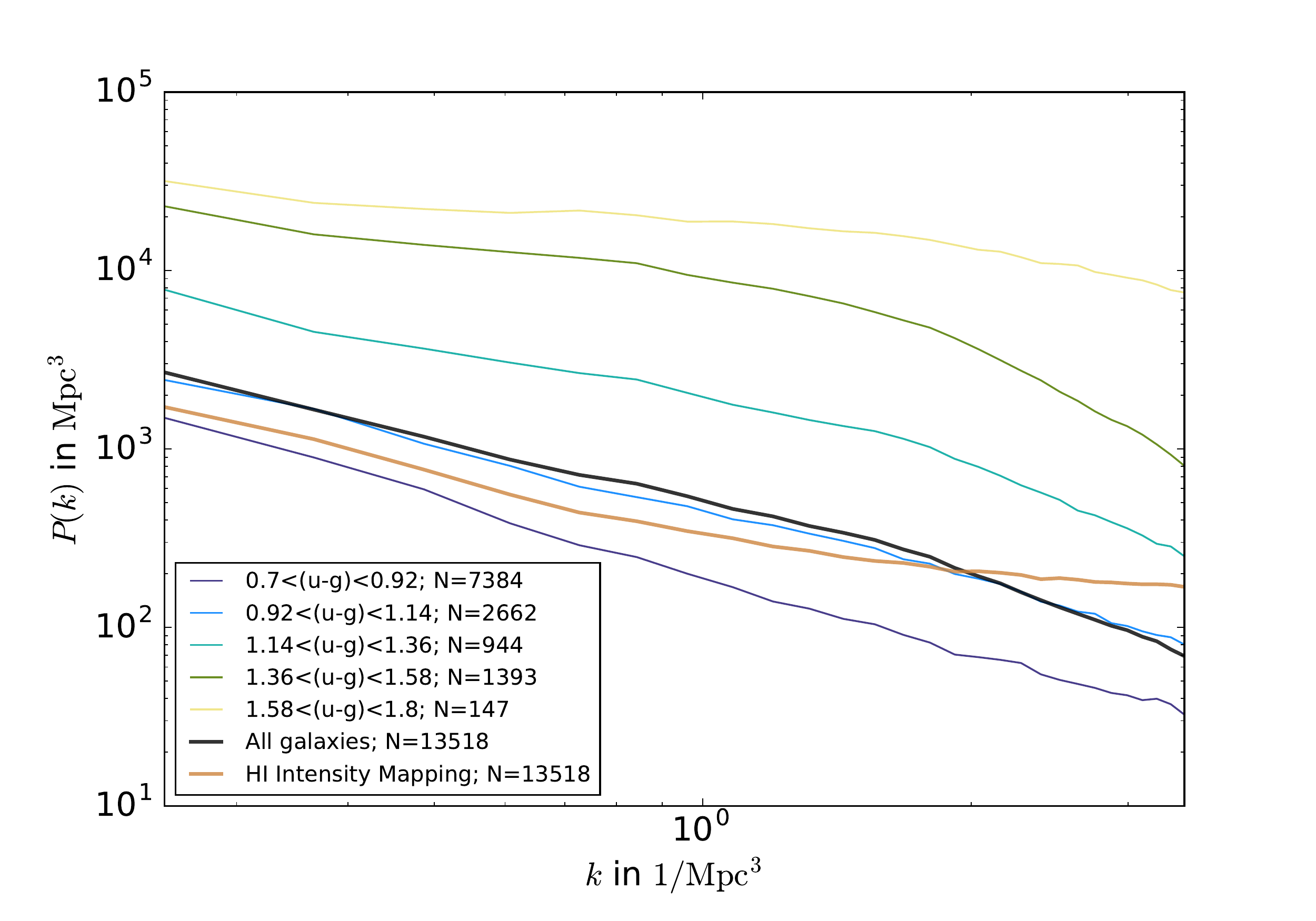}
\includegraphics[width=0.5\textwidth]{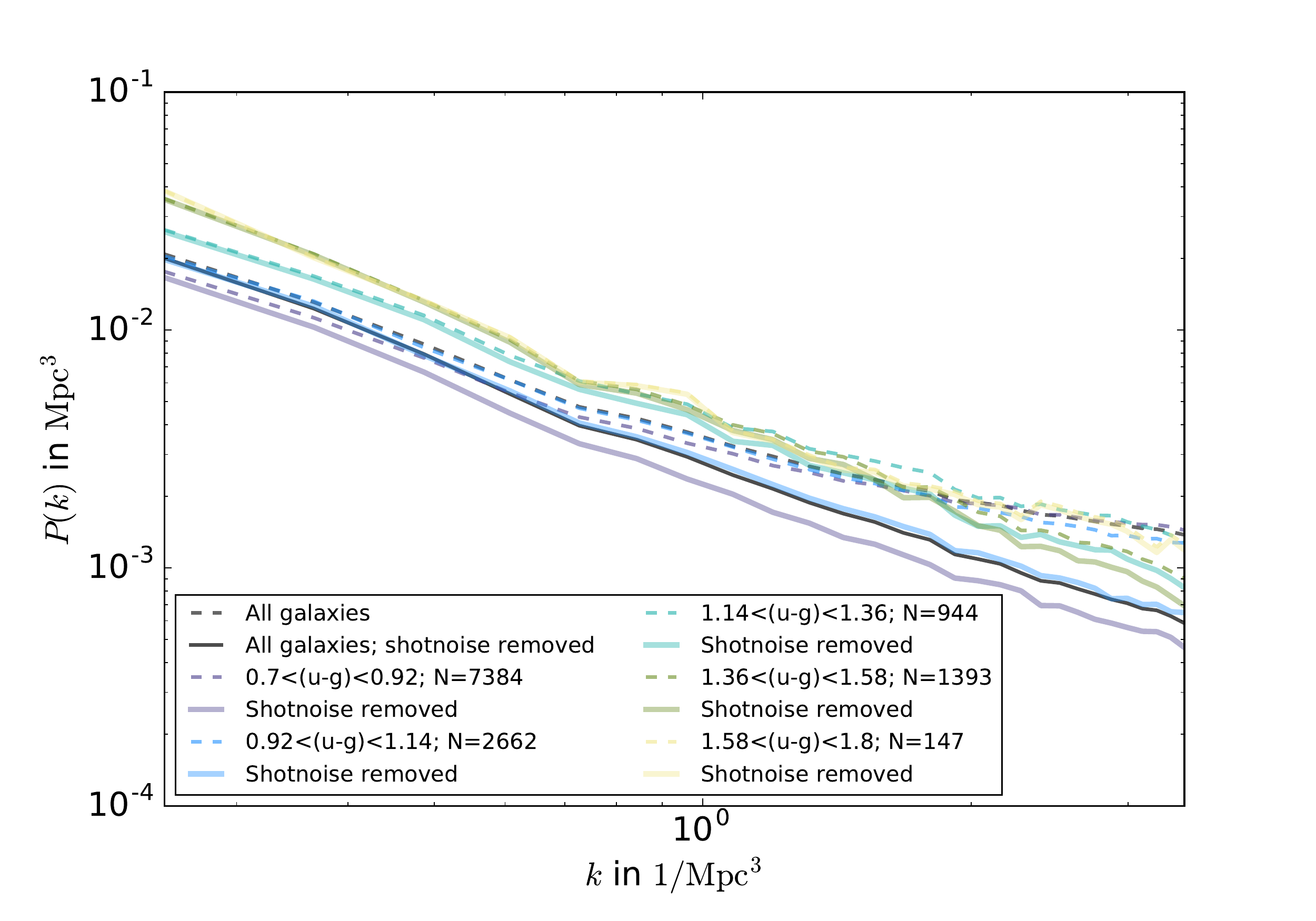}
\caption{Upper panel: we use a box of the EAGLE hydro-dynamical simulation suite at $z=0.5$ to derive the HI intensity mapping power spectrum (orange line) as well as several optical selected galaxy sample power spectra using the magnitudes in the SDSS u and g filters. The black line marks the power spectrum of all galaxies in the simulation volume. Lower panel: We cross-correlate the HI intensity maps with respective galaxy selections. The dashed lines mark the observed cross-power spectra. The solid lines have been shot-noise corrected where the shot-noise is proportional to the average HI mass in the optical galaxies.}
\label{Fig_HIPS}
\end{figure}

Understanding how HI correlates to the underlying mass of the DM halo is crucial to constraining DM properties from astrophysical observations. Observational constraints on HI abundance and clustering in the post-reionisation ($z < 6$) Universe can be divided into three categories: (1)  21 cm emission line galaxy surveys  at low redshifts ($z \sim 0-1$), (2) 21 cm intensity mapping (attempted at $z \sim 1$) measuring the integrated, unresolved emission from galaxies, and (3) higher redshift damped Lyman-$\alpha$ absorption surveys (at redshifts  $z > 1.5$). Future facilities such as the SKA will attempt to provide both galaxy surveys as well as intensity maps at moderate and high redshifts ($z > 1$), thus enhancing our understanding in this field.

The HI intensity power spectrum, $[\delta T_{\rm HI}(k,z)]^2$ \citep[as provided
by, for example,][]{battye2012}, couples contributions from (1) the
astrophysics of HI in galaxies that affects the brightness temperature and
the HI bias, and (2) the underlying DM power spectrum. \citet{hptrcar2015}
combine the astrophysical uncertainties from the available data to derive estimates of the observable HI power spectrum using current and future facilities. The astrophysics needs to be modelled effectively in order to recover the underlying cosmological parameters, and in the future also enable constraints on the DM power spectrum via HI experiments. This can be done by using a data-driven halo model framework for neutral hydrogen in the post-reionisation Universe \citep{hpar2017, hparaa2017}. The uncertainties on the astrophysical parameters are quantified using a Markov chain Monte Carlo technique applied to existing HI observations. This not only offers clues towards the baryonic gas evolution, but also enables  insights into the amount of astrophysical degradation expected in forecasting the cosmological and DM properties. 

Accurate cosmological interpretation of the HI intensity mapping power
spectrum requires profound understanding of the manner in which the HI gas
traces the underlying DM distribution.  This is most commonly expressed through the
HI bias \citep{Sarkar:2016ip, 2016arXiv160905157C}. Numerical simulations show the HI bias to scale-dependently increase for wavenumbers $k \leq 1.0$ Mpc$^{-1}$.  Figure \ref{Fig_HIPS} shows
an example of this effect using the HI intensity mapping power spectrum
derived from a $(100~{\rm Mpc})^3$ volume of the hydrodynamical EAGLE simulation \citep{Lagos:2015gpa, Crain:2016ex} at $z=0.5$ (marked in orange), comparing it to galaxy samples selected by their Sloan Digital Sky Survey (SDSS) u and g luminosities. One can also see from this Figure that high u-g luminosities show a greater amplitude in their power spectra.  This reflects the correlation between u-g luminosity and the age of the galaxy, with high u-g luminosities disproportionately selecting quiescent red galaxies which live in higher density regions than their younger blue counterparts. The total galaxy power spectrum as a tracer for the DM is also marked in black for comparison. Measuring the HI bias in future intensity mapping observations at low and high redshifts as seen by 
SKA1-Mid and SKA1-Low, respectively, will be crucial to gaining a new
understanding of how HI correlates with the underlying host DM halo mass as
well as to the properties of the host galaxy. The latter can be facilitated
by the cross-correlation of HI intensity maps with different galaxy samples,
allowing measurement of the cross-correlation coefficients of HI to galaxy
properties such as age, star-formation activity, AGN activity, and halo
mass. The shot-noise on the cross-correlation power spectrum determines the
average HI mass of the optical galaxy sample, constraining the scaling
relation of HI mass to optical galaxy tracers \citep{2017arXiv170308268W}.
An example of these effects can be seen in the lower panel of
Figure~\ref{Fig_HIPS}, which shows the cross-correlation of
$(\rm{u-g})$-magnitude selected galaxy samples with HI intensity mapping
signals given by the EAGLE simulation. If the shot-noise is not taken into
account, the cross-correlations of different galaxy selections exhibit
vastly varying scale-dependent clustering behaviour on smaller scales, which
is relieved once the shot noise is correctly removed as marked by the solid lines.

SKA1-Mid and SKA1-Low will both be equipped to perform HI intensity mapping
observations spanning $0<z<6$. The resulting HI power spectrum measurements
will allow the determination of the scale-dependence of the HI bias, as well
as the absolute amplitude of the HI bias when employing outside constraints
for the HI energy density, $\Omega_{\rm HI}$. The cross-correlations of these observations with galaxy surveys performed by Euclid or LSST will provide additional insights into the coupling of galaxy and halo properties to the HI distribution.

The DM problems on galaxy scales can be studied with the SKA precursors,  
with the help of suitable samples of galaxies and long integration times 
to attain the necessary sensitivity to detect the HI as far out as possible.
The accompanying multi-wavelength optical studies are also reaching a 
great sophistication, so that they hopefully set stringent constraints to
the stellar mass-to-light ratios of galactic disks.

\subsubsection{Thermal characteristics of dark matter}

Determining the magnitude of the DM thermal velocities will give us clues to
unveiling the nature of DM. We already know that DM cannot be hot, that is,
it cannot be mostly made up of particles with large thermal velocities such
as neutrinos, since this would change the structure formation paradigm from
bottom-up to top-down. On the other hand, the possibility of DM having
relatively small thermal velocities (i.e., warm dark matter) is not in
contradiction with  cosmological observations. Currently, the tightest
constraints come from observations of the Lyman-$\alpha$
forest~\citep{Vid_2017} with $m_{\rm WDM}>5.3$ keV at $2\sigma$ confidence, but a large parameter space remains unexplored and could in principle be investigated. 
The SKA can further constrain these WDM properties by measuring the global
21 cm evolution and power spectra in different frequency bands. The following sections summarise the effect of such thermal properties and the prospects for measurement.

\subsubsection{Warm dark matter and the HI power spectrum}
\label{wdm_hi}

The shape and amplitude of the 21 cm power spectrum at different redshifts is sensitive to the abundance, clustering and HI mass function of DM halos. It is expected therefore that the significant impact WDM has on the properties of low-mass halos will result in an observable signature. The impact of WDM on halo properties has been studied in some detail, and analytic formalisms such as the halo model have been extended to include it~\citep{Dunstan_2011}. \citet{Carucci_2015} prepared forecasts by using the results of hydrodynamic simulations with CDM and WDM and pointed out that 5 000 hours of interferometer intensity mapping observations by SKA1-Low can be used to rule out a WDM model with an effective particle mass, $m_{X}$, of 4 keV at $3\sigma$. These are competitive constraints that can complement bounds from independent probes such as those from the Lyman-$\alpha$ forest~\citep{Vid_2017}.

\begin{figure*}
\centerline{
\includegraphics[width=0.5\linewidth]{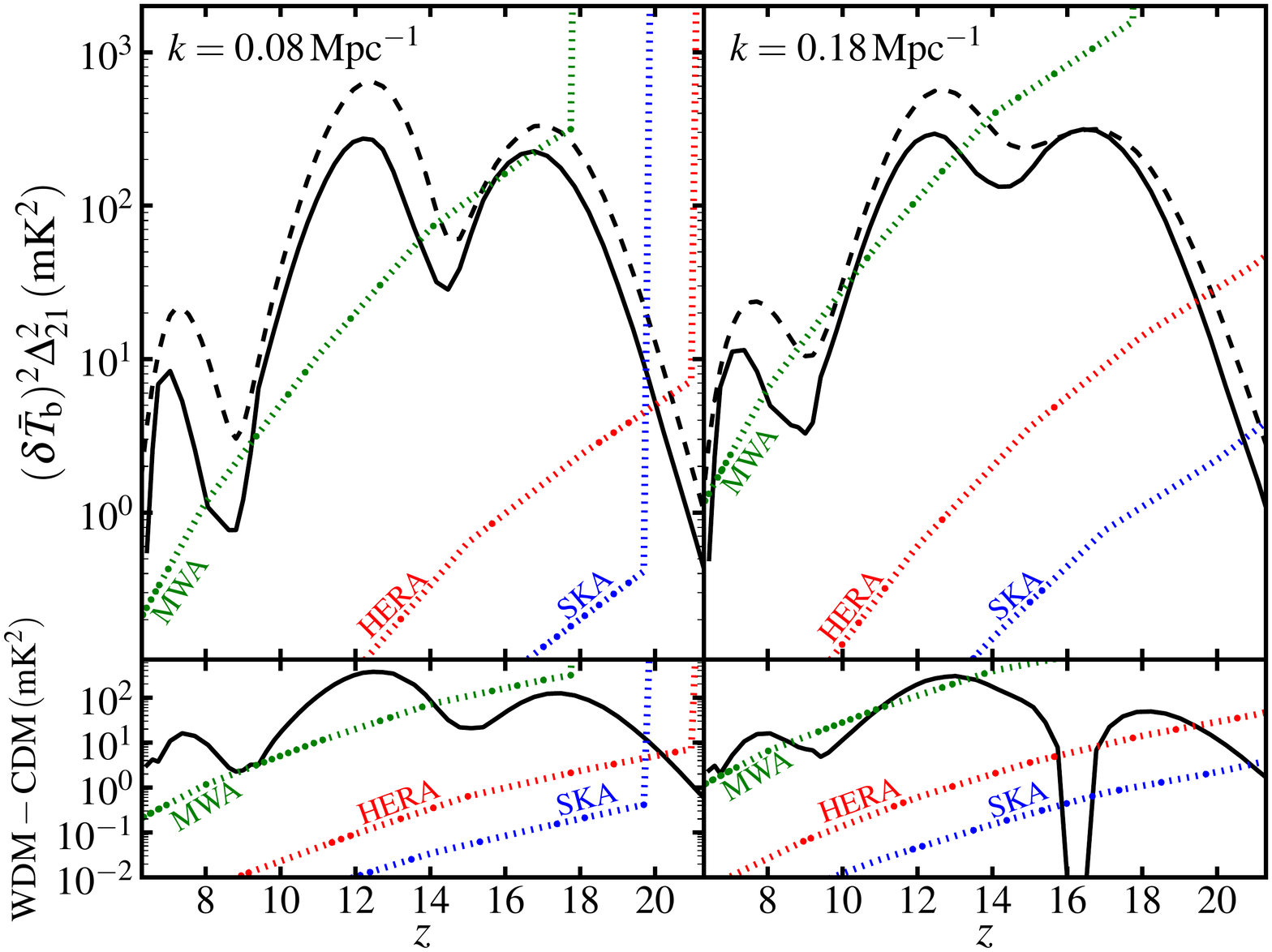}
\includegraphics[width=0.5\linewidth]{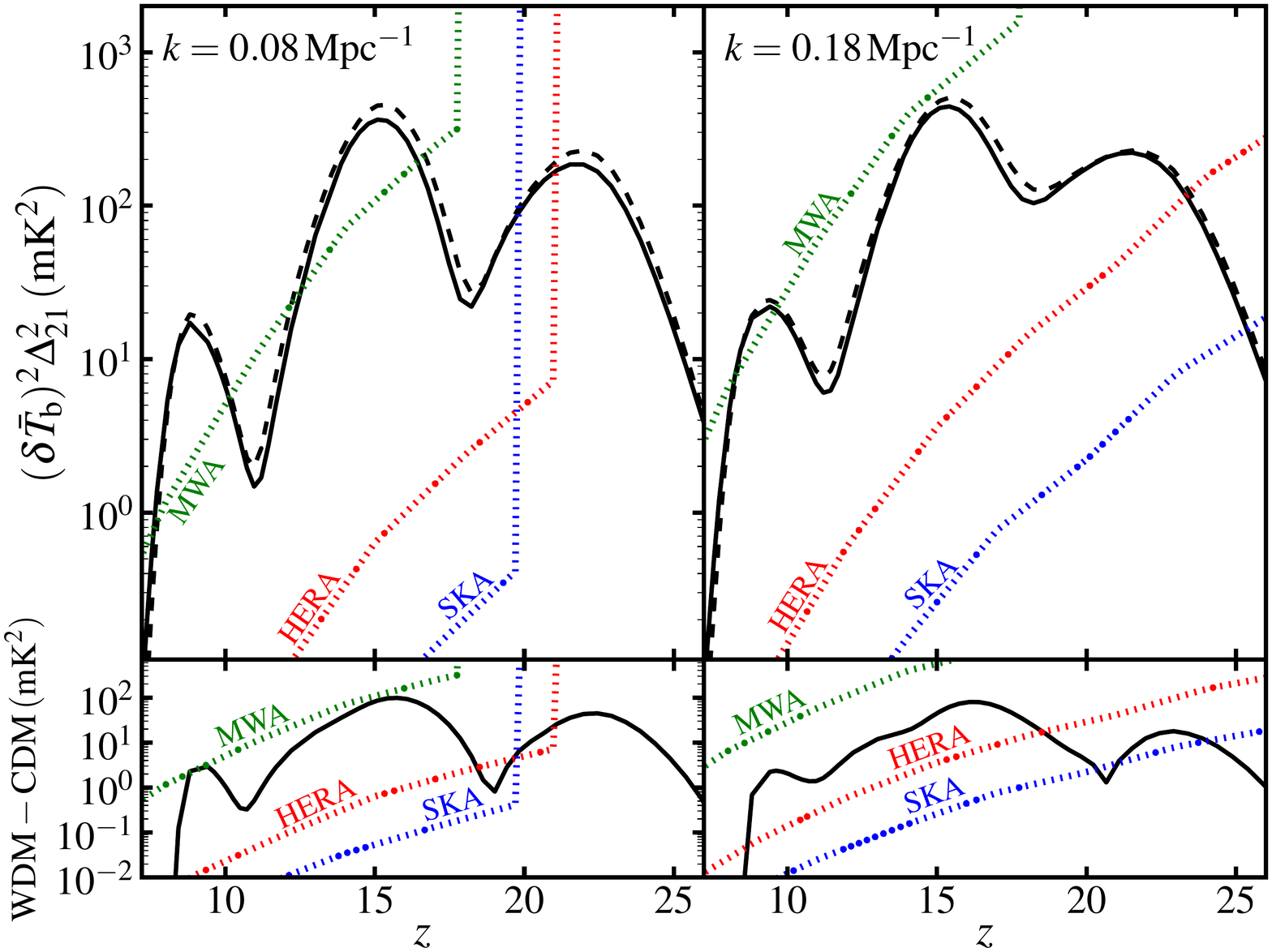}}
\caption{Evolution of the power spectrum of the brightness temperature for WDM with (left panel) $\mx=2 \, \text{keV}$ and (right panel) $\mx=4 \, \text{keV}$. The top panels show power spectra at $k=0.08,~0.18~\text{Mpc}^{-1}$ for the WDM (dashed) and the CDM model (solid). The bottom panel is the subtraction of CDM power spectrum from the WDM power spectrum, showing the difference. The dotted curves show the $1\sigma$ thermal noise power spectrum forecasts computed by~\citet{mesinger2013a} and \citet{Sitwell14} with 2 000 hours observational time. The green, red and blue lines are for MWA, HERA and SKA-Low respectively. Figure taken from~\citet{Sitwell14}.} 
\label{fig:ps_k}
\end{figure*}

To show the evolution of the power spectrum at different scales, Figure~\ref{fig:ps_k} plots the brightness temperature power spectrum as a function of redshift, $(\delta \bar{T}_b)^2\Delta_{21}^2$, for the modes $k=0.08$ Mpc$^{-1}$ and $k=0.18$ Mpc$^{-1}$. One can see a three-peak structure that (moving from right to left) is due to inhomogeneities in the coupling coefficient for scattering of Lyman-$\alpha$ photons, $x_{\alpha}$, kinetic gas temperature, $\Tk$, and the neutral fraction of hydrogen, $x_{\text{HI}}$. At their peak these inhomogeneities could enhance the power at $k=0.08,0.18$ Mpc$^{-1}$ for the WDM model by up to a factor of $2.4, 2.0\,$ for $m_{\rm X}=2\,$keV and $1.3,1.1$ for $\mx=4$ keV model. 

The current interferometric radio surveys such as MWA, HERA and SKA-Low may provide the sensitivity and noise level that are able to detect this boost of power spectrum by WDM. To forecast the capability of these new surveys to place constraints on WDM model, we plot the forecasts of $1\sigma$ thermal noise of the power spectrum computed by \citet{mesinger2013a} and \citet{Sitwell14} for 2 000 hours of observational time for MWA, HERA and SKA-Low in Figure~\ref{fig:ps_k}. One can see that, although MWA's capability is marginally able to detect the boosted power of $\mx=2\,$keV model at reionisation and X-ray heating peak locations, SKA-Low should provide a strong constraint on the excess power on these scales over the range of redshift $10<z<25$.

\subsubsection{Determining thermal properties from the epoch of reionisation}

Free-free emission from an ionised medium can produce a potentially remarkable diffuse signal, particularly at frequencies lower than $\sim 10$ GHz.
The baryonic matter variance can be calculated by integrating the power spectrum of matter density perturbations over an appropriate range of wavenumbers:
\begin{equation} 
\sigma^{2} (z) = \frac{1}{2\pi^{2}} \int P(k, z) k^{2 }dk .
\end{equation} 
In a given cosmological model, the cosmological parameters determine the level of matter density contrast. 
While the structure distribution does not depend significantly on the DM model at large scales \citep{2007Sci...317.1527G}, 
the small scales and, consequently, the amplitude of the clumping factor, $1+\sigma^2 (z)$, are particularly sensitive to the thermal properties of DM particles.

In CDM standard models, cold and essentially collisionless particles had a negligible velocity dispersion in the cosmic epochs relevant for structure formation, and the corresponding power spectrum is then 
essentially undamped up to very high wavenumbers. 
Contrariwise, in WDM models, the intrinsic thermal velocity dispersions related to the particle distribution properties could imply a substantial free streaming process, affecting clustering properties 
and suppressing power spectrum above certain wavenumbers, dependent on particle thermal history and mass. Traditional cold and hot DM particle masses are, respectively, in the range of $\sim 10 - 10^{2}$ GeV and  
$\sim$ few eV, while masses of about $1 - 10$ keV are typically considered for WDM particle candidates \citep{2008PhRvD..77d3518B}, in form of, e.g., gravitinos or sterile neutrinos. 

The evolution of the power spectrum can be described in terms of transfer function and for WDM models it is approximately given by that of the corresponding CDM model
but with with a cut-off $k_\mathrm{max}$ at high wavenumbers, i.e.\ at scales smaller than the scale of free-streaming of the WDM particle.
Values of $k_\mathrm{max}$ in the range $20$ to $10^3$ are usually considered. 
Thus, together with the amplitude of initial perturbations, $k_\mathrm{max}$ mainly determines the values the clumping factor at the relevant redshifts. 
In turn, the damping of inhomogeneities in $\Lambda$WDM models at small scales delays the growth of structures~\citep{2005PhRvD..71f3534V},  
influencing the early stages of star formation history. 

Ultimately, the non-negligible IGM density contrast, related to DM particle properties,
implies an amplification factor of $[\simeq 1 + \sigma^{2} (z)]$
of the diffuse free-free emission triggered by a specific reionisation mechanism \citep{2014MNRAS.437.2507T}.   
Therefore, other than contributing to the deep understanding of the astrophysical reionisation process,
the detailed analysis of the free-free diffuse signal represents a way to study DM properties 
exploiting their influence on the power spectrum at high wavenumbers.

\subsection{Dark Matter Searches}
\label{sec:dmsearch}

The nature of DM will remain a mystery without the clear and unequivocal
detection of its particle physics nature\footnote{For a review on DM direct
and indirect searches see, for example,~\citet{Gaskins:2016,Conrad:2017},
and see~\citet{GambitSUSY2017,GambitMSSM2017} for the accelerator-based status.}. A surprisingly common prediction of many particle physics models is that DM is not completely dark. It can either couple to standard model particles with a very weak interaction or it can self-annihilate or decay, and via cascading processes eventually end up as standard model particles such as neutrinos, photons, positrons and other antimatter elements~\citep{Bertone:2004pz}.

Such particles may be observed through different channels. Neutrinos and
photons, usually in the form of $\gamma$-rays, have zero electromagnetic
charge and consequently maintain their original trajectory. Conversely,
charged particles are effectively isotropised by their tangled propagation
in Galactic magnetic fields. The acceleration of the charged particle
products in the magnetic fields do however provide an additional detection
channel via the emission of secondary radiation such as bremsstrahlung,
Compton or synchrotron radiation. The latter can reach frequencies from MHz
to a few hundred GHz, and needs to be isolated from overwhelming and complex astrophysical backgrounds that mask the expected DM signal both morphologically and in spectral features. The DM signals often possess features that differentiate them from the backgrounds (typically, a non-power-law spectrum and a cut-off at an energy related to the DM mass). This is not observed in the data (for example from Fermi, HESS and MAGIC) and therefore relevant bounds on DM properties (particle mass, annihilation or decay rate) are the typical outcome~\citep[for example,][and references therein]{Gaskins:2016,Conrad:2017}.

The most studied option is the so-called WIMP, elementary particles with a
mass from a few GeV to several TeV, endowed with weak or sub-weak type
interactions. In particular, $\gamma$-ray observations are considered a
promising avenue to probe WIMP scenarios~\citep{Bringmann:2012ez}, but
$\gamma$-rays alone cannot usually be disentangled from the astrophysical
sources, and radio astronomy can play a crucial role in background
determination. Although DM targets (clusters, galaxies, galaxy satellites or
subhalos) appear to be invisible individually, their cumulative emission
might be detectable with advanced techniques. We will discuss possible
general signals in \S\ref{sec:DMgamma} and more specific cases in
\S\ref{sec:DarkestSources}, \S\ref{sec:one} and \S\ref{sec:MSP}. But as we
will discuss in \S\ref{sec:branons} and \S\ref{sec:weakly}, other options should also be considered.

\subsubsection{Photon fluxes from WIMP-like dark matter}\label{sec:DMgamma} \label{Sec:II}

Positrons and electrons propagating in magnetic fields will lose energy
(mainly) due to synchrotron radiation and inverse Compton
scattering~\citep{Sarazin:1999}. Other ways to lose energy, such as
Coulombian interactions or bremsstrahlung are subdominant. For the energies
and magnetic fields relevant for DM positrons in the Milky Way, the
resulting radiation has frequencies from MHz to a few hundred GHz. For
high-energy positrons and electrons propagating in a Galactic environment,
the density per unit energy ($\psi$ in units of cm$^{-3}$~GeV$^{-1}$) is well described as a purely diffusive equation~\citep{Delahaye:2007fr}:
\begin{eqnarray} \label{eqn:transport}
-K_{0}\epsilon ^{\delta}\nabla^{2}\psi-\frac{\partial}{\partial E}(b(E)\psi)=Q(\vec x,E),
\end{eqnarray}
in which the parameters $K_{0}$ and $\delta$ model the diffusion of the positrons in the Galactic magnetic field, $b(E)$ describes the loss of energy, and $\epsilon=E/1{\rm GeV}$.

The source term $Q(\vec x,E)$ contains the information on the source that injects positrons into the environment. If the only primary source of positrons is the annihilation of WIMP particles with mass $M$, the source term becomes  
\begin{eqnarray}\label{diffuse}
 Q(\vec{x},E)=\frac{1}{2}\left\langle \sigma v\right\rangle\left(\frac{\rho(\vec x)}{M_{\text{}}}\right)^2\sum_{i}\beta_{j}\frac{{\rm d}N_e^j}{{\rm d}
E}, 
\end{eqnarray}     
where $\beta_{j}$ is the branching ratio of the different annihilation channels.
The thermally averaged annihilation cross-section, $\langle\sigma v\rangle$, is mainly described by the theory explaining the WIMP physics, whereas the number of positrons and electrons produced in each decay channel per energy interval, $\text{d}\,N_\gamma^i/\text{d}\,E_{\gamma}$ involves decays and/or hadronisation of unstable products (e.g., quarks and gauge bosons) involving non-perturbative effects related to quantum chromodynamics (QCD), which can be obtained from numerical software packages such as DarkSUSY\footnote{\url{http://www.darksusy.org}} or micrOMEGAs\footnote{\url{https://lapth.cnrs.fr/micromegas/}} based on PYTHIA Monte Carlo event generator\footnote{\url{http://home.thep.lu.se/~torbjorn/Pythia.html}}. 

The power of emission is related to the positron kinetic energy by the synchrotron power
\begin{eqnarray}
 P_\mathrm{syn}=\frac{1}{4\pi \epsilon_{0}} \frac{\sqrt{3}e^{3} B}{m_{e} c} y \int^{\infty}_{y} {\rm d}\xi\, K_{5/3}(\xi) \left(\frac{\nu}{\nu_{c}}\right), 
\end{eqnarray}  
where 
\begin{eqnarray}
 \nu_{c} =\frac{3eE^{2} B}{4\pi m_{e}^{3} c^{4}},
\end{eqnarray}

\noindent is defined as a the critical frequency of the emission. In the
above equations, $B$ is the Galactic magnetic field, $m_{e}$ the mass of the
electron, $E$ the kinetic energy of the electron/positrons, $c$ the speed of
light. and $K_{5/3}(\xi)$ is the modified Bessel function of the second kind.

Taking into consideration every positron and electron with a specific energy
as a synchrotron emitter, it is necessary to sum all the possible contributions over the line-of-sight as follows:
\begin{eqnarray}
F(\nu)=\left(\frac{2}{4 \pi}\right) \int_\mathrm{l.o.s} {\rm d}l \int^{M}_{m_{e}}{\rm d}
E'\, P_{\rm syn}(\nu,E') \psi (\vec{x},E')\,.
\end{eqnarray}
\noindent $F(\nu)$ corresponds to the so-called density of radiation (which
can be measured in janskys), and $\psi (\vec{x},E')$ is the number density
of electrons/positrons previously calculated through the diffusion equation
in Equation~(\ref{diffuse}). Focussing on the Milky Way, a diffuse contribution in radio frequencies could be expected due to the disposition of DM in halos.

\subsubsection{Searches for diffuse emission in the darkest sources\label{sec:DarkestSources}}

As described in \S\ref{sec:DMgamma}, DM annihilations may cascade to
non-thermal electrons and positrons, which in turn emit radio waves as
synchrotron radiation in regions where an ambient magnetic field is present.
Therefore, a generic prediction of WIMP models is diffuse radio emission
from DM halos induced by non-gravitational interactions of DM. Its discovery
would be a significant step toward the solution of the DM mystery. The
improved sensitivity of the SKA and its precursors will allow us to reach the sensitivity required to detect the radio flux emitted by DM halos, especially at low redshifts, both in the investigation of the number counts of sources, and in their statistics across the sky.

Intriguingly, a few years ago, the ARCADE-2 collaboration reported isotropic
radio emission that is significantly brighter than the expected
contributions from known extragalactic astrophysical
sources \citep{2011ApJ...734....6S, 2014JCAP...04..008F}
and is well-fitted
by WIMP-induced emission \citep{2011PhRvL.107A1302F}. If the cosmological
signal from DM is at such level, then the contribution from particle DM in
the data from the Evolutionary Map of the Universe
\citep[EMU;][]{2011PASA...28..215N} survey on the Australian Square
Kilometre Array Pathfinder \citep[ASKAP;][]{Johnston:2008hp} should be
significant.

\begin{figure}
\vspace{-3.7cm}
\centering
\includegraphics[width=0.5\textwidth]{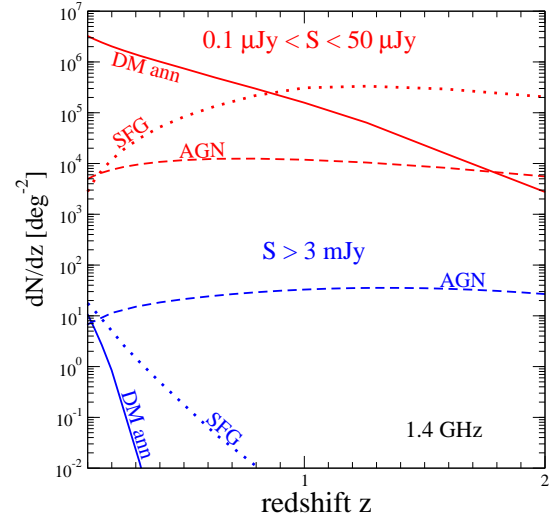}
\caption{Redshift distribution of sources for bright ($S> 3$ mJy, lower in
blue) and faint ($0.1 \mu\rm{Jy} < S < 50 \mu\rm{Jy}$, upper in red)
samples,  showing that DM annihilation will be visible in the faint source number count but not in the bright source number count. Benchmark models for 
astrophysical and DM radio sources are taken from~\citet{2012JCAP...03..033F}.}
\label{fig:dNdz_DM}
\end{figure}

Figure~\ref{fig:dNdz_DM} shows the number of sources as a function of redshift in two brightness ranges for the DM component~\citep[for model A of][which fits the ARCADE excess]{2011PhRvL.107A1302F}, compared to more mundane astrophysical sources that explain the source counts observed so far.
The emission from DM is mainly provided by faint sources at low redshift,
and the median size of the source is large ($\gtrsim$~arcmin). This implies
that previous analyses of counts or angular correlation were not sensitive
to a WIMP-induced signal. A relevant exception
is~\citet{2015MNRAS.447.2243V}, who interestingly reported a possible
deviation associated with faint extended sources. Figure~\ref{fig:dNdz_DM}
shows that while the DM contribution is subdominant in the range probed by
NRAO Very Large Array Sky Survey (NVSS, above mJy), it can become relevant
for fluxes within the sensitivity reach of the SKA and its precursors. 

The analysis of number count fluctuations in SKA data and the angular auto- and cross-correlation of the SKA density field are thus promising techniques to test the DM interpretation of the ARCADE excess and, more generally, to constrain the WIMP parameter space~\citep{2012JCAP...03..033F}.

\subsubsection{Branons as WIMP candidates}\label{Sec:III}\label{sec:branons}

Massive brane fluctuations (branons) provide an example of a DM candidate
that is detectable or constrainable with the SKA. They arise in brane-world models with low tension~\citep{2003AIPC..670..235C,PhysRevLett.90.241301,KUGO2001301}.

\begin{figure}
\centering
\includegraphics[width=9cm,height=7cm]{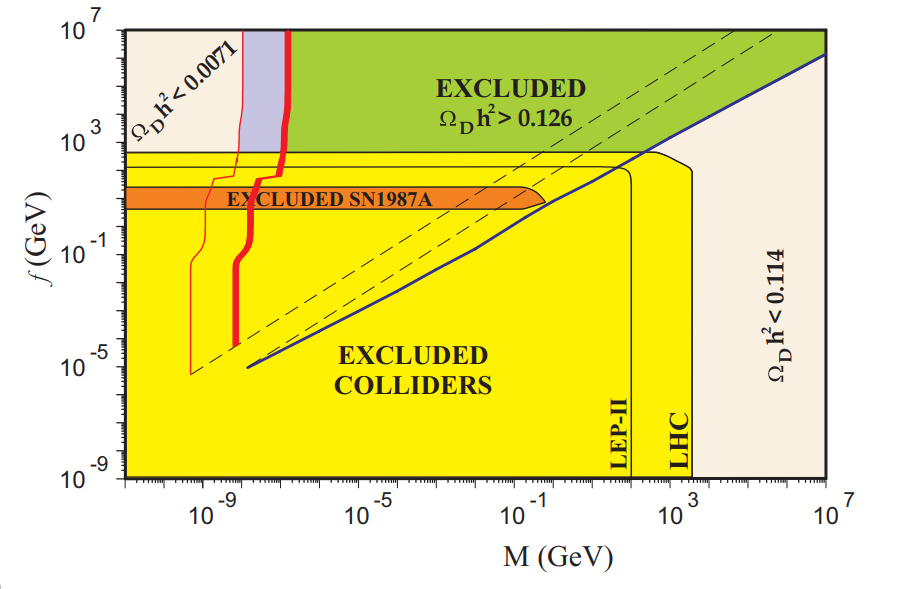}
\caption{Combined exclusion plot for a branon model with a single disformal
scalar from total and hot DM \citep[taken from][]{Cembranos:2016jun}
including constraints from LEP-II \citep{PhysRevD.67.075010,ACHARD2004145}
and LHC
\citep{PhysRevD.70.096001,Cembranos:2011cm,Landsberg:2015pka,Khachatryan:2014rwa}
single photon event, and supernova cooling \citep{Cembranos:2003fu}. The two solid (red) lines on the right are associated with the hot DM; the thicker line corresponds to the total DM range $\Omega_{D} h^{2}= 0.126-0.114$
~\citep{Ade:2015xua} and the thin line is the hot DM limit $\Omega_{D} h^{2}<0.126-0.114$. The solid (blue) line along the diagonal corresponds to CDM behaviour, and the dashed lines corresponds to $M/T_\mathrm{freeze-out} = 3$ for hot (upper curve) and cold (lower curve) DM.}
\label{fig:2}
\end{figure}

Massive branons are pseudo-scalar fields that can be understood as the pseudo-Goldstone bosons corresponding to the spontaneous breaking of translational invariance in the bulk space. The broken symmetry is precisely produced by the presence of the brane~\citep{PhysRevD.59.085009, PhysRevLett.83.3601, Dobado:2000gr}.
Branons are prevented from decaying into standard model particles by parity
invariance on the brane, but can annihilate into standard model particles,
although the coupling is suppressed by the brane tension scale. The branons
are mass degenerate, and consequently the annihilation fluxes only depend
upon two parameters of the effective theory describing the low-energy
dynamics of flexible brane-worlds, namely the brane tension scale and the
branon mass, $M$ \citep{PhysRevD.65.026005, Cembranos:2001my,PhysRevD.70.096001,PhysRevD.67.075010}.
Bounds and constraints on the model parameters from cosmology and tree-level processes in colliders are shown in Figure~\ref{fig:2} \citep{KUGO2001301,PhysRevLett.90.241301,PhysRevD.69.043509,PhysRevD.69.101304}.
Further astrophysical and cosmological bounds serve to parametrise the tension in terms of the branon mass, rendering the dynamics dependent on the mass alone \citep{KUGO2001301,PhysRevLett.90.241301,PhysRevD.69.043509,PhysRevD.69.101304}.

For branons, the thermally averaged annihilation cross-sections depend solely upon the spin and the mass of the branon \citep{Cembranos:2003fu,Cembranos:2006mj}. 
In the case of heavy branons $M \gg m_{W,\,Z}$, the main contribution to the
indirect photon flux comes from branon annihilation into bosons, $W^+W^-$
and $ZZ$. The contribution from heavy fermions, that is the annihilation
into top-antitop, can be shown to be subdominant. On the contrary, whenever
$M \ll m_{W,\,Z}$, the annihilation into $W$ or $Z$ bosons is kinematically forbidden so that the remaining channels must be considered, mainly annihilation into the heaviest possible quarks~\citep{Cembranos:2011hi}.

\begin{figure}
\centering
\includegraphics[width=6.3cm]{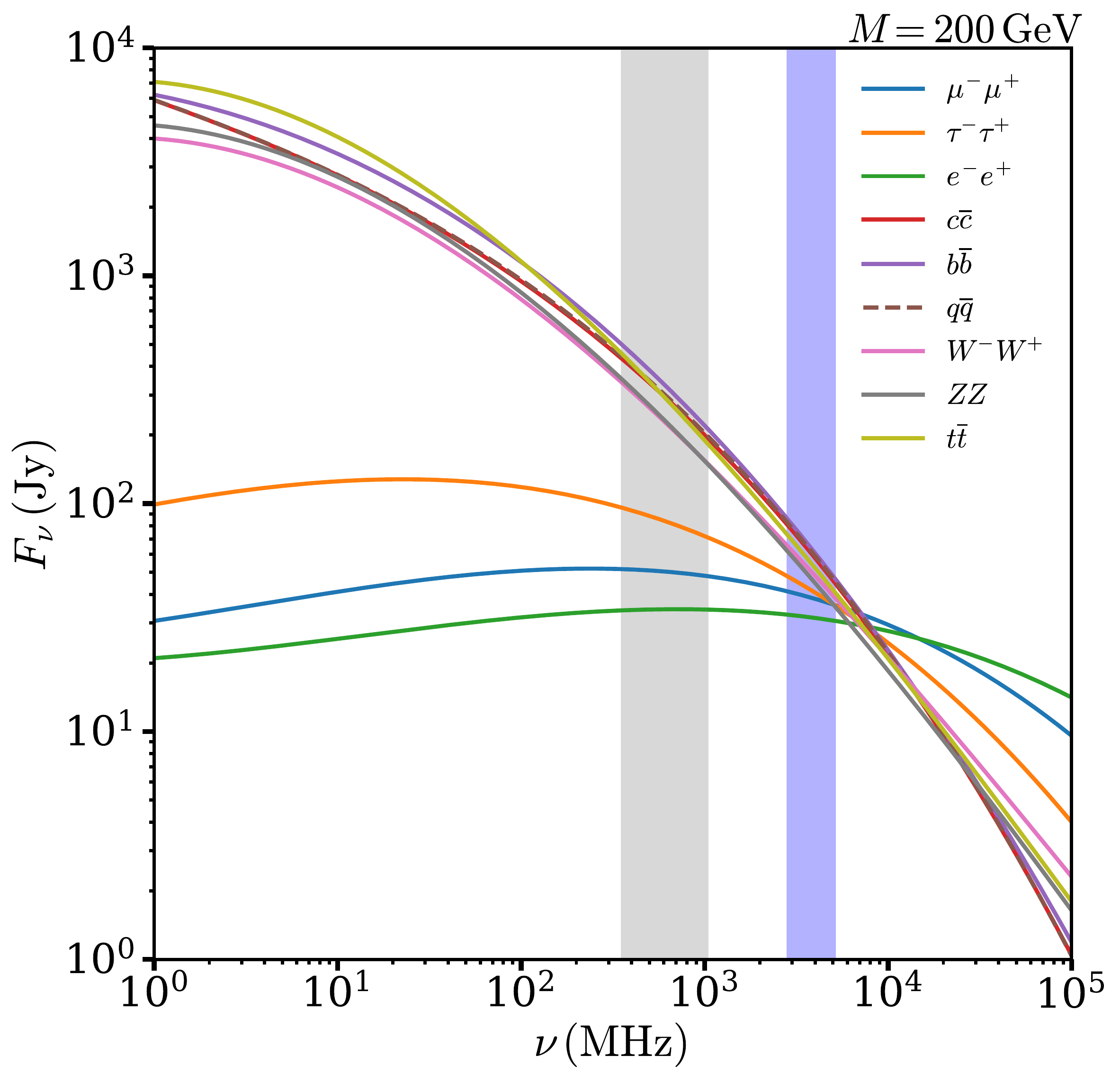}
\includegraphics[width=6.3cm]{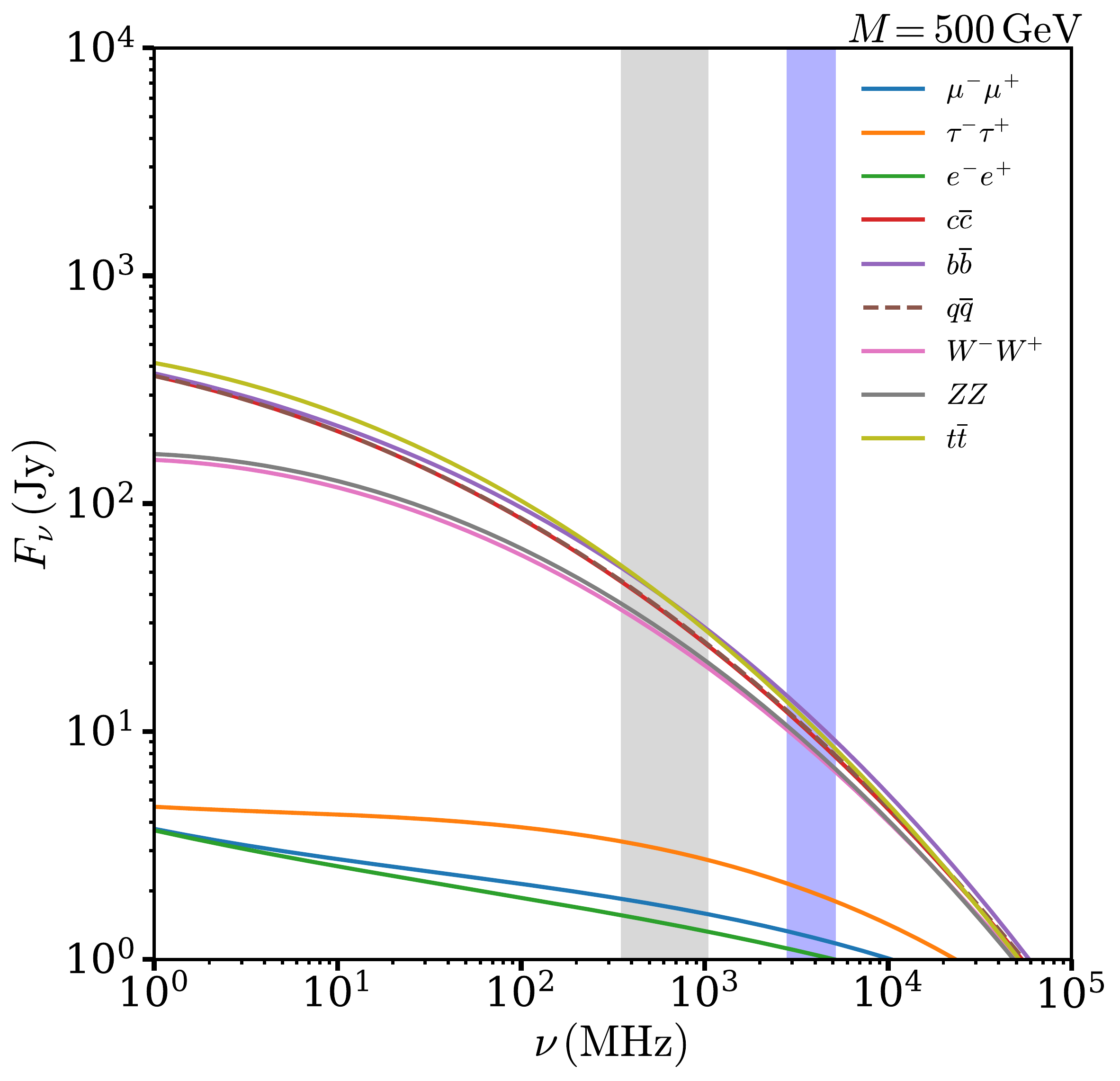}
\includegraphics[width=6.3cm]{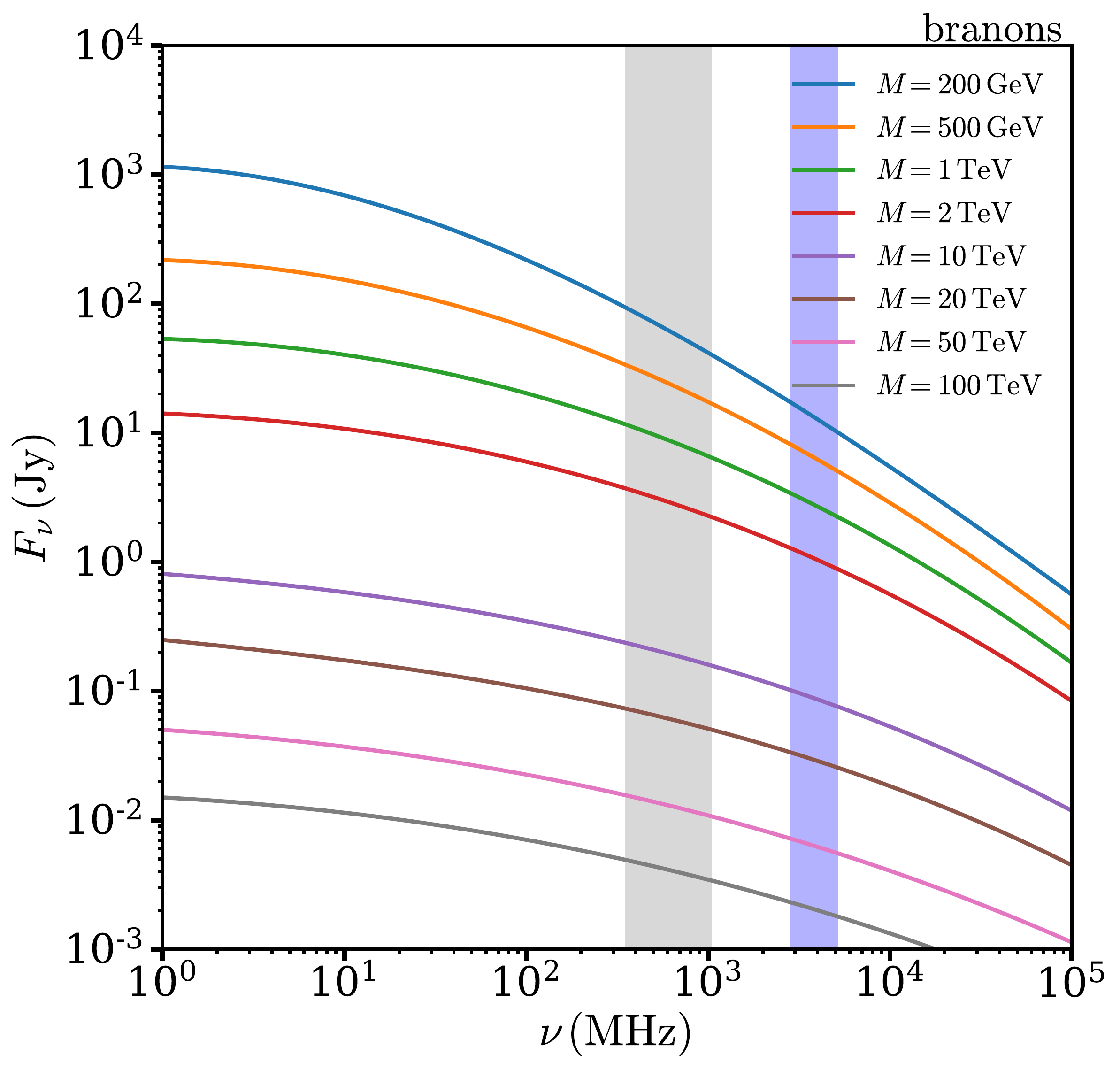}
\caption{Synchrotron density fluxes from the Galactic Centre ($l=0^{ o},b=0^{ o}$ in galactic coordinates) for different channels of branons spontaneous annihilations. Channel $q\overline{q}$ corresponds to the annihilation via $u\overline{u}$, $d\overline{d}$ and $s\overline{s}$ . The first panel represents a DM mass of $\text{M}=200$ GeV and the second one is for a mass of $\text{M}=1000$ GeV. The diffusion considered was $\text{K}_{0}=0.00595\,\text{kpc}^{2}/\text{MYr}$, $\delta=0.55$ and $\text{L}_{\text{z}}= 1$ kpc and the DM density profile used was the isothermal. The considered magnetic field is $6 \mu \text{G}$. As we can see, the synchrotron signal decreases more drastically in the case of the $W^{+}W^{-}$, $Z^{+}Z^{-}$ bosons and $q\overline{q}$, $t\overline{t}$, $b\overline{b}$ and $c\overline{c}$ quarks than the signal of the leptonic channels. The signal increases at low frequencies showing the suitable ranges to detect the signature. No boost factors are considered in this figure. In the {\it lower panel}, the signature decrease as a function of mass has been exemplified for a model with one extra dimension with a tension of $f=8.25 M^{0.75}$ \citep{PhysRevLett.90.241301,Cembranos:2003fu}}
\label{fig:4}
\end{figure}

Figure ~\ref{fig:4} shows the expected synchrotron flux densities from DM annihilations in the Galactic Centre for various annihilation channels (upper and middle panels) for a generic WIMP or different branon masses (lower panel). Consequently, the SKA would have the potential to disentangle different model-independent DM masses and the preferred annihilation channel(s) (upper and middle panels) as well as different branon candidates (lower panel) by combining different frequency ranges. For example, one can see the SKA1-Mid band-1 (0.35-1.05 GHz) and SKA1-Mid band-4 (2.80-5.18 GHz), and the sensitivity to both the DM annihilation channels and also DM masses \citep[see][]{Colafrancesco:2015ola}. In addition, since the SKA minimum detectable density flux would lie on the $\sim\mu\text{Jy}$ range \citep{2019arXiv190511154C}, such inferences would be feasible for the depicted fluxes.

In Figure~\ref{fig:4} we show the predicted fluxes for branon masses in the
range of $200$ GeV to $100$ TeV, so that the principal channels of
annihilation are via $W$ and $Z$ bosons and the top quark. Qualitatively, we
observe that the radio emission shape depends on the annihilation channel,
potentially providing information about the nature of the branon. In
addition, we observe that for branons (with a brane tension that depends on
the mass), the expected signature decreases with the mass. That means that
detectable signatures can be associated to low masses of this
extra-dimensional particle. Furthermore, this methodology of obtaining the flux density will allow us to discard regions of parameter space in the case where we observe a smaller signal than predicted by experimental results.

Finally, a monochromatic $\gamma$-ray line is expected at the energy equal
to the branon mass as a consequence of direct annihilation into photons.
This annihilation takes place in the $d$-wave channel. Consequently it is
highly suppressed, and is not detectable with current instruments
\citep[e.g., Fermi;][]{Cembranos:2011hi}. However, masses above 150~GeV can
potentially be detected with the increased sensitivity of the Cherenkov
Telescope Array, in which case cross-correlation with a synchrotron signal
from the SKA becomes a crucial test.

\subsubsection{Cross-correlation of SKA1 HI galaxies and $\gamma$-rays}\label{sec:one}

Faint sources of emission from DM might not be detectable on their own, but
they contribute a cumulative component. This method builds on a recent
proposal to use cross-correlations between a gravitational tracer of DM
(cosmic shear or galaxy clustering as proxies for the DM distribution in the
Universe) and any DM-sourced electromagnetic signals~\citep{Camera:2012cj,
Fornengo:2013rga}. In addition, it has the potential to bring redshift
information to the electromagnetic signal that is otherwise unavailable,
exploiting the different behaviour of DM emission peaking at low redshift
and the unresolved astrophysical production of the same observables more
pronounced towards intermediate redshift~\citep[for
example]{Camera:2012cj,Camera:2014rja,Fornengo:2014cya,2015ApJS..217...15X,Branchini:2016glc,Troster:2016sgf}.
Specifically, we discuss here the impact of the catalogue of HI
galaxies that will be obtained by the SKA on the cross-correlation with
Fermi-Large Area Telescope (LAT) $\gamma$-ray maps.

Currently, the vast majority of the $\gamma$-ray sky is unresolved and only a few thousand $\gamma$-ray sources are known. The two frequency regimes where, on large scales, non-thermal emission mechanisms are expected to exceed greatly any other process are the low-frequency radio band and the $\gamma$-ray range.
Radio data are thus expected to correlate with the $\gamma$-ray sky and can
be exploited to filter out information on the composition of the $\gamma$-ray background contained in unresolved $\gamma$-ray data.

\begin{figure}
\centering
\includegraphics[trim={0cm 1cm 2.5cm 9cm},clip,width=0.5\textwidth]{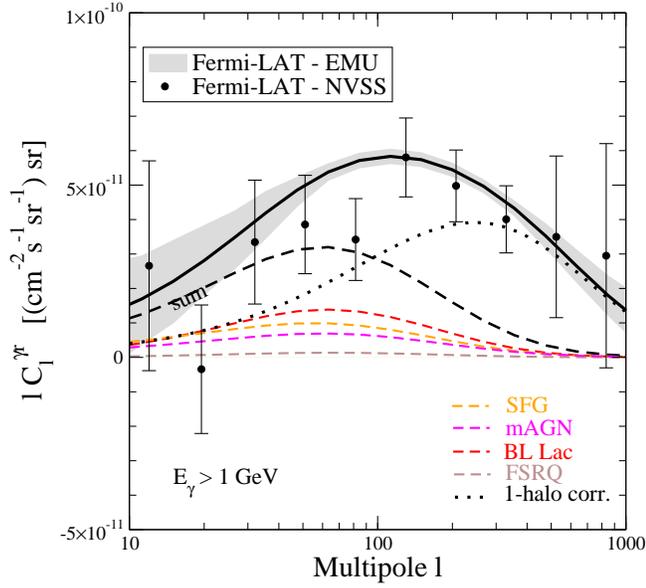}
\caption{Angular power spectrum of cross-correlation between the unresolved
$\gamma$-ray background and the distribution of radio sources. Data points
refer to the measurement performed using Fermi-LAT and NVSS data, the solid
curve shows a reference model, and the grey area reports the expected
sensitivity for the cross-correlation between EMU and Fermi-LAT data. For
details concerning data and models see, \cite{2015ApJS..217...15X}
and \cite{Cuoco:2015rfa}, respectively.}
\label{fig:CAPS_gamma}
\end{figure}

Indeed, a cross correlation between the unresolved $\gamma$-ray background seen by the Fermi-LAT telescope and the distribution of sources detected in continuum in the NVSS~\citep{Condon:1998} catalogue has been recently detected~\citep{2015ApJS..217...15X}. The angular power spectrum data are shown in Figure~\ref{fig:CAPS_gamma} together with a reference model~\citep{Cuoco:2015rfa}.
Combining this measurement with other catalogues, relevant constraints on the composition of the $\gamma$-ray background can be derived~\citep{2015ApJS..217...15X,Cuoco:2015rfa,2019arXiv190713484A}.
The improvement in sensitivity offered by ASKAP and
the SKA is dramatic, as shown in
the example of the EMU survey (grey area) in Figure~\ref{fig:CAPS_gamma}. SKA and precursor data will therefore allow us to discriminate between different explanations for the composition of the $\gamma$-ray background.

Following on from these seminal observational results and the techniques and
forecasts first proposed by \citet{Camera:2012cj,Camera:2014rja}
and~\citet{Fornengo:2013rga}, we present some preliminary results on the
cross-correlation of SKA1 HI galaxies and the $\gamma$-ray sky from Fermi.
Note that, contrary to NVSS and EMU, in this case we use HI galaxies for
which spectroscopic redshifts will be available. This will allow us to
exploit fully the tomographic-spectral approach outlined
in~\citet{Camera:2014rja}. Moreover, there is major added value in the use
of SKA1 HI galaxies, their redshift distribution peaking at low redshift and
having an extremely low shot-noise~\citep[see Fig.~4 of][]{Yahya:2014yva}.
This is the very regime where the non-gravitational DM signal is strongest.
The emission associated with WIMP decay is proportional to the DM density,
and consequently stronger at low redshift because the produced radiation is
diluted by the expansion of the Universe more rapidly than its source. The
WIMP annihilation signal, proportional to the density squared, also peaks at
low redshift since the density contrast associated with cosmic structures grows non-linearly at late times.

Specifically, we adopt an SKA1 HI galaxy survey with specifics given by
\citet{Yahya:2014yva} for the baseline configuration. We consider only
galaxies in the redshift range $0<z\le0.5$, which we further subdivide into
ten spectroscopic, narrow redshift bins. For the $\gamma$-ray angular power
spectrum, we employ the fitting formulae found by~\citet{Troster:2016sgf}
for Pass-8 Fermi-LAT events gathered over eight years through to September 2016. This is a very conservative approach, as by the time the SKA1 HI galaxy catalogue will be available, Fermi-LAT will have produced a much larger amount of data.

\begin{figure}
\centering
\includegraphics[width=\columnwidth]{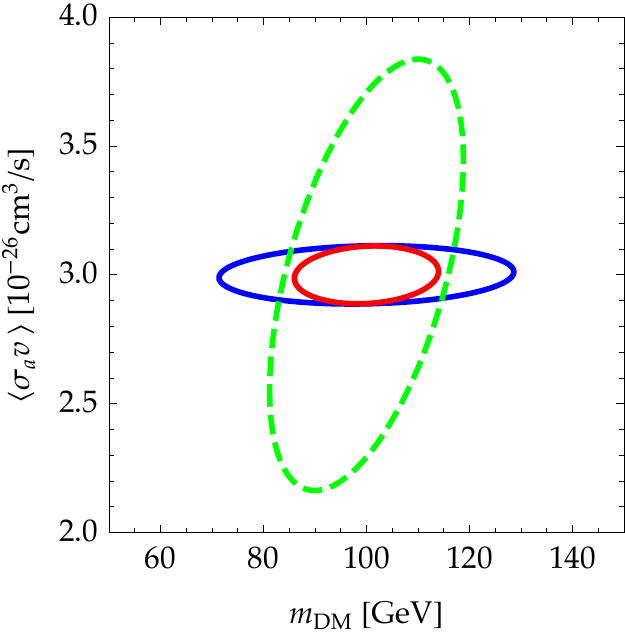}
\caption{Joint 1$\sigma$ marginal error contours on WIMP parameters for Fermi LAT $\gamma$-ray data cross-correlated with DES cosmic shear (green), SKA1 HI galaxies (blue) and their combination (red).}\label{fig:ellipse}
\end{figure}

Figure~\ref{fig:ellipse} preliminarily shows the precision with which we
will be able to reconstruct the WIMP cross-section and mass in the case of a
DM candidate with $m_{\rm DM}=100$~GeV and thermal cross-section
$3\times10^{-26}$~cm$^3$~s$^{-1}$. The green contour~\citep[the same as in
Fig.~~4 of][]{Camera:2014rja} refers to the cross-correlation of Fermi-LAT
$\gamma$ rays and cosmic shear from DES, while the blue contour depicts the
constraining power of an SKA1-Fermi joint analysis. The main take-home
message here is the high complementarity of the two techniques, the
combination of which (red ellipse) has the potential to tightly constrain {\em both}\
WIMP mass and cross-section.

\subsubsection{SKA1-Mid and Millisecond Pulsars} \label{sec:MSP}
Given the large DM density in the inner Galaxy, any WIMP annihilation signal
is expected to be particularly bright from that direction.  Interestingly, a
signal candidate has been found in $\gamma$-ray data from the
Fermi-LAT~\citep{Goodenough:2009gk, Vitale:2009hr}. The existence of an
excess emission in the GeV energy range ($\sim 100 \, \rm MeV - 10 \, \rm
GeV$), above conventional models for the diffuse $\gamma$-ray emission from
the Galaxy, is now firmly established
\citep{Abazajian:2014fta,Calore:2014xka,TheFermi-LAT:2015kwa,Daylan:2014rsa,Fermi17a}.
However, the signal may also have an astrophysical origin, e.g., from the
combined emission of thousands of millisecond pulsars in the Galactic bulge
\citep{Wang:2005ti, Abazajian:2010zy}, young pulsars \citep{OLeary:2016cwz},
or stellar remnants from disrupted globular clusters \citep{Brandt:2015}.

We briefly summarise how the sensitivity of the SKA to detect the bulge population of millisecond pulsars can be estimated.  Following~\citet{1984bens.work..234D}, the root mean square uncertainty of the flux density (in mJy) is given by
\begin{equation} 
  S_{\nu, \rm rms} = \frac{T_{\rm sys}}{G \, \sqrt{t_{\rm obs} \,  \Delta
    \nu \, n_p}} \left(\frac{W_{\rm obs}}{P-W_{\rm obs}}
\right)^{1/2} \,, 
\label{eq:Snurms}
\end{equation}
where $T_\text{sys}$ refers to the system temperature, $G$ the instrument gain, $t_\text{obs}$ the observation time, $\Delta\nu$ the bandwidth and $n_p=2$ the number of polarisations.  Furthermore, $P$ is the pulse period, and $W_\text{obs}$ the effective pulse width.  We adopt the parameters corresponding to SKA1-Mid observations \citep{Dewdney2015}.
More specifically, we adopt a central frequency of $1.67$ GHz, a $770$ MHz
bandwidth, a receiver temperature of $25$ K, and an effective gain of
8.5~K~Jy$^{-1}$ for the considered sub-array~\citep[see][for
details]{Calore:2015bsx}.  The beam full-width half-maximum is 0.77~arcmin, and we
assume 3000 synthesised beams and 20~minutes of integration time per pointing.

\begin{figure}
  \includegraphics[width=0.9\linewidth]{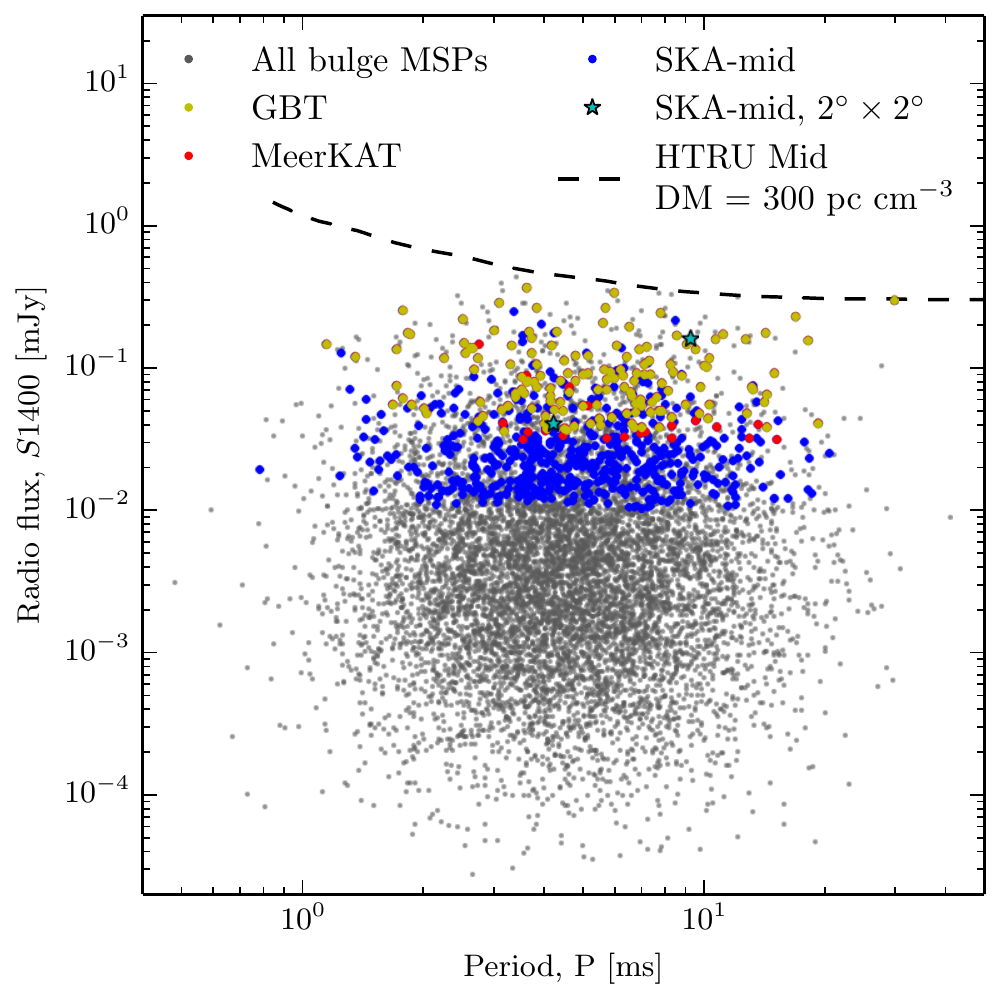}
  \caption{Distribution of the period vs.~flux density 
    of simulated bulge millisecond pulsars (\textit{grey dots}).  Sources detectable by the three 
    observational scenarios described by \cite{Calore:2015bsx} are represented by
    \textit{coloured dots}. The improvement of SKA1-Mid with respect to the GBT and MeerKAT is
    represented by the \textit{blue points}. ``SKA-Mid $2\times2$" refers to sources detectable in 
    the inner $2^\circ\times2^\circ$ about the Galactic centre.
    The \textit{dashed black line}
    is the flux sensitivity of the Parkes High Time Resolution Universe
    mid-latitude survey (for DM~$= 300$~pc~cm$^{-3}$).
    Figure adopted from~\citet{Calore:2015bsx}.}
  \label{fig:simulation}
\end{figure}

We find that a dedicated search in the region ($|\ell|<5^\circ$ and
$3^\circ<|b|<7^\circ$) plus ($|\ell|<3^\circ$ and $1^\circ<|b|<3^\circ$)
plus ($|\ell|,|b|<1^\circ$), which would take about 90 hours, can detect 207
bulge and 112 foreground sources at $10\sigma$ significance or higher.  This
number is about 7 times larger than the number of sources expected for a
similar survey with MeerKAT.  As shown in Figure \ref{fig:simulation}, the
SKA will be able to detect three times fainter sources than what is accessible by an equivalent MeerKAT survey, or by targeted observations with the Green Bank Telescope (GBT).

The large number of detections will mark significant progress for our understanding of the millisecond pulsar bulge population, and hence backgrounds for DM searches in the inner Galaxy.  First, the large number of detections will allow us to determine the number of radio-bright bulge millisecond pulsars down to $10\%$ precision, and to measure the distribution of sources and confirm that they indeed correspond to the morphology of the Fermi GeV excess. Second, the large number of measured ephemerides will be useful for searches for $\gamma$-ray pulsations in the Fermi LAT data, and likely allow us to significantly increase the part of the inner Galaxy $\gamma$-ray emission that can be directly attributed to millisecond pulsars.  Third, anticipating a better determination of a potential correlation between $\gamma$-ray and radio emission, these results can lead to relatively clear predictions for the $\gamma$-ray emission from the Galactic bulge that can then be subtracted from DM signal searches.

Finally, we stress that a non-detection of a significant number of bulge
millisecond pulsars with the SKA would practically exclude the millisecond
pulsar hypothesis as explanation for the Fermi GeV excess, and strengthen
the case for a DM signal, unless the radio emission of millisecond pulsars
is a factor 10--100 weaker than what is suggested by globular clusters,
while keeping the $\gamma$-ray emission unchanged.

\subsubsection{Extremely weakly interacting dark matter candidates}\label{sec:weakly}

Looking for radio signals could be equally (if not more) important than
searching for anomalous $\gamma$-ray production. Radio signatures have
already ruled out GeV DM particles with thermal interactions, have
constrained DM models in general \citep[e.g.,][]{Boehm:2002yz,Crocker:2010gy,Boehm:2010kg,Fornengo:2011iq,Bringmann:2014lpa,Cirelli:2016mrc}, and could further be used to probe the existence of extremely weakly interacting dark matter.

\paragraph{Black hole shadow} \label{sec:BHShadow} The DM density profile
increases in the inner regions of galaxies, but the `spikiness' of the
profile is under debate.  One hypothesis is that the DM density profile
becomes as steep as $\rho \propto r^{-7/3}$ near the central black hole,
referred to as a DM spike.  The formation and survival of DM spikes is
controversial (partly due to galaxy dynamics), however one can test their
existence if one further assumes that DM interacts (even very weakly) with
standard model particles.  Indeed, taking the specific case of heavy DM
particles annihilating into standard model particles, it was shown that the
presence of a DM spike in M87 leads to a copious production of synchrotron
emission in the frequency range and spatial region that is currently being
probed by the Event Horizon
Telescope.\footnote{http://www.eventhorizontelescope.org/}. As the
additional radiation from DM further enhances the photon ring around the
black hole shadow for any value of the self-annihilation cross section
greater than $10^{-31}$~cm$^3$~s$^{-1}$ (assuming a 10~GeV candidate), one
should be able to confirm the presence of DM spikes even in scenarios of
light p-wave annihilating DM candidates \citep{Lacroix:2016qpq}. 

\paragraph{Dark Ages} 
Alternative  DM scenarios include particles with self and/or interactions with standard model particles in the early Universe. One consequence of these interactions is the damping of the primordial DM fluctuations, leading to a small-scale cut-off in the primordial power spectrum~\citep[see for example][]{deLaix:1995vi,Boehm:2000gq,Boehm:2001hm,Boehm:2004th,Mangano:2006mp}.  
A late-time manifestation of these effects is the suppression of small-scale companions of the Milky Way as well as a smaller number of  small-structures in the Universe as a whole, which becomes even more prominent as one goes back in time \citep{Boehm:2014vja,Schewtschenko:2014fca,Schewtschenko:2015rno,Cyr-Racine:2015ihg,Vogelsberger:2015gpr,Moline:2016fdo}. By probing the Dark Ages, the SKA 
will be able to measure the primordial power spectrum at high redshift and probe a potential suppression of power due to the nature of the DM.

\paragraph{Axions} \label{sec:QCDaxions}
An alternative to the WIMP dark matter model is that some or all of the dark matter is comprised of QCD axions.  

Axion-two photon coupling in the astrophysical environment may result in an
observable DM signature, with the strength and shape of the axion signature
strongly dependent on the relative properties of the magnetic field.
Assuming that the axion comprises a substantial component of the CDM
density, conversion in a static magnetic field will produce a line profile
with central frequency principally derived from the mass of the axion, that
is in the range $0.2 - 200$ GHz.  The width and polarisation of the signal
is then dependent on the velocity distribution of the axion, the relative
movement of the Earth and the source, and the polarisation of the magnetic
field itself (see also \S\ref{sec:CR} on the SKA's ability to trace magnetic fields).  In fact, the polarisation and strength of the axion signal should trace the spatial profile of the magnetic field.  This is of particular use in extracting the axion signature from other foreground signals, since the polarisation should be perpendicular to synchrotron radiation.

\begin{figure}
\centering
\includegraphics[width=\columnwidth,trim={4.5cm 3cm 6cm 4cm},clip]{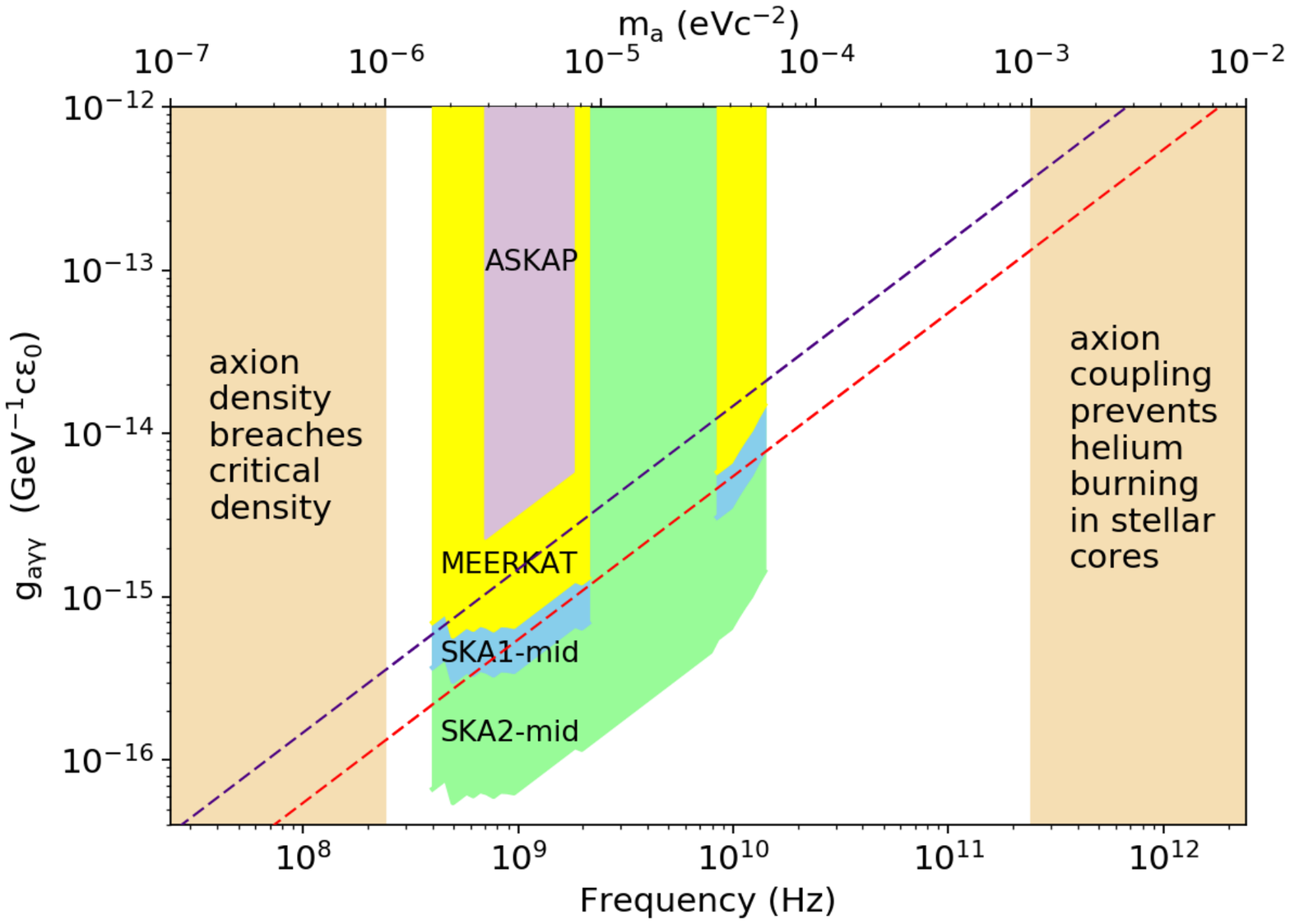}
\caption{The coupling strength that could be probed by observing the
interstellar medium across the frequency range accessible to ASKAP, MeerKAT
and SKA-Mid. The sensitivities of SKA1-Mid and SKA2-Mid (blue and green,
respectively) show considerable
improvement on the pre-cursor telescopes, ASKAP
(purple) and MeerKAT (yellow). The system
temperature of the SKA is minimised between $\sim 2 - 7$ GHz, corresponding
to an axion mass of $\sim 8.26 - 28.91\ \mu$eVc$^{-2}$ and providing a good
opportunity for detection of both KSVZ and DFSZ axions. Figure reproduced
from~\citet{2017ApJ...845L...4K} by permission of the AAS.}.
\label{fig:axion_ska}
\end{figure}

With such distinguishing features, new technologies offer the opportunity
for astrophysical observations to make a significant contribution to axion
search efforts.  Laboratory experiments searching for the axion have
received significant investment in recent years but remain sensitive only to
the most optimistic axion model and operate over small areas in frequency
space.  SKA2 in particular offers a significant improvement in the axion
coupling strength that can be probed and in the breadth of frequency space
that can be observed.  To demonstrate the potential power of the SKA and its
precursors, Figure~\ref{fig:axion_ska} \citep[taken
from][]{2017ApJ...845L...4K} shows the axion-two photon coupling strength
that could be probed with observations of the interstellar medium, the
parallel lines from the bottom-left to top-right showing the coupling
strength expected from the Kim-Shifman-Vainshtein-Zakharov (KSVZ) and
Dine-Fischler-Srednicki-Zhitnitsky (DFSZ) models \citep{1979PhRvL..43..103K,
1980NuPhB.166..493S,1981PhLB..104..199D}.  Further constraint of this
parameter space with the SKA, or indeed detection of the axion itself, would
represent an important step forward in both particle physics and cosmology.

\paragraph{Primordial Black Holes 
\label{sec:PBHs}}

\begin{figure}
\includegraphics[width=\columnwidth]{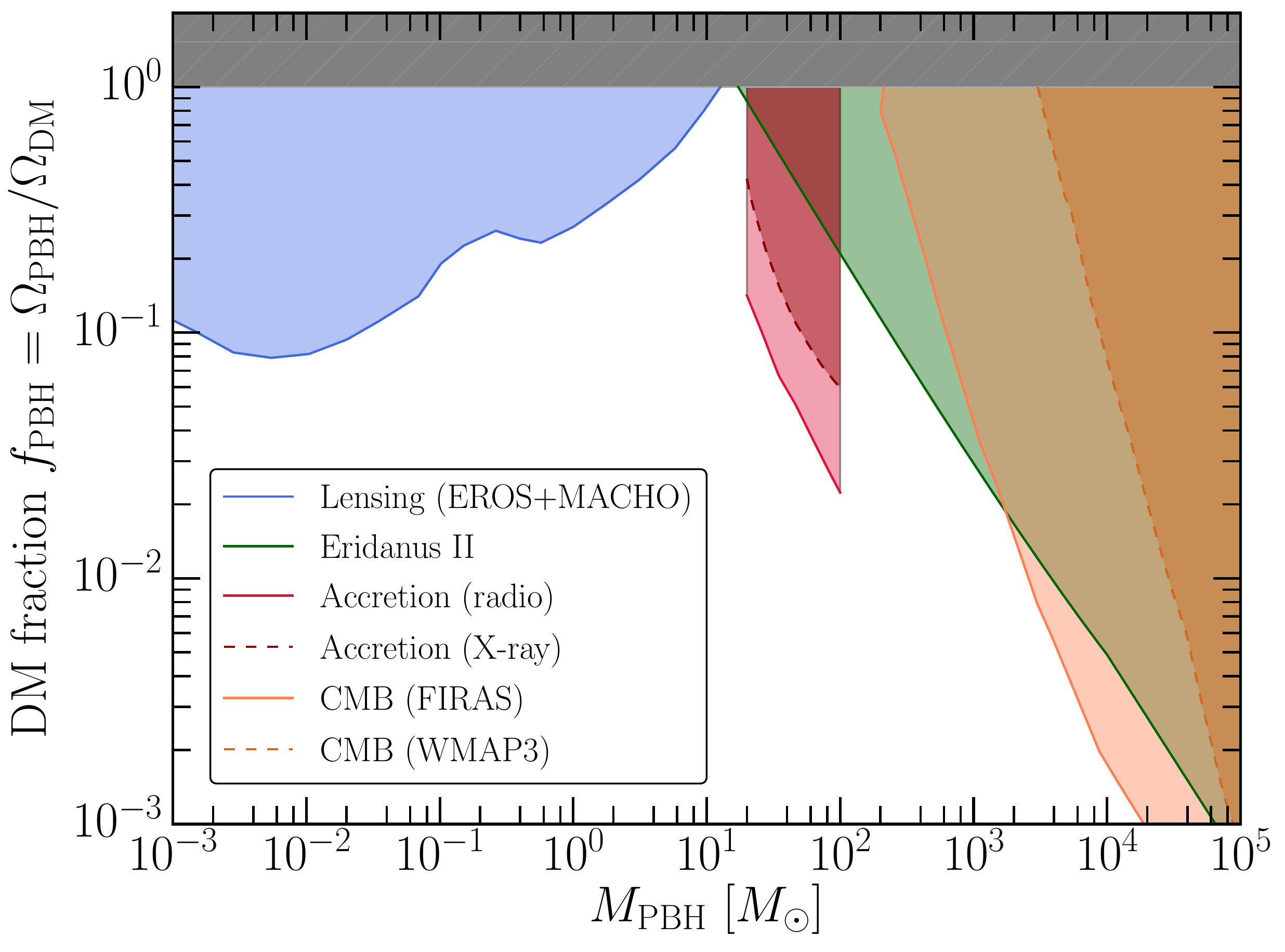}
\caption{Summary of astrophysical constraints on PBHs in the mass range $M \in [10^{-3},\,10^5]\,M_\odot$.}
\label{PBH_frac}
\end{figure}

The idea that PBHs can collapse in the early Universe out of small-scale density fluctuations (possibly originated during inflation) dates back to the early 1970s \citep[see, e.g.,][]{Hawking:1971ei,CarrHawking}. The PBH mass is of order the horizon mass at formation and can in principle span a very large range, from the Planck scale all the way up to $\mathcal{O}(10 - 100)$ M$_{\odot}$. For masses larger than $10^{-17}$ M$_{\odot}$, the PBH evaporation lifetime
(due to Hawking-Bekenstein radiation) is larger than the age of the Universe, and PBHs in that range can be a viable DM candidate, as first outlined by \cite{chapline}. However, observations have allowed astrophysical constraints to be placed on the mass ranges for which PBHs contribute significantly to the DM density, see Figure \ref{PBH_frac}~\citep{cksy,1607.06077}. 

The recent LIGO detection of gravitational waves from binary-black-hole mergers has in particular prompted a renewed interest in the PBH mass window around $10-100\ M_{\odot}$. The PBH merger rate was initially found to be
consistent with that expected where PBHs make up all of the DM in the Universe \citep{1603.00464}, but subsequent analyses have contradicted this \citep{sasaki,Ali-Haimoud:2017rtz}. SKA observations will allow this mass range to be further investigated through measurement of third-order Shapiro time delay induced by PBHs in MSP timing~\citep{schutz}, or by looking for radio signatures both in the astronomical and cosmological context. On the cosmological side, the redshifted HI line is a very interesting observable. In fact, a population of PBHs is expected to accrete gas during the Dark Ages (see \S\ref{sec:overview}) and significantly change the reionisation history of the
Universe: this effect can be probed by the SKA up to redshifts $\simeq 30$ \citep{Poulin2017}. On the astronomical side, it is possible to look for radio and X-ray sources in the sky associated with a population of PBHs distributed in the Galaxy and accreting interstellar gas.  There also remain other mass windows open in the context of DM, in particular the lunar-mass range $10^{20}-10^{24}$ g and the atomic-size range $10^{16}-10^{17}$ g. Pulsar timing with the SKA could be an important probe of the mass range $10^{22}-10^{28}$~g, even if these PBHs are highly subdominant~\citep{kashiyama}.

\begin{figure}
\begin{center}
\includegraphics[width=\columnwidth]{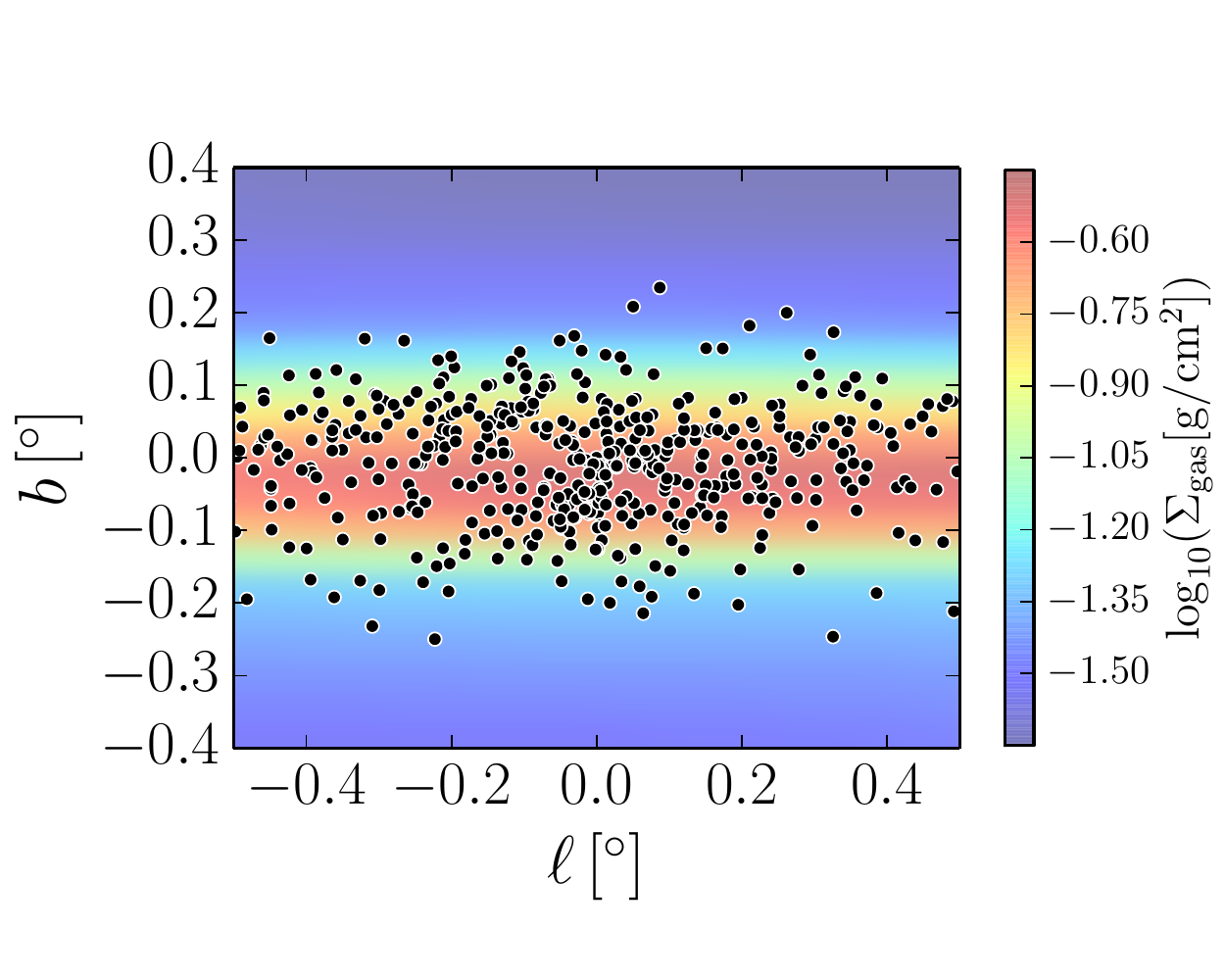}
\caption{Radio sources above the SKA1-Mid point source sensitivity, for 1000
hours of data taking, if PBHs are $\sim 1\%$ of the DM.}\label{fig:SKAsky}
\end{center}
\end{figure}

\underline{Detecting PBHs using radio sources.}
PBHs can in principle be discovered by measuring the radio and X-ray emission produced by the accretion of interstellar gas onto these objects~\citep{Gaggero2017}. By comparing the predicted number of sources with astronomical catalogues, it may be possible to further constrain the fraction of DM in the form of PBHs. Even if they represent a subdominant contribution to dark matter, the SKA may still allow their discovery. Let us now demonstrate this claim in more detail, following the ideas and the approach of \cite{Gaggero2017}. We focus on a small region of interest that includes the high density region of the Galactic ridge.  The ridge is a promising region because it both represents a peak in the density of interstellar gas and therefore the strength of accretion, and resides near the Galactic centre where the DM density is also maximised. For radio emission, we assume that a jet is launched and adopt the universal
empirical relation known as the fundamental plane \citep{Plotkin:2011dy}. We exploit this relation to compute the 5-GHz radio flux and assume a flat radio spectrum, such that $F_{5GHz}=F_{1.4GHz}$. First, we consider the projected bound for the SKA1-Mid (band 2, $0.95 - 1.76$ GHz) point-source sensitivity. Assuming PBHs of mass $\sim30M_{\odot}$ do account for all of the DM, and with a Monte Carlo simulation, we predict a detection of $\simeq 2000$ sources in our region of interest ($<1^\circ$ away from the Galactic centre) for 1 hour of exposure. In Figure~\ref{fig:SKAsky}, we show the predicted map of radio sources above the SKA1-Mid sensitivity threshold (for 1h exposure) when the DM fraction is reduced to only $1$\%. Even in this case, corresponding to a subdominant population of PBHs, SKA1-Mid can detect a large number of sources. However, a detailed calculation of the SKA1-Mid detection sensitivity, correctly accounting for other radio source backgrounds, is postponed to a future work. 

\underline{PBHs and quantum gravity.}
Different approaches to quantum gravity converge in pointing out the
possibility of instabilities of quantum-gravitational origin that can
manifest in an explosive event in a time scale shorter than the evaporation
time~\citep{Gregory:1993vy,Kol:2004pn}.
In particular, loop quantum gravity has recently provided a framework to
compute explicitly this time~\citep{Christodoulou:2016ve}. Loop quantum
gravity removes the classical curvature
singularities~\citep{Ashtekar:2005qt,Rovelli:2013osa,Corichi:2015xia}, such as the one at the black hole centre, because of quantum spacetime discreteness. The consequences of this discreteness on the dynamics can be modelled at the effective level by an effective potential that prevents the gravitational collapse from forming the singularity and triggers a bounce. The bounce connects a collapsing solution of the Einstein equation, that is the classical black hole, to an explosive expanding one, a white hole~\citep{Haggard:2014fv}, through an intermediate quantum region. This process is a typical quantum tunnelling event, and the characteristic time at which it takes place, the hole lifetime, can be as a decaying time, similar to the lifetime of conventional nuclear radioactivity.  The resulting picture is conservative in comparison to other models of non-singular black holes. The collapse still produces a horizon, but it is now a dynamical horizon with a finite lifetime, rather than a perpetual event horizon. The collapsing matter continues its fall after entering the trapping region, forming a very dense object whose further collapse is prevented by quantum pressure~\citep[referred to as a Planck Star;][]{Rovelli:2014cta}. 

While this fate should be generic for all black holes, it becomes experimentally relevant only for tiny and old black holes, being the primordial ones. The collapsing matter that forms PBHs in the radiation-dominated epoch is mainly constituted  by photons. Seen from the centre of the hole, those photons collapse through the trapping region, then expand passing through an anti-trapping region and eventually exit the white-hole horizon, always at the speed of light; the process is thus extremely fast. On the other hand, for an observer sitting outside the horizon, a huge but finite redshift stretches this time to cosmological times. This time, properly called the hole lifetime, as discussed before, has a minimal duration of $M_{\rm BH}^2$ and a maximal duration below $M_{\rm BH}^3$. In analogy with standard quantum-decay processes, one may expect the shortest possible time $M_{\rm BH}^2$ to be favoured.

Astrophysical signals produced in the explosive event associated with the
black hole decay could be  detectable directly.  
Various signals can be expected \citep{Barrau:2015uca,2018arXiv180808857B}: (1) a high-energy signal determined by the
temperature of the photons emitted,  (2) a signal determined by the size of
the hole exploding, (3) a signal in the radio due to the possible presence
of magnetic fields around the exploding hole, and (4) the emission in
gravitational waves. 
The peak of signal (2) falls at millimetre wavelengths; however,
the full distribution of events is accessible to SKA1-Mid \citep{2018PhRvD..97f6019B}. Interestingly, the signal presents a peculiar wavelength-distance relation \citep{Barrau:2014yka}, which allows it to be discriminated from other astrophysical sources, either via direct detection, or via the resulting background radiation \citep{Barrau:2015uca} that SKA intensity mapping may detect, especially with the improved sensitivity of SKA2-Mid.

Signal (3)  is fully within the frequency range of SKA1-Mid and SKA2-Mid.  The interaction of signal (1) with the ionised interstellar medium produces a radio pulse at a frequency $\sim1$~GHz
\citep{Rees:1977nat,Blandford:1977}. Interestingly, this signal has similar properties to those of fast radio bursts. In this case, the emission mechanism relies on the presence of a shell of relativistic charged particles produced in the explosion. The shell behaves as a superconductor that expels the interstellar magnetic field from a spherical volume centred on the original black hole site 
\citep{Cutchin:2016gbu}.

Finally,  PBH decay has the peculiar property of lowering the DM energy-density content of the Universe, since the decay effectively converts DM into radiation. This affects the galaxy number count in large-scale galaxy surveys, in particular measurements of galaxy clustering, galaxy lensing and redshift-space distortions \citep{Raccanelli:2017one}. The large-scale structure surveys performed by the SKA will provide key data in this respect by detecting individual galaxies in the radio continuum \citep{Jarvis:2015asa}.

\subsection{Astroparticle Physics}
\label{sec:astrop}

In this section we discuss how the SKA can constrain the masses of photons
(\S\ref{sec:photonmass}) and neutrinos  (\S\ref{sec:neutrinomass} and
\S\ref{sec:lssneutrinos}). We also consider the problem of cosmic ray
acceleration in magnetic fields, and how the SKA can improve the relatively
little knowledge we have about magnetic fields (\S\ref{sec:CR}).

\subsubsection{Constraining the neutrino mass}\label{sec:neutrinomass}
Determining the sum of the neutrino masses and their hierarchy is one of the most important tasks of modern physics. Unfortunately, setting upper bounds from laboratory experiments is very challenging. It is expected that in the near future \textsc{katrin}\footnote{https://www.katrin.kit.edu} will set an upper limit of $M_\nu=\sum_i m_{\nu_i}<0.6$ eV. A different way to determine the sum of the neutrino masses is through cosmological observables, where their very large thermal velocities (in contrast with the assumed negligible ones for CDM) produce a clear neutrino signature, in particular a suppression of power on small scales in the matter power spectrum. Understanding and measuring this effect is also important for dark energy and general relativity tests, as models of modified gravity or interacting DM/energy also lead to modifications of small scales power \citep[see, for example,][]{Wright:2017dkw}.    

Current constraints on the sum of the neutrino masses, arising by combining data from the CMB, galaxy clustering and/or the Lyman-$\alpha$ forest, are $M_\nu\lesssim0.12$ eV \citep{2014PhRvD..89j3505R,Palanque_2015, Cuesta_2016, Vagnozzi_2017}. One would naively expect that those constraints can be improved by using galaxy clustering at higher redshifts, since the available volume is much larger, the non-linear clustering effects are weaker, and the effects of dark energy will be smaller.

The possibility of using high-redshift optical galaxy surveys (combined with CMB data in order to lift parameter degeneracies) to provide precision measurements of the neutrino masses is not new~\citep[see for example][]{Takada:2006}.  However, the detection of galaxies at high redshifts becomes more difficult and expensive, and shot-noise effects may dominate. In~\citet{Takada:2006}, a space-based galaxy survey with a $300$ deg$^2$ sky coverage at redshifts $3.5<z<6.5$ (assuming a very large number density and bias of the galaxy tracers) was found to be able to measure the neutrino mass with  $\sigma(m_{\nu,{\, \rm tot}}) = 0.025$ eV combined with CMB data. 

Another possibility is to map the large scale structure of the Universe through 21 cm intensity mapping. Given the fact that neutrinos modify the abundance of halos \citep{Castorina_2014,Costanzi_2014}, their clustering \citep{Villaescusa-Navarro_2014, Castorina_2014} and also the internal halo properties such as concentration \citep{Villaescusa-Navarro_2013}, it is expected that they will also leave a signature  on the abundance and spatial distribution of cosmic HI in the post-reionisation era. This has been explicitly checked by means of hydrodynamic simulations by~\citet{Villaescusa-Navarro_2015}. The key point to understand the impact of neutrino masses on the abundance and clustering properties of HI is the fact that halos of the same mass have very similar HI content, independently of the sum of the neutrino masses. 
The results show that in cosmologies with massive neutrinos the abundance of cosmic HI will be suppressed with respect to the equivalent massless neutrino model. At the same time, the presence of massive neutrinos will make the HI more clustered. 

\begin{figure}[t]
\includegraphics[width=\columnwidth]{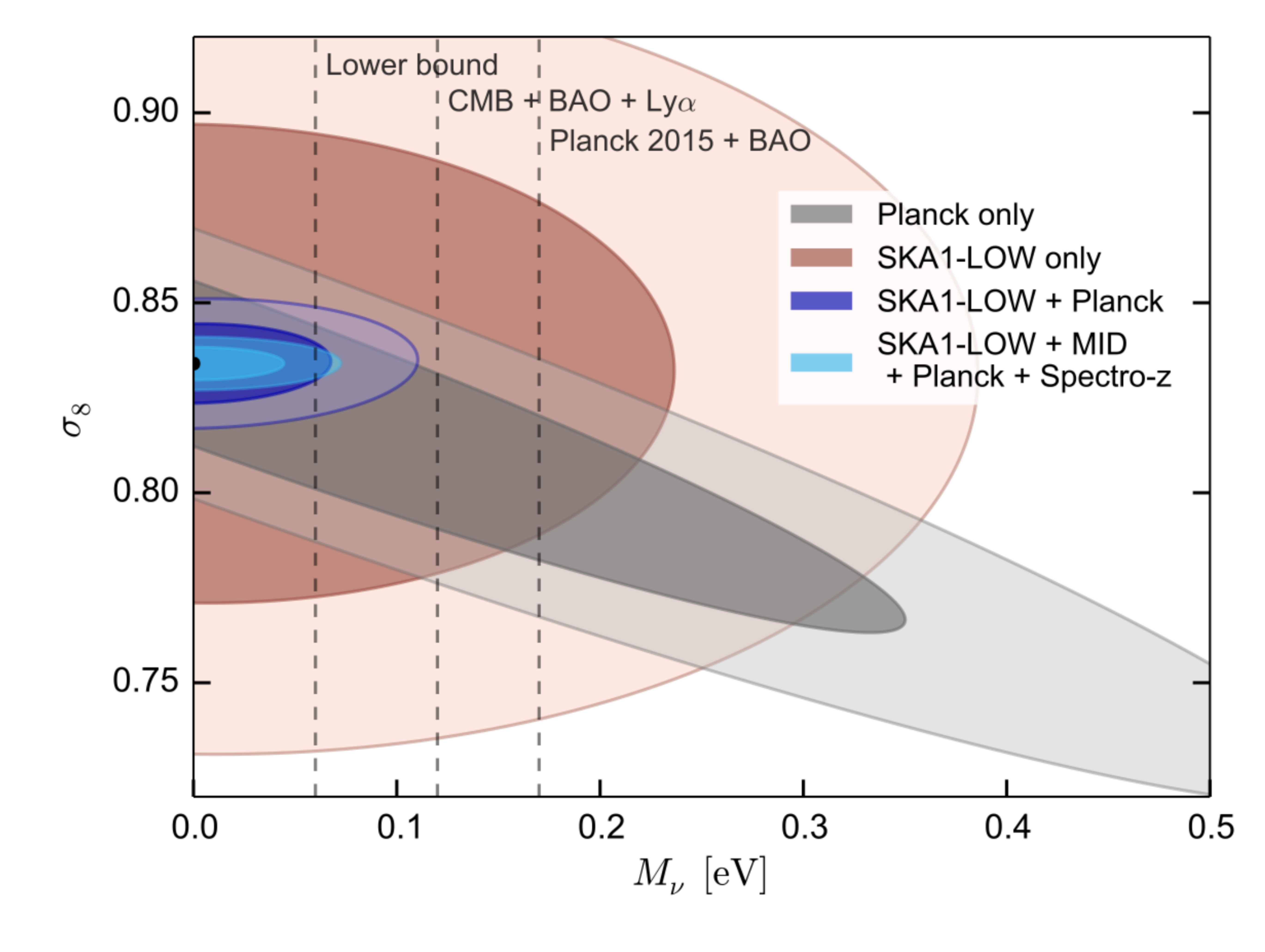}
\caption{\label{fig:mnu} Constraints on the $M_\nu-\sigma_8$ plane from Planck (grey), SKA1-low (brown), SKA1-low + Planck (dark blue) and SKA1-low + SKA1-mid + Planck + Euclid (light blue). The vertical dashed lines indicate the minimum sum of the neutrino masses from neutrino oscillations together with recent bounds from cosmological probes. Adapted from \citet{Villaescusa-Navarro_2015}.}
\end{figure}

\citet{Villaescusa-Navarro_2015} investigated the constraints that intensity mapping observations using SKA1 can place on the sum of the neutrino masses. They considered a deep survey by SKA1-low covering $\simeq20$ deg$^2$ with $10,000$ hours using interferometry over a bandwidth covering redshifts $z\in[3-6]$, and a wide SKA1-mid survey covering $20,000$ deg$^2$ over $10,000$ hours using the single-dish mode of observation from $z=0$ to $z=3$. As shown in Figure \ref{fig:mnu} the neutrino mass can be constrained with $\simeq0.09$ eV ($1\sigma$). It is important to note that, for the given observing time, the constraints from SKA1-low remain practically constant up to a survey area of $100$ deg$^2$, and similarly for SKA1-mid down to $2,000$ deg$^2$. The constraints are more sensitive to the total available observation time. By combining intensity mapping observations with data from Planck and optical galaxy surveys like Euclid the uncertainty can shrink to $\simeq0.03$ eV ($1\sigma$), which is very competitive with respect to any other probe such as galaxy clustering probes or the Lyman-$\alpha$ forest.

\subsubsection{Constraining neutrino properties with SKA voids}\label{sec:lssneutrinos}

\begin{figure}[t]
\includegraphics[width=\columnwidth,trim={0cm 0 0cm 0},clip]{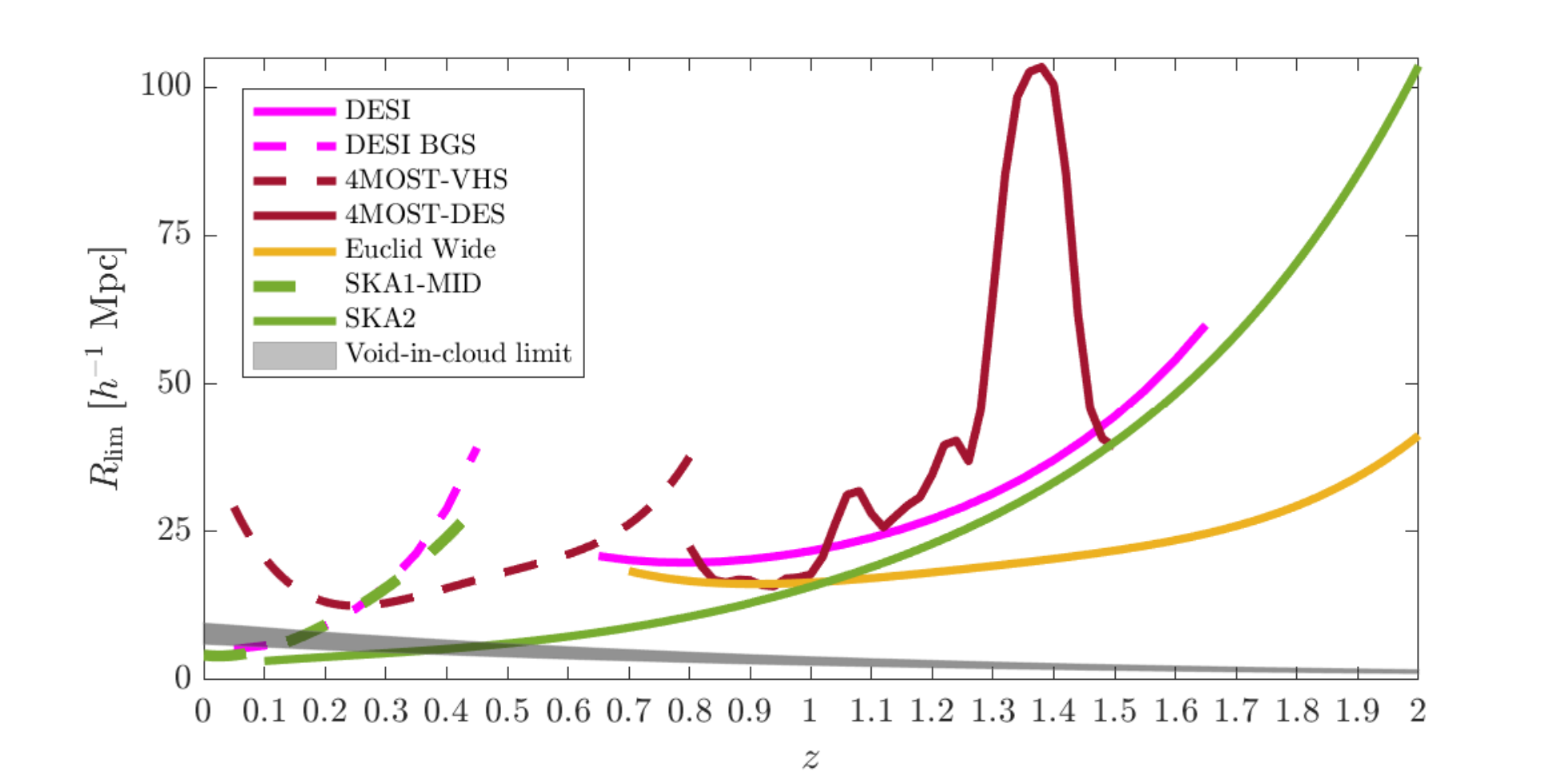}
\caption{\label{fig:voidlimrad} Expected limiting void radii for future spectroscopic galaxy surveys (not including quasars) across the corresponding survey redshift ranges. An approximate void-in-cloud limit is indicated (shaded), below which theoretical predictions are uncertain as regards to what extent voids inside overdensity clouds disappear due to halo collapse of the overdensity.}
\end{figure}

Future SKA HI galaxy surveys will offer an unprecedented spectroscopic view of both large and small scales in the cosmic web. This will allow the identification and mapping of around $10^5 - 10^6$ voids in the galaxy distribution, from the smallest to the largest voids in the Universe \citep{Sahlen:2016kzx}. Figure \ref{fig:voidlimrad} shows the expected limiting void radii for a selection of future large, spectroscopic surveys (we limit the discussion to spectroscopic surveys to minimise the impact of redshift-space systematics). The number counts \citep{Pisani:2015jha, Sahlen:2015wpc}, shapes \citep{Massara:2015msa}, RSDs \citep{Sutter:2014oca}, and lensing properties \citep{Spolyar:2013maa} of voids are examples of sensitive void probes of cosmological parameters. Voids are particularly sensitive to the normalisation and shape of the matter power spectrum, and the effects of screened theories of gravity which exhibit a modification to General Relativity in low-density environments \citep{Voivodic:2016kog}. This is because void distributions contain objects ranging from the linear to the non-linear regimes, across both scale and redshift. SKA2 will also reach well into the void-in-cloud limit across a wide range of redshifts, allowing a detailed study of this theoretically uncertain process whereby small voids disappear through the collapse of the larger overdensities within which they arise.

As a particular case for the SKA, we consider number counts of voids, and forecast cosmological parameter constraints from future SKA surveys in combination with Euclid, using the Fisher-matrix method. Massive neutrinos affect void and galaxy cluster distributions by shifting the turn-over scale in the matter power spectrum as set by the redshift of matter-radiation equality. Their free-streaming also suppresses power on the neutrino free-streaming scale (set by the neutrino masses), which significantly affects the number counts and shapes of voids \citep{Massara:2015msa} and the number counts of clusters \citep{Brandbyge:2010ge}.

\begin{figure}[t]
\includegraphics[trim={0cm 0cm 0cm 0cm},clip,width=\columnwidth]{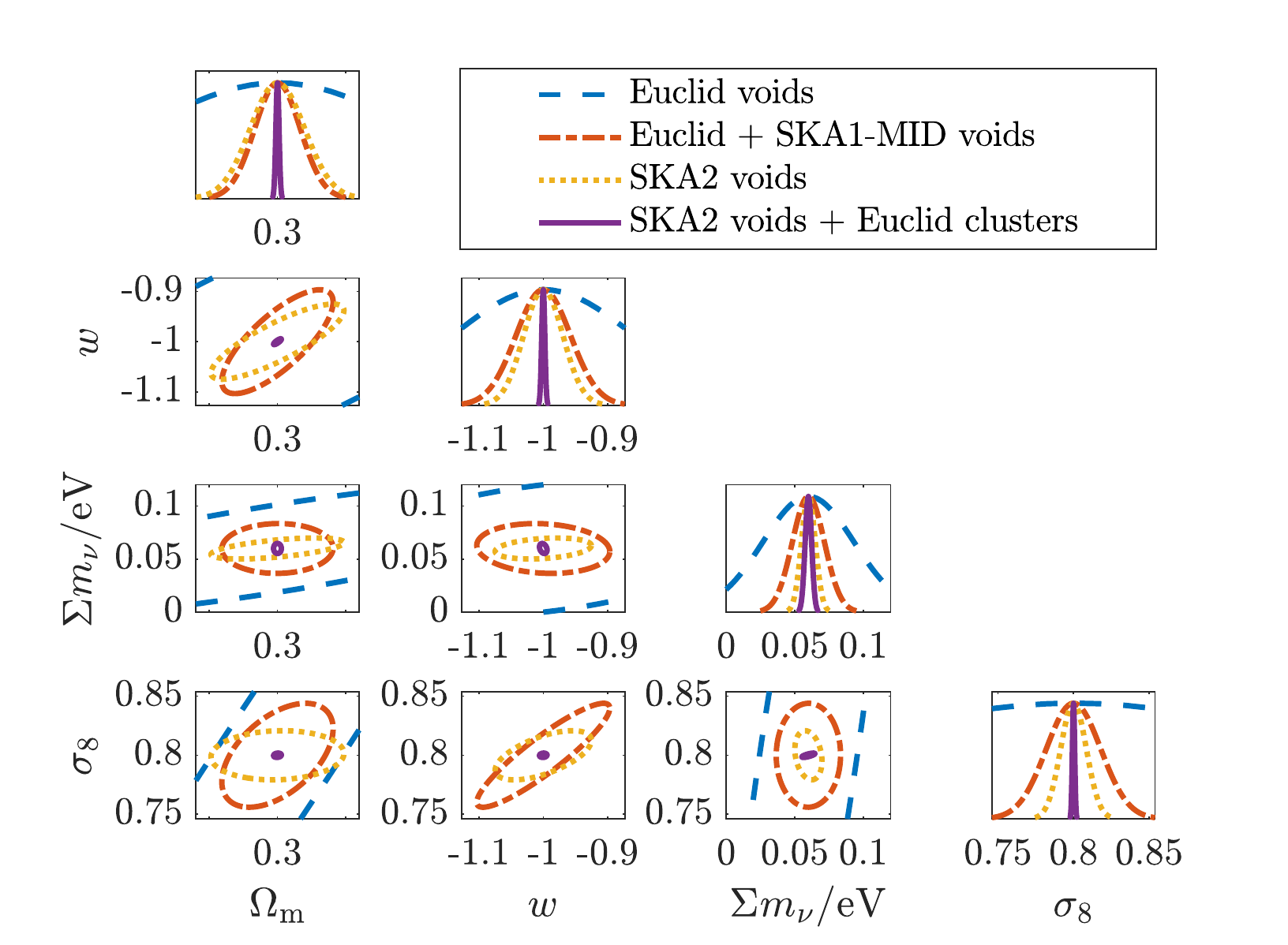}
\caption{\label{fig:voidmnuforecast} Forecast parameter constraints (95\%
confidence levels) for a flat wCDM model with massive neutrinos. Note the
considerable degeneracy breaking between the Euclid and SKA1 void samples,
and between the SKA2 void and Euclid cluster samples. SKA1-Mid covers
5000~deg$^2$, $z = 0-0.43$. SKA2 covers 30\,000 deg$^2$, $z=0.1-2$. Euclid
voids covers 15\,000 deg$^2$, $z=0.7-2$. Euclid clusters covers 15\,000
deg$^2$, $z=0.2-2$. The fiducial cosmological model is given by
$\{\Omega_{\rm m} = 0.3, w = -1, \Sigma m_{\nu} = 0.06 \,{\rm eV}, \sigma_8
= 0.8, n_{\rm s} = 0.96, h = 0.7, \Omega_{\rm b} = 0.044\}$. We have also
marginalised over uncertainty in void radius and cluster mass
\citep{Sahlen:2016kzx}, and in the theoretical void distribution function
\citep{Pisani:2015jha}.}
\end{figure}

We consider a flat wCDM cosmology with massive neutrinos described by the sum of neutrino masses $\Sigma m_{\nu}$. The void distribution is modelled following~\citet{Sahlen:2015wpc}, \citet{Sahlen:2016kzx} and \citet{2019PhRvD..99f3525S}, also taking into account the galaxy density and bias for each survey~\citep{Yahya:2014yva,Raccanelli:2015vla}. The results are shown in Figure \ref{fig:voidmnuforecast} (see caption for survey and model assumptions). The combined SKA1-Mid \&  Euclid void number counts could achieve a precision $\sigma(\Sigma m_{\nu}) = 0.02$~eV, marginalised over all 6 other parameters. No additional priors are included. The SKA2 void number counts could improve on this by a factor of two, potentially distinguishing $\Sigma m_{\nu} = 0.06$ eV from $\Sigma m_{\nu} = 0.1$ eV and allowing for a determination of the neutrino hierarchy characterised by those masses (inverted and normal respectively). By using the powerful degeneracy-breaking complementarity between clusters of galaxies and voids \citep{Sahlen:2015wpc,Sahlen:2016kzx,2019PhRvD..99f3525S}, SKA2 voids + Euclid clusters number counts could reach as low as $\sigma(\Sigma m_{\nu}) = 0.002$~eV. These forecasts are highly competitive with expectations for planned Stage IV CMB experiments/probes~\citep[for example][]{Zhen:2015yba}.

\subsubsection{Measuring the photon mass with fast radio bursts} \label{sec:photonmass}
Fast radio bursts (FRBs) are short, dispersed spikes of radio waves, typically lasting a few milliseconds at $\sim$ GHz frequencies~\citep{Lorimer:2007qn, Thornton:2013iua}. They appear to come from powerful events at cosmological distances with their cause still unknown~\citep{Katz:2016dti}. Despite our ignorance of their origin, FRBs can be used to study fundamental physics, particularly in setting upper limits on the mass of the photon~\citep{Wu:2016brq,Bonetti:2016cpo, Bonetti:2017pym, Shao:2017tuu}.

If photons are massive, the speed of light will be energy dependent (in a Lorentz-invariant theory), with high-energy photons travelling faster. Thus, low energy photons will have a time delay after they traverse a fixed distance. Because of (1) the short time duration
($\sim$\,ms), (2) the large travelling distance ($\sim$\,Gpc), and (3) the low energy of photons ($\sim\,\mu$eV). FRBs are among the best celestial objects to constrain the photon mass, $m_\gamma$~\citep{Shao:2017tuu}.
Individual sources with redshift measurements have been used to obtain a
limit of $m_\nu < {\cal O}(10^{-50})$ kg \citep{Wu:2016brq, Bonetti:2016cpo,
Bonetti:2017pym}. Nevertheless, measurements of FRB redshifts are rare \citep{Chatterjee:2017dqg} and consequently a Bayesian framework, where FRBs with and without redshift measurement equally contribute to the constraint, was used to obtain the currently best limit from the kinematics of light propagation~\citep{Shao:2017tuu}. 

In January 2017, the Commensal Real-time ASKAP Fast Transients survey (CRAFT) found a FRB in a 3.4-day pilot survey~\citep{Bannister:2017sie}. Such a survey benefits greatly from the large field of view with the phased-array-feed technology. The Canadian Hydrogen Intensity Mapping Experiment (CHIME) will also find many more FRBs \citep{2018ApJ...863...48C,2018ATel11901....1B}, while when SKA2 is operating, FRBs will be detected daily. These FRBs will contribute to an even tighter limit on the photon mass, or may even discover new physics beyond the standard model if photons are indeed massive.

\subsubsection{The non-thermal Universe} \label{sec:CR}
\begin{figure}
\begin{center}
\includegraphics[width=\columnwidth, trim=0cm 10cm 0cm 8cm, clip=true]{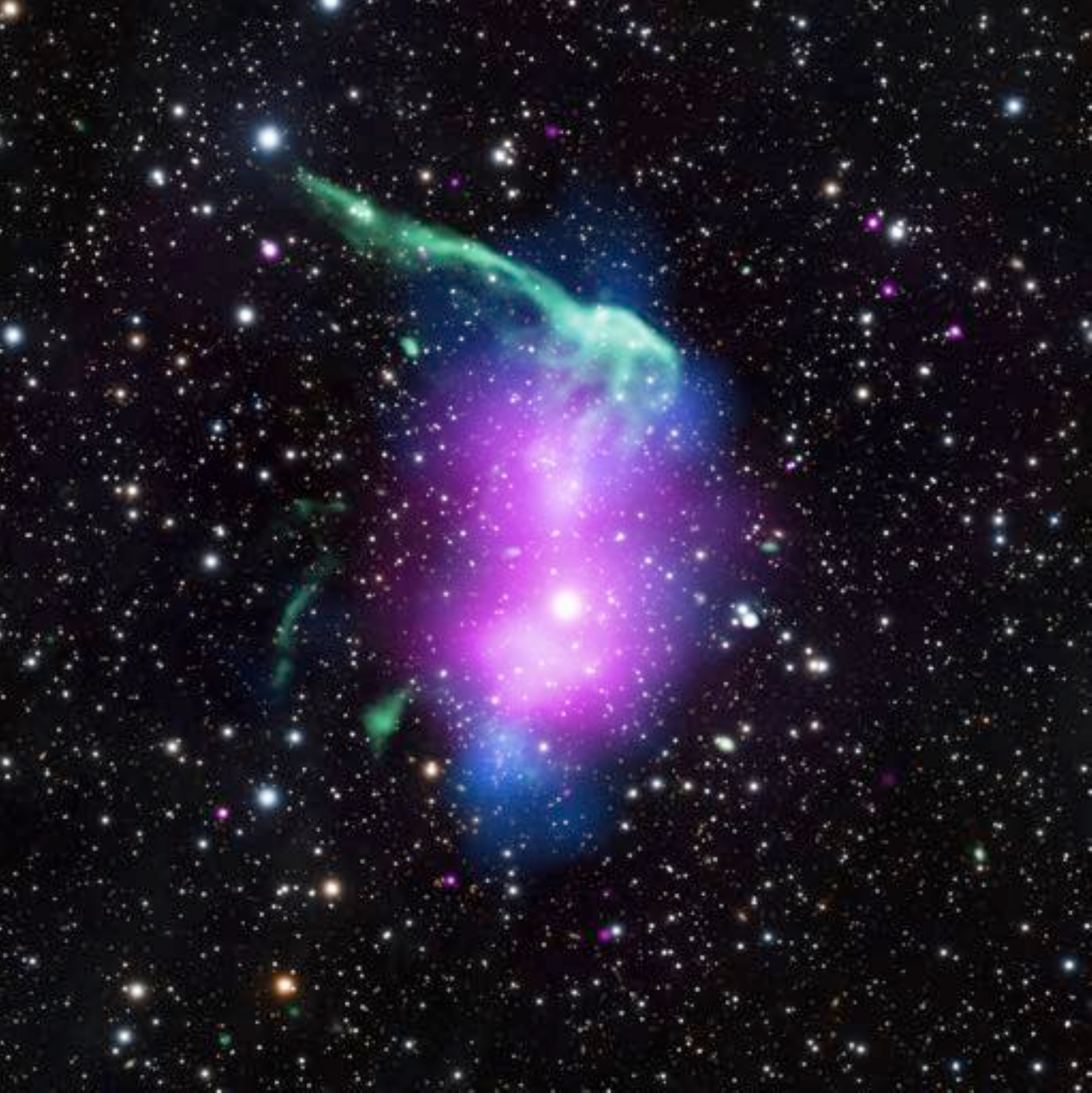}
\caption{Multi-wavelength image of the so-called `toothbrush' radio relic. The green colours show the radio image (LOFAR), the magenta the X-ray (Chandra) view and the white the optical data (Subaru) \citep[][]{2016ApJ...818..204V}.}
 \label{toothbrush}
\end{center}
\end{figure}

Continuum observations with the SKA will allow profound insights into the
non-thermal Universe. This encompasses sources of the energy density that
stem from magnetic fields, non-thermal particles (also called cosmic rays),
and turbulent motions. In spiral galaxies, these non-thermal components
dominate the total energy density. Also on larger scales, in the cosmic web
of galaxy clusters and filaments, there is ample evidence of substantial
non-thermal components. An example is shown in Figure \ref{toothbrush},
which shows a multi-wavelength image of the `toothbrush' relic, a 3-Mpc long diffuse radio source, located at the periphery of a merging galaxy cluster \citep{2016ApJ...818..204V}. 
At radio wavelengths, these non-thermal components are mostly observed via
the synchrotron emission that is produced by relativistic electrons with
Lorentz factors of a few hundred gyrating in magnetic fields (see also
\S\ref{sec:DMgamma}). However, the origins of both the magnetic fields and
the relativistic particles are unknown.

The diffuse radio sources observed in galaxy clusters span vast scales of up
to several Mpc. The short synchrotron cooling time of the electrons implies
that they must be injected in-situ by a process operating over the same
spatial scale as the source itself. The plasmas in which radio relics and
halos occur are collisionless and have number densities of $10^{-4} -
10^{-3}$~cm$^{-3}$. Moreover, the magnetic fields have strengths of around a
few $\mu$G, leading to a substantially weaker magnetic pressure than the thermal pressure. Similar conditions are unattainable in a laboratory on Earth and hence can only be studied remotely through astronomical observations. 

Cosmological simulations predict that the largest part of radio relics and halos have not been discovered yet, since their large size and low surface brightness makes them difficult to find~\citep[for example][]{2017MNRAS.470..240N,wi17}. Yet, through the study of these objects one hopes to find (1) a process that can generate magnetic fields that fills large volumes (probably) in very low-density cosmic environments, and (2) processes that can accelerate electrons to relativistic energies such that they fill an entire galaxy cluster or operate at the outskirts of clusters.  The main candidates are shock waves and turbulence, but it is not known whether low-Mach number shocks are efficient enough, or what fraction of the magnetic field is produced by the shock wave itself and what part is merely amplified via compression.

The SKA will be able to probe the cosmic filaments that are predicted to
contain most of the baryons in the Universe \citep{2001ApJ...552..473D}.
Very little is known about these filaments, since the thermal state of the baryons and their low column densities makes them very hard to observe, but they are expected to be sheathed by accretion shocks for which both the continuum and polarised emission could be detectable. Provided that the shocks in this extreme environment are at least as efficient as cluster shocks in accelerating relativistic electrons, we have predicted the flux densities of the synchrotron radiation that
could be detected by the SKA for magnetic fields of order $\sim 0.01-0.1$~$\mu$G.

\begin{figure}
\begin{center}
\includegraphics[width=\columnwidth]{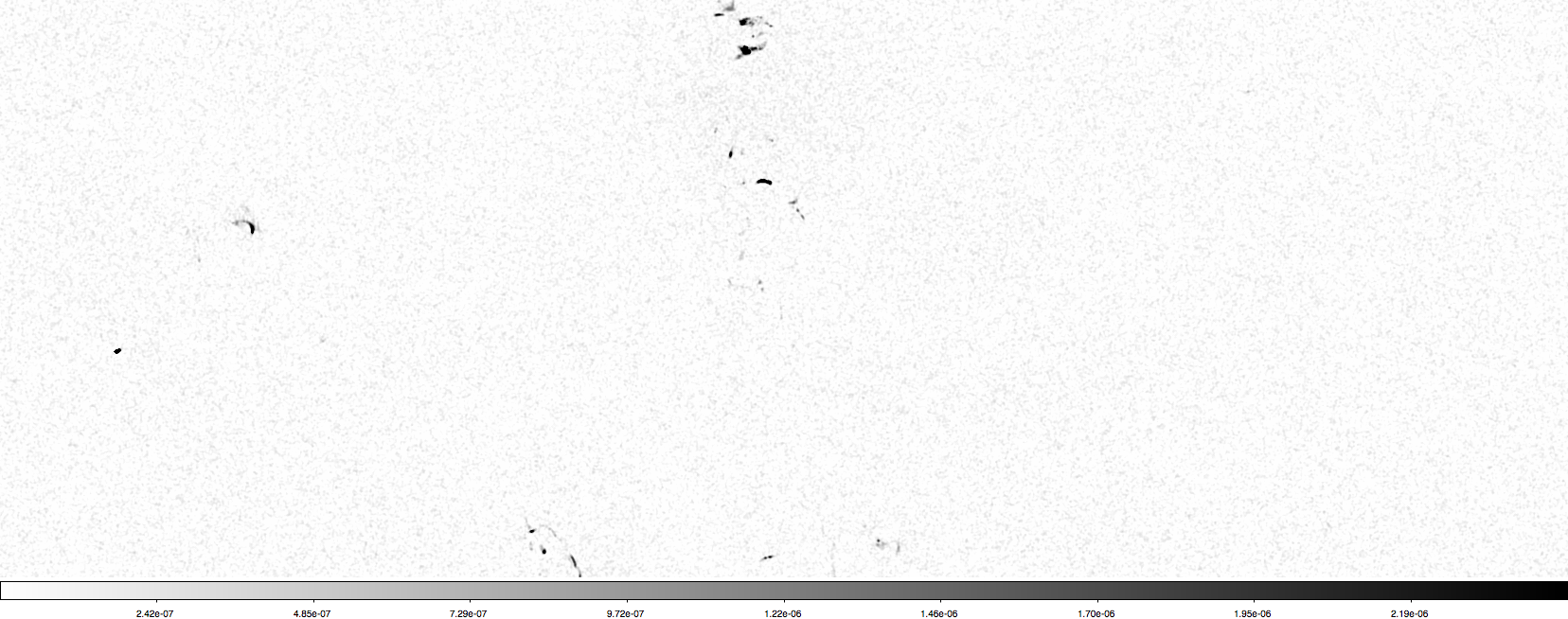}
\includegraphics[width=\columnwidth]{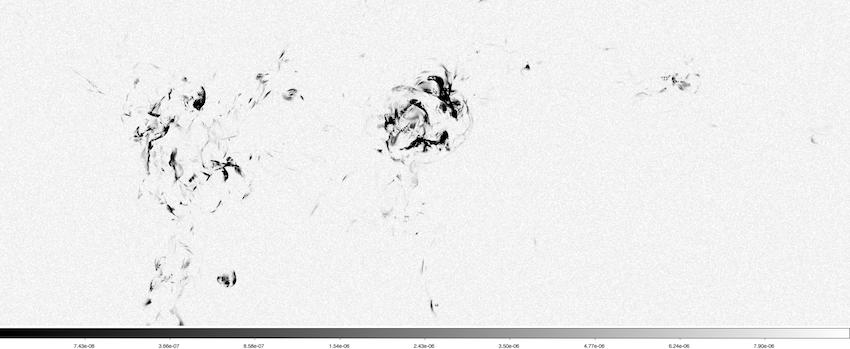}
\caption{Mock observations of a $5 \times 15$ degree area including galaxy clusters and filaments, assuming the sensitivity of the NVSS survey with the Very Large Array at $1.4$ GHz  (top) and with  the sensitivity of a survey with SKA1-Low at $110$ MHz (bottom) in units of $\rm Jy/arcsec^2$. The underlying cosmological simulations are part of the CHRONOS++ suite of MHD simulations with the ENZO code and was run on the Piz-Daint computer cluster at CSCS in Lugano \citep[][]{va14mhd}.}
 \label{ska}
\end{center}
\end{figure}

In Figure~\ref{ska} we show the predicted synchrotron signal from shock accelerated relativistic electrons in a $5 \times 15$ degree region from a cosmological simulation, as observed assuming the typical sensitivities of current and future technologies.
The region was extracted from a cosmological magnetohydrodynamic (MHD)
simulation performed with the ENZO code~\citep{enzo14}. The acceleration
efficiency of electrons by shocks is tuned to reproduce the radio emission
observed at radio relics in the intracluster medium
\citep{va15radio}.\footnote{A public repository of radio maps for the full
volumes studied by \citet{va15radio} is available at \url{http://cosmosimfrazza.myfreesites.net/radio-web}}
This mock observation illustrates the large jump in sensitivity to the
diffuse and shocked cosmic web that we expect to achieve with SKA1-Low, allowing us to greatly improve our present view of extragalactic magnetic fields beyond the innermost regions of galaxy clusters. 
There are strong indications that both radio relics and halos are in fact
made up of quite distinct subclasses of sources~\citep[for
example][]{2015MNRAS.448.2197D,2017NatAs...1E...5V}. The improved
sensitivity of the SKA will provide sufficient statistics to study and distinguish between the various particle acceleration and re-acceleration scenarios.

The SKA will be able detect cosmic filaments if the magnetic field energy
density is at the level of a few percent of the thermal energy
density~\citep{va15radio}. Observations below $\approx 200$~MHz are best suited to detect the large-scale diffuse emission produced by cosmological shock waves. These shocks are characterised by a flat emission spectrum and show flux densities  of $\sim\mu$Jy/arcsec$^2$ at low redshifts.
Because of cosmological dimming, most of the detectable radio emission is caused by structures at $z \leq 0.1$. Especially at high frequencies, the detection of nearby filaments ($z \leq 0.02$) is  difficult because of the lack of short baselines.
The range of magnetic fields that should ensure a systematic detection of the cosmic web at the periphery of galaxy clusters is in line with the magnetic field detected along an accreting group in the nearby Coma galaxy cluster \citep{bo13} and can be achieved either via a small-scale dynamo or via release of magnetic fields by nearby galaxies/AGN. Assessing which of these mechanism(s) is responsible for the  magnetisation in such rarefied cosmic environments is an exciting and important challenge for the SKA.

\subsection{Summary}
\label{sec:conclusion}

It is clear that the SKA will provide significant advancements in our
understanding of astronomical, cosmological and even particle theories.  The
wide range of opportunities set out in this section focus only on the fields
of DM and astroparticle physics and demonstrate the power of this new technology.  Just the increased observability of high-energy objects such as pulsars, binary stars and AGN, and the ability to trace back to extremely high redshift offers a wealth of new data that will contribute to a number of important and groundbreaking discoveries. 

HI intensity mapping in particular is a new and powerful tool for mapping the structure formation of the Universe. It will provide insight on the clustering properties and thermal characteristics of DM. The increased sensitivity also opens the door for DM particle searches as well as new experiments for constraining standard model particle properties such as the neutrino mass.  It may even offer the opportunity to test new quantum gravity and string theories through observations of primordial black holes and binary pulsars.

\section{Conclusions}

Physicists seek to understand the nature of matter, energy and spacetime, plus how the three of these have interacted over cosmic time.  While great strides have come from terrestrial and solar system experiments, it is increasingly clear that immense progress lies ahead through studying the cosmos.

Modern radio telescopes now have the capability to gather enormous statistical samples of celestial objects, and to make ultra-precise measurements of astrophysical effects. In this paper, we have explained the many ways in which the Square Kilometre Array will push far beyond the current frontiers in these areas, and will allow us to ask and answer new questions about cosmology, gravity, dark matter, dark energy, and more. The SKA will not just be a revolutionary facility for astronomy, but will also be an extraordinary machine for advancing fundamental physics.

\begin{acknowledgements}

The workshop ''Fundamental Physics with the Square Kilometre Array'' was made possible through the generous support of the Dunlap Institute for Astronomy and Astrophysics, the
Square Kilometre Array Organisation,
SKA South Africa,
the ARC Centre of Excellence for All-Sky Astrophysics (CAASTRO),
the International Centre for Radio Astronomy Research (ICRAR), the
National Institute for Theoretical Physics (NITheP), the
Netherlands Research School for Astronomy (NOVA), and the
NRF/DST South African Research Chairs Initiative (SARChI).

Individually the author's recognise a number of collaborators, institutions and funding bodies for their support, namely:\\

A. Weltman acknowledges financial support from the DST/NRF South African Research Chairs programme as well as the Institute for Advanced Study, Princeton and the Flatiron Institute through the Simons Foundation.

P. Bull's research was supported by an appointment to the NASA Postdoctoral Program at the Jet Propulsion Laboratory, California Institute of Technology, administered by Universities Space Research Association under contract with NASA.

S. Camera is supported by MIUR through Rita Levi Montalcini project `\textsc{prometheus} -- Probing and Relating Observables with Multi-wavelength Experiments To Help Enlightening the Universe's Structure', and by the `Departments of Excellence 2018-2022' Grant awarded by MIUR (L. 232/2016).

H. Padmanabhan's research was supported by the Tomalla Foundation.

J.R. Pritchard is pleased to acknowledge support from the European Research Council under ERC grant number 638743-FIRSTDAWN.

A. Racanelli has received funding from the People Programme (Marie Curie Actions) of the European Union H2020 Programme under REA grant agreement number 706896 (COSMOFLAGS). Funding for this work was partially provided by the Spanish MINECO under MDM-2014-0369 of ICCUB (Unidad de Excelencia `Maria de Maeztu').

E.Athanassoula and A. Bosma thank the Action Sp\'{e}cifique SKA-LOFAR of CNRS/INSU for financial support. 

For R. Barkana, this project/publication was made possible through the support of a grant from the John Templeton Foundation; the opinions expressed in this publication are those of the authors and do not necessarily reflect the views of the John Templeton Foundation. 

 G. Bertone, F. Calore, R.M.T. Connors, and D. Gaggero acknowledge collaboration with M. Lovell, S. Markoff, and E. Storm.

Camille Bonvin acknowledges partial support by INAF-Direzione Scientifica and
by ASI through ASI/INAF Agreement 2014-024-R.1 for the Planck LFI Activity of Phase E2.

M. Br\"uggen's work was strongly supported by computing resources from the Swiss National Supercomputing Centre (CSCS) under projects ID s585 and s701. F. Vazza acknowledges financial support from the European Union's Horizon 2020 research and innovation programme under the Marie-Sklodowska-Curie grant agreement no.664931 and the ERC Starting Grant "MAGCOW", no.714196. They acknowledge allocations 9059 on supercomputers at the NIC of the Forschungszentrum J\"ulich. Computations described in their work were performed using the ENZO code, which is the product of a collaborative effort of scientists at many universities and national laboratories.

 C. Burigana and T. Trombetti acknowledge partial support from the INAF PRIN SKA/CTA project FORmation and Evolution of Cosmic STructures (FORECaST) with Future Radio Surveys.

C. Weniger and F. Calore  acknowledge collaboration with F.~Donato, M.~Di Mauro and J.~W.~T.~Hessels. Their research is funded by NWO through an NWO Vidi research grant.

J.A.R. Cembranos and \'A. de la Cruz-Dombriz acknowledge financial support from the Consolider-Ingenio MULTIDARK CSD2009-00064 project. 

A. de la Cruz-Dombriz acknowledges financial support from projects Consolider-Ingenio MULTIDARK CSD2009-00064. FPA2014-53375-C2-1-P Spanish Ministry of Economy and Science, FIS2016-78859-P European Regional Development Fund and Spanish Research Agency (AEI), CA15117 CANTATA and CA16104 COST Actions EU Framework Programme Horizon 2020, CSIC I-LINK1019 Project, Spanish Ministry of Economy and Science, University of Cape Town Launching Grants Programme.

 A. Dombriz (grant numbers 99077, CSUR150628121624, IFR170131220846)  P. Dunsby, J. Larena, , Y-z Ma (grant numbers 104800, 105925) are supported by the National Research Foundation (South Africa).

M. M\'endez-Isla acknowledges financial support from the University of Cape Town Doctoral Fellowships and the Erasmus+ Alliance4Universities Mobility Programme.

 S. Camera, M. Regis and N. Fornengo are supported by: the research grant The Anisotropic Dark Universe, number CSTO161409, funded under the program CSP-UNITO Research for the Territory 2016 by Compagnia di Sanpaolo and University of Torino; the research grants TAsP (Theoretical Astroparticle Physics) and InDark (Inflation, Dark Matter and the Large-Scale Structure of the Universe) funded by the Istituto Nazionale di Fisica Nucleare (INFN). Their research is also funded by MIUR through the Rita Levi Montalcini project `\textsc{prometheus} -- Probing and Relating Observables with Multi-wavelength Experiments To Help Enlightening the Universe's Structure'. 

The contribution of S. D. Mohanty to this paper is supported by NSF Grants PHY-1505861 and HRD-0734800. 

R. Maartens is supported by the South African SKA Project and by the UK STFC, Grant ST/N000668/1.

D. Parkinson acknowledges support of the Australian Research Council through the award of a Future Fellowship [Grant No. FT130101086].

A. Pourtsidou's work for this project was partly supported by a Dennis Sciama Fellowship at the University of Portsmouth.

M. Regis acknowledges support by the Excellent Young PI Grant: ``The Particle Dark-matter Quest in the Extragalactic Sky'' funded by the University of Torino and Compagnia di San Paolo, by ``Deciphering the high-energy sky via cross correlation'' funded by Accordo Attuativo ASI-INAF n. 2017-14-H.0., and by the project ``Theoretical Astroparticle Physics (TAsP)'' funded by the INFN.

M. Sahl\'{e}n is supported by Olle Engkvist Foundation.

M. Sakellariadou is partially supported by STFC (UK) under the research grant ST/L000326/1.

T. Venumadhav acknowledges support from the Schmidt Fellowship and the Fund for Memberships in Natural Sciences at the Institute for Advanced Study.

F. Vidotto acknowledges support by grant IT956-16 of the Basque Government and by grant FIS2017-85076-P (MINECO/AEI/FEDER, UE).

The work of F. Villaescusa-Navarro is supported by the Simons Foundation.

Y. Wang is supported by the National Natural Science Foundation of China (NSFC) under Grants No. 11973024, No. 91636111, No. 11690021, No. 11503007, and ``the Fundamental Research Funds for the Central Universities'' under Grant No. 2019kfyRCPY106. 

L. Wolz is supported by an ARC Discovery Early Career Researcher Award (DE170100356). Part of her research was also supported by the Australian Research Council Centre of Excellence for All-sky Astrophysics (CAASTRO), through project number CE110001020.
 
B. M. Gaensler acknowledges support from: the Dunlap Institute of Astronomy and Astrophysics, funded through an endowment established by the David Dunlap family and the University of Toronto; the Natural Sciences and Engineering Research Council of Canada (NSERC) through grant RGPIN-2015-05948; and of the Canada Research Chairs program.

The use of the {\it Planck} Legacy
Archive\footnote{\url{http://pla.esac.esa.int/pla/}} is acknowledged. We
acknowledge use of the HEALPix\footnote{\url{http://healpix.jpl.nasa.gov/}}
\citep{2005ApJ...622..759G} software and analysis package.

\end{acknowledgements}

\bibliographystyle{pasa-mnras}
\bibliography{refs_grav,refs_astropart,refs_cosmology,refs_eor,refs_other}

\begin{thebibliography}{}
\makeatletter
\relax
\def\mn@urlcharsother{\let\do\@makeother \do\$\do\&\do\#\do\^\do\_\do\%\do\~}
\definecolor{darkblue}{rgb}{0,0,0.597656}
\def\mndoi{\begingroup\mn@urlcharsother \@ifnextchar [ {\mndoi@} {\mndoi@[]}}
\def\mndoi@[#1]#2{\def\@tempa{#1}\ifx\@tempa\@empty \href
  {http://dx.doi.org/#2} {\textcolor{darkblue}{doi:#2}}\else \href
  {http://dx.doi.org/#2} {\textcolor{darkblue}{#1}}\fi \endgroup}
\def\mn@eprint#1#2{\mn@eprint@#1:#2::\@nil}
\def\mn@eprint@arXiv#1{\href {http://arxiv.org/abs/#1} {{\tt arXiv:#1}}}
\def\mn@eprint@dblp#1{\href {http://dblp.uni-trier.de/rec/bibtex/#1.xml}
  {dblp:#1}}
\def\mn@eprint@#1:#2:#3:#4\@nil{\def\@tempa {#1}\def\@tempb {#2}\def\@tempc
  {#3}\ifx \@tempc \@empty \let \@tempc \@tempb \let \@tempb \@tempa \fi \ifx
  \@tempb \@empty \def\@tempb {arXiv}\fi \@ifundefined
  {mn@eprint@\@tempb}{\@tempb:\@tempc}{\expandafter \expandafter \csname
  mn@eprint@\@tempb\endcsname \expandafter{\@tempc}}}

\bibitem[\protect\citeauthoryear{Abazajian}{Abazajian}{2011}]{Abazajian:2010zy}
Abazajian K.~N.,  2011, \mndoi [JCAP] {10.1088/1475-7516/2011/03/010}, 1103,
  010

\bibitem[\protect\citeauthoryear{Abazajian, Canac, Horiuchi  \&
  Kaplinghat}{Abazajian et~al.}{2014}]{Abazajian:2014fta}
Abazajian K.~N.,  Canac N.,  Horiuchi S.,   Kaplinghat M.,  2014, \mndoi
  [Phys.Rev.] {10.1103/PhysRevD.90.023526}, D90, 023526

\bibitem[\protect\citeauthoryear{{Abbott} et~al.,}{{Abbott}
  et~al.}{2016a}]{2016PhRvD..93l2004A}
{Abbott} B.~P.,  et~al., 2016a, \mndoi [\prd] {10.1103/PhysRevD.93.122004},
  \href {http://adsabs.harvard.edu/abs/2016PhRvD..93l2004A} {93, 122004}

\bibitem[\protect\citeauthoryear{{Abbott} et~al.,}{{Abbott}
  et~al.}{2016b}]{2016PhRvL.116f1102A}
{Abbott} B.~P.,  et~al., 2016b, \mndoi [Physical Review Letters]
  {10.1103/PhysRevLett.116.061102}, \href
  {http://adsabs.harvard.edu/abs/2016PhRvL.116f1102A} {116, 061102}

\bibitem[\protect\citeauthoryear{Abbott et~al.}{Abbott
  et~al.}{2016c}]{TheLIGOScientific:2016src}
Abbott B.~P.,  et~al., 2016c, \mndoi [Phys. Rev. Lett.]
  {10.1103/PhysRevLett.116.221101}, 116, 221101

\bibitem[\protect\citeauthoryear{Abbott et~al.}{Abbott
  et~al.}{2017}]{Abbott:2017vtc}
Abbott B.~P.,  et~al., 2017, \mndoi [Phys. Rev. Lett.]
  {10.1103/PhysRevLett.118.221101}, 118, 221101

\bibitem[\protect\citeauthoryear{Abramo \& Leonard}{Abramo \&
  Leonard}{2013}]{Abramo:2013awa}
Abramo L.~R.,  Leonard K.~E.,  2013, \mndoi [Mon. Not. Roy. Astron. Soc.]
  {10.1093/mnras/stt465}, 432, 318

\bibitem[\protect\citeauthoryear{Achard et~al.,}{Achard
  et~al.}{2004}]{ACHARD2004145}
Achard P.,  et~al., 2004, \mndoi [Physics Letters B]
  {http://dx.doi.org/10.1016/j.physletb.2004.07.014}, 597, 145

\bibitem[\protect\citeauthoryear{{Ackermann} et~al.,}{{Ackermann}
  et~al.}{2017}]{Fermi17a}
{Ackermann} M.,  et~al., 2017, \mndoi [\apj] {10.3847/1538-4357/aa6cab}, \href
  {http://adsabs.harvard.edu/abs/2017ApJ...840...43A} {840, 43}

\bibitem[\protect\citeauthoryear{Ade et~al.}{Ade et~al.}{2016}]{Ade:2015xua}
Ade P. A.~R.,  et~al., 2016, \mndoi [Astron. Astrophys.]
  {10.1051/0004-6361/201525830}, 594, A13

\bibitem[\protect\citeauthoryear{{Adshead}, {Easther}, {Pritchard}  \&
  {Loeb}}{{Adshead} et~al.}{2011}]{1007.3748}
{Adshead} P.,  {Easther} R.,  {Pritchard} J.,   {Loeb} A.,  2011, \mndoi
  [\jcap] {10.1088/1475-7516/2011/02/021}, \href
  {http://adsabs.harvard.edu/abs/2011JCAP...02..021A} {2, 021}

\bibitem[\protect\citeauthoryear{Aghanim et~al.}{Aghanim
  et~al.}{2014}]{Aghanim:2013suk}
Aghanim N.,  et~al., 2014, \mndoi [Astron. Astrophys.]
  {10.1051/0004-6361/201321556}, 571, A27

\bibitem[\protect\citeauthoryear{{Ahn}, {Shapiro}, {Alvarez}, {Iliev}, {Martel}
   \& {Ryu}}{{Ahn} et~al.}{2006}]{2006NewAR..50..179A}
{Ahn} K.,  {Shapiro} P.~R.,  {Alvarez} M.~A.,  {Iliev} I.~T.,  {Martel} H.,
  {Ryu} D.,  2006, \mndoi [New Astron. Rev.] {10.1016/j.newar.2005.11.021},
  \href {http://adsabs.harvard.edu/abs/2006NewAR..50..179A} {50, 179}

\bibitem[\protect\citeauthoryear{{Aiola}, {Wang}, {Kosowsky}, {Kahniashvili}
  \& {Firouzjahi}}{{Aiola} et~al.}{2015}]{Aiola:2015rqa}
{Aiola} S.,  {Wang} B.,  {Kosowsky} A.,  {Kahniashvili} T.,   {Firouzjahi} H.,
  2015, \mndoi [Physical Review D] {10.1103/PhysRevD.92.063008}, \href
  {https://ui.adsabs.harvard.edu/abs/2015PhRvD..92f3008A} {92, 063008}

\bibitem[\protect\citeauthoryear{{Ajello} et~al.,}{{Ajello}
  et~al.}{2016}]{TheFermi-LAT:2015kwa}
{Ajello} M.,  et~al., 2016, \mndoi [\apj] {10.3847/0004-637X/819/1/44}, \href
  {http://adsabs.harvard.edu/abs/2016ApJ...819...44A} {819, 44}

\bibitem[\protect\citeauthoryear{Akrami, Fantaye, Shafieloo, Eriksen, Hansen
  et~al.}{Akrami et~al.}{2014}]{Akrami:2014eta}
Akrami Y.,  Fantaye Y.,  Shafieloo A.,  Eriksen H.,  Hansen F.,   et~al., 2014,
  \mndoi [Astrophys.J.] {10.1088/2041-8205/784/2/L42}, 784, L42

\bibitem[\protect\citeauthoryear{Alcaraz, Cembranos, Dobado  \& Maroto}{Alcaraz
  et~al.}{2003}]{PhysRevD.67.075010}
Alcaraz J.,  Cembranos J. A.~R.,  Dobado A.,   Maroto A.~L.,  2003, \mndoi
  [Phys. Rev. D] {10.1103/PhysRevD.67.075010}, 67, 075010

\bibitem[\protect\citeauthoryear{{Ali-Ha{\"i}moud}, {Meerburg}  \&
  {Yuan}}{{Ali-Ha{\"i}moud} et~al.}{2014}]{2014PhRvD..89h3506A}
{Ali-Ha{\"i}moud} Y.,  {Meerburg} P.~D.,   {Yuan} S.,  2014, \mndoi [\prd]
  {10.1103/PhysRevD.89.083506}, \href
  {http://adsabs.harvard.edu/abs/2014PhRvD..89h3506A} {89, 083506}

\bibitem[\protect\citeauthoryear{Ali-Ha{\"i}moud, Kovetz  \&
  Kamionkowski}{Ali-Ha{\"i}moud et~al.}{2017}]{Ali-Haimoud:2017rtz}
Ali-Ha{\"i}moud Y.,  Kovetz E.~D.,   Kamionkowski M.,  2017, \mndoi [Phys.
  Rev.] {10.1103/PhysRevD.96.123523}, D96, 123523

\bibitem[\protect\citeauthoryear{Ali et~al.,}{Ali et~al.}{2015}]{ali2015}
Ali Z.~S.,  et~al., 2015, \mndoi [The Astrophysical Journal]
  {10.1088/0004-637X/809/1/61}, 809, 61

\bibitem[\protect\citeauthoryear{{Allison} \& {Dalgarno}}{{Allison} \&
  {Dalgarno}}{1969}]{1969ApJ...158..423A}
{Allison} A.~C.,  {Dalgarno} A.,  1969, \mndoi [\apj] {10.1086/150204}, \href
  {http://adsabs.harvard.edu/abs/1969ApJ...158..423A} {158, 423}

\bibitem[\protect\citeauthoryear{Alonso \& Ferreira}{Alonso \&
  Ferreira}{2015}]{Alonso:2015sfa}
Alonso D.,  Ferreira P.~G.,  2015, \mndoi [Phys. Rev.]
  {10.1103/PhysRevD.92.063525}, D92, 063525

\bibitem[\protect\citeauthoryear{Alonso, Bull, Ferreira  \& Santos}{Alonso
  et~al.}{2015a}]{Alonso2015}
Alonso D.,  Bull P.,  Ferreira P.~G.,   Santos M.~G.,  2015a, \mndoi [Monthly
  Notices of the Royal Astronomical Society] {10.1093/mnras/stu2474}, 447, 400

\bibitem[\protect\citeauthoryear{Alonso, Bull, Ferreira, Maartens  \&
  Santos}{Alonso et~al.}{2015b}]{Alonso:2015uua}
Alonso D.,  Bull P.,  Ferreira P.~G.,  Maartens R.,   Santos M.,  2015b, \mndoi
  [Astrophys. J.] {10.1088/0004-637X/814/2/145}, 814, 145

\bibitem[\protect\citeauthoryear{Amendola, Kunz, Motta, Saltas  \&
  Sawicki}{Amendola et~al.}{2013}]{Amendola:2012ky}
Amendola L.,  Kunz M.,  Motta M.,  Saltas I.~D.,   Sawicki I.,  2013, \mndoi
  [Phys. Rev.] {10.1103/PhysRevD.87.023501}, D87, 023501

\bibitem[\protect\citeauthoryear{{Amendola} et~al.,}{{Amendola}
  et~al.}{2018}]{Amendola:2016saw}
{Amendola} L.,  et~al., 2018, \mndoi [Living Reviews in Relativity]
  {10.1007/s41114-017-0010-3}, \href
  {http://adsabs.harvard.edu/abs/2018LRR....21....2A} {21, 2}

\bibitem[\protect\citeauthoryear{{Ammazzalorso} et~al.,}{{Ammazzalorso}
  et~al.}{2019}]{2019arXiv190713484A}
{Ammazzalorso} S.,  et~al., 2019, arXiv e-prints, \href
  {https://ui.adsabs.harvard.edu/abs/2019arXiv190713484A} {p. arXiv:1907.13484}

\bibitem[\protect\citeauthoryear{Andersson \& Coley}{Andersson \&
  Coley}{2011}]{Andersson:2011za}
Andersson L.,  Coley A.,  2011, \mndoi [Class. Quant. Grav.]
  {10.1088/0264-9381/28/16/160301}, 28, 160301

\bibitem[\protect\citeauthoryear{Andr\'e et~al.}{Andr\'e
  et~al.}{2014}]{Andre:2013nfa}
Andr\'e P.,  et~al., 2014, \mndoi [JCAP] {10.1088/1475-7516/2014/02/006}, 1402,
  006

\bibitem[\protect\citeauthoryear{{Ang{\'e}lil} \& {Saha}}{{Ang{\'e}lil} \&
  {Saha}}{2010}]{2010ApJ...711..157A}
{Ang{\'e}lil} R.,  {Saha} P.,  2010, \mndoi [\apj]
  {10.1088/0004-637X/711/1/157}, \href
  {http://adsabs.harvard.edu/abs/2010ApJ...711..157A} {711, 157}

\bibitem[\protect\citeauthoryear{{Ang{\'e}lil} \& {Saha}}{{Ang{\'e}lil} \&
  {Saha}}{2014}]{2014MNRAS.444.3780A}
{Ang{\'e}lil} R.,  {Saha} P.,  2014, \mndoi [\mnras] {10.1093/mnras/stu1686},
  \href {http://adsabs.harvard.edu/abs/2014MNRAS.444.3780A} {444, 3780}

\bibitem[\protect\citeauthoryear{{Ang{\'e}lil}, {Saha}  \&
  {Merritt}}{{Ang{\'e}lil} et~al.}{2010}]{2010ApJ...720.1303A}
{Ang{\'e}lil} R.,  {Saha} P.,   {Merritt} D.,  2010, \mndoi [\apj]
  {10.1088/0004-637X/720/2/1303}, \href
  {http://adsabs.harvard.edu/abs/2010ApJ...720.1303A} {720, 1303}

\bibitem[\protect\citeauthoryear{{Archibald} et~al.,}{{Archibald}
  et~al.}{2018}]{2018arXiv180702059A}
{Archibald} A.~M.,  et~al., 2018, Nature, \href
  {http://adsabs.harvard.edu/abs/2018arXiv180702059A} {559, 73}

\bibitem[\protect\citeauthoryear{Arzoumanian et~al.}{Arzoumanian
  et~al.}{2016}]{Arzoumanian:2015liz}
Arzoumanian Z.,  et~al., 2016, \mndoi [Astrophys. J.]
  {10.3847/0004-637X/821/1/13}, 821, 13

\bibitem[\protect\citeauthoryear{Asad et~al.,}{Asad et~al.}{2015}]{Asad2015}
Asad K. M.~B.,  et~al., 2015, \mndoi [Monthly Notices of the Royal Astronomical
  Society] {10.1093/mnras/stv1107}, 451, 3709

\bibitem[\protect\citeauthoryear{Asad et~al.,}{Asad et~al.}{2016}]{Asad2016}
Asad K. M.~B.,  et~al., 2016, \mndoi [Monthly Notices of the Royal Astronomical
  Society] {10.1093/mnras/stw1863}, 462, 4482

\bibitem[\protect\citeauthoryear{{Asad}, {Koopmans}, {Jeli{\'c}}, {de Bruyn},
  {Pandey}  \& {Gehlot}}{{Asad} et~al.}{2018}]{Asad2017}
{Asad} K.~M.~B.,  {Koopmans} L.~V.~E.,  {Jeli{\'c}} V.,  {de Bruyn} A.~G.,
  {Pandey} V.~N.,   {Gehlot} B.~K.,  2018, \mndoi [\mnras]
  {10.1093/mnras/sty258}, \href
  {http://adsabs.harvard.edu/abs/2018MNRAS.476.3051A} {476, 3051}

\bibitem[\protect\citeauthoryear{Ashtekar \& Bojowald}{Ashtekar \&
  Bojowald}{2006}]{Ashtekar:2005qt}
Ashtekar A.,  Bojowald M.,  2006, \mndoi [Class. Quant. Grav.]
  {10.1088/0264-9381/23/2/008}, 23, 391

\bibitem[\protect\citeauthoryear{{Athron} et~al.,}{{Athron}
  et~al.}{2017a}]{GambitSUSY2017}
{Athron} P.,  et~al., 2017a, \mndoi [European Physical Journal C]
  {10.1140/epjc/s10052-017-5167-0}, \href
  {http://adsabs.harvard.edu/abs/2017EPJC...77..824A} {77, 824}

\bibitem[\protect\citeauthoryear{{Athron} et~al.,}{{Athron}
  et~al.}{2017b}]{GambitMSSM2017}
{Athron} P.,  et~al., 2017b, \mndoi [European Physical Journal C]
  {10.1140/epjc/s10052-017-5196-8}, \href
  {http://adsabs.harvard.edu/abs/2017EPJC...77..879A} {77, 879}

\bibitem[\protect\citeauthoryear{{BICEP2/Keck Collaboration}
  et~al.,}{{BICEP2/Keck Collaboration} et~al.}{2015}]{2015PhRvL.114j1301B}
{BICEP2/Keck Collaboration} et~al., 2015, \mndoi [Physical Review Letters]
  {10.1103/PhysRevLett.114.101301}, \href
  {http://adsabs.harvard.edu/abs/2015PhRvL.114j1301B} {114, 101301}

\bibitem[\protect\citeauthoryear{Babak \& Sesana}{Babak \&
  Sesana}{2012}]{Babak:2011mr}
Babak S.,  Sesana A.,  2012, \mndoi [Phys. Rev.] {10.1103/PhysRevD.85.044034},
  D85, 044034

\bibitem[\protect\citeauthoryear{{Bacon}, {Andrianomena}, {Clarkson}, {Bolejko}
   \& {Maartens}}{{Bacon} et~al.}{2014}]{Bacon:2014uja}
{Bacon} D.~J.,  {Andrianomena} S.,  {Clarkson} C.,  {Bolejko} K.,   {Maartens}
  R.,  2014, \mndoi [\mnras] {10.1093/mnras/stu1270}, \href
  {http://adsabs.harvard.edu/abs/2014MNRAS.443.1900B} {443, 1900}

\bibitem[\protect\citeauthoryear{{Bacon} et~al.,}{{Bacon}
  et~al.}{2015}]{2015aska.confE.145B}
{Bacon} D.,  et~al., 2015, Advancing Astrophysics with the Square Kilometre
  Array (AASKA14), \href {http://adsabs.harvard.edu/abs/2015aska.confE.145B}
  {p.~145}

\bibitem[\protect\citeauthoryear{Baker \& Bull}{Baker \&
  Bull}{2015}]{Baker:2015bva}
Baker T.,  Bull P.,  2015, \mndoi [Astrophys. J.]
  {10.1088/0004-637X/811/2/116}, 811, 116

\bibitem[\protect\citeauthoryear{Baker, Ferreira  \& Skordis}{Baker
  et~al.}{2014a}]{Baker:2013hia}
Baker T.,  Ferreira P.~G.,   Skordis C.,  2014a, \mndoi [Phys. Rev.]
  {10.1103/PhysRevD.89.024026}, D89, 024026

\bibitem[\protect\citeauthoryear{Baker, Ferreira, Leonard  \& Motta}{Baker
  et~al.}{2014b}]{Baker:2014zva}
Baker T.,  Ferreira P.~G.,  Leonard C.~D.,   Motta M.,  2014b, \mndoi [Phys.
  Rev.] {10.1103/PhysRevD.90.124030}, D90, 124030

\bibitem[\protect\citeauthoryear{Baker, Psaltis  \& Skordis}{Baker
  et~al.}{2015}]{Baker:2014zba}
Baker T.,  Psaltis D.,   Skordis C.,  2015, \mndoi [Astrophys. J.]
  {10.1088/0004-637X/802/1/63}, 802, 63

\bibitem[\protect\citeauthoryear{Bando, Kugo, Noguchi  \& Yoshioka}{Bando
  et~al.}{1999}]{PhysRevLett.83.3601}
Bando M.,  Kugo T.,  Noguchi T.,   Yoshioka K.,  1999, \mndoi [Phys. Rev.
  Lett.] {10.1103/PhysRevLett.83.3601}, 83, 3601

\bibitem[\protect\citeauthoryear{Bannister et~al.}{Bannister
  et~al.}{2017}]{Bannister:2017sie}
Bannister K.,  et~al., 2017, \mndoi [Astrophys. J.] {10.3847/2041-8213/aa71ff,
  10.3847/2041-8213}, 841, L12

\bibitem[\protect\citeauthoryear{Bardeen, Steinhardt  \& Turner}{Bardeen
  et~al.}{1983}]{Bardeen:1983qw}
Bardeen J.~M.,  Steinhardt P.~J.,   Turner M.~S.,  1983, \mndoi [Phys. Rev. D]
  {10.1103/PhysRevD.28.679}, 28, 679

\bibitem[\protect\citeauthoryear{Barkana}{Barkana}{2018}]{barkana2018a}
Barkana R.,  2018, Nature, 555, 71 EP

\bibitem[\protect\citeauthoryear{{Barkana} \& {Loeb}}{{Barkana} \&
  {Loeb}}{2001}]{astro-ph/0010468}
{Barkana} R.,  {Loeb} A.,  2001, \mndoi [\physrep]
  {10.1016/S0370-1573(01)00019-9}, \href
  {http://adsabs.harvard.edu/abs/2001PhR...349..125B} {349, 125}

\bibitem[\protect\citeauthoryear{{Barkana} \& {Loeb}}{{Barkana} \&
  {Loeb}}{2005a}]{2005ApJ...624L..65B}
{Barkana} R.,  {Loeb} A.,  2005a, \mndoi [\apjl] {10.1086/430599}, \href
  {http://adsabs.harvard.edu/abs/2005ApJ...624L..65B} {624, L65}

\bibitem[\protect\citeauthoryear{{Barkana} \& {Loeb}}{{Barkana} \&
  {Loeb}}{2005b}]{2005ApJ...626....1B}
{Barkana} R.,  {Loeb} A.,  2005b, \mndoi [\apj] {10.1086/429954}, \href
  {http://adsabs.harvard.edu/abs/2005ApJ...626....1B} {626, 1}

\bibitem[\protect\citeauthoryear{{Barkana}, {Haiman}  \& {Ostriker}}{{Barkana}
  et~al.}{2001}]{Barkarna01}
{Barkana} R.,  {Haiman} Z.,   {Ostriker} J.~P.,  2001, \mndoi [\apj]
  {10.1086/322393}, \href {http://adsabs.harvard.edu/abs/2001ApJ...558..482B}
  {558, 482}

\bibitem[\protect\citeauthoryear{{Barkana}, {Outmezguine}, {Redigol}  \&
  {Volansky}}{{Barkana} et~al.}{2018}]{barkana2018b}
{Barkana} R.,  {Outmezguine} N.~J.,  {Redigol} D.,   {Volansky} T.,  2018,
  \mndoi [Physical Review D] {10.1103/PhysRevD.98.103005}, \href
  {https://ui.adsabs.harvard.edu/abs/2018PhRvD..98j3005B} {98, 103005}

\bibitem[\protect\citeauthoryear{Barrau, Rovelli  \& Vidotto}{Barrau
  et~al.}{2014}]{Barrau:2014yka}
Barrau A.,  Rovelli C.,   Vidotto F.,  2014, \mndoi [Phys. Rev.]
  {10.1103/PhysRevD.90.127503}, D90, 127503

\bibitem[\protect\citeauthoryear{Barrau, Bolliet, Vidotto  \& Weimer}{Barrau
  et~al.}{2016}]{Barrau:2015uca}
Barrau A.,  Bolliet B.,  Vidotto F.,   Weimer C.,  2016, \mndoi [JCAP]
  {10.1088/1475-7516/2016/02/022}, 1602, 022

\bibitem[\protect\citeauthoryear{{Barrau}, {Martineau}  \& {Moulin}}{{Barrau}
  et~al.}{2018a}]{2018arXiv180808857B}
{Barrau} A.,  {Martineau} K.,   {Moulin} F.,  2018a, \mndoi [Universe]
  {10.3390/universe4100102}, \href
  {https://ui.adsabs.harvard.edu/abs/2018Univ....4..102B} {4, 102}

\bibitem[\protect\citeauthoryear{{Barrau}, {Moulin}  \& {Martineau}}{{Barrau}
  et~al.}{2018b}]{2018PhRvD..97f6019B}
{Barrau} A.,  {Moulin} F.,   {Martineau} K.,  2018b, \mndoi [\prd]
  {10.1103/PhysRevD.97.066019}, \href
  {http://adsabs.harvard.edu/abs/2018PhRvD..97f6019B} {97, 066019}

\bibitem[\protect\citeauthoryear{{Barry}, {Beardsley}, {Byrne}, {Hazelton},
  {Morales}, {Pober}  \& {Sullivan}}{{Barry} et~al.}{2019}]{barry2019}
{Barry} N.,  {Beardsley} A.~P.,  {Byrne} R.,  {Hazelton} B.,  {Morales} M.~F.,
  {Pober} J.~C.,   {Sullivan} I.,  2019, \mndoi [Publications of the
  Astronomical Society of Australia] {10.1017/pasa.2019.21}, \href
  {https://ui.adsabs.harvard.edu/abs/2019PASA...36...26B} {36, e026}

\bibitem[\protect\citeauthoryear{Bartelmann \& Schneider}{Bartelmann \&
  Schneider}{2001}]{Bartelmann:1999yn}
Bartelmann M.,  Schneider P.,  2001, \mndoi [Phys. Rept.]
  {10.1016/S0370-1573(00)00082-X}, 340, 291

\bibitem[\protect\citeauthoryear{{Bartlett} \& {Stebbins}}{{Bartlett} \&
  {Stebbins}}{1991}]{1991ApJ...371....8B}
{Bartlett} J.~G.,  {Stebbins} A.,  1991, \mndoi [\apj] {10.1086/169865}, \href
  {http://adsabs.harvard.edu/abs/1991ApJ...371....8B} {371, 8}

\bibitem[\protect\citeauthoryear{{Bartolo}, {Komatsu}, {Matarrese}  \&
  {Riotto}}{{Bartolo} et~al.}{2004}]{astro-ph/0406398}
{Bartolo} N.,  {Komatsu} E.,  {Matarrese} S.,   {Riotto} A.,  2004, \mndoi
  [\physrep] {10.1016/j.physrep.2004.08.022}, \href
  {http://adsabs.harvard.edu/abs/2004PhR...402..103B} {402, 103}

\bibitem[\protect\citeauthoryear{{Bates} et~al.,}{{Bates}
  et~al.}{2011}]{2011MNRAS.416.2455B}
{Bates} S.~D.,  et~al., 2011, \mndoi [\mnras]
  {10.1111/j.1365-2966.2011.18416.x}, \href
  {http://adsabs.harvard.edu/abs/2011MNRAS.416.2455B} {416, 2455}

\bibitem[\protect\citeauthoryear{Battye, Davies  \& Weller}{Battye
  et~al.}{2004}]{Battye:2004re}
Battye R.~A.,  Davies R.~D.,   Weller J.,  2004, \mndoi [Mon. Not. Roy. Astron.
  Soc.] {10.1111/j.1365-2966.2004.08416.x}, 355, 1339

\bibitem[\protect\citeauthoryear{{Battye}, {Browne}, {Dickinson}, {Heron},
  {Maffei}  \& {Pourtsidou}}{{Battye} et~al.}{2013}]{battye2012}
{Battye} R.~A.,  {Browne} I.~W.~A.,  {Dickinson} C.,  {Heron} G.,  {Maffei} B.,
    {Pourtsidou} A.,  2013, \mndoi [\mnras] {10.1093/mnras/stt1082}, \href
  {http://adsabs.harvard.edu/abs/2013MNRAS.434.1239B} {434, 1239}

\bibitem[\protect\citeauthoryear{{Baumann}}{{Baumann}}{2009}]{0907.5424}
{Baumann} D.,  2009, preprint, \href
  {http://adsabs.harvard.edu/abs/2009arXiv0907.5424B} {} (\mn@eprint {arXiv}
  {0907.5424})

\bibitem[\protect\citeauthoryear{{Beardsley} et~al.,}{{Beardsley}
  et~al.}{2016}]{Beardsley2016}
{Beardsley} A.~P.,  et~al., 2016, \mndoi [\apj] {10.3847/1538-4357/833/1/102},
  \href {http://adsabs.harvard.edu/abs/2016ApJ...833..102B} {833, 102}

\bibitem[\protect\citeauthoryear{{Begeman}}{{Begeman}}{1989}]{1989A&A...223...47B}
{Begeman} K.~G.,  1989, \aap, \href
  {http://adsabs.harvard.edu/abs/1989A%26A...223...47B} {223, 47}

\bibitem[\protect\citeauthoryear{Bekenstein}{Bekenstein}{2004}]{Bekenstein:2004ne}
Bekenstein J.~D.,  2004, \mndoi [Phys. Rev.] {10.1103/PhysRevD.70.083509,
  10.1103/PhysRevD.71.069901}, D70, 083509

\bibitem[\protect\citeauthoryear{Bengaly, Bernui, Alcaniz, Xavier  \&
  Novaes}{Bengaly et~al.}{2017}]{Bengaly:2016amk}
Bengaly C. A.~P.,  Bernui A.,  Alcaniz J.~S.,  Xavier H.~S.,   Novaes C.~P.,
  2017, \mndoi [Mon. Not. Roy. Astron. Soc.] {10.1093/mnras/stw2268}, 464, 768

\bibitem[\protect\citeauthoryear{Bernardi et~al.,}{Bernardi
  et~al.}{2009}]{Bernardi2009}
Bernardi G.,  et~al., 2009, \mndoi [Astronomy and Astrophysics]
  {10.1051/0004-6361/200911627}, 500, 965

\bibitem[\protect\citeauthoryear{Bernardi et~al.,}{Bernardi
  et~al.}{2010}]{Bernardi2010}
Bernardi G.,  et~al., 2010, \mndoi [Astronomy and Astrophysics]
  {10.1051/0004-6361/200913420}, 522, A67

\bibitem[\protect\citeauthoryear{{Bertacca}, {Raccanelli}, {Piattella},
  {Pietrobon}, {Bartolo}, {Matarrese}  \& {Giannantonio}}{{Bertacca}
  et~al.}{2011}]{bertacca11}
{Bertacca} D.,  {Raccanelli} A.,  {Piattella} O.~F.,  {Pietrobon} D.,
  {Bartolo} N.,  {Matarrese} S.,   {Giannantonio} T.,  2011, \mndoi [\jcap]
  {10.1088/1475-7516/2011/03/039}, \href
  {http://adsabs.harvard.edu/abs/2011JCAP...03..039B} {3, 039}

\bibitem[\protect\citeauthoryear{Bertone, Hooper  \& Silk}{Bertone
  et~al.}{2005}]{Bertone:2004pz}
Bertone G.,  Hooper D.,   Silk J.,  2005, \mndoi [Phys. Rept.]
  {10.1016/j.physrep.2004.08.031}, 405, 279

\bibitem[\protect\citeauthoryear{Bester, Larena  \& Bishop}{Bester
  et~al.}{2015}]{Bester:2015gla}
Bester H.~L.,  Larena J.,   Bishop N.~T.,  2015, \mndoi [Mon. Not. Roy. Astron.
  Soc.] {10.1093/mnras/stv1672}, 453, 2364

\bibitem[\protect\citeauthoryear{Bester, Larena  \& Bishop}{Bester
  et~al.}{2017}]{Bester:2017lrt}
Bester H.~L.,  Larena J.,   Bishop N.~T.,  2017, preprint (\mn@eprint {arXiv}
  {1705.00994})

\bibitem[\protect\citeauthoryear{{Bird}, {Cholis}, {Mu{\~n}oz},
  {Ali-Ha{\"i}moud}, {Kamionkowski}, {Kovetz}, {Raccanelli}  \& {Riess}}{{Bird}
  et~al.}{2016}]{1603.00464}
{Bird} S.,  {Cholis} I.,  {Mu{\~n}oz} J.~B.,  {Ali-Ha{\"i}moud} Y.,
  {Kamionkowski} M.,  {Kovetz} E.~D.,  {Raccanelli} A.,   {Riess} A.~G.,  2016,
  \mndoi [Physical Review Letters] {10.1103/PhysRevLett.116.201301}, \href
  {http://adsabs.harvard.edu/abs/2016PhRvL.116t1301B} {116, 201301}

\bibitem[\protect\citeauthoryear{Blandford}{Blandford}{1977}]{Blandford:1977}
Blandford R.~D.,  1977, Mon. Not. R. Astron. Soc., 181, 489

\bibitem[\protect\citeauthoryear{{Blandford} \& {Teukolsky}}{{Blandford} \&
  {Teukolsky}}{1976}]{1976ApJ...205..580B}
{Blandford} R.,  {Teukolsky} S.~A.,  1976, \mndoi [\apj] {10.1086/154315},
  \href {http://adsabs.harvard.edu/abs/1976ApJ...205..580B} {205, 580}

\bibitem[\protect\citeauthoryear{Boehm \& Schaeffer}{Boehm \&
  Schaeffer}{2005}]{Boehm:2004th}
Boehm C.,  Schaeffer R.,  2005, \mndoi [Astron. Astrophys.]
  {10.1051/0004-6361:20042238}, 438, 419

\bibitem[\protect\citeauthoryear{Boehm, Fayet  \& Schaeffer}{Boehm
  et~al.}{2001}]{Boehm:2000gq}
Boehm C.,  Fayet P.,   Schaeffer R.,  2001, \mndoi [Phys. Lett.]
  {10.1016/S0370-2693(01)01060-7}, B518, 8

\bibitem[\protect\citeauthoryear{Boehm, Riazuelo, Hansen  \& Schaeffer}{Boehm
  et~al.}{2002}]{Boehm:2001hm}
Boehm C.,  Riazuelo A.,  Hansen S.~H.,   Schaeffer R.,  2002, \mndoi [Phys.
  Rev.] {10.1103/PhysRevD.66.083505}, D66, 083505

\bibitem[\protect\citeauthoryear{Boehm, Ensslin  \& Silk}{Boehm
  et~al.}{2004}]{Boehm:2002yz}
Boehm C.,  Ensslin T.~A.,   Silk J.,  2004, \mndoi [J. Phys.]
  {10.1088/0954-3899/30/3/004}, G30, 279

\bibitem[\protect\citeauthoryear{{Boehm}, {Silk}  \& {Ensslin}}{{Boehm}
  et~al.}{2010}]{Boehm:2010kg}
{Boehm} C.,  {Silk} J.,   {Ensslin} T.,  2010, arXiv e-prints, \href
  {https://ui.adsabs.harvard.edu/abs/2010arXiv1008.5175B} {p. arXiv:1008.5175}

\bibitem[\protect\citeauthoryear{Boehm, Schewtschenko, Wilkinson, Baugh  \&
  Pascoli}{Boehm et~al.}{2014}]{Boehm:2014vja}
Boehm C.,  Schewtschenko J.~A.,  Wilkinson R.~J.,  Baugh C.~M.,   Pascoli S.,
  2014, \mndoi [Mon. Not. Roy. Astron. Soc.] {10.1093/mnrasl/slu115}, 445, L31

\bibitem[\protect\citeauthoryear{{Bolejko}, {Clarkson}, {Maartens}, {Bacon},
  {Meures}  \& {Beynon}}{{Bolejko} et~al.}{2013}]{Bolejko:2012uj}
{Bolejko} K.,  {Clarkson} C.,  {Maartens} R.,  {Bacon} D.,  {Meures} N.,
  {Beynon} E.,  2013, \mndoi [Physical Review Letters]
  {10.1103/PhysRevLett.110.021302}, \href
  {http://adsabs.harvard.edu/abs/2013PhRvL.110b1302B} {110, 021302}

\bibitem[\protect\citeauthoryear{{Bonafede}, {Vazza}, {Br{\"u}ggen}, {Murgia},
  {Govoni}, {Feretti}, {Giovannini}  \& {Ogrean}}{{Bonafede}
  et~al.}{2013}]{bo13}
{Bonafede} A.,  {Vazza} F.,  {Br{\"u}ggen} M.,  {Murgia} M.,  {Govoni} F.,
  {Feretti} L.,  {Giovannini} G.,   {Ogrean} G.,  2013, \mndoi [\mnras]
  {10.1093/mnras/stt960}, \href
  {http://adsabs.harvard.edu/abs/2013MNRAS.433.3208B} {433, 3208}

\bibitem[\protect\citeauthoryear{Bonaldi \& Brown}{Bonaldi \&
  Brown}{2015}]{Bonaldi2015}
Bonaldi A.,  Brown M.~L.,  2015, \mndoi [Monthly Notices of the Royal
  Astronomical Society] {10.1093/mnras/stu2601}, 447, 1973

\bibitem[\protect\citeauthoryear{{Bonaldi}, {Harrison}, {Camera}  \&
  {Brown}}{{Bonaldi} et~al.}{2016}]{2016MNRAS.463.3686B}
{Bonaldi} A.,  {Harrison} I.,  {Camera} S.,   {Brown} M.~L.,  2016, \mndoi
  [\mnras] {10.1093/mnras/stw2104}, \href
  {http://adsabs.harvard.edu/abs/2016MNRAS.463.3686B} {463, 3686}

\bibitem[\protect\citeauthoryear{Bonetti, Ellis, Mavromatos, Sakharov,
  Sarkisyan-Grinbaum  \& Spallicci}{Bonetti et~al.}{2016}]{Bonetti:2016cpo}
Bonetti L.,  Ellis J.,  Mavromatos N.~E.,  Sakharov A.~S.,  Sarkisyan-Grinbaum
  E. K.~G.,   Spallicci A. D. A.~M.,  2016, \mndoi [Phys. Lett.]
  {10.1016/j.physletb.2016.04.035}, B757, 548

\bibitem[\protect\citeauthoryear{Bonetti, Ellis, Mavromatos, Sakharov,
  Sarkisyan-Grinbaum  \& Spallicci}{Bonetti et~al.}{2017}]{Bonetti:2017pym}
Bonetti L.,  Ellis J.,  Mavromatos N.~E.,  Sakharov A.~S.,  Sarkisyan-Grinbaum
  E.~K.,   Spallicci A. D. A.~M.,  2017, \mndoi [Phys. Lett.]
  {10.1016/j.physletb.2017.03.014}, B768, 326

\bibitem[\protect\citeauthoryear{{Bonvin}}{{Bonvin}}{2008}]{Bonvin:2008ni}
{Bonvin} C.,  2008, \mndoi [\prd] {10.1103/PhysRevD.78.123530}, \href
  {http://adsabs.harvard.edu/abs/2008PhRvD..78l3530B} {78, 123530}

\bibitem[\protect\citeauthoryear{Bonvin \& Durrer}{Bonvin \&
  Durrer}{2011}]{Bonvin:2011bg}
Bonvin C.,  Durrer R.,  2011, \mndoi [Phys. Rev.] {10.1103/PhysRevD.84.063505},
  D84, 063505

\bibitem[\protect\citeauthoryear{Bonvin, Clarkson, Durrer, Maartens  \&
  Umeh}{Bonvin et~al.}{2015}]{Bonvin:2015kea}
Bonvin C.,  Clarkson C.,  Durrer R.,  Maartens R.,   Umeh O.,  2015, \mndoi
  [JCAP] {10.1088/1475-7516/2015/07/040}, 1507, 040

\bibitem[\protect\citeauthoryear{{Bonvin}, {Andrianomena}, {Bacon}, {Clarkson},
  {Maartens}, {Moloi}  \& {Bull}}{{Bonvin} et~al.}{2017}]{Bonvin:2016dze}
{Bonvin} C.,  {Andrianomena} S.,  {Bacon} D.,  {Clarkson} C.,  {Maartens} R.,
  {Moloi} T.,   {Bull} P.,  2017, \mndoi [\mnras] {10.1093/mnras/stx2049},
  \href {http://adsabs.harvard.edu/abs/2017MNRAS.472.3936B} {472, 3936}

\bibitem[\protect\citeauthoryear{{Book} \& {Flanagan}}{{Book} \&
  {Flanagan}}{2011}]{Book:2011}
{Book} L.~G.,  {Flanagan} {\'E}.~{\'E}.,  2011, \mndoi [\prd]
  {10.1103/PhysRevD.83.024024}, \href
  {http://adsabs.harvard.edu/abs/2011PhRvD..83b4024B} {83, 024024}

\bibitem[\protect\citeauthoryear{{Book}, {Kamionkowski}  \& {Schmidt}}{{Book}
  et~al.}{2012}]{2012PhRvL.108u1301B}
{Book} L.,  {Kamionkowski} M.,   {Schmidt} F.,  2012, \mndoi [Physical Review
  Letters] {10.1103/PhysRevLett.108.211301}, \href
  {http://adsabs.harvard.edu/abs/2012PhRvL.108u1301B} {108, 211301}

\bibitem[\protect\citeauthoryear{{Bosma}}{{Bosma}}{1981a}]{1981AJ.....86.1791B}
{Bosma} A.,  1981a, \mndoi [\aj] {10.1086/113062}, \href
  {http://adsabs.harvard.edu/abs/1981AJ.....86.1791B} {86, 1791}

\bibitem[\protect\citeauthoryear{{Bosma}}{{Bosma}}{1981b}]{1981AJ.....86.1825B}
{Bosma} A.,  1981b, \mndoi [\aj] {10.1086/113063}, \href
  {http://adsabs.harvard.edu/abs/1981AJ.....86.1825B} {86, 1825}

\bibitem[\protect\citeauthoryear{{Bosma}}{{Bosma}}{2017}]{2017ASSL..434..209B}
{Bosma} A.,  2017, in {Knapen} J.~H.,  {Lee} J.~C.,   {Gil de Paz} A.,  eds,
  Astrophysics and Space Science Library Vol. 434, Outskirts of Galaxies.
  p.~209

\bibitem[\protect\citeauthoryear{Bourke et~al.,}{Bourke
  et~al.}{2015}]{aaska2015}
Bourke T.~L.,  et~al., eds, 2015, Advancing Astrophysics with the Square
  Kilometre Array Square Kilometre Array Organisation, Jodrell Bank, UK

\bibitem[\protect\citeauthoryear{Bowman, Morales  \& Hewitt}{Bowman
  et~al.}{2006}]{Bowman2006}
Bowman J.~D.,  Morales M.~F.,   Hewitt J.~N.,  2006, \mndoi [The Astrophysical
  Journal] {10.1086/498703}, 638, 20

\bibitem[\protect\citeauthoryear{Bowman, Morales  \& Hewitt}{Bowman
  et~al.}{2008}]{Bowman2008}
Bowman J.~D.,  Morales M.~F.,   Hewitt J.~N.,  2008, \mndoi [The Astrophysical
  Journal] {10.1088/0004-637X/695/1/183}, 695, 183

\bibitem[\protect\citeauthoryear{Bowman, Rogers, Monsalve, Mozdzen  \&
  Mahesh}{Bowman et~al.}{2018}]{bowman2018}
Bowman J.~D.,  Rogers A. E.~E.,  Monsalve R.~A.,  Mozdzen T.~J.,   Mahesh N.,
  2018, Nature, 555, 67 EP

\bibitem[\protect\citeauthoryear{{Boyanovsky}, {de Vega}  \&
  {Sanchez}}{{Boyanovsky} et~al.}{2008}]{2008PhRvD..77d3518B}
{Boyanovsky} D.,  {de Vega} H.~J.,   {Sanchez} N.~G.,  2008, \mndoi [\prd]
  {10.1103/PhysRevD.77.043518}, \href
  {http://adsabs.harvard.edu/abs/2008PhRvD..77d3518B} {77, 043518}

\bibitem[\protect\citeauthoryear{{Boyle} et~al.,}{{Boyle}
  et~al.}{2018}]{2018ATel11901....1B}
{Boyle} P.~C.,  et~al., 2018, The Astronomer's Telegram, \href
  {http://adsabs.harvard.edu/abs/2018ATel11901....1B} {11901}

\bibitem[\protect\citeauthoryear{{Braginsky}, {Kardashev}, {Polnarev}  \&
  {Novikov}}{{Braginsky} et~al.}{1990}]{Braginsky:1990}
{Braginsky} V.~B.,  {Kardashev} N.~S.,  {Polnarev} A.~G.,   {Novikov} I.~D.,
  1990, Nuovo Cimento B Serie, \href
  {http://adsabs.harvard.edu/abs/1990NCimB.105.1141B} {105, 1141}

\bibitem[\protect\citeauthoryear{Branchini, Camera, Cuoco, Fornengo, Regis,
  Viel  \& Xia}{Branchini et~al.}{2017}]{Branchini:2016glc}
Branchini E.,  Camera S.,  Cuoco A.,  Fornengo N.,  Regis M.,  Viel M.,   Xia
  J.-Q.,  2017, \mndoi [Astrophys. J. Suppl.] {10.3847/1538-4365/228/1/8}, 228,
  8

\bibitem[\protect\citeauthoryear{Brandbyge, Hannestad, Haugb{\o}lle  \&
  Wong}{Brandbyge et~al.}{2010}]{Brandbyge:2010ge}
Brandbyge J.,  Hannestad S.,  Haugb{\o}lle T.,   Wong Y. Y.~Y.,  2010, \mndoi
  [JCAP] {10.1088/1475-7516/2010/09/014}, 1009, 014

\bibitem[\protect\citeauthoryear{{Brandenberger}, {Danos}, {Hern{\'a}ndez}  \&
  {Holder}}{{Brandenberger} et~al.}{2010}]{2010JCAP...12..028B}
{Brandenberger} R.~H.,  {Danos} R.~J.,  {Hern{\'a}ndez} O.~F.,   {Holder}
  G.~P.,  2010, \mndoi [\jcap] {10.1088/1475-7516/2010/12/028}, \href
  {http://adsabs.harvard.edu/abs/2010JCAP...12..028B} {12, 028}

\bibitem[\protect\citeauthoryear{{Brandt} \& {Kocsis}}{{Brandt} \&
  {Kocsis}}{2015}]{Brandt:2015}
{Brandt} T.~D.,  {Kocsis} B.,  2015, \mndoi [\apj]
  {10.1088/0004-637X/812/1/15}, \href
  {http://adsabs.harvard.edu/abs/2015ApJ...812...15B} {812, 15}

\bibitem[\protect\citeauthoryear{Braun}{Braun}{2017}]{braun2017}
Braun R.,  2017, Anticipated SKA1 Science Performance.
Square Kilometre Array Organisation, Jodrell Bank, UK

\bibitem[\protect\citeauthoryear{{Breysse}, {Ali-Ha{\"i}moud}  \&
  {Hirata}}{{Breysse} et~al.}{2018}]{2018PhRvD..98d3520B}
{Breysse} P.~C.,  {Ali-Ha{\"i}moud} Y.,   {Hirata} C.~M.,  2018, \mndoi [\prd]
  {10.1103/PhysRevD.98.043520}, \href
  {http://adsabs.harvard.edu/abs/2018PhRvD..98d3520B} {98, 043520}

\bibitem[\protect\citeauthoryear{Bringmann \& Weniger}{Bringmann \&
  Weniger}{2012}]{Bringmann:2012ez}
Bringmann T.,  Weniger C.,  2012, \mndoi [Phys. Dark Univ.]
  {10.1016/j.dark.2012.10.005}, 1, 194

\bibitem[\protect\citeauthoryear{Bringmann, Vollmann  \& Weniger}{Bringmann
  et~al.}{2014}]{Bringmann:2014lpa}
Bringmann T.,  Vollmann M.,   Weniger C.,  2014, \mndoi [Phys. Rev.]
  {10.1103/PhysRevD.90.123001}, D90, 123001

\bibitem[\protect\citeauthoryear{{Brown} \& {Battye}}{{Brown} \&
  {Battye}}{2011}]{2011MNRAS.410.2057B}
{Brown} M.~L.,  {Battye} R.~A.,  2011, \mndoi [\mnras]
  {10.1111/j.1365-2966.2010.17583.x}, \href
  {http://adsabs.harvard.edu/abs/2011MNRAS.410.2057B} {410, 2057}

\bibitem[\protect\citeauthoryear{Brown et~al.}{Brown
  et~al.}{2015}]{Brown:2015ucq}
Brown M.~L.,  et~al., 2015, PoS, AASKA14, 023

\bibitem[\protect\citeauthoryear{{Bryan} et~al.,}{{Bryan}
  et~al.}{2014}]{enzo14}
{Bryan} G.~L.,  et~al., 2014, \mndoi [\apjs] {10.1088/0067-0049/211/2/19},
  \href {http://adsabs.harvard.edu/abs/2014ApJS..211...19B} {211, 19}

\bibitem[\protect\citeauthoryear{Bull}{Bull}{2016}]{Bull:2015lja}
Bull P.,  2016, \mndoi [Astrophys. J.] {10.3847/0004-637X/817/1/26}, 817, 26

\bibitem[\protect\citeauthoryear{Bull \& Clifton}{Bull \&
  Clifton}{2012}]{Bull:2012zx}
Bull P.,  Clifton T.,  2012, \mndoi [Phys. Rev.] {10.1103/PhysRevD.85.103512},
  D85, 103512

\bibitem[\protect\citeauthoryear{Bull, Clifton  \& Ferreira}{Bull
  et~al.}{2012}]{Bull:2011wi}
Bull P.,  Clifton T.,   Ferreira P.~G.,  2012, \mndoi [Phys. Rev.]
  {10.1103/PhysRevD.85.024002}, D85, 024002

\bibitem[\protect\citeauthoryear{Bull, Ferreira, Patel  \& Santos}{Bull
  et~al.}{2015}]{Bull:2014rha}
Bull P.,  Ferreira P.~G.,  Patel P.,   Santos M.~G.,  2015, \mndoi [Astrophys.
  J.] {10.1088/0004-637X/803/1/21}, 803, 21

\bibitem[\protect\citeauthoryear{{Bull} et~al.,}{{Bull}
  et~al.}{2016}]{Bull2015}
{Bull} P.,  et~al., 2016, \mndoi [Physics of the Dark Universe]
  {10.1016/j.dark.2016.02.001}, \href
  {http://adsabs.harvard.edu/abs/2016PDU....12...56B} {12, 56}

\bibitem[\protect\citeauthoryear{{Burigana}, {Popa}, {Salvaterra}, {Schneider},
  {Choudhury}  \& {Ferrara}}{{Burigana} et~al.}{2008}]{2008MNRAS.385..404B}
{Burigana} C.,  {Popa} L.~A.,  {Salvaterra} R.,  {Schneider} R.,  {Choudhury}
  T.~R.,   {Ferrara} A.,  2008, \mndoi [\mnras]
  {10.1111/j.1365-2966.2008.12845.x}, \href
  {http://adsabs.harvard.edu/abs/2008MNRAS.385..404B} {385, 404}

\bibitem[\protect\citeauthoryear{{Burigana} et~al.,}{{Burigana}
  et~al.}{2015}]{2015aska.confE.149B}
{Burigana} C.,  et~al., 2015, Advancing Astrophysics with the Square Kilometre
  Array (AASKA14), \href {http://adsabs.harvard.edu/abs/2015aska.confE.149B}
  {p.~149}

\bibitem[\protect\citeauthoryear{{Burns} et~al.,}{{Burns}
  et~al.}{2012}]{burns2012}
{Burns} J.~O.,  et~al., 2012, \mndoi [Advances in Space Research]
  {10.1016/j.asr.2011.10.014}, \href
  {http://adsabs.harvard.edu/abs/2012AdSpR..49..433B} {49, 433}

\bibitem[\protect\citeauthoryear{{CHIME/FRB Collaboration} et~al.,}{{CHIME/FRB
  Collaboration} et~al.}{2018}]{2018ApJ...863...48C}
{CHIME/FRB Collaboration} et~al., 2018, \mndoi [\apj]
  {10.3847/1538-4357/aad188}, \href
  {http://adsabs.harvard.edu/abs/2018ApJ...863...48C} {863, 48}

\bibitem[\protect\citeauthoryear{Caldwell}{Caldwell}{2002}]{Caldwell:1999ew}
Caldwell R.~R.,  2002, \mndoi [Phys. Lett.] {10.1016/S0370-2693(02)02589-3},
  B545, 23

\bibitem[\protect\citeauthoryear{Calore, Cholis  \& Weniger}{Calore
  et~al.}{2015}]{Calore:2014xka}
Calore F.,  Cholis I.,   Weniger C.,  2015, \mndoi [JCAP]
  {10.1088/1475-7516/2015/03/038}, 1503, 038

\bibitem[\protect\citeauthoryear{Calore, Di~Mauro, Donato, Hessels  \&
  Weniger}{Calore et~al.}{2016}]{Calore:2015bsx}
Calore F.,  Di~Mauro M.,  Donato F.,  Hessels J. W.~T.,   Weniger C.,  2016,
  \mndoi [Astrophys. J.] {10.3847/0004-637X/827/2/143}, 827, 143

\bibitem[\protect\citeauthoryear{Camera \& Nishizawa}{Camera \&
  Nishizawa}{2013}]{Camera:2013xfa}
Camera S.,  Nishizawa A.,  2013, \mndoi [Phys. Rev. Lett.]
  {10.1103/PhysRevLett.110.151103}, 110, 151103

\bibitem[\protect\citeauthoryear{Camera, Santos, Bacon, Jarvis, McAlpine,
  Norris, Raccanelli  \& Rottgering}{Camera et~al.}{2012}]{Camera:2012ez}
Camera S.,  Santos M.~G.,  Bacon D.~J.,  Jarvis M.~J.,  McAlpine K.,  Norris
  R.~P.,  Raccanelli A.,   Rottgering H.,  2012, \mndoi [Mon. Not. Roy. Astron.
  Soc.] {10.1111/j.1365-2966.2012.22073.x}, 427, 2079

\bibitem[\protect\citeauthoryear{{Camera}, {Santos}, {Ferreira}  \&
  {Ferramacho}}{{Camera} et~al.}{2013a}]{Camera2013}
{Camera} S.,  {Santos} M.~G.,  {Ferreira} P.~G.,   {Ferramacho} L.,  2013a,
  \mndoi [Physical Review Letters] {10.1103/PhysRevLett.111.171302}, \href
  {http://adsabs.harvard.edu/abs/2013PhRvL.111q1302C} {111, 171302}

\bibitem[\protect\citeauthoryear{Camera, Fornasa, Fornengo  \& Regis}{Camera
  et~al.}{2013b}]{Camera:2012cj}
Camera S.,  Fornasa M.,  Fornengo N.,   Regis M.,  2013b, \mndoi [Astrophys.
  J.] {10.1088/2041-8205/771/1/L5}, 771, L5

\bibitem[\protect\citeauthoryear{Camera, Santos  \& Maartens}{Camera
  et~al.}{2015a}]{Camera:2014bwa}
Camera S.,  Santos M.~G.,   Maartens R.,  2015a, \mndoi [Mon. Not. Roy. Astron.
  Soc.] {10.1093/mnras/stv040}, 448, 1035

\bibitem[\protect\citeauthoryear{Camera, Maartens  \& Santos}{Camera
  et~al.}{2015b}]{Camera:2014sba}
Camera S.,  Maartens R.,   Santos M.~G.,  2015b, \mndoi [Mon. Not. Roy. Astron.
  Soc.] {10.1093/mnrasl/slv069}, 451, L80

\bibitem[\protect\citeauthoryear{Camera, Carbone, Fedeli  \& Moscardini}{Camera
  et~al.}{2015c}]{Camera:2014dia}
Camera S.,  Carbone C.,  Fedeli C.,   Moscardini L.,  2015c, \mndoi [Phys.
  Rev.] {10.1103/PhysRevD.91.043533}, D91, 043533

\bibitem[\protect\citeauthoryear{Camera, Fornasa, Fornengo  \& Regis}{Camera
  et~al.}{2015d}]{Camera:2014rja}
Camera S.,  Fornasa M.,  Fornengo N.,   Regis M.,  2015d, \mndoi [JCAP]
  {10.1088/1475-7516/2015/06/029}, 1506, 029

\bibitem[\protect\citeauthoryear{Camera et~al.}{Camera
  et~al.}{2015e}]{Camera:2015fsa}
Camera S.,  et~al., 2015e, PoS, AASKA14, 025

\bibitem[\protect\citeauthoryear{{Camera}, {Harrison}, {Bonaldi}  \&
  {Brown}}{{Camera} et~al.}{2017}]{2017MNRAS.464.4747C}
{Camera} S.,  {Harrison} I.,  {Bonaldi} A.,   {Brown} M.~L.,  2017, \mndoi
  [\mnras] {10.1093/mnras/stw2688}, \href
  {http://adsabs.harvard.edu/abs/2017MNRAS.464.4747C} {464, 4747}

\bibitem[\protect\citeauthoryear{Capozziello, Lambiase, Sakellariadou  \&
  Stabile}{Capozziello et~al.}{2015}]{Capozziello:2014mea}
Capozziello S.,  Lambiase G.,  Sakellariadou M.,   Stabile A.,  2015, \mndoi
  [Phys. Rev.] {10.1103/PhysRevD.91.044012}, D91, 044012

\bibitem[\protect\citeauthoryear{{Carr}}{{Carr}}{2005}]{astro-ph/0511743}
{Carr} B.~J.,  2005, preprint, \href
  {http://adsabs.harvard.edu/abs/2005astro.ph.11743C} {} (\mn@eprint {arXiv}
  {astro-ph/0511743})

\bibitem[\protect\citeauthoryear{Carr \& Hawking}{Carr \&
  Hawking}{1974}]{CarrHawking}
Carr B.~J.,  Hawking S.~W.,  1974, Mon. Not. Roy. Astron. Soc., 168

\bibitem[\protect\citeauthoryear{{Carr}, {Kohri}, {Sendouda}  \&
  {Yokoyama}}{{Carr} et~al.}{2010}]{cksy}
{Carr} B.~J.,  {Kohri} K.,  {Sendouda} Y.,   {Yokoyama} J.,  2010, \mndoi
  [\prd] {10.1103/PhysRevD.81.104019}, \href
  {http://adsabs.harvard.edu/abs/2010PhRvD..81j4019C} {81, 104019}

\bibitem[\protect\citeauthoryear{{Carr}, {K{\"u}hnel}  \& {Sandstad}}{{Carr}
  et~al.}{2016}]{1607.06077}
{Carr} B.,  {K{\"u}hnel} F.,   {Sandstad} M.,  2016, \mndoi [\prd]
  {10.1103/PhysRevD.94.083504}, \href
  {http://adsabs.harvard.edu/abs/2016PhRvD..94h3504C} {94, 083504}

\bibitem[\protect\citeauthoryear{{Carucci}, {Villaescusa-Navarro}, {Viel}  \&
  {Lapi}}{{Carucci} et~al.}{2015}]{Carucci_2015}
{Carucci} I.~P.,  {Villaescusa-Navarro} F.,  {Viel} M.,   {Lapi} A.,  2015,
  \mndoi [\jcap] {10.1088/1475-7516/2015/07/047}, \href
  {http://adsabs.harvard.edu/abs/2015JCAP...07..047C} {7, 047}

\bibitem[\protect\citeauthoryear{{Castorina} \&
  {Villaescusa-Navarro}}{{Castorina} \&
  {Villaescusa-Navarro}}{2017}]{2016arXiv160905157C}
{Castorina} E.,  {Villaescusa-Navarro} F.,  2017, \mndoi [Monthly Notices of
  the Royal Astronomical Society] {10.1093/mnras/stx1599}, \href
  {https://ui.adsabs.harvard.edu/abs/2017MNRAS.471.1788C} {471, 1788}

\bibitem[\protect\citeauthoryear{{Castorina}, {Sefusatti}, {Sheth},
  {Villaescusa-Navarro}  \& {Viel}}{{Castorina} et~al.}{2014}]{Castorina_2014}
{Castorina} E.,  {Sefusatti} E.,  {Sheth} R.~K.,  {Villaescusa-Navarro} F.,
  {Viel} M.,  2014, \mndoi [\jcap] {10.1088/1475-7516/2014/02/049}, \href
  {http://adsabs.harvard.edu/abs/2014JCAP...02..049C} {2, 49}

\bibitem[\protect\citeauthoryear{Cembranos \& Maroto}{Cembranos \&
  Maroto}{2016}]{Cembranos:2016jun}
Cembranos J. A.~R.,  Maroto A.~L.,  2016, \mndoi [Int. J. Mod. Phys.]
  {10.1142/S0217751X16300155}, 31, 1630015

\bibitem[\protect\citeauthoryear{Cembranos, Dobado  \& Maroto}{Cembranos
  et~al.}{2001a}]{Cembranos:2001my}
Cembranos J. A.~R.,  Dobado A.,   Maroto A.~L.,  2001a, in {Lepton and photon
  interactions at high energies. Proceedings, 20th International Symposium, LP
  2001, Rome, Italy, July 23-28, 2001}.  (\mn@eprint {arXiv} {hep-ph/0107155})

\bibitem[\protect\citeauthoryear{Cembranos, Dobado  \& Maroto}{Cembranos
  et~al.}{2001b}]{PhysRevD.65.026005}
Cembranos J. A.~R.,  Dobado A.,   Maroto A.~L.,  2001b, \mndoi [Phys. Rev. D]
  {10.1103/PhysRevD.65.026005}, 65, 026005

\bibitem[\protect\citeauthoryear{Cembranos, Dobado  \& Maroto}{Cembranos
  et~al.}{2003a}]{PhysRevLett.90.241301}
Cembranos J. A.~R.,  Dobado A.,   Maroto A.~L.,  2003a, \mndoi [Phys. Rev.
  Lett.] {10.1103/PhysRevLett.90.241301}, 90, 241301

\bibitem[\protect\citeauthoryear{{Cembranos}, {Dobado}  \&
  {Maroto}}{{Cembranos} et~al.}{2003b}]{2003AIPC..670..235C}
{Cembranos} J.~A.~R.,  {Dobado} A.,   {Maroto} A.~L.,  2003b, in {Cotti} U.,
  {Mondrag{\'o}n} M.,   {Tavares-Velasco} G.,  eds,  American Institute of
  Physics Conference Series Vol. 670, Particles and Fields. pp 235--242,
  \mndoi{10.1063/1.1594340}

\bibitem[\protect\citeauthoryear{Cembranos, Dobado  \& Maroto}{Cembranos
  et~al.}{2003c}]{Cembranos:2003fu}
Cembranos J. A.~R.,  Dobado A.,   Maroto A.~L.,  2003c, \mndoi [Phys. Rev.]
  {10.1103/PhysRevD.68.103505}, D68, 103505

\bibitem[\protect\citeauthoryear{Cembranos, Dobado  \& Maroto}{Cembranos
  et~al.}{2004}]{PhysRevD.70.096001}
Cembranos J. A.~R.,  Dobado A.,   Maroto A.~L.,  2004, \mndoi [Phys. Rev. D]
  {10.1103/PhysRevD.70.096001}, 70, 096001

\bibitem[\protect\citeauthoryear{Cembranos, Dobado  \& Maroto}{Cembranos
  et~al.}{2006}]{Cembranos:2006mj}
Cembranos J. A.~R.,  Dobado A.,   Maroto A.~L.,  2006, in {Recent developments
  in theoretical and experimental general relativity, gravitation and
  relativistic field theories. Proceedings, 11th Marcel Grossmann Meeting,
  MG11, Berlin, Germany, July 23-29, 2006. Pt. A-C}. pp 2851--2853

\bibitem[\protect\citeauthoryear{Cembranos, Diaz-Cruz  \& Prado}{Cembranos
  et~al.}{2011}]{Cembranos:2011cm}
Cembranos J. A.~R.,  Diaz-Cruz J.~L.,   Prado L.,  2011, \mndoi [Phys. Rev.]
  {10.1103/PhysRevD.84.083522}, D84, 083522

\bibitem[\protect\citeauthoryear{Cembranos, de~la Cruz-Dombriz, Gammaldi  \&
  Maroto}{Cembranos et~al.}{2012}]{Cembranos:2011hi}
Cembranos J. A.~R.,  de~la Cruz-Dombriz A.,  Gammaldi V.,   Maroto A.~L.,
  2012, \mndoi [Phys. Rev.] {10.1103/PhysRevD.85.043505}, D85, 043505

\bibitem[\protect\citeauthoryear{{Cembranos}, {de la Cruz-Dombriz}, {Gammaldi}
  \& {Mendez-Isla}}{{Cembranos} et~al.}{2019}]{2019arXiv190511154C}
{Cembranos} J.~A.~R.,  {de la Cruz-Dombriz} A.,  {Gammaldi} V.,   {Mendez-Isla}
  M.,  2019, arXiv e-prints, \href
  {https://ui.adsabs.harvard.edu/abs/2019arXiv190511154C} {p. arXiv:1905.11154}

\bibitem[\protect\citeauthoryear{{Challinor} \& {Lewis}}{{Challinor} \&
  {Lewis}}{2011}]{2011PhRvD..84d3516C}
{Challinor} A.,  {Lewis} A.,  2011, \mndoi [\prd] {10.1103/PhysRevD.84.043516},
  \href {http://adsabs.harvard.edu/abs/2011PhRvD..84d3516C} {84, 043516}

\bibitem[\protect\citeauthoryear{Chamseddine, Connes  \& Marcolli}{Chamseddine
  et~al.}{2007}]{Chamseddine:2006ep}
Chamseddine A.~H.,  Connes A.,   Marcolli M.,  2007, \mndoi [Adv. Theor. Math.
  Phys.] {10.4310/ATMP.2007.v11.n6.a3}, 11, 991

\bibitem[\protect\citeauthoryear{Chang, Pen, Peterson  \& McDonald}{Chang
  et~al.}{2008}]{Chang:2007xk}
Chang T.-C.,  Pen U.-L.,  Peterson J.~B.,   McDonald P.,  2008, \mndoi [Phys.
  Rev. Lett.] {10.1103/PhysRevLett.100.091303}, 100, 091303

\bibitem[\protect\citeauthoryear{{Chapline}}{{Chapline}}{1975}]{chapline}
{Chapline} G.~F.,  1975, \mndoi [\nat] {10.1038/253251a0}, \href
  {http://adsabs.harvard.edu/abs/1975Natur.253..251C} {253, 251}

\bibitem[\protect\citeauthoryear{Chapman et~al.,}{Chapman
  et~al.}{2012}]{Chapman2012}
Chapman E.,  et~al., 2012, \mndoi [Monthly Notices of the Royal Astronomical
  Society] {10.1111/j.1365-2966.2012.21065.x}, 423, 2518

\bibitem[\protect\citeauthoryear{Chapman et~al.,}{Chapman
  et~al.}{2013}]{Chapman2013}
Chapman E.,  et~al., 2013, \mndoi [Monthly Notices of the Royal Astronomical
  Society] {10.1093/mnras/sts333}, 429, 165

\bibitem[\protect\citeauthoryear{Chapman et~al.,}{Chapman
  et~al.}{2015}]{Chapman2015}
Chapman E.,  et~al., 2015, Advancing Astrophysics with the Square Kilometre
  Array (AASKA14), p.~5

\bibitem[\protect\citeauthoryear{Chapman, Zaroubi, Abdalla, Dulwich, Jeli{\'c}
  \& Mort}{Chapman et~al.}{2016}]{Chapman2016}
Chapman E.,  Zaroubi S.,  Abdalla F.~B.,  Dulwich F.,  Jeli{\'c} V.,   Mort B.,
   2016, \mndoi [Monthly Notices of the Royal Astronomical Society]
  {10.1093/mnras/stw161}, 458, 2928

\bibitem[\protect\citeauthoryear{Chatterjee et~al.}{Chatterjee
  et~al.}{2017}]{Chatterjee:2017dqg}
Chatterjee S.,  et~al., 2017, \mndoi [Nature] {10.1038/nature20797}, 541, 58

\bibitem[\protect\citeauthoryear{{Chennamangalam} \&
  {Lorimer}}{{Chennamangalam} \& {Lorimer}}{2014}]{2014MNRAS.440L..86C}
{Chennamangalam} J.,  {Lorimer} D.~R.,  2014, \mndoi [\mnras]
  {10.1093/mnrasl/slu025}, \href
  {http://adsabs.harvard.edu/abs/2014MNRAS.440L..86C} {440, L86}

\bibitem[\protect\citeauthoryear{{Chluba} \& {Sunyaev}}{{Chluba} \&
  {Sunyaev}}{2012}]{2012MNRAS.419.1294C}
{Chluba} J.,  {Sunyaev} R.~A.,  2012, \mndoi [\mnras]
  {10.1111/j.1365-2966.2011.19786.x}, \href
  {http://adsabs.harvard.edu/abs/2012MNRAS.419.1294C} {419, 1294}

\bibitem[\protect\citeauthoryear{{Cho} \& {Lazarian}}{{Cho} \&
  {Lazarian}}{2002}]{2002ApJ...575L..63C}
{Cho} J.,  {Lazarian} A.,  2002, \mndoi [\apjl] {10.1086/342722}, \href
  {http://adsabs.harvard.edu/abs/2002ApJ...575L..63C} {575, L63}

\bibitem[\protect\citeauthoryear{{Christodoulou}, {Rovelli}, {Speziale}  \&
  {Vilensky}}{{Christodoulou} et~al.}{2016}]{Christodoulou:2016ve}
{Christodoulou} M.,  {Rovelli} C.,  {Speziale} S.,   {Vilensky} I.,  2016,
  \mndoi [Physical Review D] {10.1103/PhysRevD.94.084035}, \href
  {https://ui.adsabs.harvard.edu/abs/2016PhRvD..94h4035C} {94, 084035}

\bibitem[\protect\citeauthoryear{Cirelli \& Taoso}{Cirelli \&
  Taoso}{2016}]{Cirelli:2016mrc}
Cirelli M.,  Taoso M.,  2016, \mndoi [JCAP] {10.1088/1475-7516/2016/07/041},
  1607, 041

\bibitem[\protect\citeauthoryear{{Clark}, {Dutta}, {Gao}, {Strigari}  \&
  {Watson}}{{Clark} et~al.}{2017}]{clark2017}
{Clark} S.~J.,  {Dutta} B.,  {Gao} Y.,  {Strigari} L.~E.,   {Watson} S.,  2017,
  \mndoi [\prd] {10.1103/PhysRevD.95.083006}, \href
  {http://adsabs.harvard.edu/abs/2017PhRvD..95h3006C} {95, 083006}

\bibitem[\protect\citeauthoryear{Clarkson}{Clarkson}{2012}]{Clarkson:2012bg}
Clarkson C.,  2012, \mndoi [Comptes Rendus Physique]
  {10.1016/j.crhy.2012.04.005}, 13, 682

\bibitem[\protect\citeauthoryear{Clarkson \& Maartens}{Clarkson \&
  Maartens}{2010}]{Clarkson:2010uz}
Clarkson C.,  Maartens R.,  2010, \mndoi [Class. Quant. Grav.]
  {10.1088/0264-9381/27/12/124008}, 27, 124008

\bibitem[\protect\citeauthoryear{Clarkson, Ellis, Faltenbacher, Maartens, Umeh
  \& Uzan}{Clarkson et~al.}{2012}]{Clarkson:2011br}
Clarkson C.,  Ellis G. F.~R.,  Faltenbacher A.,  Maartens R.,  Umeh O.,   Uzan
  J.-P.,  2012, \mndoi [Mon. Not. Roy. Astron. Soc.]
  {10.1111/j.1365-2966.2012.21750.x}, 426, 1121

\bibitem[\protect\citeauthoryear{Clifton, Clarkson  \& Bull}{Clifton
  et~al.}{2012a}]{Clifton:2011sn}
Clifton T.,  Clarkson C.,   Bull P.,  2012a, \mndoi [Phys. Rev. Lett.]
  {10.1103/PhysRevLett.109.051303}, 109, 051303

\bibitem[\protect\citeauthoryear{Clifton, Ferreira, Padilla  \&
  Skordis}{Clifton et~al.}{2012b}]{Clifton:2011jh}
Clifton T.,  Ferreira P.~G.,  Padilla A.,   Skordis C.,  2012b, \mndoi [Phys.
  Rept.] {10.1016/j.physrep.2012.01.001}, 513, 1

\bibitem[\protect\citeauthoryear{{Cohen}, {Fialkov}, {Barkana}  \&
  {Lotem}}{{Cohen} et~al.}{2017}]{2016arXiv160902312C}
{Cohen} A.,  {Fialkov} A.,  {Barkana} R.,   {Lotem} M.,  2017, \mndoi [\mnras]
  {10.1093/mnras/stx2065}, \href
  {http://adsabs.harvard.edu/abs/2017MNRAS.472.1915C} {472, 1915}

\bibitem[\protect\citeauthoryear{Colafrancesco, Regis, Marchegiani, Beck, Beck,
  Zechlin, Lobanov  \& Horns}{Colafrancesco
  et~al.}{2015}]{Colafrancesco:2015ola}
Colafrancesco S.,  Regis M.,  Marchegiani P.,  Beck G.,  Beck R.,  Zechlin H.,
  Lobanov A.,   Horns D.,  2015, PoS, AASKA14, 100

\bibitem[\protect\citeauthoryear{{Cole} \& {Byrnes}}{{Cole} \&
  {Byrnes}}{2018}]{Byrnes3}
{Cole} P.~S.,  {Byrnes} C.~T.,  2018, \mndoi [\jcap]
  {10.1088/1475-7516/2018/02/019}, \href
  {http://adsabs.harvard.edu/abs/2018JCAP...02..019C} {2, 019}

\bibitem[\protect\citeauthoryear{{Condon}, {Cotton}, {Greisen}, {Yin},
  {Perley}, {Taylor}  \& {Broderick}}{{Condon} et~al.}{1998}]{Condon:1998}
{Condon} J.~J.,  {Cotton} W.~D.,  {Greisen} E.~W.,  {Yin} Q.~F.,  {Perley}
  R.~A.,  {Taylor} G.~B.,   {Broderick} J.~J.,  1998, \mndoi [\aj]
  {10.1086/300337}, \href {http://adsabs.harvard.edu/abs/1998AJ....115.1693C}
  {115, 1693}

\bibitem[\protect\citeauthoryear{{Condon} et~al.,}{{Condon}
  et~al.}{2012}]{2012ApJ...758...23C}
{Condon} J.~J.,  et~al., 2012, \mndoi [\apj] {10.1088/0004-637X/758/1/23},
  \href {http://adsabs.harvard.edu/abs/2012ApJ...758...23C} {758, 23}

\bibitem[\protect\citeauthoryear{Conrad \& Reimer}{Conrad \&
  Reimer}{2017}]{Conrad:2017}
Conrad J.,  Reimer O.,  2017, Nat Phys, 13, 224

\bibitem[\protect\citeauthoryear{{Cooray}}{{Cooray}}{2006}]{2006PhRvL..97z1301C}
{Cooray} A.,  2006, \mndoi [Physical Review Letters]
  {10.1103/PhysRevLett.97.261301}, \href
  {http://adsabs.harvard.edu/abs/2006PhRvL..97z1301C} {97, 261301}

\bibitem[\protect\citeauthoryear{Cordes \& Shannon}{Cordes \&
  Shannon}{2010}]{Cordes:2010fh}
Cordes J.~M.,  Shannon R.~M.,  2010, preprint (\mn@eprint {arXiv} {1010.3785})

\bibitem[\protect\citeauthoryear{Corichi \& Singh}{Corichi \&
  Singh}{2016}]{Corichi:2015xia}
Corichi A.,  Singh P.,  2016, \mndoi [Class. Quant. Grav.]
  {10.1088/0264-9381/33/5/055006}, 33, 055006

\bibitem[\protect\citeauthoryear{{Costanzi}, {Villaescusa-Navarro}, {Viel},
  {Xia}, {Borgani}, {Castorina}  \& {Sefusatti}}{{Costanzi}
  et~al.}{2013}]{Costanzi_2014}
{Costanzi} M.,  {Villaescusa-Navarro} F.,  {Viel} M.,  {Xia} J.-Q.,  {Borgani}
  S.,  {Castorina} E.,   {Sefusatti} E.,  2013, \mndoi [\jcap]
  {10.1088/1475-7516/2013/12/012}, \href
  {http://adsabs.harvard.edu/abs/2013JCAP...12..012C} {12, 12}

\bibitem[\protect\citeauthoryear{{Crain} et~al.,}{{Crain}
  et~al.}{2017}]{Crain:2016ex}
{Crain} R.~A.,  et~al., 2017, \mndoi [Monthly Notices of the Royal Astronomical
  Society] {10.1093/mnras/stw2586}, \href
  {https://ui.adsabs.harvard.edu/abs/2017MNRAS.464.4204C} {464, 4204}

\bibitem[\protect\citeauthoryear{{Crittenden} \& {Turok}}{{Crittenden} \&
  {Turok}}{1996}]{crittenden96}
{Crittenden} R.~G.,  {Turok} N.,  1996, \mndoi [Physical Review Letters]
  {10.1103/PhysRevLett.76.575}, \href
  {http://adsabs.harvard.edu/abs/1996PhRvL..76..575C} {76, 575}

\bibitem[\protect\citeauthoryear{Crocker, Bell, Balazs  \& Jones}{Crocker
  et~al.}{2010}]{Crocker:2010gy}
Crocker R.~M.,  Bell N.~F.,  Balazs C.,   Jones D.~I.,  2010, \mndoi [Phys.
  Rev.] {10.1103/PhysRevD.81.063516}, D81, 063516

\bibitem[\protect\citeauthoryear{{Cuesta}, {Niro}  \& {Verde}}{{Cuesta}
  et~al.}{2016}]{Cuesta_2016}
{Cuesta} A.~J.,  {Niro} V.,   {Verde} L.,  2016, \mndoi [Physics of the Dark
  Universe] {10.1016/j.dark.2016.04.005}, \href
  {http://adsabs.harvard.edu/abs/2016PDU....13...77C} {13, 77}

\bibitem[\protect\citeauthoryear{Cuoco, Xia, Regis, Branchini, Fornengo  \&
  Viel}{Cuoco et~al.}{2015}]{Cuoco:2015rfa}
Cuoco A.,  Xia J.-Q.,  Regis M.,  Branchini E.,  Fornengo N.,   Viel M.,  2015,
  \mndoi [Astrophys. J. Suppl.] {10.1088/0067-0049/221/2/29}, 221, 29

\bibitem[\protect\citeauthoryear{{Curran}}{{Curran}}{2007}]{Curran:2007}
{Curran} S.,  2007, in {Lobanov} A.~P.,  {Zensus} J.~A.,  {Cesarsky} C.,
  {Diamond} P.~J.,  eds, Exploring the Cosmic Frontier: Astrophysical
  Instruments for the 21st Century. Springer-Verlag, p.~91,
  \mndoi{10.1007/978-3-540-39756-4_28}

\bibitem[\protect\citeauthoryear{Curran, Kanekar  \& Darling}{Curran
  et~al.}{2004}]{Curran:2004mg}
Curran S.~J.,  Kanekar N.,   Darling J.~K.,  2004, \mndoi [New Astron. Rev.]
  {10.1016/j.newar.2004.09.004}, 48, 1095

\bibitem[\protect\citeauthoryear{Cutchin, Simonetti, Ellingson, Larracuente  \&
  Kavic}{Cutchin et~al.}{2016}]{Cutchin:2016gbu}
Cutchin S.~E.,  Simonetti J.~H.,  Ellingson S.~W.,  Larracuente A.~S.,   Kavic
  M.~J.,  2016, \mndoi [Publ. Astron. Soc. Pac.] {10.1086/684195}, 127, 1269

\bibitem[\protect\citeauthoryear{Cutler \& Holz}{Cutler \&
  Holz}{2009}]{Cutler:2009qv}
Cutler C.,  Holz D.~E.,  2009, \mndoi [Phys. Rev.]
  {10.1103/PhysRevD.80.104009}, D80, 104009

\bibitem[\protect\citeauthoryear{Cyr-Racine, Sigurdson, Zavala, Bringmann,
  Vogelsberger  \& Pfrommer}{Cyr-Racine et~al.}{2016}]{Cyr-Racine:2015ihg}
Cyr-Racine F.-Y.,  Sigurdson K.,  Zavala J.,  Bringmann T.,  Vogelsberger M.,
  Pfrommer C.,  2016, \mndoi [Phys. Rev.] {10.1103/PhysRevD.93.123527}, D93,
  123527

\bibitem[\protect\citeauthoryear{{Dalal}, {Dor{\'e}}, {Huterer}  \&
  {Shirokov}}{{Dalal} et~al.}{2008}]{0710.4560}
{Dalal} N.,  {Dor{\'e}} O.,  {Huterer} D.,   {Shirokov} A.,  2008, \mndoi
  [\prd] {10.1103/PhysRevD.77.123514}, \href
  {http://adsabs.harvard.edu/abs/2008PhRvD..77l3514D} {77, 123514}

\bibitem[\protect\citeauthoryear{{Damour} \& {Deruelle}}{{Damour} \&
  {Deruelle}}{1986}]{1986AIHS...44..263D}
{Damour} T.,  {Deruelle} N.,  1986, Ann.~Inst.~Henri Poincar{\'e}
  Phys.~Th{\'e}or., \href {http://adsabs.harvard.edu/abs/1986AIHS...44..263D}
  {44, 263}

\bibitem[\protect\citeauthoryear{Damour \& Esposito-Farese}{Damour \&
  Esposito-Farese}{1992}]{Damour:1992ah}
Damour T.,  Esposito-Farese G.,  1992, \mndoi [Phys. Rev.]
  {10.1103/PhysRevD.46.4128}, D46, 4128

\bibitem[\protect\citeauthoryear{Damour \& Esposito-Farese}{Damour \&
  Esposito-Farese}{1996}]{Damour:1996ke}
Damour T.,  Esposito-Farese G.,  1996, \mndoi [Phys. Rev.]
  {10.1103/PhysRevD.54.1474}, D54, 1474

\bibitem[\protect\citeauthoryear{Damour \& Schaefer}{Damour \&
  Schaefer}{1991}]{Damour:1991rq}
Damour T.,  Schaefer G.,  1991, \mndoi [Phys. Rev. Lett.]
  {10.1103/PhysRevLett.66.2549}, 66, 2549

\bibitem[\protect\citeauthoryear{Damour \& Taylor}{Damour \&
  Taylor}{1992}]{Damour:1991rd}
Damour T.,  Taylor J.~H.,  1992, \mndoi [Phys. Rev.]
  {10.1103/PhysRevD.45.1840}, D45, 1840

\bibitem[\protect\citeauthoryear{Damour, Gibbons  \& Taylor}{Damour
  et~al.}{1988}]{Damour:1988zz}
Damour T.,  Gibbons G.~W.,   Taylor J.~H.,  1988, \mndoi [Phys. Rev. Lett.]
  {10.1103/PhysRevLett.61.1151}, 61, 1151

\bibitem[\protect\citeauthoryear{Damour, Piazza  \& Veneziano}{Damour
  et~al.}{2002}]{Damour:2002mi}
Damour T.,  Piazza F.,   Veneziano G.,  2002, \mndoi [Phys. Rev. Lett.]
  {10.1103/PhysRevLett.89.081601}, 89, 081601

\bibitem[\protect\citeauthoryear{{Dav{\'e}} et~al.,}{{Dav{\'e}}
  et~al.}{2001}]{2001ApJ...552..473D}
{Dav{\'e}} R.,  et~al., 2001, \mndoi [\apj] {10.1086/320548}, \href
  {http://adsabs.harvard.edu/abs/2001ApJ...552..473D} {552, 473}

\bibitem[\protect\citeauthoryear{{Daylan}, {Finkbeiner}, {Hooper}, {Linden},
  {Portillo}, {Rodd}  \& {Slatyer}}{{Daylan} et~al.}{2016}]{Daylan:2014rsa}
{Daylan} T.,  {Finkbeiner} D.~P.,  {Hooper} D.,  {Linden} T.,  {Portillo}
  S.~K.~N.,  {Rodd} N.~L.,   {Slatyer} T.~R.,  2016, \mndoi [Physics of the
  Dark Universe] {10.1016/j.dark.2015.12.005}, \href
  {http://adsabs.harvard.edu/abs/2016PDU....12....1D} {12, 1}

\bibitem[\protect\citeauthoryear{{De Zotti} et~al.,}{{De Zotti}
  et~al.}{2018}]{2016arXiv160907263D}
{De Zotti} G.,  et~al., 2018, \mndoi [\jcap] {10.1088/1475-7516/2018/04/020},
  \href {http://adsabs.harvard.edu/abs/2018JCAP...04..020D} {4, 020}

\bibitem[\protect\citeauthoryear{{DeBoer} et~al.}{{DeBoer}
  et~al.}{2017}]{DeBoer2017}
{DeBoer} D.~R.,  et~al., 2017, \mndoi [\pasp]
  {10.1088/1538-3873/129/974/045001}, \href
  {http://adsabs.harvard.edu/abs/2017PASP..129d5001D} {129, 045001}

\bibitem[\protect\citeauthoryear{Delahaye, Lineros, Donato, Fornengo  \&
  Salati}{Delahaye et~al.}{2008}]{Delahaye:2007fr}
Delahaye T.,  Lineros R.,  Donato F.,  Fornengo N.,   Salati P.,  2008, \mndoi
  [Phys. Rev.] {10.1103/PhysRevD.77.063527}, D77, 063527

\bibitem[\protect\citeauthoryear{{Demetroullas} \& {Brown}}{{Demetroullas} \&
  {Brown}}{2016}]{2016MNRAS.456.3100D}
{Demetroullas} C.,  {Brown} M.~L.,  2016, \mndoi [\mnras]
  {10.1093/mnras/stv2876}, \href
  {http://adsabs.harvard.edu/abs/2016MNRAS.456.3100D} {456, 3100}

\bibitem[\protect\citeauthoryear{{Deneva}, {Cordes}  \& {Lazio}}{{Deneva}
  et~al.}{2009}]{2009ApJ...702L.177D}
{Deneva} J.~S.,  {Cordes} J.~M.,   {Lazio} T.~J.~W.,  2009, \mndoi [\apjl]
  {10.1088/0004-637X/702/2/L177}, \href
  {http://adsabs.harvard.edu/abs/2009ApJ...702L.177D} {702, L177}

\bibitem[\protect\citeauthoryear{{Dewdney}, {Turner}, {Millenaar}, {McCool},
  {Lazio}  \& {Cornwell}}{{Dewdney} et~al.}{2013}]{Dewdney2015}
{Dewdney} P.~E.,  {Turner} W.,  {Millenaar} R.,  {McCool} R.,  {Lazio} J.,
  {Cornwell} T.~J.,  2013, SKA-TEL-SKO-DD001, Revlsion 01

\bibitem[\protect\citeauthoryear{Dewdney, Turner, Braun, Santander-Vela,
  Waterson  \& Tan}{Dewdney et~al.}{2016}]{dewdney2016}
Dewdney P.,  Turner W.,  Braun R.,  Santander-Vela J.,  Waterson M.,   Tan
  G.-H.,  2016, SKA1 System Baseline Design.
Square Kilometre Array Organisation, Jodrell Bank, UK

\bibitem[\protect\citeauthoryear{{Dewey}, {Stokes}, {Segelstein}, {Taylor}  \&
  {Weisberg}}{{Dewey} et~al.}{1984}]{1984bens.work..234D}
{Dewey} R.,  {Stokes} G.,  {Segelstein} D.,  {Taylor} J.,   {Weisberg} J.,
  1984, in {Reynolds} S.~P.,  {Stinebring} D.~R.,  eds, Birth and Evolution of
  Neutron Stars: Issues Raised by Millisecond Pulsars. p.~234

\bibitem[\protect\citeauthoryear{Di~Porto, Amendola  \& Branchini}{Di~Porto
  et~al.}{2012}]{DiPorto:2012ey}
Di~Porto C.,  Amendola L.,   Branchini E.,  2012, \mndoi [Mon. Not. Roy.
  Astron. Soc.] {10.1111/j.1745-3933.2012.01265.x}, 423, L97

\bibitem[\protect\citeauthoryear{Dillon et~al.,}{Dillon
  et~al.}{2015}]{Dillon2015}
Dillon J.~S.,  et~al., 2015, \mndoi [Physical Review D]
  {10.1103/PhysRevD.91.123011}, 91, 123011

\bibitem[\protect\citeauthoryear{{Dine}, {Fischler}  \& {Srednicki}}{{Dine}
  et~al.}{1981}]{1981PhLB..104..199D}
{Dine} M.,  {Fischler} W.,   {Srednicki} M.,  1981, \mndoi [Physics Letters B]
  {10.1016/0370-2693(81)90590-6}, \href
  {http://adsabs.harvard.edu/abs/1981PhLB..104..199D} {104, 199}

\bibitem[\protect\citeauthoryear{Dirac}{Dirac}{1937}]{Dirac:1937ti}
Dirac P. A.~M.,  1937, \mndoi [Nature] {10.1038/139323a0}, 139, 323

\bibitem[\protect\citeauthoryear{Dobado \& Maroto}{Dobado \&
  Maroto}{2001}]{Dobado:2000gr}
Dobado A.,  Maroto A.~L.,  2001, \mndoi [Nucl. Phys.]
  {10.1016/S0550-3213(00)00574-5}, B592, 203

\bibitem[\protect\citeauthoryear{{Dossett}, {Ishak}, {Parkinson}  \&
  {Davis}}{{Dossett} et~al.}{2015}]{2015PhRvD..92b3003D}
{Dossett} J.~N.,  {Ishak} M.,  {Parkinson} D.,   {Davis} T.~M.,  2015, \mndoi
  [\prd] {10.1103/PhysRevD.92.023003}, \href
  {http://adsabs.harvard.edu/abs/2015PhRvD..92b3003D} {92, 023003}

\bibitem[\protect\citeauthoryear{{Dunstan}, {Abazajian}, {Polisensky}  \&
  {Ricotti}}{{Dunstan} et~al.}{2011}]{Dunstan_2011}
{Dunstan} R.~M.,  {Abazajian} K.~N.,  {Polisensky} E.,   {Ricotti} M.,  2011,
  preprint, \href {http://adsabs.harvard.edu/abs/2011arXiv1109.6291D} {}
  (\mn@eprint {arXiv} {1109.6291})

\bibitem[\protect\citeauthoryear{{Eatough} et~al.,}{{Eatough}
  et~al.}{2013}]{2013Natur.501..391E}
{Eatough} R.~P.,  et~al., 2013, \mndoi [\nat] {10.1038/nature12499}, \href
  {http://adsabs.harvard.edu/abs/2013Natur.501..391E} {501, 391}

\bibitem[\protect\citeauthoryear{{Eatough} et~al.,}{{Eatough}
  et~al.}{2015}]{2015aska.confE..45E}
{Eatough} R.,  et~al., 2015, Advancing Astrophysics with the Square Kilometre
  Array (AASKA14), \href {http://adsabs.harvard.edu/abs/2015aska.confE..45E}
  {p.~45}

\bibitem[\protect\citeauthoryear{Edwards, Hobbs  \& Manchester}{Edwards
  et~al.}{2006}]{Edwards:2006zg}
Edwards R.~T.,  Hobbs G.~B.,   Manchester R.~N.,  2006, \mndoi [Mon. Not. Roy.
  Astron. Soc.] {10.1111/j.1365-2966.2006.10870.x}, 372, 1549

\bibitem[\protect\citeauthoryear{Elghozi, Mavromatos, Sakellariadou  \&
  Yusaf}{Elghozi et~al.}{2016}]{Elghozi:2015jka}
Elghozi T.,  Mavromatos N.~E.,  Sakellariadou M.,   Yusaf M.~F.,  2016, \mndoi
  [JCAP] {10.1088/1475-7516/2016/02/060}, 1602, 060

\bibitem[\protect\citeauthoryear{Elghozi, Mavromatos  \& Sakellariadou}{Elghozi
  et~al.}{2017}]{Elghozi:2016wzb}
Elghozi T.,  Mavromatos N.~E.,   Sakellariadou M.,  2017, \mndoi [Eur. Phys.
  J.] {10.1140/epjc/s10052-017-4998-z}, C77, 445

\bibitem[\protect\citeauthoryear{Ellis, Siemens  \& Creighton}{Ellis
  et~al.}{2012}]{Ellis:2012zv}
Ellis J.~A.,  Siemens X.,   Creighton J. D.~E.,  2012, \mndoi [Astrophys. J.]
  {10.1088/0004-637X/756/2/175}, 756, 175

\bibitem[\protect\citeauthoryear{Eriksen, Banday, Gorski, Hansen  \&
  Lilje}{Eriksen et~al.}{2007}]{Eriksen:2007pc}
Eriksen H.~K.,  Banday A.,  Gorski K.,  Hansen F.,   Lilje P.,  2007, \mndoi
  [Astrophys.J.] {10.1086/518091}, 660, L81

\bibitem[\protect\citeauthoryear{{Fauvet} et~al.,}{{Fauvet}
  et~al.}{2011}]{2011A&A...526A.145F}
{Fauvet} L.,  et~al., 2011, \mndoi [\aap] {10.1051/0004-6361/201014492}, \href
  {http://adsabs.harvard.edu/abs/2011A%26A...526A.145F} {526, A145}

\bibitem[\protect\citeauthoryear{{Fauvet}, {Mac{\'{\i}}as-P{\'e}rez}  \&
  {D{\'e}sert}}{{Fauvet} et~al.}{2012}]{2012APh....36...57F}
{Fauvet} L.,  {Mac{\'{\i}}as-P{\'e}rez} J.~F.,   {D{\'e}sert} F.~X.,  2012,
  \mndoi [Astroparticle Physics] {10.1016/j.astropartphys.2012.04.013}, \href
  {http://adsabs.harvard.edu/abs/2012APh....36...57F} {36, 57}

\bibitem[\protect\citeauthoryear{February, Larena, Smith  \& Clarkson}{February
  et~al.}{2010}]{February:2009pv}
February S.,  Larena J.,  Smith M.,   Clarkson C.,  2010, \mndoi [Mon. Not.
  Roy. Astron. Soc.] {10.1111/j.1365-2966.2010.16627.x}, 405, 2231

\bibitem[\protect\citeauthoryear{February, Clarkson  \& Maartens}{February
  et~al.}{2013}]{February:2012fp}
February S.,  Clarkson C.,   Maartens R.,  2013, \mndoi [JCAP]
  {10.1088/1475-7516/2013/03/023}, 1303, 023

\bibitem[\protect\citeauthoryear{Ferramacho, Santos, Jarvis  \&
  Camera}{Ferramacho et~al.}{2014}]{Ferramacho:2014pua}
Ferramacho L.~D.,  Santos M.~G.,  Jarvis M.~J.,   Camera S.,  2014, \mndoi
  [Mon. Not. Roy. Astron. Soc.] {10.1093/mnras/stu1015}, 442, 2511

\bibitem[\protect\citeauthoryear{Ferreras, Sakellariadou  \& Yusaf}{Ferreras
  et~al.}{2008}]{Ferreras:2007kw}
Ferreras I.,  Sakellariadou M.,   Yusaf M.~F.,  2008, \mndoi [Phys. Rev. Lett.]
  {10.1103/PhysRevLett.100.031302}, 100, 031302

\bibitem[\protect\citeauthoryear{Ferreras, Mavromatos, Sakellariadou  \&
  Yusaf}{Ferreras et~al.}{2009}]{Ferreras:2009rv}
Ferreras I.,  Mavromatos N.~E.,  Sakellariadou M.,   Yusaf M.~F.,  2009, \mndoi
  [Phys. Rev.] {10.1103/PhysRevD.80.103506}, D80, 103506

\bibitem[\protect\citeauthoryear{{Fialkov} \& {Loeb}}{{Fialkov} \&
  {Loeb}}{2013}]{2013JCAP...11..066F}
{Fialkov} A.,  {Loeb} A.,  2013, \mndoi [\jcap]
  {10.1088/1475-7516/2013/11/066}, \href
  {http://adsabs.harvard.edu/abs/2013JCAP...11..066F} {11, 066}

\bibitem[\protect\citeauthoryear{{Fialkov}, {Barkana}, {Visbal},
  {Tseliakhovich}  \& {Hirata}}{{Fialkov} et~al.}{2013}]{2013MNRAS.432.2909F}
{Fialkov} A.,  {Barkana} R.,  {Visbal} E.,  {Tseliakhovich} D.,   {Hirata}
  C.~M.,  2013, \mndoi [\mnras] {10.1093/mnras/stt650}, \href
  {http://adsabs.harvard.edu/abs/2013MNRAS.432.2909F} {432, 2909}

\bibitem[\protect\citeauthoryear{{Fialkov}, {Barkana}  \& {Visbal}}{{Fialkov}
  et~al.}{2014}]{2014Natur.506..197F}
{Fialkov} A.,  {Barkana} R.,   {Visbal} E.,  2014, \mndoi [\nat]
  {10.1038/nature12999}, \href
  {http://adsabs.harvard.edu/abs/2014Natur.506..197F} {506, 197}

\bibitem[\protect\citeauthoryear{{Fialkov}, {Barkana}  \& {Cohen}}{{Fialkov}
  et~al.}{2015}]{2015PhRvL.114j1303F}
{Fialkov} A.,  {Barkana} R.,   {Cohen} A.,  2015, \mndoi [Physical Review
  Letters] {10.1103/PhysRevLett.114.101303}, \href
  {http://adsabs.harvard.edu/abs/2015PhRvL.114j1303F} {114, 101303}

\bibitem[\protect\citeauthoryear{{Field}}{{Field}}{1958}]{1958PIRE...46..240F}
{Field} G.~B.,  1958, \mndoi [Proceedings of the IRE]
  {10.1109/JRPROC.1958.286741}, \href
  {http://adsabs.harvard.edu/abs/1958PIRE...46..240F} {46, 240}

\bibitem[\protect\citeauthoryear{{Finelli} et~al.,}{{Finelli}
  et~al.}{2018}]{2016arXiv161208270C}
{Finelli} F.,  et~al., 2018, \mndoi [\jcap] {10.1088/1475-7516/2018/04/016},
  \href {http://adsabs.harvard.edu/abs/2018JCAP...04..016F} {4, 016}

\bibitem[\protect\citeauthoryear{{Fixsen}, {Cheng}, {Gales}, {Mather}, {Shafer}
   \& {Wright}}{{Fixsen} et~al.}{1996}]{1996ApJ...473..576F}
{Fixsen} D.~J.,  {Cheng} E.~S.,  {Gales} J.~M.,  {Mather} J.~C.,  {Shafer}
  R.~A.,   {Wright} E.~L.,  1996, \mndoi [\apj] {10.1086/178173}, \href
  {http://adsabs.harvard.edu/abs/1996ApJ...473..576F} {473, 576}

\bibitem[\protect\citeauthoryear{Fleury, Clarkson  \& Maartens}{Fleury
  et~al.}{2017}]{Fleury:2016fda}
Fleury P.,  Clarkson C.,   Maartens R.,  2017, \mndoi [JCAP]
  {10.1088/1475-7516/2017/03/062}, 1703, 062

\bibitem[\protect\citeauthoryear{{Fomalont} \& {Reid}}{{Fomalont} \&
  {Reid}}{2004}]{2004NewAR..48.1473F}
{Fomalont} E.,  {Reid} M.,  2004, \mndoi [New Astronomy Reviews]
  {10.1016/j.newar.2004.09.037}, \href
  {http://adsabs.harvard.edu/abs/2004NewAR..48.1473F} {48, 1473}

\bibitem[\protect\citeauthoryear{Fonseca, Camera, Santos  \& Maartens}{Fonseca
  et~al.}{2015}]{Fonseca:2015laa}
Fonseca J.,  Camera S.,  Santos M.,   Maartens R.,  2015, \mndoi [Astrophys.
  J.] {10.1088/2041-8205/812/2/L22}, 812, L22

\bibitem[\protect\citeauthoryear{Fonseca, Maartens  \& Santos}{Fonseca
  et~al.}{2017}]{Fonseca:2016xvi}
Fonseca J.,  Maartens R.,   Santos M.~G.,  2017, \mndoi [Mon. Not. Roy. Astron.
  Soc.] {10.1093/mnras/stw3248}, 466, 2780

\bibitem[\protect\citeauthoryear{Fornengo \& Regis}{Fornengo \&
  Regis}{2014}]{Fornengo:2013rga}
Fornengo N.,  Regis M.,  2014, \mndoi [Front. Physics]
  {10.3389/fphy.2014.00006}, 2, 6

\bibitem[\protect\citeauthoryear{{Fornengo}, {Lineros}, {Regis}  \&
  {Taoso}}{{Fornengo} et~al.}{2011}]{2011PhRvL.107A1302F}
{Fornengo} N.,  {Lineros} R.,  {Regis} M.,   {Taoso} M.,  2011, \mndoi
  [Physical Review Letters] {10.1103/PhysRevLett.107.271302}, \href
  {http://adsabs.harvard.edu/abs/2011PhRvL.107A1302F} {107, 271302}

\bibitem[\protect\citeauthoryear{{Fornengo}, {Lineros}, {Regis}  \&
  {Taoso}}{{Fornengo} et~al.}{2012a}]{2012JCAP...03..033F}
{Fornengo} N.,  {Lineros} R.,  {Regis} M.,   {Taoso} M.,  2012a, \mndoi [JCAP]
  {10.1088/1475-7516/2012/03/033}, \href
  {http://adsabs.harvard.edu/abs/2012JCAP...03..033F} {3, 033}

\bibitem[\protect\citeauthoryear{Fornengo, Lineros, Regis  \& Taoso}{Fornengo
  et~al.}{2012b}]{Fornengo:2011iq}
Fornengo N.,  Lineros R.~A.,  Regis M.,   Taoso M.,  2012b, \mndoi [JCAP]
  {10.1088/1475-7516/2012/01/005}, 1201, 005

\bibitem[\protect\citeauthoryear{{Fornengo}, {Lineros}, {Regis}  \&
  {Taoso}}{{Fornengo} et~al.}{2014}]{2014JCAP...04..008F}
{Fornengo} N.,  {Lineros} R.~A.,  {Regis} M.,   {Taoso} M.,  2014, \mndoi
  [JCAP] {10.1088/1475-7516/2014/04/008}, \href
  {http://adsabs.harvard.edu/abs/2014JCAP...04..008F} {4, 008}

\bibitem[\protect\citeauthoryear{Fornengo, Perotto, Regis  \& Camera}{Fornengo
  et~al.}{2015}]{Fornengo:2014cya}
Fornengo N.,  Perotto L.,  Regis M.,   Camera S.,  2015, \mndoi [Astrophys. J.]
  {10.1088/2041-8205/802/1/L1}, 802, L1

\bibitem[\protect\citeauthoryear{{Foster} \& {Backer}}{{Foster} \&
  {Backer}}{1990}]{Foster:1990}
{Foster} R.~S.,  {Backer} D.~C.,  1990, \mndoi [\apj] {10.1086/169195}, \href
  {http://adsabs.harvard.edu/abs/1990ApJ...361..300F} {361, 300}

\bibitem[\protect\citeauthoryear{Freire et~al.,}{Freire
  et~al.}{2012}]{Freire:2012mg}
Freire P. C.~C.,  et~al., 2012, \mndoi [Mon. Not. Roy. Astron. Soc.]
  {10.1111/j.1365-2966.2012.21253.x}, 423, 3328

\bibitem[\protect\citeauthoryear{{Furlanetto}}{{Furlanetto}}{2006}]{2006MNRAS.371..867F}
{Furlanetto} S.~R.,  2006, \mndoi [\mnras] {10.1111/j.1365-2966.2006.10725.x},
  \href {http://adsabs.harvard.edu/abs/2006MNRAS.371..867F} {371, 867}

\bibitem[\protect\citeauthoryear{{Furlanetto}, {Zaldarriaga}  \&
  {Hernquist}}{{Furlanetto} et~al.}{2004a}]{2004ApJ...613....1F}
{Furlanetto} S.~R.,  {Zaldarriaga} M.,   {Hernquist} L.,  2004a, \mndoi [\apj]
  {10.1086/423025}, \href {http://adsabs.harvard.edu/abs/2004ApJ...613....1F}
  {613, 1}

\bibitem[\protect\citeauthoryear{{Furlanetto}, {Zaldarriaga}  \&
  {Hernquist}}{{Furlanetto} et~al.}{2004b}]{2004ApJ...613...16F}
{Furlanetto} S.~R.,  {Zaldarriaga} M.,   {Hernquist} L.,  2004b, \mndoi [\apj]
  {10.1086/423028}, \href {http://adsabs.harvard.edu/abs/2004ApJ...613...16F}
  {613, 16}

\bibitem[\protect\citeauthoryear{{Furlanetto}, {Oh}  \&
  {Pierpaoli}}{{Furlanetto} et~al.}{2006}]{furlanetto2006dm}
{Furlanetto} S.~R.,  {Oh} S.~P.,   {Pierpaoli} E.,  2006, \mndoi [\prd]
  {10.1103/PhysRevD.74.103502}, \href
  {http://adsabs.harvard.edu/abs/2006PhRvD..74j3502F} {74, 103502}

\bibitem[\protect\citeauthoryear{{Furlanetto} et~al.,}{{Furlanetto}
  et~al.}{2009}]{2009astro2010S..82F}
{Furlanetto} S.~R.,  et~al., 2009, in astro2010: The Astronomy and Astrophysics
  Decadal Survey.  (\mn@eprint {arXiv} {0902.3259})

\bibitem[\protect\citeauthoryear{Gaggero et~al.}{Gaggero
  et~al.}{2017}]{Gaggero2017}
Gaggero D.,  et~al., 2017, \mndoi [Phys. Rev. Lett.]
  {10.1103/PhysRevLett.118.241101}, 118, 241101

\bibitem[\protect\citeauthoryear{{Gao} \& {Theuns}}{{Gao} \&
  {Theuns}}{2007}]{2007Sci...317.1527G}
{Gao} L.,  {Theuns} T.,  2007, \mndoi [Science] {10.1126/science.1146676},
  \href {http://adsabs.harvard.edu/abs/2007Sci...317.1527G} {317, 1527}

\bibitem[\protect\citeauthoryear{Garcia-Bellido \& Haugboelle}{Garcia-Bellido
  \& Haugboelle}{2008}]{GarciaBellido:2008nz}
Garcia-Bellido J.,  Haugboelle T.,  2008, \mndoi [JCAP]
  {10.1088/1475-7516/2008/04/003}, 0804, 003

\bibitem[\protect\citeauthoryear{{Gaskins}}{{Gaskins}}{2016}]{Gaskins:2016}
{Gaskins} J.~M.,  2016, \mndoi [Contemporary Physics]
  {10.1080/00107514.2016.1175160}, \href
  {http://adsabs.harvard.edu/abs/2016ConPh..57..496G} {57, 496}

\bibitem[\protect\citeauthoryear{Geng, Zhang  \& Zhang}{Geng
  et~al.}{2014}]{Geng:2014ypa}
Geng J.-J.,  Zhang J.-F.,   Zhang X.,  2014, \mndoi [JCAP]
  {10.1088/1475-7516/2014/12/018}, 1412, 018

\bibitem[\protect\citeauthoryear{{Germani} \& {Prokopec}}{{Germani} \&
  {Prokopec}}{2017}]{Germani}
{Germani} C.,  {Prokopec} T.,  2017, \mndoi [Physics of the Dark Universe]
  {10.1016/j.dark.2017.09.001}, \href
  {http://adsabs.harvard.edu/abs/2017PDU....18....6G} {18, 6}

\bibitem[\protect\citeauthoryear{{Gervasi}, {Zannoni}, {Tartari}, {Boella}  \&
  {Sironi}}{{Gervasi} et~al.}{2008}]{2008ApJ...688...24G}
{Gervasi} M.,  {Zannoni} M.,  {Tartari} A.,  {Boella} G.,   {Sironi} G.,  2008,
  \mndoi [\apj] {10.1086/592134}, \href
  {http://adsabs.harvard.edu/abs/2008ApJ...688...24G} {688, 24}

\bibitem[\protect\citeauthoryear{Ghosh, Bharadwaj, Ali  \& Chengalur}{Ghosh
  et~al.}{2011}]{Ghosh2011}
Ghosh A.,  Bharadwaj S.,  Ali S.~S.,   Chengalur J.~N.,  2011, \mndoi [Monthly
  Notices of the Royal Astronomical Society]
  {10.1111/j.1365-2966.2011.19649.x}, 418, 2584

\bibitem[\protect\citeauthoryear{Ghosh, Koopmans, Chapman  \& Jeli{\'c}}{Ghosh
  et~al.}{2015}]{Ghosh2015}
Ghosh A.,  Koopmans L. V.~E.,  Chapman E.,   Jeli{\'c} V.,  2015, \mndoi
  [Monthly Notices of the Royal Astronomical Society] {10.1093/mnras/stv1355},
  452, 1587

\bibitem[\protect\citeauthoryear{{Giannantonio}, {Scranton}, {Crittenden},
  {Nichol}, {Boughn}, {Myers}  \& {Richards}}{{Giannantonio}
  et~al.}{2008a}]{giannantonio08a}
{Giannantonio} T.,  {Scranton} R.,  {Crittenden} R.~G.,  {Nichol} R.~C.,
  {Boughn} S.~P.,  {Myers} A.~D.,   {Richards} G.~T.,  2008a, \mndoi [\prd]
  {10.1103/PhysRevD.77.123520}, \href
  {http://adsabs.harvard.edu/abs/2008PhRvD..77l3520G} {77, 123520}

\bibitem[\protect\citeauthoryear{{Giannantonio}, {Song}  \&
  {Koyama}}{{Giannantonio} et~al.}{2008b}]{giannantonio08dgp}
{Giannantonio} T.,  {Song} Y.-S.,   {Koyama} K.,  2008b, \mndoi [\prd]
  {10.1103/PhysRevD.78.044017}, \href
  {http://adsabs.harvard.edu/abs/2008PhRvD..78d4017G} {78, 044017}

\bibitem[\protect\citeauthoryear{{Giannantonio}, {Crittenden}, {Nichol}  \&
  {Ross}}{{Giannantonio} et~al.}{2012}]{giannantonio12}
{Giannantonio} T.,  {Crittenden} R.,  {Nichol} R.,   {Ross} A.~J.,  2012,
  \mndoi [\mnras] {10.1111/j.1365-2966.2012.21896.x}, \href
  {http://adsabs.harvard.edu/abs/2012MNRAS.426.2581G} {426, 2581}

\bibitem[\protect\citeauthoryear{{Gleyzes}}{{Gleyzes}}{2017}]{Gleyzes:2017kpi}
{Gleyzes} J.,  2017, \mndoi [\prd] {10.1103/PhysRevD.96.063516}, \href
  {http://adsabs.harvard.edu/abs/2017PhRvD..96f3516G} {96, 063516}

\bibitem[\protect\citeauthoryear{{Godfrey} et~al.,}{{Godfrey}
  et~al.}{2012}]{2012PASA...29...42G}
{Godfrey} L.~E.~H.,  et~al., 2012, \mndoi [\pasa] {10.1071/AS11050}, \href
  {http://adsabs.harvard.edu/abs/2012PASA...29...42G} {29, 42}

\bibitem[\protect\citeauthoryear{{Goodenough} \& {Hooper}}{{Goodenough} \&
  {Hooper}}{2009}]{Goodenough:2009gk}
{Goodenough} L.,  {Hooper} D.,  2009, preprint, \href
  {http://adsabs.harvard.edu/abs/2009arXiv0910.2998G} {} (\mn@eprint {arXiv}
  {0910.2998})

\bibitem[\protect\citeauthoryear{Gordon}{Gordon}{2007}]{Gordon:2006ag}
Gordon C.,  2007, \mndoi [Astrophys.J.] {10.1086/510511}, 656, 636

\bibitem[\protect\citeauthoryear{Gordon, Hu, Huterer  \& Crawford}{Gordon
  et~al.}{2005}]{Gordon:2005ai}
Gordon C.,  Hu W.,  Huterer D.,   Crawford T.~M.,  2005, \mndoi [Phys.Rev.]
  {10.1103/PhysRevD.72.103002}, D72, 103002

\bibitem[\protect\citeauthoryear{{G{\'o}rski}, {Hivon}, {Banday}, {Wandelt},
  {Hansen}, {Reinecke}  \& {Bartelmann}}{{G{\'o}rski}
  et~al.}{2005}]{2005ApJ...622..759G}
{G{\'o}rski} K.~M.,  {Hivon} E.,  {Banday} A.~J.,  {Wandelt} B.~D.,  {Hansen}
  F.~K.,  {Reinecke} M.,   {Bartelmann} M.,  2005, \mndoi [\apj]
  {10.1086/427976}, \href {http://adsabs.harvard.edu/abs/2005ApJ...622..759G}
  {622, 759}

\bibitem[\protect\citeauthoryear{Graham et~al.,}{Graham
  et~al.}{2015}]{Graham:2015gma}
Graham M.~J.,  et~al., 2015, \mndoi [Nature] {10.1038/nature14143}, 518, 74

\bibitem[\protect\citeauthoryear{{Green} \& {Liddle}}{{Green} \&
  {Liddle}}{1999}]{astro-ph/9901268}
{Green} A.~M.,  {Liddle} A.~R.,  1999, \mndoi [\prd]
  {10.1103/PhysRevD.60.063509}, \href
  {http://adsabs.harvard.edu/abs/1999PhRvD..60f3509G} {60, 063509}

\bibitem[\protect\citeauthoryear{Gregory \& Laflamme}{Gregory \&
  Laflamme}{1993}]{Gregory:1993vy}
Gregory R.,  Laflamme R.,  1993, \mndoi [Phys. Rev. Lett.]
  {10.1103/PhysRevLett.70.2837}, 70, 2837

\bibitem[\protect\citeauthoryear{{Grieb} et~al.,}{{Grieb}
  et~al.}{2017}]{Grieb:2016uuo}
{Grieb} J.~N.,  et~al., 2017, \mndoi [\mnras] {10.1093/mnras/stw3384}, \href
  {http://adsabs.harvard.edu/abs/2017MNRAS.467.2085G} {467, 2085}

\bibitem[\protect\citeauthoryear{{Haggard} \& {Rovelli}}{{Haggard} \&
  {Rovelli}}{2015}]{Haggard:2014fv}
{Haggard} H.~M.,  {Rovelli} C.,  2015, \mndoi [\prd]
  {10.1103/PhysRevD.92.104020}, \href
  {http://adsabs.harvard.edu/abs/2015PhRvD..92j4020H} {92, 104020}

\bibitem[\protect\citeauthoryear{Hansen, Banday  \& Gorski}{Hansen
  et~al.}{2004}]{Hansen:2004vq}
Hansen F.~K.,  Banday A.,   Gorski K.,  2004, \mndoi [Mon.Not.Roy.Astron.Soc.]
  {10.1111/j.1365-2966.2004.08229.x}, 354, 641

\bibitem[\protect\citeauthoryear{Hanson \& Lewis}{Hanson \&
  Lewis}{2009}]{Hanson:2009gu}
Hanson D.,  Lewis A.,  2009, \mndoi [Phys.Rev.] {10.1103/PhysRevD.80.063004},
  D80, 063004

\bibitem[\protect\citeauthoryear{{Harker}, {Pritchard}, {Burns}  \&
  {Bowman}}{{Harker} et~al.}{2012}]{harker}
{Harker} G.~J.~A.,  {Pritchard} J.~R.,  {Burns} J.~O.,   {Bowman} J.~D.,  2012,
  \mndoi [\mnras] {10.1111/j.1365-2966.2011.19766.x}, \href
  {http://adsabs.harvard.edu/abs/2012MNRAS.419.1070H} {419, 1070}

\bibitem[\protect\citeauthoryear{{Harrison}, {Camera}, {Zuntz}  \&
  {Brown}}{{Harrison} et~al.}{2016}]{2016MNRAS.463.3674H}
{Harrison} I.,  {Camera} S.,  {Zuntz} J.,   {Brown} M.~L.,  2016, \mndoi
  [\mnras] {10.1093/mnras/stw2082}, \href
  {http://adsabs.harvard.edu/abs/2016MNRAS.463.3674H} {463, 3674}

\bibitem[\protect\citeauthoryear{Harrison, Lochner  \& Brown}{Harrison
  et~al.}{2017}]{Harrison:2017pcu}
Harrison I.,  Lochner M.,   Brown M.~L.,  2017, preprint (\mn@eprint {arXiv}
  {1704.08278})

\bibitem[\protect\citeauthoryear{{Haslam}, {Salter}, {Stoffel}  \&
  {Wilson}}{{Haslam} et~al.}{1982}]{1982A&AS...47....1H}
{Haslam} C.~G.~T.,  {Salter} C.~J.,  {Stoffel} H.,   {Wilson} W.~E.,  1982,
  \aaps, \href {http://adsabs.harvard.edu/abs/1982A%26AS...47....1H} {47, 1}

\bibitem[\protect\citeauthoryear{Hawking}{Hawking}{1971}]{Hawking:1971ei}
Hawking S.,  1971, Mon. Not. Roy. Astron. Soc., 152, 75

\bibitem[\protect\citeauthoryear{{Hayasaki} \& {Loeb}}{{Hayasaki} \&
  {Loeb}}{2016}]{Hayasaki:2015qoa}
{Hayasaki} K.,  {Loeb} A.,  2016, \mndoi [Scientific Reports]
  {10.1038/srep35629}, \href
  {http://adsabs.harvard.edu/abs/2016NatSR...635629H} {6, 35629}

\bibitem[\protect\citeauthoryear{Hazelton, Morales  \& Sullivan}{Hazelton
  et~al.}{2013}]{Hazelton2013}
Hazelton B.~J.,  Morales M.~F.,   Sullivan I.~S.,  2013, \mndoi [The
  Astrophysical Journal] {10.1088/0004-637X/770/2/156}, 770, 156

\bibitem[\protect\citeauthoryear{Hellings \& Downs}{Hellings \&
  Downs}{1983}]{Hellings:1983fr}
Hellings R.~W.,  Downs G.~S.,  1983, \mndoi [Astrophys. J.] {10.1086/183954},
  265, L39

\bibitem[\protect\citeauthoryear{Hellwing, Barreira, Frenk, Li  \&
  Cole}{Hellwing et~al.}{2014}]{Hellwing:2014nma}
Hellwing W.~A.,  Barreira A.,  Frenk C.~S.,  Li B.,   Cole S.,  2014, \mndoi
  [Phys. Rev. Lett.] {10.1103/PhysRevLett.112.221102}, 112, 221102

\bibitem[\protect\citeauthoryear{{Hirata}}{{Hirata}}{2006}]{Hirata06}
{Hirata} C.~M.,  2006, \mndoi [\mnras] {10.1111/j.1365-2966.2005.09949.x},
  \href {http://adsabs.harvard.edu/abs/2006MNRAS.367..259H} {367, 259}

\bibitem[\protect\citeauthoryear{Hirata}{Hirata}{2009}]{Hirata:2009ar}
Hirata C.~M.,  2009, \mndoi [JCAP] {10.1088/1475-7516/2009/09/011}, 0909, 011

\bibitem[\protect\citeauthoryear{{Hirata}, {Mishra}  \& {Venumadhav}}{{Hirata}
  et~al.}{2018}]{2017arXiv170703513H}
{Hirata} C.~M.,  {Mishra} A.,   {Venumadhav} T.,  2018, \mndoi [\prd]
  {10.1103/PhysRevD.97.103521}, \href
  {http://adsabs.harvard.edu/abs/2018PhRvD..97j3521H} {97, 103521}

\bibitem[\protect\citeauthoryear{{Ho}, {Hirata}, {Padmanabhan}, {Seljak}  \&
  {Bahcall}}{{Ho} et~al.}{2008}]{ho08}
{Ho} S.,  {Hirata} C.,  {Padmanabhan} N.,  {Seljak} U.,   {Bahcall} N.,  2008,
  \mndoi [\prd] {10.1103/PhysRevD.78.043519}, \href
  {http://adsabs.harvard.edu/abs/2008PhRvD..78d3519H} {78, 043519}

\bibitem[\protect\citeauthoryear{Hobbs \& Dai}{Hobbs \&
  Dai}{2017}]{Hobbs:2017oam}
Hobbs G.,  Dai S.,  2017, preprint (\mn@eprint {arXiv} {1707.01615})

\bibitem[\protect\citeauthoryear{{Hobbs}, {Edwards}  \& {Manchester}}{{Hobbs}
  et~al.}{2006}]{2006MNRAS.369..655H}
{Hobbs} G.~B.,  {Edwards} R.~T.,   {Manchester} R.~N.,  2006, \mndoi [\mnras]
  {10.1111/j.1365-2966.2006.10302.x}, \href
  {http://adsabs.harvard.edu/abs/2006MNRAS.369..655H} {369, 655}

\bibitem[\protect\citeauthoryear{{Hogan} \& {Rees}}{{Hogan} \&
  {Rees}}{1979}]{1979MNRAS.188..791H}
{Hogan} C.~J.,  {Rees} M.~J.,  1979, \mnras, \href
  {http://adsabs.harvard.edu/abs/1979MNRAS.188..791H} {188, 791}

\bibitem[\protect\citeauthoryear{{Hu}}{{Hu}}{2001}]{Hu:2001tn}
{Hu} W.,  2001, \mndoi [\apjl] {10.1086/323253}, \href
  {http://adsabs.harvard.edu/abs/2001ApJ...557L..79H} {557, L79}

\bibitem[\protect\citeauthoryear{{Huff}, {Krause}, {Eifler}, {George}  \&
  {Schlegel}}{{Huff} et~al.}{2013}]{2013arXiv1311.1489H}
{Huff} E.~M.,  {Krause} E.,  {Eifler} T.,  {George} M.~R.,   {Schlegel} D.,
  2013, preprint, \href {http://adsabs.harvard.edu/abs/2013arXiv1311.1489H} {}
  (\mn@eprint {arXiv} {1311.1489})

\bibitem[\protect\citeauthoryear{Hutter, Dayal, M\"{u}ller  \& Trott}{Hutter
  et~al.}{2017}]{0004-637X-836-2-176}
Hutter A.,  Dayal P.,  M\"{u}ller V.,   Trott C.~M.,  2017, The Astrophysical
  Journal, 836, 176

\bibitem[\protect\citeauthoryear{{Ir{\v s}i{\v c}} et~al.,}{{Ir{\v s}i{\v c}}
  et~al.}{2017}]{Vid_2017}
{Ir{\v s}i{\v c}} V.,  et~al., 2017, \mndoi [\prd]
  {10.1103/PhysRevD.96.023522}, \href
  {http://adsabs.harvard.edu/abs/2017PhRvD..96b3522I} {96, 023522}

\bibitem[\protect\citeauthoryear{{Ishino} et~al.,}{{Ishino}
  et~al.}{2016}]{2016SPIE.9904E..0XI}
{Ishino} H.,  et~al., 2016, in Space Telescopes and Instrumentation 2016:
  Optical, Infrared, and Millimeter Wave. p. 99040X, \mndoi{10.1117/12.2231995}

\bibitem[\protect\citeauthoryear{Ivarsen, Bull, Llinares  \& Mota}{Ivarsen
  et~al.}{2016}]{Ivarsen:2016xre}
Ivarsen M.~F.,  Bull P.,  Llinares C.,   Mota D.~F.,  2016, \mndoi [Astron.
  Astrophys.] {10.1051/0004-6361/201628604}, 595, A40

\bibitem[\protect\citeauthoryear{Jacobs et~al.,}{Jacobs
  et~al.}{2015}]{jacobs2015}
Jacobs D.~C.,  et~al., 2015, \mndoi [The Astrophysical Journal]
  {10.1088/0004-637X/801/1/51}, 801, 51

\bibitem[\protect\citeauthoryear{{Jacobs} et~al.,}{{Jacobs}
  et~al.}{2016}]{Jacobs2016}
{Jacobs} D.~C.,  et~al., 2016, \mndoi [\apj] {10.3847/0004-637X/825/2/114},
  \href {https://ui.adsabs.harvard.edu/abs/2016ApJ...825..114J} {825, 114}

\bibitem[\protect\citeauthoryear{{Jaffe}}{{Jaffe}}{2004}]{Jaffe:2004}
{Jaffe} A.~H.,  2004, \mndoi [New Astronomy Reviews]
  {10.1016/j.newar.2004.09.018}, \href
  {http://adsabs.harvard.edu/abs/2004NewAR..48.1483J} {48, 1483}

\bibitem[\protect\citeauthoryear{Janssen et~al.}{Janssen
  et~al.}{2015}]{Janssen:2014dka}
Janssen G.,  et~al., 2015, PoS, AASKA14, 037

\bibitem[\protect\citeauthoryear{{Jarvis}, {Bacon}, {Blake}, {Brown},
  {Lindsay}, {Raccanelli}, {Santos}  \& {Schwarz}}{{Jarvis}
  et~al.}{2015a}]{2015aska.confE..18J}
{Jarvis} M.,  {Bacon} D.,  {Blake} C.,  {Brown} M.,  {Lindsay} S.,
  {Raccanelli} A.,  {Santos} M.,   {Schwarz} D.~J.,  2015a, Advancing
  Astrophysics with the Square Kilometre Array (AASKA14), \href
  {http://adsabs.harvard.edu/abs/2015aska.confE..18J} {p.~18}

\bibitem[\protect\citeauthoryear{Jarvis, Bacon, Blake, Brown, Lindsay,
  Raccanelli, Santos  \& Schwarz}{Jarvis et~al.}{2015b}]{Jarvis:2015asa}
Jarvis M.,  Bacon D.,  Blake C.,  Brown M.,  Lindsay S.,  Raccanelli A.,
  Santos M.,   Schwarz D.~J.,  2015b, PoS, AASKA14, 018

\bibitem[\protect\citeauthoryear{Jeannerot, Rocher  \& Sakellariadou}{Jeannerot
  et~al.}{2003}]{Jeannerot:2003qv}
Jeannerot R.,  Rocher J.,   Sakellariadou M.,  2003, \mndoi [Phys. Rev.]
  {10.1103/PhysRevD.68.103514}, D68, 103514

\bibitem[\protect\citeauthoryear{Jeli{\'c} et~al.,}{Jeli{\'c}
  et~al.}{2014}]{Jelic2014}
Jeli{\'c} V.,  et~al., 2014, \mndoi [Astronomy and Astrophysics]
  {10.1051/0004-6361/201423998}, 568, A101

\bibitem[\protect\citeauthoryear{{Jensen}, {Zackrisson}, {Pelckmans},
  {Binggeli}, {Ausmees}  \& {Lundholm}}{{Jensen}
  et~al.}{2016}]{2016ApJ...827....5J}
{Jensen} H.,  {Zackrisson} E.,  {Pelckmans} K.,  {Binggeli} C.,  {Ausmees} K.,
   {Lundholm} U.,  2016, \mndoi [\apj] {10.3847/0004-637X/827/1/5}, \href
  {http://adsabs.harvard.edu/abs/2016ApJ...827....5J} {827, 5}

\bibitem[\protect\citeauthoryear{Jeong \& Schmidt}{Jeong \&
  Schmidt}{2012}]{CosmicRulers1}
Jeong D.,  Schmidt F.,  2012, \mndoi [Phys. Rev.] {10.1103/PhysRevD.86.083512},
  D86, 083512

\bibitem[\protect\citeauthoryear{Jeong, Schmidt  \& Hirata}{Jeong
  et~al.}{2012}]{Jeong:2011as}
Jeong D.,  Schmidt F.,   Hirata C.~M.,  2012, \mndoi [\prd]
  {10.1103/PhysRevD.85.023504}, 85, 023504

\bibitem[\protect\citeauthoryear{{Johannsen}}{{Johannsen}}{2016}]{2016CQGra..33k3001J}
{Johannsen} T.,  2016, \mndoi [Classical and Quantum Gravity]
  {10.1088/0264-9381/33/11/113001}, \href
  {http://adsabs.harvard.edu/abs/2016CQGra..33k3001J} {33, 113001}

\bibitem[\protect\citeauthoryear{Johnston \& Wall}{Johnston \&
  Wall}{2008}]{Johnston:2008hp}
Johnston S.,  Wall J.,  2008, \mndoi [Exper. Astron.]
  {10.1007/s10686-008-9124-7}, 22, 151

\bibitem[\protect\citeauthoryear{{Joudaki}, {Dor{\'e}}, {Ferramacho},
  {Kaplinghat}  \& {Santos}}{{Joudaki} et~al.}{2011}]{joudaki2011}
{Joudaki} S.,  {Dor{\'e}} O.,  {Ferramacho} L.,  {Kaplinghat} M.,   {Santos}
  M.~G.,  2011, \mndoi [Physical Review Letters]
  {10.1103/PhysRevLett.107.131304}, \href
  {http://adsabs.harvard.edu/abs/2011PhRvL.107m1304J} {107, 131304}

\bibitem[\protect\citeauthoryear{Joyce, Jain, Khoury  \& Trodden}{Joyce
  et~al.}{2015}]{Joyce:2014kja}
Joyce A.,  Jain B.,  Khoury J.,   Trodden M.,  2015, \mndoi [Phys. Rept.]
  {10.1016/j.physrep.2014.12.002}, 568, 1

\bibitem[\protect\citeauthoryear{{Kaiser} \& {Hudson}}{{Kaiser} \&
  {Hudson}}{2015}]{Kaiser:2014jca}
{Kaiser} N.,  {Hudson} M.~J.,  2015, \mndoi [\mnras] {10.1093/mnras/stv693},
  \href {http://adsabs.harvard.edu/abs/2015MNRAS.450..883K} {450, 883}

\bibitem[\protect\citeauthoryear{{Kaiser} \& {Jaffe}}{{Kaiser} \&
  {Jaffe}}{1997}]{Kaiser:1997}
{Kaiser} N.,  {Jaffe} A.,  1997, \mndoi [\apj] {10.1086/304357}, \href
  {http://adsabs.harvard.edu/abs/1997ApJ...484..545K} {484, 545}

\bibitem[\protect\citeauthoryear{{Kashiyama} \& {Seto}}{{Kashiyama} \&
  {Seto}}{2012}]{kashiyama}
{Kashiyama} K.,  {Seto} N.,  2012, \mndoi [\mnras]
  {10.1111/j.1365-2966.2012.21935.x}, \href
  {http://adsabs.harvard.edu/abs/2012MNRAS.426.1369K} {426, 1369}

\bibitem[\protect\citeauthoryear{Katz}{Katz}{2016}]{Katz:2016dti}
Katz J.~I.,  2016, \mndoi [Mod. Phys. Lett.] {10.1142/S0217732316300135}, A31,
  1630013

\bibitem[\protect\citeauthoryear{Kehl, Wex, Kramer  \& Liu}{Kehl
  et~al.}{2017}]{Kehl:2016mgp}
Kehl M.~S.,  Wex N.,  Kramer M.,   Liu K.,  2017, in {Proceedings, 14th Marcel
  Grossmann Meeting on Recent Developments in Theoretical and Experimental
  General Relativity, Astrophysics, and Relativistic Field Theories (MG14) (In
  4 Volumes): Rome, Italy, July 12-18, 2015}. pp 1860--1865

\bibitem[\protect\citeauthoryear{{Kelley} \& {Quinn}}{{Kelley} \&
  {Quinn}}{2017}]{2017ApJ...845L...4K}
{Kelley} K.,  {Quinn} P.~J.,  2017, \mndoi [\apjl] {10.3847/2041-8213/aa808d},
  \href {http://adsabs.harvard.edu/abs/2017ApJ...845L...4K} {845, L4}

\bibitem[\protect\citeauthoryear{{Kern}, {Liu}, {Parsons}, {Mesinger}  \&
  {Greig}}{{Kern} et~al.}{2017}]{2017arXiv170504688K}
{Kern} N.~S.,  {Liu} A.,  {Parsons} A.~R.,  {Mesinger} A.,   {Greig} B.,  2017,
  \mndoi [\apj] {10.3847/1538-4357/aa8bb4}, \href
  {http://adsabs.harvard.edu/abs/2017ApJ...848...23K} {848, 23}

\bibitem[\protect\citeauthoryear{Khachatryan et~al.}{Khachatryan
  et~al.}{2016}]{Khachatryan:2014rwa}
Khachatryan V.,  et~al., 2016, \mndoi [Phys. Lett.]
  {10.1016/j.physletb.2016.01.057}, B755, 102

\bibitem[\protect\citeauthoryear{Khatri \& Wandelt}{Khatri \&
  Wandelt}{2007}]{Khatri:2007yv}
Khatri R.,  Wandelt B.~D.,  2007, \mndoi [Phys. Rev. Lett.]
  {10.1103/PhysRevLett.98.111301}, 98, 111301

\bibitem[\protect\citeauthoryear{Khoury \& Weltman}{Khoury \&
  Weltman}{2004a}]{Khoury:2003aq}
Khoury J.,  Weltman A.,  2004a, \mndoi [Phys. Rev. Lett.]
  {10.1103/PhysRevLett.93.171104}, 93, 171104

\bibitem[\protect\citeauthoryear{Khoury \& Weltman}{Khoury \&
  Weltman}{2004b}]{Khoury:2003rn}
Khoury J.,  Weltman A.,  2004b, \mndoi [Phys. Rev.]
  {10.1103/PhysRevD.69.044026}, D69, 044026

\bibitem[\protect\citeauthoryear{{Kim}}{{Kim}}{1979}]{1979PhRvL..43..103K}
{Kim} J.~E.,  1979, \mndoi [Physical Review Letters]
  {10.1103/PhysRevLett.43.103}, \href
  {http://adsabs.harvard.edu/abs/1979PhRvL..43..103K} {43, 103}

\bibitem[\protect\citeauthoryear{Kim, Linder, Edelstein  \& Erskine}{Kim
  et~al.}{2015}]{Kim:2014uha}
Kim A.~G.,  Linder E.~V.,  Edelstein J.,   Erskine D.,  2015, \mndoi
  [Astropart. Phys.] {10.1016/j.astropartphys.2014.09.004}, 62, 195

\bibitem[\protect\citeauthoryear{Kl{\"o}ckner et~al.,}{Kl{\"o}ckner
  et~al.}{2015}]{Klockner:2015rqa}
Kl{\"o}ckner H.-R.,  et~al., 2015, PoS, AASKA14, 027

\bibitem[\protect\citeauthoryear{Koda et~al.,}{Koda
  et~al.}{2014}]{Koda:2013eya}
Koda J.,  et~al., 2014, \mndoi [Mon. Not. Roy. Astron. Soc.]
  {10.1093/mnras/stu1610}, 445, 4267

\bibitem[\protect\citeauthoryear{{Kogut}}{{Kogut}}{1996}]{1996astro.ph..7100K}
{Kogut} A.,  1996, preprint, \href
  {http://adsabs.harvard.edu/abs/1996astro.ph..7100K} {} (\mn@eprint {arXiv}
  {astro-ph/9607100})

\bibitem[\protect\citeauthoryear{Kogut et~al.}{Kogut
  et~al.}{1993}]{Kogut:1993ag}
Kogut A.,  et~al., 1993, \mndoi [Astrophys. J.] {10.1086/173453}, 419, 1

\bibitem[\protect\citeauthoryear{Kol \& Sorkin}{Kol \&
  Sorkin}{2004}]{Kol:2004pn}
Kol B.,  Sorkin E.,  2004, \mndoi [Class. Quant. Grav.]
  {10.1088/0264-9381/21/21/003}, 21, 4793

\bibitem[\protect\citeauthoryear{{Komatsu}}{{Komatsu}}{2010}]{1003.6097}
{Komatsu} E.,  2010, \mndoi [Classical and Quantum Gravity]
  {10.1088/0264-9381/27/12/124010}, \href
  {http://adsabs.harvard.edu/abs/2010CQGra..27l4010K} {27, 124010}

\bibitem[\protect\citeauthoryear{{Koopmans} et~al.,}{{Koopmans}
  et~al.}{2015}]{2015aska.confE...1K}
{Koopmans} L.,  et~al., 2015, Advancing Astrophysics with the Square Kilometre
  Array (AASKA14), \href {http://adsabs.harvard.edu/abs/2015aska.confE...1K}
  {p.~1}

\bibitem[\protect\citeauthoryear{{Kovetz}, {Raccanelli}  \& {Rahman}}{{Kovetz}
  et~al.}{2017}]{Kovetz:2016}
{Kovetz} E.~D.,  {Raccanelli} A.,   {Rahman} M.,  2017, \mndoi [\mnras]
  {10.1093/mnras/stx691}, \href
  {http://adsabs.harvard.edu/abs/2017MNRAS.468.3650K} {468, 3650}

\bibitem[\protect\citeauthoryear{Kramer}{Kramer}{2016}]{Kramer:2016kwa}
Kramer M.,  2016, \mndoi [Int. J. Mod. Phys.] {10.1142/S0218271816300299}, D25,
  1630029

\bibitem[\protect\citeauthoryear{Kramer \& Champion}{Kramer \&
  Champion}{2013}]{Kramer:2013kea}
Kramer M.,  Champion D.~J.,  2013, \mndoi [Class. Quant. Grav.]
  {10.1088/0264-9381/30/22/224009}, 30, 224009

\bibitem[\protect\citeauthoryear{Kramer, Backer, Cordes, Lazio, Stappers  \&
  Johnston}{Kramer et~al.}{2004}]{Kramer:2004hd}
Kramer M.,  Backer D.~C.,  Cordes J.~M.,  Lazio T. J.~W.,  Stappers B.~W.,
  Johnston S.,  2004, \mndoi [New Astron. Rev.] {10.1016/j.newar.2004.09.020},
  48, 993

\bibitem[\protect\citeauthoryear{Kramer et~al.}{Kramer
  et~al.}{2006}]{Kramer:2006nb}
Kramer M.,  et~al., 2006, \mndoi [Science] {10.1126/science.1132305}, 314, 97

\bibitem[\protect\citeauthoryear{Kugo \& Yoshioka}{Kugo \&
  Yoshioka}{2001}]{KUGO2001301}
Kugo T.,  Yoshioka K.,  2001, \mndoi [Nuclear Physics B]
  {http://dx.doi.org/10.1016/S0550-3213(00)00645-3}, 594, 301

\bibitem[\protect\citeauthoryear{{Kuhlen}, {Madau}  \& {Montgomery}}{{Kuhlen}
  et~al.}{2006}]{2006ApJ...637L...1K}
{Kuhlen} M.,  {Madau} P.,   {Montgomery} R.,  2006, \mndoi [\apjl]
  {10.1086/500548}, \href {http://adsabs.harvard.edu/abs/2006ApJ...637L...1K}
  {637, L1}

\bibitem[\protect\citeauthoryear{{LSST Science Collaboration} et~al.,}{{LSST
  Science Collaboration} et~al.}{2009}]{2009arXiv0912.0201L}
{LSST Science Collaboration} et~al., 2009, preprint, \href
  {http://adsabs.harvard.edu/abs/2009arXiv0912.0201L} {} (\mn@eprint {arXiv}
  {0912.0201})

\bibitem[\protect\citeauthoryear{{La Porta}, {Burigana}, {Reich}  \&
  {Reich}}{{La Porta} et~al.}{2008}]{2008A&A...479..641L}
{La Porta} L.,  {Burigana} C.,  {Reich} W.,   {Reich} P.,  2008, \mndoi [\aap]
  {10.1051/0004-6361:20078435}, \href
  {http://adsabs.harvard.edu/abs/2008A%26A...479..641L} {479, 641}

\bibitem[\protect\citeauthoryear{{Lacroix}, {Karami}, {Broderick}, {Silk}  \&
  {B{\AA}`hm}}{{Lacroix} et~al.}{2017}]{Lacroix:2016qpq}
{Lacroix} T.,  {Karami} M.,  {Broderick} A.~E.,  {Silk} J.,   {B{\AA}`hm} C.,
  2017, \mndoi [\prd] {10.1103/PhysRevD.96.063008}, \href
  {http://adsabs.harvard.edu/abs/2017PhRvD..96f3008L} {96, 063008}

\bibitem[\protect\citeauthoryear{{Lagos} et~al.,}{{Lagos}
  et~al.}{2015}]{Lagos:2015gpa}
{Lagos} C.~d.~P.,  et~al., 2015, \mndoi [\mnras] {10.1093/mnras/stv1488}, \href
  {http://adsabs.harvard.edu/abs/2015MNRAS.452.3815L} {452, 3815}

\bibitem[\protect\citeauthoryear{{Lagos}, {Baker}, {Ferreira}  \&
  {Noller}}{{Lagos} et~al.}{2016}]{2016JCAP...08..007L}
{Lagos} M.,  {Baker} T.,  {Ferreira} P.~G.,   {Noller} J.,  2016, \mndoi
  [\jcap] {10.1088/1475-7516/2016/08/007}, \href
  {http://adsabs.harvard.edu/abs/2016JCAP...08..007L} {8, 007}

\bibitem[\protect\citeauthoryear{Lambiase, Sakellariadou  \& Stabile}{Lambiase
  et~al.}{2013}]{Lambiase:2013dai}
Lambiase G.,  Sakellariadou M.,   Stabile A.,  2013, \mndoi [JCAP]
  {10.1088/1475-7516/2013/12/020}, 1312, 020

\bibitem[\protect\citeauthoryear{Lambiase, Sakellariadou, Stabile  \&
  Stabile}{Lambiase et~al.}{2015}]{Lambiase:2015yia}
Lambiase G.,  Sakellariadou M.,  Stabile A.,   Stabile A.,  2015, \mndoi [JCAP]
  {10.1088/1475-7516/2015/07/003}, 1507, 003

\bibitem[\protect\citeauthoryear{Landsberg}{Landsberg}{2015}]{Landsberg:2015pka}
Landsberg G.,  2015, \mndoi [Mod. Phys. Lett.] {10.1142/S0217732315400179},
  A30, 1540017

\bibitem[\protect\citeauthoryear{Lazaridis et~al.}{Lazaridis
  et~al.}{2009}]{Lazaridis:2009kq}
Lazaridis K.,  et~al., 2009, \mndoi [Mon. Not. R. Astron. Soc.]
  {10.1111/j.1365-2966.2009.15481.x}, 400, 805

\bibitem[\protect\citeauthoryear{{Lewis} \& {Challinor}}{{Lewis} \&
  {Challinor}}{2007}]{2007PhRvD..76h3005L}
{Lewis} A.,  {Challinor} A.,  2007, \mndoi [\prd] {10.1103/PhysRevD.76.083005},
  \href {http://adsabs.harvard.edu/abs/2007PhRvD..76h3005L} {76, 083005}

\bibitem[\protect\citeauthoryear{{Linder}}{{Linder}}{1986}]{Linder:1986}
{Linder} E.~V.,  1986, \mndoi [\prd] {10.1103/PhysRevD.34.1759}, \href
  {http://adsabs.harvard.edu/abs/1986PhRvD..34.1759L} {34, 1759}

\bibitem[\protect\citeauthoryear{{Linder}}{{Linder}}{1988}]{Linder:1988}
{Linder} E.~V.,  1988, \mndoi [\apj] {10.1086/166269}, \href
  {http://adsabs.harvard.edu/abs/1988ApJ...328...77L} {328, 77}

\bibitem[\protect\citeauthoryear{Line, Webster, Pindor, Mitchell  \&
  Trott}{Line et~al.}{2017}]{Line2017}
Line J. L.~B.,  Webster R.~L.,  Pindor B.,  Mitchell D.~A.,   Trott C.~M.,
  2017, \mndoi [Publications of the Astronomical Society of Australia]
  {10.1017/pasa.2016.58}, 34, e003

\bibitem[\protect\citeauthoryear{Liu \& Tegmark}{Liu \&
  Tegmark}{2011}]{Liu2011}
Liu A.,  Tegmark M.,  2011, \mndoi [Physical Review D]
  {10.1103/PhysRevD.83.103006}, 83

\bibitem[\protect\citeauthoryear{{Liu}, {Wex}, {Kramer}, {Cordes}  \&
  {Lazio}}{{Liu} et~al.}{2012}]{2012ApJ...747....1L}
{Liu} K.,  {Wex} N.,  {Kramer} M.,  {Cordes} J.~M.,   {Lazio} T.~J.~W.,  2012,
  \mndoi [\apj] {10.1088/0004-637X/747/1/1}, \href
  {http://adsabs.harvard.edu/abs/2012ApJ...747....1L} {747, 1}

\bibitem[\protect\citeauthoryear{{Liu}, {Parsons}  \& {Trott}}{{Liu}
  et~al.}{2014a}]{Liu2014a}
{Liu} A.,  {Parsons} A.~R.,   {Trott} C.~M.,  2014a, \mndoi [\prd]
  {10.1103/PhysRevD.90.023018}, \href
  {http://adsabs.harvard.edu/abs/2014PhRvD..90b3018L} {90, 023018}

\bibitem[\protect\citeauthoryear{{Liu}, {Parsons}  \& {Trott}}{{Liu}
  et~al.}{2014b}]{Liu2014}
{Liu} A.,  {Parsons} A.~R.,   {Trott} C.~M.,  2014b, \mndoi [\prd]
  {10.1103/PhysRevD.90.023019}, \href
  {http://adsabs.harvard.edu/abs/2014PhRvD..90b3019L} {90, 023019}

\bibitem[\protect\citeauthoryear{{Liu}, {Pritchard}, {Allison}, {Parsons},
  {Seljak}  \& {Sherwin}}{{Liu} et~al.}{2016}]{2016PhRvD..93d3013L}
{Liu} A.,  {Pritchard} J.~R.,  {Allison} R.,  {Parsons} A.~R.,  {Seljak} U.,
  {Sherwin} B.~D.,  2016, \mndoi [\prd] {10.1103/PhysRevD.93.043013}, \href
  {http://adsabs.harvard.edu/abs/2016PhRvD..93d3013L} {93, 043013}

\bibitem[\protect\citeauthoryear{Loeb}{Loeb}{1998}]{Loeb:1998bu}
Loeb A.,  1998, \mndoi [Astrophys. J.] {10.1086/311375}, 499, L111

\bibitem[\protect\citeauthoryear{{Loeb} \& {Zaldarriaga}}{{Loeb} \&
  {Zaldarriaga}}{2004}]{2004PhRvL..92u1301L}
{Loeb} A.,  {Zaldarriaga} M.,  2004, \mndoi [Physical Review Letters]
  {10.1103/PhysRevLett.92.211301}, \href
  {http://adsabs.harvard.edu/abs/2004PhRvL..92u1301L} {92, 211301}

\bibitem[\protect\citeauthoryear{{Lombriser}, {Yoo}  \& {Koyama}}{{Lombriser}
  et~al.}{2013}]{Lombriser2013}
{Lombriser} L.,  {Yoo} J.,   {Koyama} K.,  2013, \mndoi [\prd]
  {10.1103/PhysRevD.87.104019}, \href
  {http://adsabs.harvard.edu/abs/2013PhRvD..87j4019L} {87, 104019}

\bibitem[\protect\citeauthoryear{Lorimer, Bailes, McLaughlin, Narkevic  \&
  Crawford}{Lorimer et~al.}{2007}]{Lorimer:2007qn}
Lorimer D.~R.,  Bailes M.,  McLaughlin M.~A.,  Narkevic D.~J.,   Crawford F.,
  2007, \mndoi [Science] {10.1126/science.1147532}, 318, 777

\bibitem[\protect\citeauthoryear{Lovelock}{Lovelock}{1971}]{Lovelock:1971yv}
Lovelock D.,  1971, \mndoi [J. Math. Phys.] {10.1063/1.1665613}, 12, 498

\bibitem[\protect\citeauthoryear{{Lu}, {Do}, {Ghez}, {Morris}, {Yelda}  \&
  {Matthews}}{{Lu} et~al.}{2013}]{2013ApJ...764..155L}
{Lu} J.~R.,  {Do} T.,  {Ghez} A.~M.,  {Morris} M.~R.,  {Yelda} S.,   {Matthews}
  K.,  2013, \mndoi [\apj] {10.1088/0004-637X/764/2/155}, \href
  {http://adsabs.harvard.edu/abs/2013ApJ...764..155L} {764, 155}

\bibitem[\protect\citeauthoryear{{Ma} \& {Bertschinger}}{{Ma} \&
  {Bertschinger}}{1995}]{astro-ph/9506072}
{Ma} C.-P.,  {Bertschinger} E.,  1995, \mndoi [\apj] {10.1086/176550}, \href
  {http://adsabs.harvard.edu/abs/1995ApJ...455....7M} {455, 7}

\bibitem[\protect\citeauthoryear{Maartens}{Maartens}{2011}]{Maartens:2011yx}
Maartens R.,  2011, \mndoi [Phil. Trans. Roy. Soc. Lond.]
  {10.1098/rsta.2011.0289}, A369, 5115

\bibitem[\protect\citeauthoryear{Macfadyen \& Milosavljevic}{Macfadyen \&
  Milosavljevic}{2008}]{Macfadyen:2006jx}
Macfadyen A.~I.,  Milosavljevic M.,  2008, \mndoi [Astrophys. J.]
  {10.1086/523869}, 672, 83

\bibitem[\protect\citeauthoryear{{Mack} \& {Wesley}}{{Mack} \&
  {Wesley}}{2008}]{2008arXiv0805.1531M}
{Mack} K.~J.,  {Wesley} D.~H.,  2008, preprint, \href
  {http://adsabs.harvard.edu/abs/2008arXiv0805.1531M} {} (\mn@eprint {arXiv}
  {0805.1531})

\bibitem[\protect\citeauthoryear{{Macquart}, {Kanekar}, {Frail}  \&
  {Ransom}}{{Macquart} et~al.}{2010}]{2010ApJ...715..939M}
{Macquart} J.-P.,  {Kanekar} N.,  {Frail} D.~A.,   {Ransom} S.~M.,  2010,
  \mndoi [\apj] {10.1088/0004-637X/715/2/939}, \href
  {http://adsabs.harvard.edu/abs/2010ApJ...715..939M} {715, 939}

\bibitem[\protect\citeauthoryear{{Madau} \& {Dickinson}}{{Madau} \&
  {Dickinson}}{2014}]{2014ARA&A..52..415M}
{Madau} P.,  {Dickinson} M.,  2014, \mndoi [\araa]
  {10.1146/annurev-astro-081811-125615}, \href
  {http://adsabs.harvard.edu/abs/2014ARA%26A..52..415M} {52, 415}

\bibitem[\protect\citeauthoryear{{Madau}, {Meiksin}  \& {Rees}}{{Madau}
  et~al.}{1997}]{1997ApJ...475..429M}
{Madau} P.,  {Meiksin} A.,   {Rees} M.~J.,  1997, \apj, \href
  {http://adsabs.harvard.edu/abs/1997ApJ...475..429M} {475, 429}

\bibitem[\protect\citeauthoryear{Manchester}{Manchester}{2015}]{Manchester:2015mda}
Manchester R.~N.,  2015, \mndoi [Int. J. Mod. Phys.]
  {10.1142/S0218271815300189}, D24, 1530018

\bibitem[\protect\citeauthoryear{Manchester et~al.}{Manchester
  et~al.}{2013}]{Manchester:2012za}
Manchester R.~N.,  et~al., 2013, \mndoi [Publ. Astron. Soc. Austral.]
  {10.1017/pasa.2012.017}, 30, 17

\bibitem[\protect\citeauthoryear{Mangano, Melchiorri, Serra, Cooray  \&
  Kamionkowski}{Mangano et~al.}{2006}]{Mangano:2006mp}
Mangano G.,  Melchiorri A.,  Serra P.,  Cooray A.,   Kamionkowski M.,  2006,
  \mndoi [Phys. Rev.] {10.1103/PhysRevD.74.043517}, D74, 043517

\bibitem[\protect\citeauthoryear{Maroto}{Maroto}{2004a}]{PhysRevD.69.043509}
Maroto A.~L.,  2004a, \mndoi [Phys. Rev. D] {10.1103/PhysRevD.69.043509}, 69,
  043509

\bibitem[\protect\citeauthoryear{Maroto}{Maroto}{2004b}]{PhysRevD.69.101304}
Maroto A.~L.,  2004b, \mndoi [Phys. Rev. D] {10.1103/PhysRevD.69.101304}, 69,
  101304

\bibitem[\protect\citeauthoryear{{Martinelli}, {Pandolfi}, {Martins}  \&
  {Vielzeuf}}{{Martinelli} et~al.}{2012}]{2012PhRvD..86l3001M}
{Martinelli} M.,  {Pandolfi} S.,  {Martins} C.~J.~A.~P.,   {Vielzeuf} P.~E.,
  2012, \mndoi [\prd] {10.1103/PhysRevD.86.123001}, \href
  {http://adsabs.harvard.edu/abs/2012PhRvD..86l3001M} {86, 123001}

\bibitem[\protect\citeauthoryear{Massara, Villaescusa-Navarro, Viel  \&
  Sutter}{Massara et~al.}{2015}]{Massara:2015msa}
Massara E.,  Villaescusa-Navarro F.,  Viel M.,   Sutter P.~M.,  2015, \mndoi
  [JCAP] {10.1088/1475-7516/2015/11/018}, 1511, 018

\bibitem[\protect\citeauthoryear{{Massardi}, {Bonaldi}, {Negrello},
  {Ricciardi}, {Raccanelli}  \& {de Zotti}}{{Massardi}
  et~al.}{2010}]{massardi10}
{Massardi} M.,  {Bonaldi} A.,  {Negrello} M.,  {Ricciardi} S.,  {Raccanelli}
  A.,   {de Zotti} G.,  2010, \mndoi [\mnras]
  {10.1111/j.1365-2966.2010.16305.x}, \href
  {http://adsabs.harvard.edu/abs/2010MNRAS.404..532M} {404, 532}

\bibitem[\protect\citeauthoryear{{Masui} \& {Pen}}{{Masui} \&
  {Pen}}{2010}]{2010PhRvL.105p1302M}
{Masui} K.~W.,  {Pen} U.-L.,  2010, \mndoi [Physical Review Letters]
  {10.1103/PhysRevLett.105.161302}, \href
  {http://adsabs.harvard.edu/abs/2010PhRvL.105p1302M} {105, 161302}

\bibitem[\protect\citeauthoryear{Masui et~al.}{Masui
  et~al.}{2013a}]{Masui:2012zc}
Masui K.~W.,  et~al., 2013a, \mndoi [Astrophys. J.]
  {10.1088/2041-8205/763/1/L20}, 763, L20

\bibitem[\protect\citeauthoryear{Masui et~al.,}{Masui
  et~al.}{2013b}]{2013ApJ...763L..20M}
Masui K.~W.,  et~al., 2013b, \mnras, 763, L20

\bibitem[\protect\citeauthoryear{Mavromatos \& Sakellariadou}{Mavromatos \&
  Sakellariadou}{2007}]{Mavromatos:2007sp}
Mavromatos N.,  Sakellariadou M.,  2007, \mndoi [Phys. Lett.]
  {10.1016/j.physletb.2007.07.004}, B652, 97

\bibitem[\protect\citeauthoryear{Mavromatos, Sakellariadou  \&
  Yusaf}{Mavromatos et~al.}{2009}]{Mavromatos:2009xh}
Mavromatos N.~E.,  Sakellariadou M.,   Yusaf M.~F.,  2009, \mndoi [Phys. Rev.]
  {10.1103/PhysRevD.79.081301}, D79, 081301

\bibitem[\protect\citeauthoryear{Mavromatos, Sakellariadou  \&
  Yusaf}{Mavromatos et~al.}{2013}]{Mavromatos:2012ha}
Mavromatos N.~E.,  Sakellariadou M.,   Yusaf M.~F.,  2013, \mndoi [JCAP]
  {10.1088/1475-7516/2013/03/015}, 1303, 015

\bibitem[\protect\citeauthoryear{McLaughlin}{McLaughlin}{2013}]{McLaughlin:2013ira}
McLaughlin M.~A.,  2013, \mndoi [Class. Quant. Grav.]
  {10.1088/0264-9381/30/22/224008}, 30, 224008

\bibitem[\protect\citeauthoryear{{McQuinn} \& {O'Leary}}{{McQuinn} \&
  {O'Leary}}{2012}]{2012ApJ...760....3M}
{McQuinn} M.,  {O'Leary} R.~M.,  2012, \mndoi [\apj]
  {10.1088/0004-637X/760/1/3}, \href
  {http://adsabs.harvard.edu/abs/2012ApJ...760....3M} {760, 3}

\bibitem[\protect\citeauthoryear{McQuinn, Zahn, Zaldarriaga, Hernquist  \&
  Furlanetto}{McQuinn et~al.}{2006}]{McQuinn2006}
McQuinn M.,  Zahn O.,  Zaldarriaga M.,  Hernquist L.,   Furlanetto S.~R.,
  2006, \mndoi [The Astrophysical Journal] {10.1086/505167}, 653, 815

\bibitem[\protect\citeauthoryear{{Mellema} et~al.,}{{Mellema}
  et~al.}{2013}]{2013ExA....36..235M}
{Mellema} G.,  et~al., 2013, \mndoi [Experimental Astronomy]
  {10.1007/s10686-013-9334-5}, \href
  {http://adsabs.harvard.edu/abs/2013ExA....36..235M} {36, 235}

\bibitem[\protect\citeauthoryear{{Merritt}, {Alexander}, {Mikkola}  \&
  {Will}}{{Merritt} et~al.}{2010}]{2010PhRvD..81f2002M}
{Merritt} D.,  {Alexander} T.,  {Mikkola} S.,   {Will} C.~M.,  2010, \mndoi
  [\prd] {10.1103/PhysRevD.81.062002}, \href
  {http://adsabs.harvard.edu/abs/2010PhRvD..81f2002M} {81, 062002}

\bibitem[\protect\citeauthoryear{{Mesinger} \& {Furlanetto}}{{Mesinger} \&
  {Furlanetto}}{2007}]{2007ApJ...669..663M}
{Mesinger} A.,  {Furlanetto} S.,  2007, \mndoi [\apj] {10.1086/521806}, \href
  {http://adsabs.harvard.edu/abs/2007ApJ...669..663M} {669, 663}

\bibitem[\protect\citeauthoryear{{Mesinger}, {Furlanetto}  \& {Cen}}{{Mesinger}
  et~al.}{2011}]{Mesinger11}
{Mesinger} A.,  {Furlanetto} S.,   {Cen} R.,  2011, \mndoi [\mnras]
  {10.1111/j.1365-2966.2010.17731.x}, \href
  {http://adsabs.harvard.edu/abs/2011MNRAS.411..955M} {411, 955}

\bibitem[\protect\citeauthoryear{{Mesinger}, {Ewall-Wice}  \&
  {Hewitt}}{{Mesinger} et~al.}{2014}]{mesinger2013a}
{Mesinger} A.,  {Ewall-Wice} A.,   {Hewitt} J.,  2014, \mndoi [\mnras]
  {10.1093/mnras/stu125}, \href
  {http://adsabs.harvard.edu/abs/2014MNRAS.439.3262M} {439, 3262}

\bibitem[\protect\citeauthoryear{{Meszaros}}{{Meszaros}}{1974}]{Meszaros}
{Meszaros} P.,  1974, \aap, \href
  {http://adsabs.harvard.edu/abs/1974A%26A....37..225M} {37, 225}

\bibitem[\protect\citeauthoryear{{Metcalf} \& {White}}{{Metcalf} \&
  {White}}{2009}]{Metcalf:2009}
{Metcalf} R.~B.,  {White} S.~D.~M.,  2009, \mndoi [\mnras]
  {10.1111/j.1365-2966.2009.14401.x}, \href
  {http://adsabs.harvard.edu/abs/2009MNRAS.394..704M} {394, 704}

\bibitem[\protect\citeauthoryear{{Mishra} \& {Hirata}}{{Mishra} \&
  {Hirata}}{2018}]{2017arXiv170703514M}
{Mishra} A.,  {Hirata} C.~M.,  2018, \mndoi [\prd]
  {10.1103/PhysRevD.97.103522}, \href
  {http://adsabs.harvard.edu/abs/2018PhRvD..97j3522M} {97, 103522}

\bibitem[\protect\citeauthoryear{Mo \& White}{Mo \& White}{1996}]{Mo:1995}
Mo H.~J.,  White S. D.~M.,  1996, \mndoi [Mon. Not. Roy. Astron. Soc.]
  {10.1093/mnras/282.2.347}, 282, 347

\bibitem[\protect\citeauthoryear{Moline, Schewtschenko, Palomares-Ruiz, Boehm
  \& Baugh}{Moline et~al.}{2016}]{Moline:2016fdo}
Moline A.,  Schewtschenko J.~A.,  Palomares-Ruiz S.,  Boehm C.,   Baugh C.~M.,
  2016, \mndoi [JCAP] {10.1088/1475-7516/2016/08/069}, 1608, 069

\bibitem[\protect\citeauthoryear{{Moore} et~al.,}{{Moore}
  et~al.}{2017}]{Moore2015}
{Moore} D.~F.,  et~al., 2017, \mndoi [The Astrophysical Journal]
  {10.3847/1538-4357/836/2/154}, \href
  {https://ui.adsabs.harvard.edu/abs/2017ApJ...836..154M} {836, 154}

\bibitem[\protect\citeauthoryear{{Moresco}}{{Moresco}}{2015}]{moresco2015}
{Moresco} M.,  2015, \mndoi [MNRAS] {10.1093/mnrasl/slv037}, \href
  {http://adsabs.harvard.edu/abs/2015MNRAS.450L..16M} {450, L16}

\bibitem[\protect\citeauthoryear{{Moresco}, {Jimenez}, {Cimatti}  \&
  {Pozzetti}}{{Moresco} et~al.}{2011}]{moresco2011}
{Moresco} M.,  {Jimenez} R.,  {Cimatti} A.,   {Pozzetti} L.,  2011, \mndoi
  [JCAP] {10.1088/1475-7516/2011/03/045}, \href
  {http://adsabs.harvard.edu/abs/2011JCAP...03..045M} {3, 45}

\bibitem[\protect\citeauthoryear{{Mu{\~n}oz} \& {Loeb}}{{Mu{\~n}oz} \&
  {Loeb}}{2018}]{munoz2018}
{Mu{\~n}oz} J.~B.,  {Loeb} A.,  2018, Nature, 557, 684

\bibitem[\protect\citeauthoryear{{Mu{\~n}oz}, {Ali-Ha{\"i}moud}  \&
  {Kamionkowski}}{{Mu{\~n}oz} et~al.}{2015}]{2015PhRvD..92h3508M}
{Mu{\~n}oz} J.~B.,  {Ali-Ha{\"i}moud} Y.,   {Kamionkowski} M.,  2015, \mndoi
  [\prd] {10.1103/PhysRevD.92.083508}, \href
  {http://adsabs.harvard.edu/abs/2015PhRvD..92h3508M} {92, 083508}

\bibitem[\protect\citeauthoryear{{Mu{\~n}oz}, {Kovetz}, {Dai}  \&
  {Kamionkowski}}{{Mu{\~n}oz} et~al.}{2016}]{Munoz}
{Mu{\~n}oz} J.~B.,  {Kovetz} E.~D.,  {Dai} L.,   {Kamionkowski} M.,  2016,
  \mndoi [Physical Review Letters] {10.1103/PhysRevLett.117.091301}, \href
  {http://adsabs.harvard.edu/abs/2016PhRvL.117i1301M} {117, 091301}

\bibitem[\protect\citeauthoryear{{Mu{\~n}oz}, {Kovetz}, {Raccanelli},
  {Kamionkowski}  \& {Silk}}{{Mu{\~n}oz} et~al.}{2017}]{Munoz:2017}
{Mu{\~n}oz} J.~B.,  {Kovetz} E.~D.,  {Raccanelli} A.,  {Kamionkowski} M.,
  {Silk} J.,  2017, \mndoi [\jcap] {10.1088/1475-7516/2017/05/032}, \href
  {http://adsabs.harvard.edu/abs/2017JCAP...05..032M} {5, 032}

\bibitem[\protect\citeauthoryear{Mukhanov}{Mukhanov}{1985}]{Mukhanov:1985rz}
Mukhanov V.~F.,  1985, JETP Lett., 41, 493

\bibitem[\protect\citeauthoryear{{Murphy}, {Malec}  \& {Prochaska}}{{Murphy}
  et~al.}{2016}]{murphy2016}
{Murphy} M.~T.,  {Malec} A.~L.,   {Prochaska} J.~X.,  2016, \mndoi [\mnras]
  {10.1093/mnras/stw1482}, \href
  {https://ui.adsabs.harvard.edu/abs/2016MNRAS.461.2461M} {461, 2461}

\bibitem[\protect\citeauthoryear{{Murray}, {Trott}  \& {Jordan}}{{Murray}
  et~al.}{2017}]{Murray2017}
{Murray} S.~G.,  {Trott} C.~M.,   {Jordan} C.~H.,  2017, \mndoi [\apj]
  {10.3847/1538-4357/aa7d0a}, \href
  {http://adsabs.harvard.edu/abs/2017ApJ...845....7M} {845, 7}

\bibitem[\protect\citeauthoryear{{Narayanan}, {Spergel}, {Dav{\'e}}  \&
  {Ma}}{{Narayanan} et~al.}{2000}]{Narayanan01}
{Narayanan} V.~K.,  {Spergel} D.~N.,  {Dav{\'e}} R.,   {Ma} C.-P.,  2000,
  \mndoi [\apjl] {10.1086/317269}, \href
  {http://adsabs.harvard.edu/abs/2000ApJ...543L.103N} {543, L103}

\bibitem[\protect\citeauthoryear{{Naselsky} \& {Chiang}}{{Naselsky} \&
  {Chiang}}{2004}]{2004MNRAS.347..795N}
{Naselsky} P.,  {Chiang} L.-Y.,  2004, \mndoi [\mnras]
  {10.1111/j.1365-2966.2004.07250.x}, \href
  {http://adsabs.harvard.edu/abs/2004MNRAS.347..795N} {347, 795}

\bibitem[\protect\citeauthoryear{Nelson, Ochoa  \& Sakellariadou}{Nelson
  et~al.}{2010a}]{Nelson:2010ru}
Nelson W.,  Ochoa J.,   Sakellariadou M.,  2010a, \mndoi [Phys. Rev. Lett.]
  {10.1103/PhysRevLett.105.101602}, 105, 101602

\bibitem[\protect\citeauthoryear{Nelson, Ochoa  \& Sakellariadou}{Nelson
  et~al.}{2010b}]{Nelson:2010rt}
Nelson W.,  Ochoa J.,   Sakellariadou M.,  2010b, \mndoi [Phys. Rev.]
  {10.1103/PhysRevD.82.085021}, D82, 085021

\bibitem[\protect\citeauthoryear{{Nishizawa}}{{Nishizawa}}{2014}]{nishizawa14}
{Nishizawa} A.~J.,  2014, \mndoi [Progress of Theoretical and Experimental
  Physics] {10.1093/ptep/ptu062}, \href
  {http://adsabs.harvard.edu/abs/2014PTEP.2014fB110N} {2014, 06B110}

\bibitem[\protect\citeauthoryear{{Nolta} et~al.,}{{Nolta}
  et~al.}{2004}]{nolta04}
{Nolta} M.~R.,  et~al., 2004, \mndoi [\apj] {10.1086/386536}, \href
  {http://adsabs.harvard.edu/abs/2004ApJ...608...10N} {608, 10}

\bibitem[\protect\citeauthoryear{{Nordtvedt}}{{Nordtvedt}}{1987}]{Nordtvedt:1987}
{Nordtvedt} K.,  1987, \mndoi [\apj] {10.1086/165603}, \href
  {http://adsabs.harvard.edu/abs/1987ApJ...320..871N} {320, 871}

\bibitem[\protect\citeauthoryear{Nordtvedt}{Nordtvedt}{1990}]{Nordtvedt:1990zz}
Nordtvedt K.,  1990, \mndoi [Phys. Rev. Lett.] {10.1103/PhysRevLett.65.953},
  65, 953

\bibitem[\protect\citeauthoryear{{Norris} et~al.,}{{Norris}
  et~al.}{2011}]{2011PASA...28..215N}
{Norris} R.~P.,  et~al., 2011, \mndoi [\pasa] {10.1071/AS11021}, \href
  {http://adsabs.harvard.edu/abs/2011PASA...28..215N} {28, 215}

\bibitem[\protect\citeauthoryear{{Nuza}, {Gelszinnis}, {Hoeft}  \&
  {Yepes}}{{Nuza} et~al.}{2017}]{2017MNRAS.470..240N}
{Nuza} S.~E.,  {Gelszinnis} J.,  {Hoeft} M.,   {Yepes} G.,  2017, \mndoi
  [\mnras] {10.1093/mnras/stx1109}, \href
  {http://adsabs.harvard.edu/abs/2017MNRAS.470..240N} {470, 240}

\bibitem[\protect\citeauthoryear{{O'Leary}, {Kistler}, {Kerr}  \&
  {Dexter}}{{O'Leary} et~al.}{2016}]{OLeary:2016cwz}
{O'Leary} R.~M.,  {Kistler} M.~D.,  {Kerr} M.,   {Dexter} J.,  2016, arXiv
  e-prints, \href {https://ui.adsabs.harvard.edu/abs/2016arXiv160105797O} {p.
  arXiv:1601.05797}

\bibitem[\protect\citeauthoryear{Offringa et~al.,}{Offringa
  et~al.}{2016}]{Offringa2016}
Offringa A.~R.,  et~al., 2016, \mndoi [Monthly Notices of the Royal
  Astronomical Society] {10.1093/mnras/stw310}, 458, 1057

\bibitem[\protect\citeauthoryear{{Oguri}}{{Oguri}}{2016}]{Oguri:2016}
{Oguri} M.,  2016, \mndoi [\prd] {10.1103/PhysRevD.93.083511}, \href
  {http://adsabs.harvard.edu/abs/2016PhRvD..93h3511O} {93, 083511}

\bibitem[\protect\citeauthoryear{{Oh}}{{Oh}}{1999}]{1999ApJ...527...16O}
{Oh} S.~P.,  1999, \mndoi [\apj] {10.1086/308077}, \href
  {http://adsabs.harvard.edu/abs/1999ApJ...527...16O} {527, 16}

\bibitem[\protect\citeauthoryear{{Oosterloo}, {Fraternali}  \&
  {Sancisi}}{{Oosterloo} et~al.}{2007}]{Oost07}
{Oosterloo} T.,  {Fraternali} F.,   {Sancisi} R.,  2007, \mndoi [\aj]
  {10.1086/520332}, \href {http://adsabs.harvard.edu/abs/2007AJ....134.1019O}
  {134, 1019}

\bibitem[\protect\citeauthoryear{{Paciga} et~al.,}{{Paciga}
  et~al.}{2013}]{paciga2013}
{Paciga} G.,  et~al., 2013, \mndoi [\mnras] {10.1093/mnras/stt753}, \href
  {http://adsabs.harvard.edu/abs/2013MNRAS.433..639P} {433, 639}

\bibitem[\protect\citeauthoryear{{Padmanabhan} \& {Refregier}}{{Padmanabhan} \&
  {Refregier}}{2017}]{hpar2017}
{Padmanabhan} H.,  {Refregier} A.,  2017, \mndoi [\mnras]
  {10.1093/mnras/stw2706}, \href
  {http://adsabs.harvard.edu/abs/2017MNRAS.464.4008P} {464, 4008}

\bibitem[\protect\citeauthoryear{{Padmanabhan}, {Choudhury}  \&
  {Refregier}}{{Padmanabhan} et~al.}{2015}]{hptrcar2015}
{Padmanabhan} H.,  {Choudhury} T.~R.,   {Refregier} A.,  2015, \mndoi [\mnras]
  {10.1093/mnras/stu2702}, \href
  {http://adsabs.harvard.edu/abs/2015MNRAS.447.3745P} {447, 3745}

\bibitem[\protect\citeauthoryear{{Padmanabhan}, {Refregier}  \&
  {Amara}}{{Padmanabhan} et~al.}{2017}]{hparaa2017}
{Padmanabhan} H.,  {Refregier} A.,   {Amara} A.,  2017, \mndoi [\mnras]
  {10.1093/mnras/stx979}, \href
  {http://adsabs.harvard.edu/abs/2017MNRAS.469.2323P} {469, 2323}

\bibitem[\protect\citeauthoryear{{Palanque-Delabrouille}
  et~al.,}{{Palanque-Delabrouille} et~al.}{2015}]{Palanque_2015}
{Palanque-Delabrouille} N.,  et~al., 2015, \mndoi [\jcap]
  {10.1088/1475-7516/2015/11/011}, \href
  {http://adsabs.harvard.edu/abs/2015JCAP...11..011P} {11, 011}

\bibitem[\protect\citeauthoryear{Pan \& Knox}{Pan \& Knox}{2015}]{Zhen:2015yba}
Pan Z.,  Knox L.,  2015, \mndoi [Mon. Not. Roy. Astron. Soc.]
  {10.1093/mnras/stv2164}, 454, 3200

\bibitem[\protect\citeauthoryear{Park}{Park}{2007}]{par_space}
Park E.-K.,  2007, DMSAG Report on the Direct Detection and Study of Dark
  Matter, \url
  {https://science.energy.gov/~/media/hep/pdf/files/pdfs/dmsagreportjuly18_2007.pdf}

\bibitem[\protect\citeauthoryear{Parkinson, Bassett  \& Barrow}{Parkinson
  et~al.}{2004}]{Parkinson:2003kf}
Parkinson D.,  Bassett B.~A.,   Barrow J.~D.,  2004, \mndoi [Phys. Lett.]
  {10.1016/j.physletb.2003.10.081}, B578, 235

\bibitem[\protect\citeauthoryear{{Parsons} et~al.,}{{Parsons}
  et~al.}{2014}]{parsons2014}
{Parsons} A.~R.,  et~al., 2014, \mndoi [\apj] {10.1088/0004-637X/788/2/106},
  \href {http://adsabs.harvard.edu/abs/2014ApJ...788..106P} {788, 106}

\bibitem[\protect\citeauthoryear{Patil et~al.,}{Patil et~al.}{2017}]{Patil2017}
Patil A.~H.,  et~al., 2017, \mndoi [The Astrophysical Journal]
  {10.3847/1538-4357/aa63e7}, 838, 65

\bibitem[\protect\citeauthoryear{{Patra}, {Subrahmanyan}, {Raghunathan}  \&
  {Udaya Shankar}}{{Patra} et~al.}{2013}]{Patra2013}
{Patra} N.,  {Subrahmanyan} R.,  {Raghunathan} A.,   {Udaya Shankar} N.,  2013,
  \mndoi [Experimental Astronomy] {10.1007/s10686-013-9336-3}, \href
  {http://adsabs.harvard.edu/abs/2013ExA....36..319P} {36, 319}

\bibitem[\protect\citeauthoryear{{Paumard} et~al.,}{{Paumard}
  et~al.}{2006}]{2006ApJ...643.1011P}
{Paumard} T.,  et~al., 2006, \mndoi [\apj] {10.1086/503273}, \href
  {http://adsabs.harvard.edu/abs/2006ApJ...643.1011P} {643, 1011}

\bibitem[\protect\citeauthoryear{Pen \& Broderick}{Pen \&
  Broderick}{2014}]{Pen:2013qva}
Pen U.-L.,  Broderick A.~E.,  2014, \mndoi [Mon. Not. Roy. Astron. Soc.]
  {10.1093/mnras/stu1919}, 445, 3370

\bibitem[\protect\citeauthoryear{Petiteau, Babak, Sesana  \& de
  Araújo}{Petiteau et~al.}{2013}]{Petiteau:2012zq}
Petiteau A.,  Babak S.,  Sesana A.,   de Araújo M.,  2013, \mndoi [Phys. Rev.]
  {10.1103/PhysRevD.87.064036}, D87, 064036

\bibitem[\protect\citeauthoryear{{Pfahl} \& {Loeb}}{{Pfahl} \&
  {Loeb}}{2004}]{2004ApJ...615..253P}
{Pfahl} E.,  {Loeb} A.,  2004, \mndoi [\apj] {10.1086/423975}, \href
  {http://adsabs.harvard.edu/abs/2004ApJ...615..253P} {615, 253}

\bibitem[\protect\citeauthoryear{{Pietrobon}, {Balbi}  \&
  {Marinucci}}{{Pietrobon} et~al.}{2006}]{pietrobon06}
{Pietrobon} D.,  {Balbi} A.,   {Marinucci} D.,  2006, \mndoi [\prd]
  {10.1103/PhysRevD.74.043524}, \href
  {http://adsabs.harvard.edu/abs/2006PhRvD..74d3524P} {74, 043524}

\bibitem[\protect\citeauthoryear{{Pillepich}, {Porciani}  \&
  {Matarrese}}{{Pillepich} et~al.}{2007}]{2007ApJ...662....1P}
{Pillepich} A.,  {Porciani} C.,   {Matarrese} S.,  2007, \mndoi [\apj]
  {10.1086/517963}, \href {http://adsabs.harvard.edu/abs/2007ApJ...662....1P}
  {662, 1}

\bibitem[\protect\citeauthoryear{Pindor, Wyithe, Mitchell, Ord, Wayth  \&
  Greenhill}{Pindor et~al.}{2011}]{Pindor2011}
Pindor B.,  Wyithe J. S.~B.,  Mitchell D.~A.,  Ord S.~M.,  Wayth R.~B.,
  Greenhill L.~J.,  2011, \mndoi [Publications of the Astronomical Society of
  Australia] {10.1071/AS10023}, 28, 46

\bibitem[\protect\citeauthoryear{Pisani, Sutter, Hamaus, Alizadeh, Biswas,
  Wandelt  \& Hirata}{Pisani et~al.}{2015}]{Pisani:2015jha}
Pisani A.,  Sutter P.~M.,  Hamaus N.,  Alizadeh E.,  Biswas R.,  Wandelt B.~D.,
    Hirata C.~M.,  2015, \mndoi [Phys. Rev.] {10.1103/PhysRevD.92.083531}, D92,
  083531

\bibitem[\protect\citeauthoryear{{Planck Collaboration} et~al.,}{{Planck
  Collaboration} et~al.}{2014a}]{planckisw}
{Planck Collaboration} et~al., 2014a, \mndoi [\aap]
  {10.1051/0004-6361/201321526}, \href
  {http://adsabs.harvard.edu/abs/2014A%26A...571A..19P} {571, A19}

\bibitem[\protect\citeauthoryear{{Planck Collaboration} et~al.,}{{Planck
  Collaboration} et~al.}{2014b}]{Ade:2013nlj}
{Planck Collaboration} et~al., 2014b, \mndoi [\aap]
  {10.1051/0004-6361/201321534}, \href
  {http://adsabs.harvard.edu/abs/2014A%26A...571A..23P} {571, A23}

\bibitem[\protect\citeauthoryear{{Planck Collaboration} et~al.,}{{Planck
  Collaboration} et~al.}{2016a}]{2016A&A...586A.133P}
{Planck Collaboration} et~al., 2016a, \mndoi [\aap]
  {10.1051/0004-6361/201425034}, \href
  {http://adsabs.harvard.edu/abs/2016A%26A...586A.133P} {586, A133}

\bibitem[\protect\citeauthoryear{{Planck Collaboration} et~al.,}{{Planck
  Collaboration} et~al.}{2016b}]{2016A&A...594A..10P}
{Planck Collaboration} et~al., 2016b, \mndoi [\aap]
  {10.1051/0004-6361/201525967}, \href
  {http://adsabs.harvard.edu/abs/2016A%26A...594A..10P} {594, A10}

\bibitem[\protect\citeauthoryear{{Planck Collaboration} et~al.,}{{Planck
  Collaboration} et~al.}{2016c}]{Ade:2015sjc}
{Planck Collaboration} et~al., 2016c, \mndoi [\aap]
  {10.1051/0004-6361/201526681}, \href
  {http://adsabs.harvard.edu/abs/2016A%26A...594A..16P} {594, A16}

\bibitem[\protect\citeauthoryear{{Planck Collaboration} et~al.,}{{Planck
  Collaboration} et~al.}{2016d}]{planckfnl}
{Planck Collaboration} et~al., 2016d, \mndoi [\aap]
  {10.1051/0004-6361/201525836}, \href
  {http://adsabs.harvard.edu/abs/2016A%26A...594A..17P} {594, A17}

\bibitem[\protect\citeauthoryear{{Planck Collaboration} et~al.,}{{Planck
  Collaboration} et~al.}{2016e}]{1502.02114}
{Planck Collaboration} et~al., 2016e, \mndoi [\aap]
  {10.1051/0004-6361/201525898}, \href
  {http://adsabs.harvard.edu/abs/2016A%26A...594A..20P} {594, A20}

\bibitem[\protect\citeauthoryear{{Planck Collaboration} et~al.,}{{Planck
  Collaboration} et~al.}{2016f}]{planckisw2015}
{Planck Collaboration} et~al., 2016f, \mndoi [\aap]
  {10.1051/0004-6361/201525831}, \href
  {http://adsabs.harvard.edu/abs/2016A%26A...594A..21P} {594, A21}

\bibitem[\protect\citeauthoryear{{Planck Collaboration} et~al.,}{{Planck
  Collaboration} et~al.}{2016g}]{2016A&A...594A..25P}
{Planck Collaboration} et~al., 2016g, \mndoi [\aap]
  {10.1051/0004-6361/201526803}, \href
  {http://adsabs.harvard.edu/abs/2016A%26A...594A..25P} {594, A25}

\bibitem[\protect\citeauthoryear{{Planck Collaboration} et~al.,}{{Planck
  Collaboration} et~al.}{2018}]{2018arXiv180706209P}
{Planck Collaboration} et~al., 2018, preprint, \href
  {http://adsabs.harvard.edu/abs/2018arXiv180706209P} {} (\mn@eprint {arXiv}
  {1807.06209})

\bibitem[\protect\citeauthoryear{Plotkin, Markoff, Kelly, Koerding  \&
  Anderson}{Plotkin et~al.}{2012}]{Plotkin:2011dy}
Plotkin R.~M.,  Markoff S.,  Kelly B.~C.,  Koerding E.,   Anderson S.~F.,
  2012, \mndoi [Mon. Not. Roy. Astron. Soc.]
  {10.1111/j.1365-2966.2011.19689.x}, 419, 267

\bibitem[\protect\citeauthoryear{{Pober} et~al.,}{{Pober}
  et~al.}{2015}]{pober2015}
{Pober} J.~C.,  et~al., 2015, \mndoi [\apj] {10.1088/0004-637X/809/1/62}, \href
  {http://adsabs.harvard.edu/abs/2015ApJ...809...62P} {809, 62}

\bibitem[\protect\citeauthoryear{{Polisensky} \& {Ricotti}}{{Polisensky} \&
  {Ricotti}}{2011}]{Polisensky11}
{Polisensky} E.,  {Ricotti} M.,  2011, \mndoi [\prd]
  {10.1103/PhysRevD.83.043506}, \href
  {http://adsabs.harvard.edu/abs/2011PhRvD..83d3506P} {83, 043506}

\bibitem[\protect\citeauthoryear{{Ponente}, {Diego}, {Sheth}, {Burigana},
  {Knollmann}  \& {Ascasibar}}{{Ponente} et~al.}{2011}]{2011MNRAS.410.2353P}
{Ponente} P.~P.,  {Diego} J.~M.,  {Sheth} R.~K.,  {Burigana} C.,  {Knollmann}
  S.~R.,   {Ascasibar} Y.,  2011, \mndoi [\mnras]
  {10.1111/j.1365-2966.2010.17611.x}, \href
  {http://adsabs.harvard.edu/abs/2011MNRAS.410.2353P} {410, 2353}

\bibitem[\protect\citeauthoryear{{Poulin}, {Lesgourgues}  \&
  {Serpico}}{{Poulin} et~al.}{2017}]{Poulin2017}
{Poulin} V.,  {Lesgourgues} J.,   {Serpico} P.~D.,  2017, \mndoi [Journal of
  Cosmology and Astroparticle Physics] {10.1088/1475-7516/2017/03/043}, \href
  {http://adsabs.harvard.edu/abs/2017JCAP...03..043P} {3, 043}

\bibitem[\protect\citeauthoryear{{Pourtsidou}}{{Pourtsidou}}{2016a}]{Pourtsidou}
{Pourtsidou} A.,  2016a, preprint, \href
  {http://adsabs.harvard.edu/abs/2016arXiv161205138P} {} (\mn@eprint {arXiv}
  {1612.05138})

\bibitem[\protect\citeauthoryear{Pourtsidou}{Pourtsidou}{2016b}]{Pourtsidou:2015ksn}
Pourtsidou A.,  2016b, \mndoi [Mon. Not. Roy. Astron. Soc.]
  {10.1093/mnras/stw1406}, 461, 1457

\bibitem[\protect\citeauthoryear{{Pourtsidou} \& {Metcalf}}{{Pourtsidou} \&
  {Metcalf}}{2014}]{PourtsidouMetcalf:2014}
{Pourtsidou} A.,  {Metcalf} R.~B.,  2014, \mndoi [\mnras]
  {10.1093/mnrasl/slt175}, \href
  {http://adsabs.harvard.edu/abs/2014MNRAS.439L..36P} {439, L36}

\bibitem[\protect\citeauthoryear{Pourtsidou, Bacon, Crittenden  \&
  Metcalf}{Pourtsidou et~al.}{2016}]{Pourtsidou:2015mia}
Pourtsidou A.,  Bacon D.,  Crittenden R.,   Metcalf R.~B.,  2016, \mndoi [Mon.
  Not. Roy. Astron. Soc.] {10.1093/mnras/stw658}, 459, 863

\bibitem[\protect\citeauthoryear{{Pourtsidou}, {Bacon}  \&
  {Crittenden}}{{Pourtsidou} et~al.}{2017}]{Pourtsidou:2016dzn}
{Pourtsidou} A.,  {Bacon} D.,   {Crittenden} R.,  2017, \mndoi [\mnras]
  {10.1093/mnras/stx1479}, \href
  {http://adsabs.harvard.edu/abs/2017MNRAS.470.4251P} {470, 4251}

\bibitem[\protect\citeauthoryear{{Prandoni} \& {Seymour}}{{Prandoni} \&
  {Seymour}}{2015}]{2015aska.confE..67P}
{Prandoni} I.,  {Seymour} N.,  2015, Advancing Astrophysics with the Square
  Kilometre Array (AASKA14), \href
  {http://adsabs.harvard.edu/abs/2015aska.confE..67P} {p.~67}

\bibitem[\protect\citeauthoryear{{Prandoni}, {Gregorini}, {Parma}, {de Ruiter},
  {Vettolani}, {Wieringa}  \& {Ekers}}{{Prandoni}
  et~al.}{2001}]{2001A&A...365..392P}
{Prandoni} I.,  {Gregorini} L.,  {Parma} P.,  {de Ruiter} H.~R.,  {Vettolani}
  G.,  {Wieringa} M.~H.,   {Ekers} R.~D.,  2001, \mndoi [\aap]
  {10.1051/0004-6361:20000142}, \href
  {http://adsabs.harvard.edu/abs/2001A%26A...365..392P} {365, 392}

\bibitem[\protect\citeauthoryear{{Pritchard} \& {Furlanetto}}{{Pritchard} \&
  {Furlanetto}}{2007}]{2007MNRAS.376.1680P}
{Pritchard} J.~R.,  {Furlanetto} S.~R.,  2007, \mndoi [\mnras]
  {10.1111/j.1365-2966.2007.11519.x}, \href
  {http://adsabs.harvard.edu/abs/2007MNRAS.376.1680P} {376, 1680}

\bibitem[\protect\citeauthoryear{{Pritchard} \& {Loeb}}{{Pritchard} \&
  {Loeb}}{2008}]{pritchard2008}
{Pritchard} J.~R.,  {Loeb} A.,  2008, \mndoi [\prd]
  {10.1103/PhysRevD.78.103511}, \href
  {http://adsabs.harvard.edu/abs/2008PhRvD..78j3511P} {78, 103511}

\bibitem[\protect\citeauthoryear{{Pritchard} \& {Loeb}}{{Pritchard} \&
  {Loeb}}{2010}]{2010PhRvD..82b3006P}
{Pritchard} J.~R.,  {Loeb} A.,  2010, \mndoi [\prd]
  {10.1103/PhysRevD.82.023006}, \href
  {http://adsabs.harvard.edu/abs/2010PhRvD..82b3006P} {82, 023006}

\bibitem[\protect\citeauthoryear{{Pritchard} \& {Loeb}}{{Pritchard} \&
  {Loeb}}{2012}]{Pritchard12}
{Pritchard} J.~R.,  {Loeb} A.,  2012, \mndoi [Reports on Progress in Physics]
  {10.1088/0034-4885/75/8/086901}, \href
  {http://adsabs.harvard.edu/abs/2012RPPh...75h6901P} {75, 086901}

\bibitem[\protect\citeauthoryear{{Pritchard} et~al.,}{{Pritchard}
  et~al.}{2015}]{2015aska.confE..12P}
{Pritchard} J.,  et~al., 2015, Advancing Astrophysics with the Square Kilometre
  Array (AASKA14), \href {http://adsabs.harvard.edu/abs/2015aska.confE..12P}
  {p.~12}

\bibitem[\protect\citeauthoryear{{Prochaska} \& {Wolfe}}{{Prochaska} \&
  {Wolfe}}{2009}]{2009ApJ...696.1543P}
{Prochaska} J.~X.,  {Wolfe} A.~M.,  2009, \mndoi [\apj]
  {10.1088/0004-637X/696/2/1543}, \href
  {http://adsabs.harvard.edu/abs/2009ApJ...696.1543P} {696, 1543}

\bibitem[\protect\citeauthoryear{{Procopio} et~al.,}{{Procopio}
  et~al.}{2017}]{Procopio2017}
{Procopio} P.,  et~al., 2017, \mndoi [\pasa] {10.1017/pasa.2017.26}, \href
  {http://adsabs.harvard.edu/abs/2017PASA...34...33P} {34, e033}

\bibitem[\protect\citeauthoryear{Psaltis}{Psaltis}{2008}]{Psaltis:2008bb}
Psaltis D.,  2008, \mndoi [Living Rev. Rel.] {10.12942/lrr-2008-9}, 11, 9

\bibitem[\protect\citeauthoryear{{Psaltis}, {Wex}  \& {Kramer}}{{Psaltis}
  et~al.}{2016}]{2016ApJ...818..121P}
{Psaltis} D.,  {Wex} N.,   {Kramer} M.,  2016, \mndoi [\apj]
  {10.3847/0004-637X/818/2/121}, \href
  {http://adsabs.harvard.edu/abs/2016ApJ...818..121P} {818, 121}

\bibitem[\protect\citeauthoryear{Pullen \& Kamionkowski}{Pullen \&
  Kamionkowski}{2007}]{Pullen:2007tu}
Pullen A.~R.,  Kamionkowski M.,  2007, \mndoi [Phys. Rev.]
  {10.1103/PhysRevD.76.103529}, D76, 103529

\bibitem[\protect\citeauthoryear{{Pullen}, {Chang}, {Dor{\'e}}  \&
  {Lidz}}{{Pullen} et~al.}{2013}]{Pullen:2012}
{Pullen} A.~R.,  {Chang} T.-C.,  {Dor{\'e}} O.,   {Lidz} A.,  2013, \mndoi
  [\apj] {10.1088/0004-637X/768/1/15}, \href
  {http://adsabs.harvard.edu/abs/2013ApJ...768...15P} {768, 15}

\bibitem[\protect\citeauthoryear{Pullen, Alam  \& Ho}{Pullen
  et~al.}{2015}]{Pullen:2014fva}
Pullen A.~R.,  Alam S.,   Ho S.,  2015, \mndoi [Mon. Not. Roy. Astron. Soc.]
  {10.1093/mnras/stv554}, 449, 4326

\bibitem[\protect\citeauthoryear{Pullen, Alam, He  \& Ho}{Pullen
  et~al.}{2016}]{Pullen:2015vtb}
Pullen A.~R.,  Alam S.,  He S.,   Ho S.,  2016, \mndoi [Mon. Not. Roy. Astron.
  Soc.] {10.1093/mnras/stw1249}, 460, 4098

\bibitem[\protect\citeauthoryear{{Raccanelli}}{{Raccanelli}}{2017}]{Raccanelli:2016GW}
{Raccanelli} A.,  2017, \mndoi [\mnras] {10.1093/mnras/stx835}, \href
  {http://adsabs.harvard.edu/abs/2017MNRAS.469..656R} {469, 656}

\bibitem[\protect\citeauthoryear{{Raccanelli}, {Bonaldi}, {Negrello},
  {Matarrese}, {Tormen}  \& {de Zotti}}{{Raccanelli}
  et~al.}{2008}]{raccanelli08}
{Raccanelli} A.,  {Bonaldi} A.,  {Negrello} M.,  {Matarrese} S.,  {Tormen} G.,
   {de Zotti} G.,  2008, \mndoi [\mnras] {10.1111/j.1365-2966.2008.13189.x},
  \href {http://adsabs.harvard.edu/abs/2008MNRAS.386.2161R} {386, 2161}

\bibitem[\protect\citeauthoryear{{Raccanelli} et~al.,}{{Raccanelli}
  et~al.}{2015}]{Raccanelli:2014ISW}
{Raccanelli} A.,  et~al., 2015, \mndoi [\jcap] {10.1088/1475-7516/2015/01/042},
  \href {http://adsabs.harvard.edu/abs/2015JCAP...01..042R} {1, 042}

\bibitem[\protect\citeauthoryear{{Raccanelli}, {Kovetz}, {Dai}  \&
  {Kamionkowski}}{{Raccanelli} et~al.}{2016a}]{Raccanelli:21cmISW}
{Raccanelli} A.,  {Kovetz} E.,  {Dai} L.,   {Kamionkowski} M.,  2016a, \mndoi
  [\prd] {10.1103/PhysRevD.93.083512}, \href
  {http://adsabs.harvard.edu/abs/2016PhRvD..93h3512R} {93, 083512}

\bibitem[\protect\citeauthoryear{Raccanelli, Kovetz, Bird, Cholis  \&
  Munoz}{Raccanelli et~al.}{2016b}]{Raccanelli:2016PBH}
Raccanelli A.,  Kovetz E.~D.,  Bird S.,  Cholis I.,   Munoz J.~B.,  2016b,
  \mndoi [Phys. Rev.] {10.1103/PhysRevD.94.023516}, D94, 023516

\bibitem[\protect\citeauthoryear{Raccanelli, Montanari, Bertacca, Doré  \&
  Durrer}{Raccanelli et~al.}{2016c}]{Raccanelli:2015vla}
Raccanelli A.,  Montanari F.,  Bertacca D.,  Doré O.,   Durrer R.,  2016c,
  \mndoi [JCAP] {10.1088/1475-7516/2016/05/009}, 1605, 009

\bibitem[\protect\citeauthoryear{{Raccanelli}, {Shiraishi}, {Bartolo},
  {Bertacca}, {Liguori}, {Matarrese}, {Norris}  \& {Parkinson}}{{Raccanelli}
  et~al.}{2017}]{Raccanelli17}
{Raccanelli} A.,  {Shiraishi} M.,  {Bartolo} N.,  {Bertacca} D.,  {Liguori} M.,
   {Matarrese} S.,  {Norris} R.~P.,   {Parkinson} D.,  2017, \mndoi [Physics of
  the Dark Universe] {10.1016/j.dark.2016.10.006}, \href
  {http://adsabs.harvard.edu/abs/2017PDU....15...35R} {15, 35}

\bibitem[\protect\citeauthoryear{{Raccanelli}, {Vidotto}  \&
  {Verde}}{{Raccanelli} et~al.}{2018}]{Raccanelli:2017one}
{Raccanelli} A.,  {Vidotto} F.,   {Verde} L.,  2018, \mndoi [\jcap]
  {10.1088/1475-7516/2018/08/003}, \href
  {http://adsabs.harvard.edu/abs/2018JCAP...08..003R} {8, 003}

\bibitem[\protect\citeauthoryear{Ransom et~al.}{Ransom
  et~al.}{2014}]{Ransom:2014xla}
Ransom S.~M.,  et~al., 2014, \mndoi [Nature] {10.1038/nature12917}, 505, 520

\bibitem[\protect\citeauthoryear{{Raveri}, {Bull}, {Silvestri}  \&
  {Pogosian}}{{Raveri} et~al.}{2017}]{Raveri:2017qvt}
{Raveri} M.,  {Bull} P.,  {Silvestri} A.,   {Pogosian} L.,  2017, \mndoi
  [Physical Review D] {10.1103/PhysRevD.96.083509}, \href
  {https://ui.adsabs.harvard.edu/abs/2017PhRvD..96h3509R} {96, 083509}

\bibitem[\protect\citeauthoryear{{Rea} et~al.,}{{Rea}
  et~al.}{2013}]{2013ApJ...775L..34R}
{Rea} N.,  et~al., 2013, \mndoi [\apjl] {10.1088/2041-8205/775/2/L34}, \href
  {http://adsabs.harvard.edu/abs/2013ApJ...775L..34R} {775, L34}

\bibitem[\protect\citeauthoryear{Rees}{Rees}{1977}]{Rees:1977nat}
Rees M.,  1977, Nature, 266, 333

\bibitem[\protect\citeauthoryear{Remazeilles, Dickinson, Banday, Bigot-Sazy  \&
  Ghosh}{Remazeilles et~al.}{2015}]{Remazeilles2015}
Remazeilles M.,  Dickinson C.,  Banday A.~J.,  Bigot-Sazy M.-A.,   Ghosh T.,
  2015, \mndoi [Monthly Notices of the Royal Astronomical Society]
  {10.1093/mnras/stv1274}, 451, 4311

\bibitem[\protect\citeauthoryear{Reyes, Mandelbaum, Seljak, Baldauf, Gunn,
  Lombriser  \& Smith}{Reyes et~al.}{2010}]{Reyes:2010tr}
Reyes R.,  Mandelbaum R.,  Seljak U.,  Baldauf T.,  Gunn J.~E.,  Lombriser L.,
   Smith R.~E.,  2010, \mndoi [Nature] {10.1038/nature08857}, 464, 256

\bibitem[\protect\citeauthoryear{Ricciardi et~al.,}{Ricciardi
  et~al.}{2010}]{Ricciardi2010}
Ricciardi S.,  et~al., 2010, \mndoi [Monthly Notices of the Royal Astronomical
  Society] {10.1111/j.1365-2966.2010.16819.x}, 406, 1644

\bibitem[\protect\citeauthoryear{{Riemer-S{\o}rensen}, {Parkinson}  \&
  {Davis}}{{Riemer-S{\o}rensen} et~al.}{2014}]{2014PhRvD..89j3505R}
{Riemer-S{\o}rensen} S.,  {Parkinson} D.,   {Davis} T.~M.,  2014, \mndoi [\prd]
  {10.1103/PhysRevD.89.103505}, \href
  {http://adsabs.harvard.edu/abs/2014PhRvD..89j3505R} {89, 103505}

\bibitem[\protect\citeauthoryear{{Romeo}, {Metcalf}  \& {Pourtsidou}}{{Romeo}
  et~al.}{2018}]{2017arXiv170801235R}
{Romeo} A.,  {Metcalf} R.~B.,   {Pourtsidou} A.,  2018, \mndoi [\mnras]
  {10.1093/mnras/stx2733}, \href
  {http://adsabs.harvard.edu/abs/2018MNRAS.474.1787R} {474, 1787}

\bibitem[\protect\citeauthoryear{{Rosenband} et~al.,}{{Rosenband}
  et~al.}{2008}]{Rosenband:2008}
{Rosenband} T.,  et~al., 2008, \mndoi [Science] {10.1126/science.1154622},
  \href {http://adsabs.harvard.edu/abs/2008Sci...319.1808R} {319, 1808}

\bibitem[\protect\citeauthoryear{Rovelli \& Vidotto}{Rovelli \&
  Vidotto}{2013}]{Rovelli:2013osa}
Rovelli C.,  Vidotto F.,  2013, \mndoi [Phys.Rev.Lett.]
  {10.1103/PhysRevLett.111.091303}, 111, 091303

\bibitem[\protect\citeauthoryear{Rovelli \& Vidotto}{Rovelli \&
  Vidotto}{2014}]{Rovelli:2014cta}
Rovelli C.,  Vidotto F.,  2014, \mndoi [Int. J. Mod. Phys.]
  {10.1142/S0218271814420267}, D23, 1442026

\bibitem[\protect\citeauthoryear{{SKA Science Working Group }}{{SKA Science
  Working Group }}{2011}]{SKA:2011}
{SKA Science Working Group } 2011, The Square Kilometre Array design reference
  mission: SKA phase 1, \url
  {https://www.skatelescope.org/uploaded/18714_SKA1DesRefMission.pdf}

\bibitem[\protect\citeauthoryear{Saadeh, Feeney, Pontzen, Peiris  \&
  McEwen}{Saadeh et~al.}{2016}]{Saadeh:2016sak}
Saadeh D.,  Feeney S.~M.,  Pontzen A.,  Peiris H.~V.,   McEwen J.~D.,  2016,
  \mndoi [Phys. Rev. Lett.] {10.1103/PhysRevLett.117.131302}, 117, 131302

\bibitem[\protect\citeauthoryear{{Sachs} \& {Wolfe}}{{Sachs} \&
  {Wolfe}}{1967}]{sachs67}
{Sachs} R.~K.,  {Wolfe} A.~M.,  1967, \mndoi [\apj] {10.1086/148982}, \href
  {http://adsabs.harvard.edu/abs/1967ApJ...147...73S} {147, 73}

\bibitem[\protect\citeauthoryear{{Sahl{\'e}n}}{{Sahl{\'e}n}}{2019}]{2019PhRvD..99f3525S}
{Sahl{\'e}n} M.,  2019, \mndoi [\prd] {10.1103/PhysRevD.99.063525}, \href
  {https://ui.adsabs.harvard.edu/abs/2019PhRvD..99f3525S} {99, 063525}

\bibitem[\protect\citeauthoryear{Sahl\'{e}n \& Silk}{Sahl\'{e}n \&
  Silk}{2018}]{Sahlen:2016kzx}
Sahl\'{e}n M.,  Silk J.,  2018, \mndoi [Phys. Rev.]
  {10.1103/PhysRevD.97.103504}, D97, 103504

\bibitem[\protect\citeauthoryear{Sahl\'{e}n, Zubeldia  \& Silk}{Sahl\'{e}n
  et~al.}{2016}]{Sahlen:2015wpc}
Sahl\'{e}n M.,  Zubeldia I.,   Silk J.,  2016, \mndoi [Astrophys. J.]
  {10.3847/2041-8205/820/1/L7}, 820, L7

\bibitem[\protect\citeauthoryear{{Saintonge} et~al.,}{{Saintonge}
  et~al.}{2016}]{2016MNRAS.462.1749S}
{Saintonge} A.,  et~al., 2016, \mndoi [\mnras] {10.1093/mnras/stw1715}, \href
  {http://adsabs.harvard.edu/abs/2016MNRAS.462.1749S} {462, 1749}

\bibitem[\protect\citeauthoryear{{Saiyad Ali}, {Bharadwaj}  \&
  {Pandey}}{{Saiyad Ali} et~al.}{2006}]{2006MNRAS.366..213S}
{Saiyad Ali} S.,  {Bharadwaj} S.,   {Pandey} S.~K.,  2006, \mndoi [\mnras]
  {10.1111/j.1365-2966.2005.09847.x}, \href
  {http://adsabs.harvard.edu/abs/2006MNRAS.366..213S} {366, 213}

\bibitem[\protect\citeauthoryear{{Salvaterra} \& {Burigana}}{{Salvaterra} \&
  {Burigana}}{2002}]{2002MNRAS.336..592S}
{Salvaterra} R.,  {Burigana} C.,  2002, \mndoi [\mnras]
  {10.1046/j.1365-8711.2002.05784.x}, \href
  {http://adsabs.harvard.edu/abs/2002MNRAS.336..592S} {336, 592}

\bibitem[\protect\citeauthoryear{{Sandage}}{{Sandage}}{1962}]{1962ApJ...136..319S}
{Sandage} A.,  1962, \mndoi [\apj] {10.1086/147385}, \href
  {http://adsabs.harvard.edu/abs/1962ApJ...136..319S} {136, 319}

\bibitem[\protect\citeauthoryear{{Santos} et~al.,}{{Santos}
  et~al.}{2015}]{Santos:2015vi}
{Santos} M.,  et~al., 2015, Advancing Astrophysics with the Square Kilometre
  Array (AASKA14), \href {http://adsabs.harvard.edu/abs/2015aska.confE..19S}
  {p.~19}

\bibitem[\protect\citeauthoryear{{Sarazin}}{{Sarazin}}{1999}]{Sarazin:1999}
{Sarazin} C.~L.,  1999, \mndoi [\apj] {10.1086/307501}, \href
  {http://adsabs.harvard.edu/abs/1999ApJ...520..529S} {520, 529}

\bibitem[\protect\citeauthoryear{{Sarkar}, {Bharadwaj}  \&
  {Anathpindika}}{{Sarkar} et~al.}{2016}]{Sarkar:2016ip}
{Sarkar} D.,  {Bharadwaj} S.,   {Anathpindika} S.,  2016, \mndoi [Monthly
  Notices of the Royal Astronomical Society] {10.1093/mnras/stw1111}, \href
  {https://ui.adsabs.harvard.edu/abs/2016MNRAS.460.4310S} {460, 4310}

\bibitem[\protect\citeauthoryear{{Sasaki}, {Suyama}, {Tanaka}  \&
  {Yokoyama}}{{Sasaki} et~al.}{2016}]{sasaki}
{Sasaki} M.,  {Suyama} T.,  {Tanaka} T.,   {Yokoyama} S.,  2016, \mndoi
  [Physical Review Letters] {10.1103/PhysRevLett.117.061101}, \href
  {http://adsabs.harvard.edu/abs/2016PhRvL.117f1101S} {117, 061101}

\bibitem[\protect\citeauthoryear{Schewtschenko, Wilkinson, Baugh, Boehm  \&
  Pascoli}{Schewtschenko et~al.}{2015}]{Schewtschenko:2014fca}
Schewtschenko J.~A.,  Wilkinson R.~J.,  Baugh C.~M.,  Boehm C.,   Pascoli S.,
  2015, \mndoi [Mon. Not. Roy. Astron. Soc.] {10.1093/mnras/stv431}, 449, 3587

\bibitem[\protect\citeauthoryear{Schewtschenko, Baugh, Wilkinson, Boehm,
  Pascoli  \& Sawala}{Schewtschenko et~al.}{2016}]{Schewtschenko:2015rno}
Schewtschenko J.~A.,  Baugh C.~M.,  Wilkinson R.~J.,  Boehm C.,  Pascoli S.,
  Sawala T.,  2016, \mndoi [Mon. Not. Roy. Astron. Soc.]
  {10.1093/mnras/stw1078}, 461, 2282

\bibitem[\protect\citeauthoryear{Schmidt \& Jeong}{Schmidt \&
  Jeong}{2012}]{CosmicRulers2}
Schmidt F.,  Jeong D.,  2012, \mndoi [Phys. Rev.] {10.1103/PhysRevD.86.083513},
  D86, 083513

\bibitem[\protect\citeauthoryear{{Schneider}, {Salvaterra}, {Choudhury},
  {Ferrara}, {Burigana}  \& {Popa}}{{Schneider}
  et~al.}{2008}]{2008MNRAS.384.1525S}
{Schneider} R.,  {Salvaterra} R.,  {Choudhury} T.~R.,  {Ferrara} A.,
  {Burigana} C.,   {Popa} L.~A.,  2008, \mndoi [\mnras]
  {10.1111/j.1365-2966.2007.12801.x}, \href
  {http://adsabs.harvard.edu/abs/2008MNRAS.384.1525S} {384, 1525}

\bibitem[\protect\citeauthoryear{{Sch{\"o}n}, {Mack}, {Avram}, {Wyithe}  \&
  {Barberio}}{{Sch{\"o}n} et~al.}{2015}]{2015MNRAS.451.2840S}
{Sch{\"o}n} S.,  {Mack} K.~J.,  {Avram} C.~A.,  {Wyithe} J.~S.~B.,   {Barberio}
  E.,  2015, \mndoi [\mnras] {10.1093/mnras/stv1056}, \href
  {http://adsabs.harvard.edu/abs/2015MNRAS.451.2840S} {451, 2840}

\bibitem[\protect\citeauthoryear{{Schutz} \& {Liu}}{{Schutz} \&
  {Liu}}{2017}]{schutz}
{Schutz} K.,  {Liu} A.,  2017, \mndoi [\prd] {10.1103/PhysRevD.95.023002},
  \href {http://adsabs.harvard.edu/abs/2017PhRvD..95b3002S} {95, 023002}

\bibitem[\protect\citeauthoryear{Schwarz et~al.,}{Schwarz
  et~al.}{2015}]{Schwarz:2015pqa}
Schwarz D.~J.,  et~al., 2015, PoS, AASKA14, 032

\bibitem[\protect\citeauthoryear{{Seiffert} et~al.,}{{Seiffert}
  et~al.}{2011}]{2011ApJ...734....6S}
{Seiffert} M.,  et~al., 2011, \mndoi [\apj] {10.1088/0004-637X/734/1/6}, \href
  {http://adsabs.harvard.edu/abs/2011ApJ...734....6S} {734, 6}

\bibitem[\protect\citeauthoryear{{Sekiguchi}, {Takahashi}, {Tashiro}  \&
  {Yokoyama}}{{Sekiguchi} et~al.}{2018}]{Sekiguchi:2017}
{Sekiguchi} T.,  {Takahashi} T.,  {Tashiro} H.,   {Yokoyama} S.,  2018, \mndoi
  [\jcap] {10.1088/1475-7516/2018/02/053}, \href
  {http://adsabs.harvard.edu/abs/2018JCAP...02..053S} {2, 053}

\bibitem[\protect\citeauthoryear{Seljak}{Seljak}{2009}]{Seljak:2008xr}
Seljak U.,  2009, \mndoi [Phys. Rev. Lett.] {10.1103/PhysRevLett.102.021302},
  102, 021302

\bibitem[\protect\citeauthoryear{Shannon \& Cordes}{Shannon \&
  Cordes}{2010}]{Shannon:2010bv}
Shannon R.~M.,  Cordes J.~M.,  2010, \mndoi [Astrophys. J.]
  {10.1088/0004-637X/725/2/1607}, 725, 1607

\bibitem[\protect\citeauthoryear{Shannon \& Johnston}{Shannon \&
  Johnston}{2013}]{Shannon:2013hla}
Shannon R.~M.,  Johnston S.,  2013, \mndoi [Mon. Not. Roy. Astron. Soc.]
  {10.1093/mnrasl/slt088}, 435, 29

\bibitem[\protect\citeauthoryear{Shao}{Shao}{2014}]{Shao:2014oha}
Shao L.,  2014, \mndoi [Phys. Rev. Lett.] {10.1103/PhysRevLett.112.111103},
  112, 111103

\bibitem[\protect\citeauthoryear{Shao}{Shao}{2016}]{Shao:2016ubu}
Shao L.,  2016, \mndoi [Phys. Rev.] {10.1103/PhysRevD.93.084023}, D93, 084023

\bibitem[\protect\citeauthoryear{Shao \& Wex}{Shao \& Wex}{2012}]{Shao:2012eg}
Shao L.,  Wex N.,  2012, \mndoi [Class. Quant. Grav.]
  {10.1088/0264-9381/29/21/215018}, 29, 215018

\bibitem[\protect\citeauthoryear{Shao \& Wex}{Shao \& Wex}{2016}]{Shao:2016ezh}
Shao L.,  Wex N.,  2016, \mndoi [Sci. China Phys. Mech. Astron.]
  {10.1007/s11433-016-0087-6}, 59, 699501

\bibitem[\protect\citeauthoryear{Shao \& Zhang}{Shao \&
  Zhang}{2017}]{Shao:2017tuu}
Shao L.,  Zhang B.,  2017, \mndoi [Phys. Rev.] {10.1103/PhysRevD.95.123010},
  D95, 123010

\bibitem[\protect\citeauthoryear{Shao, Caballero, Kramer, Wex, Champion  \&
  Jessner}{Shao et~al.}{2013}]{Shao:2013wga}
Shao L.,  Caballero R.~N.,  Kramer M.,  Wex N.,  Champion D.~J.,   Jessner A.,
  2013, \mndoi [Class. Quant. Grav.] {10.1088/0264-9381/30/16/165019}, 30,
  165019

\bibitem[\protect\citeauthoryear{Shao et~al.}{Shao et~al.}{2015}]{Shao:2014wja}
Shao L.,  et~al., 2015, PoS, AASKA14, 042

\bibitem[\protect\citeauthoryear{{Shao}, {Sennett}, {Buonanno}, {Kramer}  \&
  {Wex}}{{Shao} et~al.}{2017}]{Shao:2017gwu}
{Shao} L.,  {Sennett} N.,  {Buonanno} A.,  {Kramer} M.,   {Wex} N.,  2017,
  \mndoi [Physical Review X] {10.1103/PhysRevX.7.041025}, \href
  {http://adsabs.harvard.edu/abs/2017PhRvX...7d1025S} {7, 041025}

\bibitem[\protect\citeauthoryear{{Shapiro}, {Mao}, {Iliev}, {Mellema}, {Datta},
  {Ahn}  \& {Koda}}{{Shapiro} et~al.}{2013}]{2013PhRvL.110o1301S}
{Shapiro} P.~R.,  {Mao} Y.,  {Iliev} I.~T.,  {Mellema} G.,  {Datta} K.~K.,
  {Ahn} K.,   {Koda} J.,  2013, \mndoi [Physical Review Letters]
  {10.1103/PhysRevLett.110.151301}, \href
  {http://adsabs.harvard.edu/abs/2013PhRvL.110o1301S} {110, 151301}

\bibitem[\protect\citeauthoryear{Shaver, Windhorst, Madau  \& {de
  Bruyn}}{Shaver et~al.}{1999}]{Shaver1999}
Shaver P.~A.,  Windhorst R.~A.,  Madau P.,   {de Bruyn} A.~G.,  1999, Astronomy
  and Astrophysics, 345, 380

\bibitem[\protect\citeauthoryear{{Shifman}, {Vainshtein}  \&
  {Zakharov}}{{Shifman} et~al.}{1980}]{1980NuPhB.166..493S}
{Shifman} M.~A.,  {Vainshtein} A.~I.,   {Zakharov} V.~I.,  1980, \mndoi
  [Nuclear Physics B] {10.1016/0550-3213(80)90209-6}, \href
  {http://adsabs.harvard.edu/abs/1980NuPhB.166..493S} {166, 493}

\bibitem[\protect\citeauthoryear{{Shiraishi}, {Tashiro}  \&
  {Ichiki}}{{Shiraishi} et~al.}{2014}]{shiraishi2014}
{Shiraishi} M.,  {Tashiro} H.,   {Ichiki} K.,  2014, \mndoi [\prd]
  {10.1103/PhysRevD.89.103522}, \href
  {http://adsabs.harvard.edu/abs/2014PhRvD..89j3522S} {89, 103522}

\bibitem[\protect\citeauthoryear{{Shiraishi}, {Mu{\~n}oz}, {Kamionkowski}  \&
  {Raccanelli}}{{Shiraishi} et~al.}{2016}]{Shiraishi:2016}
{Shiraishi} M.,  {Mu{\~n}oz} J.~B.,  {Kamionkowski} M.,   {Raccanelli} A.,
  2016, \mndoi [\prd] {10.1103/PhysRevD.93.103506}, \href
  {http://adsabs.harvard.edu/abs/2016PhRvD..93j3506S} {93, 103506}

\bibitem[\protect\citeauthoryear{Sims, Lentati, Alexander  \& Carilli}{Sims
  et~al.}{2016}]{Sims2016a}
Sims P.~H.,  Lentati L.,  Alexander P.,   Carilli C.~L.,  2016, \mndoi [Monthly
  Notices of the Royal Astronomical Society] {10.1093/mnras/stw1768}, 462, 3069

\bibitem[\protect\citeauthoryear{{Sims}, {Lentati}, {Pober}, {Carilli},
  {Hobson}, {Alexander}  \& {Sutter}}{{Sims} et~al.}{2017}]{Lentati2017}
{Sims} P.~H.,  {Lentati} L.,  {Pober} J.~C.,  {Carilli} C.,  {Hobson} M.~P.,
  {Alexander} P.,   {Sutter} P.,  2017, arXiv e-prints, \href
  {https://ui.adsabs.harvard.edu/abs/2017arXiv170103384S} {p. arXiv:1701.03384}

\bibitem[\protect\citeauthoryear{{Singal} et~al.,}{{Singal}
  et~al.}{2011}]{2011ApJ...730..138S}
{Singal} J.,  et~al., 2011, \mndoi [\apj] {10.1088/0004-637X/730/2/138}, \href
  {http://adsabs.harvard.edu/abs/2011ApJ...730..138S} {730, 138}

\bibitem[\protect\citeauthoryear{{Sitwell}, {Mesinger}, {Ma}  \&
  {Sigurdson}}{{Sitwell} et~al.}{2014}]{Sitwell14}
{Sitwell} M.,  {Mesinger} A.,  {Ma} Y.-Z.,   {Sigurdson} K.,  2014, \mndoi
  [\mnras] {10.1093/mnras/stt2392}, \href
  {http://adsabs.harvard.edu/abs/2014MNRAS.438.2664S} {438, 2664}

\bibitem[\protect\citeauthoryear{{Smits}, {Kramer}, {Stappers}, {Lorimer},
  {Cordes}  \& {Faulkner}}{{Smits} et~al.}{2009}]{2009A&A...493.1161S}
{Smits} R.,  {Kramer} M.,  {Stappers} B.,  {Lorimer} D.~R.,  {Cordes} J.,
  {Faulkner} A.,  2009, \mndoi [\aap] {10.1051/0004-6361:200810383}, \href
  {http://cdsads.u-strasbg.fr/abs/2009A%26A...493.1161S} {493, 1161}

\bibitem[\protect\citeauthoryear{Smits et~al.}{Smits
  et~al.}{2011}]{Smits:2011technote}
Smits R.,  et~al., 2011, SKA document No. WP2-040.030.010-TD-003

\bibitem[\protect\citeauthoryear{{Sneden}, {McWilliam}, {Preston}, {Cowan},
  {Burris}  \& {Armosky}}{{Sneden} et~al.}{1996}]{sneden1996}
{Sneden} C.,  {McWilliam} A.,  {Preston} G.~W.,  {Cowan} J.~J.,  {Burris}
  D.~L.,   {Armosky} B.~J.,  1996, \mndoi [\apj] {10.1086/177656}, \href
  {http://adsabs.harvard.edu/abs/1996ApJ...467..819S} {467, 819}

\bibitem[\protect\citeauthoryear{Spolyar, Sahl\'{e}n  \& Silk}{Spolyar
  et~al.}{2013}]{Spolyar:2013maa}
Spolyar D.,  Sahl\'{e}n M.,   Silk J.,  2013, \mndoi [Phys. Rev. Lett.]
  {10.1103/PhysRevLett.111.241103}, 111, 241103

\bibitem[\protect\citeauthoryear{{Springel} et~al.,}{{Springel}
  et~al.}{2005}]{astro-ph/0504097}
{Springel} V.,  et~al., 2005, \mndoi [\nat] {10.1038/nature03597}, \href
  {http://adsabs.harvard.edu/abs/2005Natur.435..629S} {435, 629}

\bibitem[\protect\citeauthoryear{Springob, Masters, Haynes, Giovanelli  \&
  Marinoni}{Springob et~al.}{2007}]{Springob:2007vb}
Springob C.~M.,  Masters K.~L.,  Haynes M.~P.,  Giovanelli R.,   Marinoni C.,
  2007, \mndoi [Astrophys. J. Suppl.] {10.1088/0067-0049/182/1/474}, 172, 599

\bibitem[\protect\citeauthoryear{{Srianand}, {Chand}, {Petitjean}  \&
  {Aracil}}{{Srianand} et~al.}{2004}]{srianand2004}
{Srianand} R.,  {Chand} H.,  {Petitjean} P.,   {Aracil} B.,  2004, \mndoi
  [Physical Review Letters] {10.1103/PhysRevLett.92.121302}, \href
  {http://adsabs.harvard.edu/abs/2004PhRvL..92l1302S} {92, 121302}

\bibitem[\protect\citeauthoryear{Stairs}{Stairs}{2003}]{Stairs:2003eg}
Stairs I.~H.,  2003, \mndoi [Living Rev. Rel.] {10.12942/lrr-2003-5}, 6, 5

\bibitem[\protect\citeauthoryear{{Strong} \& {Moskalenko}}{{Strong} \&
  {Moskalenko}}{1998}]{1998ApJ...509..212S}
{Strong} A.~W.,  {Moskalenko} I.~V.,  1998, \mndoi [\apj] {10.1086/306470},
  \href {http://adsabs.harvard.edu/abs/1998ApJ...509..212S} {509, 212}

\bibitem[\protect\citeauthoryear{{Sun} \& {Reich}}{{Sun} \&
  {Reich}}{2009}]{2009A&A...507.1087S}
{Sun} X.~H.,  {Reich} W.,  2009, \mndoi [\aap] {10.1051/0004-6361/200912539},
  \href {http://adsabs.harvard.edu/abs/2009A%26A...507.1087S} {507, 1087}

\bibitem[\protect\citeauthoryear{{Sun} \& {Reich}}{{Sun} \&
  {Reich}}{2010}]{2010RAA....10.1287S}
{Sun} X.-H.,  {Reich} W.,  2010, \mndoi [Research in Astronomy and
  Astrophysics] {10.1088/1674-4527/10/12/009}, \href
  {http://adsabs.harvard.edu/abs/2010RAA....10.1287S} {10, 1287}

\bibitem[\protect\citeauthoryear{{Sun}, {Reich}, {Waelkens}  \&
  {En{\ss}lin}}{{Sun} et~al.}{2008}]{2008A&A...477..573S}
{Sun} X.~H.,  {Reich} W.,  {Waelkens} A.,   {En{\ss}lin} T.~A.,  2008, \mndoi
  [\aap] {10.1051/0004-6361:20078671}, \href
  {http://adsabs.harvard.edu/abs/2008A%26A...477..573S} {477, 573}

\bibitem[\protect\citeauthoryear{Sundrum}{Sundrum}{1999}]{PhysRevD.59.085009}
Sundrum R.,  1999, \mndoi [Phys. Rev. D] {10.1103/PhysRevD.59.085009}, 59,
  085009

\bibitem[\protect\citeauthoryear{{Sunyaev} \& {Khatri}}{{Sunyaev} \&
  {Khatri}}{2013}]{sunyaevkhatri2013}
{Sunyaev} R.~A.,  {Khatri} R.,  2013, \mndoi [International Journal of Modern
  Physics D] {10.1142/S0218271813300140}, \href
  {http://adsabs.harvard.edu/abs/2013IJMPD..2230014S} {22, 1330014}

\bibitem[\protect\citeauthoryear{{Sunyaev} \& {Zeldovich}}{{Sunyaev} \&
  {Zeldovich}}{1970}]{1970Ap&SS...7...20S}
{Sunyaev} R.~A.,  {Zeldovich} Y.~B.,  1970, \mndoi [\apss]
  {10.1007/BF00653472}, \href
  {http://adsabs.harvard.edu/abs/1970Ap%26SS...7...20S} {7, 20}

\bibitem[\protect\citeauthoryear{{Sunyaev} \& {Zeldovich}}{{Sunyaev} \&
  {Zeldovich}}{1975}]{1975MNRAS.171..375S}
{Sunyaev} R.~A.,  {Zeldovich} I.~B.,  1975, \mndoi [\mnras]
  {10.1093/mnras/171.2.375}, \href
  {http://adsabs.harvard.edu/abs/1975MNRAS.171..375S} {171, 375}

\bibitem[\protect\citeauthoryear{Sutter, Pisani, Wandelt  \& Weinberg}{Sutter
  et~al.}{2014}]{Sutter:2014oca}
Sutter P.~M.,  Pisani A.,  Wandelt B.~D.,   Weinberg D.~H.,  2014, \mndoi [Mon.
  Not. Roy. Astron. Soc.] {10.1093/mnras/stu1392}, 443, 2983

\bibitem[\protect\citeauthoryear{{Suzuki} et~al.,}{{Suzuki}
  et~al.}{2012}]{suzuki2012}
{Suzuki} N.,  et~al., 2012, \mndoi [\apj] {10.1088/0004-637X/746/1/85}, \href
  {http://adsabs.harvard.edu/abs/2012ApJ...746...85S} {746, 85}

\bibitem[\protect\citeauthoryear{Switzer et~al.,}{Switzer
  et~al.}{2013}]{2013MNRAS.434L..46S}
Switzer E.~R.,  et~al., 2013, \mnras, 434, L46

\bibitem[\protect\citeauthoryear{{Takada}, {Komatsu}  \& {Futamase}}{{Takada}
  et~al.}{2006}]{Takada:2006}
{Takada} M.,  {Komatsu} E.,   {Futamase} T.,  2006, \mndoi [\prd]
  {10.1103/PhysRevD.73.083520}, \href
  {http://adsabs.harvard.edu/abs/2006PhRvD..73h3520T} {73, 083520}

\bibitem[\protect\citeauthoryear{Taylor}{Taylor}{1992}]{Taylor:1992kea}
Taylor J.~H.,  1992, \mndoi [Phil. Trans. A. Math. Phys. Eng. Sci.]
  {10.1098/rsta.1992.0088}, 341, 117

\bibitem[\protect\citeauthoryear{Tegmark, Hamilton, Strauss, Vogeley  \&
  Szalay}{Tegmark et~al.}{1998}]{Tegmark:1997}
Tegmark M.,  Hamilton A. J.~S.,  Strauss M.~A.,  Vogeley M.~S.,   Szalay A.~S.,
   1998, \mndoi [Astrophys. J.] {10.1086/305663}, 499, 555

\bibitem[\protect\citeauthoryear{Thornton et~al.}{Thornton
  et~al.}{2013}]{Thornton:2013iua}
Thornton D.,  et~al., 2013, \mndoi [Science] {10.1126/science.1236789}, 341, 53

\bibitem[\protect\citeauthoryear{{Trombetti} \& {Burigana}}{{Trombetti} \&
  {Burigana}}{2014}]{2014MNRAS.437.2507T}
{Trombetti} T.,  {Burigana} C.,  2014, \mndoi [\mnras] {10.1093/mnras/stt2063},
  \href {http://adsabs.harvard.edu/abs/2014MNRAS.437.2507T} {437, 2507}

\bibitem[\protect\citeauthoryear{{Tr\"oster} et~al.}{{Tr\"oster}
  et~al.}{2017}]{Troster:2016sgf}
{Tr\"oster} T.,  et~al., 2017, \mndoi [Mon. Not. Roy. Astron. Soc.]
  {10.1093/mnras/stx365}, 467, 2706

\bibitem[\protect\citeauthoryear{Trott et~al.,}{Trott
  et~al.}{2016}]{Trott2016a}
Trott C.~M.,  et~al., 2016, \mndoi [The Astrophysical Journal]
  {10.3847/0004-637X/818/2/139}, 818, 1

\bibitem[\protect\citeauthoryear{{Tseliakhovich} \& {Hirata}}{{Tseliakhovich}
  \& {Hirata}}{2010}]{2010PhRvD..82h3520T}
{Tseliakhovich} D.,  {Hirata} C.,  2010, \mndoi [\prd]
  {10.1103/PhysRevD.82.083520}, \href
  {http://adsabs.harvard.edu/abs/2010PhRvD..82h3520T} {82, 083520}

\bibitem[\protect\citeauthoryear{{Tucci} \& {Toffolatti}}{{Tucci} \&
  {Toffolatti}}{2012}]{TucciToffolatti2012}
{Tucci} M.,  {Toffolatti} L.,  2012, \mndoi [Advances in Astronomy]
  {10.1155/2012/624987}, \href
  {http://adsabs.harvard.edu/abs/2012AdAst2012E..52T} {2012, 624987}

\bibitem[\protect\citeauthoryear{Tully \& Fisher}{Tully \&
  Fisher}{1977}]{Tully:1977fu}
Tully R.~B.,  Fisher J.~R.,  1977, Astron. Astrophys., 54, 661

\bibitem[\protect\citeauthoryear{Uzan}{Uzan}{2011}]{Uzan:2010pm}
Uzan J.-P.,  2011, \mndoi [Living Rev. Rel.] {10.12942/lrr-2011-2}, 14, 2

\bibitem[\protect\citeauthoryear{{Vagnozzi}, {Giusarma}, {Mena}, {Freese},
  {Gerbino}, {Ho}  \& {Lattanzi}}{{Vagnozzi} et~al.}{2017}]{Vagnozzi_2017}
{Vagnozzi} S.,  {Giusarma} E.,  {Mena} O.,  {Freese} K.,  {Gerbino} M.,  {Ho}
  S.,   {Lattanzi} M.,  2017, \mndoi [\prd] {10.1103/PhysRevD.96.123503}, \href
  {http://adsabs.harvard.edu/abs/2017PhRvD..96l3503V} {96, 123503}

\bibitem[\protect\citeauthoryear{{Vald{\'e}s}, {Evoli}, {Mesinger}, {Ferrara}
  \& {Yoshida}}{{Vald{\'e}s} et~al.}{2013}]{2013MNRAS.429.1705V}
{Vald{\'e}s} M.,  {Evoli} C.,  {Mesinger} A.,  {Ferrara} A.,   {Yoshida} N.,
  2013, \mndoi [\mnras] {10.1093/mnras/sts458}, \href
  {http://adsabs.harvard.edu/abs/2013MNRAS.429.1705V} {429, 1705}

\bibitem[\protect\citeauthoryear{{Vazza}, {Br{\"u}ggen}, {Gheller}  \&
  {Wang}}{{Vazza} et~al.}{2014}]{va14mhd}
{Vazza} F.,  {Br{\"u}ggen} M.,  {Gheller} C.,   {Wang} P.,  2014, \mndoi
  [\mnras] {10.1093/mnras/stu1896}, \href
  {http://adsabs.harvard.edu/abs/2014MNRAS.445.3706V} {445, 3706}

\bibitem[\protect\citeauthoryear{{Vazza}, {Ferrari}, {Br{\"u}ggen}, {Bonafede},
  {Gheller}  \& {Wang}}{{Vazza} et~al.}{2015}]{va15radio}
{Vazza} F.,  {Ferrari} C.,  {Br{\"u}ggen} M.,  {Bonafede} A.,  {Gheller} C.,
  {Wang} P.,  2015, \mndoi [\aap] {10.1051/0004-6361/201526228}, \href
  {http://adsabs.harvard.edu/abs/2015A%26A...580A.119V} {580, A119}

\bibitem[\protect\citeauthoryear{Venumadhav, Oklop\ifmmode \check{c}\else
  \v{c}\fi{}i\ifmmode~\acute{c}\else \'{c}\fi{}, Gluscevic, Mishra  \&
  Hirata}{Venumadhav et~al.}{2017}]{PhysRevD.95.083010}
Venumadhav T.,  Oklop\ifmmode \check{c}\else \v{c}\fi{}i\ifmmode~\acute{c}\else
  \'{c}\fi{} A.,  Gluscevic V.,  Mishra A.,   Hirata C.~M.,  2017, \mndoi
  [\prd] {10.1103/PhysRevD.95.083010}, 95, 083010

\bibitem[\protect\citeauthoryear{Verbiest et~al.}{Verbiest
  et~al.}{2016}]{Verbiest:2016vem}
Verbiest J. P.~W.,  et~al., 2016, \mndoi [Mon. Not. Roy. Astron. Soc.]
  {10.1093/mnras/stw347}, 458, 1267

\bibitem[\protect\citeauthoryear{{Vernstrom}, {Norris}, {Scott}  \&
  {Wall}}{{Vernstrom} et~al.}{2015}]{2015MNRAS.447.2243V}
{Vernstrom} T.,  {Norris} R.~P.,  {Scott} D.,   {Wall} J.~V.,  2015, \mndoi
  [\mnras] {10.1093/mnras/stu2595}, \href
  {http://adsabs.harvard.edu/abs/2015MNRAS.447.2243V} {447, 2243}

\bibitem[\protect\citeauthoryear{{Viel}}{{Viel}}{2005}]{Viel05}
{Viel} M.,  2005, in {Williams} P.,  {Shu} C.-G.,   {Menard} B.,  eds, IAU
  Colloq. 199: Probing Galaxies through Quasar Absorption Lines. pp 255--260,
  \mndoi{10.1017/S1743921305002681}

\bibitem[\protect\citeauthoryear{{Viel}, {Lesgourgues}, {Haehnelt}, {Matarrese}
   \& {Riotto}}{{Viel} et~al.}{2005}]{2005PhRvD..71f3534V}
{Viel} M.,  {Lesgourgues} J.,  {Haehnelt} M.~G.,  {Matarrese} S.,   {Riotto}
  A.,  2005, \mndoi [\prd] {10.1103/PhysRevD.71.063534}, \href
  {http://adsabs.harvard.edu/abs/2005PhRvD..71f3534V} {71, 063534}

\bibitem[\protect\citeauthoryear{{Viel}, {Becker}, {Bolton}, {Haehnelt},
  {Rauch}  \& {Sargent}}{{Viel} et~al.}{2008}]{Viel08}
{Viel} M.,  {Becker} G.~D.,  {Bolton} J.~S.,  {Haehnelt} M.~G.,  {Rauch} M.,
  {Sargent} W.~L.~W.,  2008, \mndoi [Physical Review Letters]
  {10.1103/PhysRevLett.100.041304}, \href
  {http://adsabs.harvard.edu/abs/2008PhRvL.100d1304V} {100, 041304}

\bibitem[\protect\citeauthoryear{{Villaescusa-Navarro}, {Bird},
  {Pe{\~n}a-Garay}  \& {Viel}}{{Villaescusa-Navarro}
  et~al.}{2013}]{Villaescusa-Navarro_2013}
{Villaescusa-Navarro} F.,  {Bird} S.,  {Pe{\~n}a-Garay} C.,   {Viel} M.,  2013,
  \mndoi [\jcap] {10.1088/1475-7516/2013/03/019}, \href
  {http://adsabs.harvard.edu/abs/2013JCAP...03..019V} {3, 19}

\bibitem[\protect\citeauthoryear{{Villaescusa-Navarro}, {Marulli}, {Viel},
  {Branchini}, {Castorina}, {Sefusatti}  \& {Saito}}{{Villaescusa-Navarro}
  et~al.}{2014}]{Villaescusa-Navarro_2014}
{Villaescusa-Navarro} F.,  {Marulli} F.,  {Viel} M.,  {Branchini} E.,
  {Castorina} E.,  {Sefusatti} E.,   {Saito} S.,  2014, \mndoi [\jcap]
  {10.1088/1475-7516/2014/03/011}, \href
  {http://adsabs.harvard.edu/abs/2014JCAP...03..011V} {3, 11}

\bibitem[\protect\citeauthoryear{{Villaescusa-Navarro}, {Bull}  \&
  {Viel}}{{Villaescusa-Navarro} et~al.}{2015}]{Villaescusa-Navarro_2015}
{Villaescusa-Navarro} F.,  {Bull} P.,   {Viel} M.,  2015, \mndoi [\apj]
  {10.1088/0004-637X/814/2/146}, \href
  {http://adsabs.harvard.edu/abs/2015ApJ...814..146V} {814, 146}

\bibitem[\protect\citeauthoryear{{Visbal}, {Barkana}, {Fialkov},
  {Tseliakhovich}  \& {Hirata}}{{Visbal} et~al.}{2012}]{2012Natur.487...70V}
{Visbal} E.,  {Barkana} R.,  {Fialkov} A.,  {Tseliakhovich} D.,   {Hirata}
  C.~M.,  2012, \mndoi [\nat] {10.1038/nature11177}, \href
  {http://adsabs.harvard.edu/abs/2012Natur.487...70V} {487, 70}

\bibitem[\protect\citeauthoryear{{Vitale} \& {Morselli}}{{Vitale} \&
  {Morselli}}{2009}]{Vitale:2009hr}
{Vitale} V.,  {Morselli} A.,  2009, arXiv e-prints, \href
  {https://ui.adsabs.harvard.edu/abs/2009arXiv0912.3828V} {p. arXiv:0912.3828}

\bibitem[\protect\citeauthoryear{Vogelsberger, Zavala, Cyr-Racine, Pfrommer,
  Bringmann  \& Sigurdson}{Vogelsberger et~al.}{2016}]{Vogelsberger:2015gpr}
Vogelsberger M.,  Zavala J.,  Cyr-Racine F.-Y.,  Pfrommer C.,  Bringmann T.,
  Sigurdson K.,  2016, \mndoi [Mon. Not. Roy. Astron. Soc.]
  {10.1093/mnras/stw1076}, 460, 1399

\bibitem[\protect\citeauthoryear{Voivodic, Lima, Llinares  \& Mota}{Voivodic
  et~al.}{2017}]{Voivodic:2016kog}
Voivodic R.,  Lima M.,  Llinares C.,   Mota D.~F.,  2017, \mndoi [Phys. Rev.]
  {10.1103/PhysRevD.95.024018}, D95, 024018

\bibitem[\protect\citeauthoryear{{Voytek}, {Natarajan}, {J{\'a}uregui
  Garc{\'{\i}}a}, {Peterson}  \& {L{\'o}pez-Cruz}}{{Voytek}
  et~al.}{2014}]{voytek2014}
{Voytek} T.~C.,  {Natarajan} A.,  {J{\'a}uregui Garc{\'{\i}}a} J.~M.,
  {Peterson} J.~B.,   {L{\'o}pez-Cruz} O.,  2014, \mndoi [\apjl]
  {10.1088/2041-8205/782/1/L9}, \href
  {http://adsabs.harvard.edu/abs/2014ApJ...782L...9V} {782, L9}

\bibitem[\protect\citeauthoryear{{Waelkens}, {Jaffe}, {Reinecke}, {Kitaura}  \&
  {En{\ss}lin}}{{Waelkens} et~al.}{2009}]{2009A&A...495..697W}
{Waelkens} A.,  {Jaffe} T.,  {Reinecke} M.,  {Kitaura} F.~S.,   {En{\ss}lin}
  T.~A.,  2009, \mndoi [\aap] {10.1051/0004-6361:200810564}, \href
  {http://adsabs.harvard.edu/abs/2009A%26A...495..697W} {495, 697}

\bibitem[\protect\citeauthoryear{{Wands}}{{Wands}}{2010}]{1004.0818}
{Wands} D.,  2010, \mndoi [Classical and Quantum Gravity]
  {10.1088/0264-9381/27/12/124002}, \href
  {http://adsabs.harvard.edu/abs/2010CQGra..27l4002W} {27, 124002}

\bibitem[\protect\citeauthoryear{Wang}{Wang}{2005}]{Wang:2005ti}
Wang W.,  2005, Chin.J.Astron.Astrophys.

\bibitem[\protect\citeauthoryear{Wang}{Wang}{2015}]{Wang:2015bsa}
Wang Y.,  2015, \mndoi [J. Phys. Conf. Ser.] {10.1088/1742-6596/610/1/012019},
  610, 012019

\bibitem[\protect\citeauthoryear{Wang \& Mohanty}{Wang \&
  Mohanty}{2017}]{PhysRevLett.118.151104}
Wang Y.,  Mohanty S.~D.,  2017, \mndoi [Phys. Rev. Lett.]
  {10.1103/PhysRevLett.118.151104}, 118, 151104

\bibitem[\protect\citeauthoryear{Wang, Mohanty  \& Jenet}{Wang
  et~al.}{2014}]{Wang:2014ava}
Wang Y.,  Mohanty S.~D.,   Jenet F.~A.,  2014, \mndoi [Astrophys. J.]
  {10.1088/0004-637X/795/1/96}, 795, 96

\bibitem[\protect\citeauthoryear{Wang, Mohanty  \& Jenet}{Wang
  et~al.}{2015}]{Wang:2015nqa}
Wang Y.,  Mohanty S.~D.,   Jenet F.~A.,  2015, \mndoi [Astrophys. J.]
  {10.1088/0004-637X/815/2/125}, 815, 125

\bibitem[\protect\citeauthoryear{Wang, Mohanty  \& Qian}{Wang
  et~al.}{2017}]{Wang:2017jxw}
Wang Y.,  Mohanty S.~D.,   Qian Y.-Q.,  2017, \mndoi [J. Phys. Conf. Ser.]
  {10.1088/1742-6596/840/1/012058}, 840, 012058

\bibitem[\protect\citeauthoryear{Webb, Murphy, Flambaum, Dzuba, Barrow,
  Churchill, Prochaska  \& Wolfe}{Webb et~al.}{2001}]{Webb:2000mn}
Webb J.~K.,  Murphy M.~T.,  Flambaum V.~V.,  Dzuba V.~A.,  Barrow J.~D.,
  Churchill C.~W.,  Prochaska J.~X.,   Wolfe A.~M.,  2001, \mndoi [Phys. Rev.
  Lett.] {10.1103/PhysRevLett.87.091301}, 87, 091301

\bibitem[\protect\citeauthoryear{Webb, King, Murphy, Flambaum, Carswell  \&
  Bainbridge}{Webb et~al.}{2011}]{Webb:2010hc}
Webb J.~K.,  King J.~A.,  Murphy M.~T.,  Flambaum V.~V.,  Carswell R.~F.,
  Bainbridge M.~B.,  2011, \mndoi [Phys. Rev. Lett.]
  {10.1103/PhysRevLett.107.191101}, 107, 191101

\bibitem[\protect\citeauthoryear{Wex}{Wex}{2014}]{Wex:2014nva}
Wex N.,  2014, preprint (\mn@eprint {arXiv} {1402.5594})

\bibitem[\protect\citeauthoryear{Wex \& Kopeikin}{Wex \&
  Kopeikin}{1999}]{Wex:1998wt}
Wex N.,  Kopeikin S.,  1999, \mndoi [Astrophys. J.] {10.1086/306933}, 514, 388

\bibitem[\protect\citeauthoryear{{White} \& {Hu}}{{White} \&
  {Hu}}{1997}]{astro-ph/9609105}
{White} M.,  {Hu} W.,  1997, \aap, \href
  {http://adsabs.harvard.edu/abs/1997A%26A...321....8W} {321, 8}

\bibitem[\protect\citeauthoryear{Will}{Will}{2014}]{Will:2014kxa}
Will C.~M.,  2014, \mndoi [Living Rev. Rel.] {10.12942/lrr-2014-4}, 17, 4

\bibitem[\protect\citeauthoryear{Williams, Turyshev  \& Boggs}{Williams
  et~al.}{2004}]{Williams:2004qba}
Williams J.~G.,  Turyshev S.~G.,   Boggs D.~H.,  2004, \mndoi [Phys. Rev.
  Lett.] {10.1103/PhysRevLett.93.261101}, 93, 261101

\bibitem[\protect\citeauthoryear{{Wilson}, {Rohlfs}  \&
  {H{\"u}ttemeister}}{{Wilson} et~al.}{2013}]{2013tra..book.....W}
{Wilson} T.~L.,  {Rohlfs} K.,   {H{\"u}ttemeister} S.,  2013, {Tools of Radio
  Astronomy}.
Springer-Verlag

\bibitem[\protect\citeauthoryear{{Wittor}, {Vazza}  \& {Br{\"u}ggen}}{{Wittor}
  et~al.}{2017}]{wi17}
{Wittor} D.,  {Vazza} F.,   {Br{\"u}ggen} M.,  2017, \mndoi [\mnras]
  {10.1093/mnras/stw2631}, \href
  {http://adsabs.harvard.edu/abs/2017MNRAS.464.4448W} {464, 4448}

\bibitem[\protect\citeauthoryear{Wolz et~al.}{Wolz
  et~al.}{2017a}]{Wolz:2015lwa}
Wolz L.,  et~al., 2017a, \mndoi [Mon. Not. Roy. Astron. Soc.]
  {10.1093/mnras/stw2556}, 464, 4938

\bibitem[\protect\citeauthoryear{{Wolz}, {Blake}  \& {Wyithe}}{{Wolz}
  et~al.}{2017b}]{2017arXiv170308268W}
{Wolz} L.,  {Blake} C.,   {Wyithe} J.~S.~B.,  2017b, \mndoi [\mnras]
  {10.1093/mnras/stx1388}, \href
  {http://adsabs.harvard.edu/abs/2017MNRAS.470.3220W} {470, 3220}

\bibitem[\protect\citeauthoryear{{Wouthuysen}}{{Wouthuysen}}{1952}]{1952AJ.....57R..31W}
{Wouthuysen} S.~A.,  1952, \mndoi [\aj] {10.1086/106661}, \href
  {http://adsabs.harvard.edu/abs/1952AJ.....57R..31W} {57, 31}

\bibitem[\protect\citeauthoryear{{Wright}, {Winther}  \& {Koyama}}{{Wright}
  et~al.}{2017}]{Wright:2017dkw}
{Wright} B.~S.,  {Winther} H.~A.,   {Koyama} K.,  2017, \mndoi [\jcap]
  {10.1088/1475-7516/2017/10/054}, \href
  {http://adsabs.harvard.edu/abs/2017JCAP...10..054W} {10, 054}

\bibitem[\protect\citeauthoryear{Wu et~al.,}{Wu et~al.}{2016}]{Wu:2016brq}
Wu X.-F.,  et~al., 2016, \mndoi [Astrophys. J.] {10.3847/2041-8205/822/1/L15},
  822, L15

\bibitem[\protect\citeauthoryear{{Xia}, {Cuoco}, {Branchini}  \& {Viel}}{{Xia}
  et~al.}{2015}]{2015ApJS..217...15X}
{Xia} J.-Q.,  {Cuoco} A.,  {Branchini} E.,   {Viel} M.,  2015, \mndoi
  [Astrophys. J.s] {10.1088/0067-0049/217/1/15}, \href
  {http://adsabs.harvard.edu/abs/2015ApJS..217...15X} {217, 15}

\bibitem[\protect\citeauthoryear{Yahya, Bull, Santos, Silva, Maartens, Okouma
  \& Bassett}{Yahya et~al.}{2015}]{Yahya:2014yva}
Yahya S.,  Bull P.,  Santos M.~G.,  Silva M.,  Maartens R.,  Okouma P.,
  Bassett B.,  2015, \mndoi [Mon. Not. Roy. Astron. Soc.]
  {10.1093/mnras/stv695}, 450, 2251

\bibitem[\protect\citeauthoryear{Yatawatta et~al.,}{Yatawatta
  et~al.}{2013}]{Yatawatta2013}
Yatawatta S.,  et~al., 2013, \mndoi [Astronomy and Astrophysics]
  {10.1051/0004-6361/201220874}, 550, A136

\bibitem[\protect\citeauthoryear{{Yoo}, {Hamaus}, {Seljak}  \&
  {Zaldarriaga}}{{Yoo} et~al.}{2012}]{2012PhRvD..86f3514Y}
{Yoo} J.,  {Hamaus} N.,  {Seljak} U.,   {Zaldarriaga} M.,  2012, \mndoi [\prd]
  {10.1103/PhysRevD.86.063514}, \href
  {http://adsabs.harvard.edu/abs/2012PhRvD..86f3514Y} {86, 063514}

\bibitem[\protect\citeauthoryear{{Young} \& {Byrnes}}{{Young} \&
  {Byrnes}}{2015}]{Byrnes1}
{Young} S.,  {Byrnes} C.~T.,  2015, \mndoi [\prd] {10.1103/PhysRevD.91.083521},
  \href {http://adsabs.harvard.edu/abs/2015PhRvD..91h3521Y} {91, 083521}

\bibitem[\protect\citeauthoryear{{Young}, {Regan}  \& {Byrnes}}{{Young}
  et~al.}{2016}]{Byrnes2}
{Young} S.,  {Regan} D.,   {Byrnes} C.~T.,  2016, \mndoi [\jcap]
  {10.1088/1475-7516/2016/02/029}, \href
  {http://adsabs.harvard.edu/abs/2016JCAP...02..029Y} {2, 029}

\bibitem[\protect\citeauthoryear{{Yu}, {Zhang}  \& {Pen}}{{Yu}
  et~al.}{2014}]{Chime2014}
{Yu} H.-R.,  {Zhang} T.-J.,   {Pen} U.-L.,  2014, \mndoi [Physical Review
  Letters] {10.1103/PhysRevLett.113.041303}, \href
  {http://adsabs.harvard.edu/abs/2014PhRvL.113d1303Y} {113, 041303}

\bibitem[\protect\citeauthoryear{{Zahn} \& {Zaldarriaga}}{{Zahn} \&
  {Zaldarriaga}}{2006}]{Zahn:2005ap}
{Zahn} O.,  {Zaldarriaga} M.,  2006, \mndoi [\apj] {10.1086/508916}, \href
  {http://adsabs.harvard.edu/abs/2006ApJ...653..922Z} {653, 922}

\bibitem[\protect\citeauthoryear{{Zhang} \& {Iorio}}{{Zhang} \&
  {Iorio}}{2017}]{2017ApJ...834..198Z}
{Zhang} F.,  {Iorio} L.,  2017, \mndoi [\apj] {10.3847/1538-4357/834/2/198},
  \href {http://adsabs.harvard.edu/abs/2017ApJ...834..198Z} {834, 198}

\bibitem[\protect\citeauthoryear{{Zhang} \& {Saha}}{{Zhang} \&
  {Saha}}{2017}]{2017ApJ...849...33Z}
{Zhang} F.,  {Saha} P.,  2017, \mndoi [\apj] {10.3847/1538-4357/aa8f47}, \href
  {http://adsabs.harvard.edu/abs/2017ApJ...849...33Z} {849, 33}

\bibitem[\protect\citeauthoryear{Zhang, Liguori, Bean  \& Dodelson}{Zhang
  et~al.}{2007}]{Zhang:2007nk}
Zhang P.,  Liguori M.,  Bean R.,   Dodelson S.,  2007, \mndoi [Phys. Rev.
  Lett.] {10.1103/PhysRevLett.99.141302}, 99, 141302

\bibitem[\protect\citeauthoryear{{Zhang}, {Lu}  \& {Yu}}{{Zhang}
  et~al.}{2014}]{2014ApJ...784..106Z}
{Zhang} F.,  {Lu} Y.,   {Yu} Q.,  2014, \mndoi [\apj]
  {10.1088/0004-637X/784/2/106}, \href
  {http://adsabs.harvard.edu/abs/2014ApJ...784..106Z} {784, 106}

\bibitem[\protect\citeauthoryear{{Zhang}, {Lu}  \& {Yu}}{{Zhang}
  et~al.}{2015}]{2015ApJ...809..127Z}
{Zhang} F.,  {Lu} Y.,   {Yu} Q.,  2015, \mndoi [\apj]
  {10.1088/0004-637X/809/2/127}, \href
  {http://adsabs.harvard.edu/abs/2015ApJ...809..127Z} {809, 127}

\bibitem[\protect\citeauthoryear{Zhu et~al.,}{Zhu et~al.}{2015}]{Zhu:2015tua}
Zhu X.-J.,  et~al., 2015, \mndoi [Mon. Not. Roy. Astron. Soc.]
  {10.1093/mnras/stv381}, 449, 1650

\bibitem[\protect\citeauthoryear{{Zhu} et~al.,}{{Zhu}
  et~al.}{2019}]{Zhu:2018etc}
{Zhu} W.~W.,  et~al., 2019, \mndoi [Monthly Notices of the Royal Astronomical
  Society] {10.1093/mnras/sty2905}, \href
  {https://ui.adsabs.harvard.edu/abs/2019MNRAS.482.3249Z} {482, 3249}

\bibitem[\protect\citeauthoryear{Zibin}{Zibin}{2011}]{Zibin:2011ma}
Zibin J.~P.,  2011, \mndoi [Phys. Rev.] {10.1103/PhysRevD.84.123508}, D84,
  123508

\bibitem[\protect\citeauthoryear{{Zuntz} et~al.,}{{Zuntz}
  et~al.}{2015}]{2015A&C....12...45Z}
{Zuntz} J.,  et~al., 2015, \mndoi [Astronomy and Computing]
  {10.1016/j.ascom.2015.05.005}, \href
  {http://adsabs.harvard.edu/abs/2015A%26C....12...45Z} {12, 45}

\bibitem[\protect\citeauthoryear{{Zygelman}}{{Zygelman}}{2005}]{Zygelman05}
{Zygelman} B.,  2005, \mndoi [Astrophysical Journal] {10.1086/427682}, \href
  {http://adsabs.harvard.edu/abs/2005ApJ...622.1356Z} {622, 1356}

\bibitem[\protect\citeauthoryear{de Cesare \& Sakellariadou}{de~Cesare \&
  Sakellariadou}{2017}]{deCesare:2016axk}
de Cesare M.,  Sakellariadou M.,  2017, \mndoi [Phys. Lett.]
  {10.1016/j.physletb.2016.10.051}, B764, 49

\bibitem[\protect\citeauthoryear{de Cesare, Lizzi  \& Sakellariadou}{de~Cesare
  et~al.}{2016}]{deCesare:2016dnp}
de Cesare M.,  Lizzi F.,   Sakellariadou M.,  2016, \mndoi [Phys. Lett.]
  {10.1016/j.physletb.2016.07.015}, B760, 498

\bibitem[\protect\citeauthoryear{{de Gasperin}, {Ogrean}, {van Weeren},
  {Dawson}, {Br{\"u}ggen}, {Bonafede}  \& {Simionescu}}{{de Gasperin}
  et~al.}{2015}]{2015MNRAS.448.2197D}
{de Gasperin} F.,  {Ogrean} G.~A.,  {van Weeren} R.~J.,  {Dawson} W.~A.,
  {Br{\"u}ggen} M.,  {Bonafede} A.,   {Simionescu} A.,  2015, \mndoi [\mnras]
  {10.1093/mnras/stv129}, \href
  {http://adsabs.harvard.edu/abs/2015MNRAS.448.2197D} {448, 2197}

\bibitem[\protect\citeauthoryear{de Laix, Scherrer  \& Schaefer}{de~Laix
  et~al.}{1995}]{deLaix:1995vi}
de Laix A.~A.,  Scherrer R.~J.,   Schaefer R.~K.,  1995, \mndoi [Astrophys. J.]
  {10.1086/176322}, 452, 495

\bibitem[\protect\citeauthoryear{{van Albada}, {Bahcall}, {Begeman}  \&
  {Sancisi}}{{van Albada} et~al.}{1985}]{1985ApJ...295..305V}
{van Albada} T.~S.,  {Bahcall} J.~N.,  {Begeman} K.,   {Sancisi} R.,  1985,
  \mndoi [\apj] {10.1086/163375}, \href
  {http://adsabs.harvard.edu/abs/1985ApJ...295..305V} {295, 305}

\bibitem[\protect\citeauthoryear{{van Weeren} et~al.,}{{van Weeren}
  et~al.}{2016}]{2016ApJ...818..204V}
{van Weeren} R.~J.,  et~al., 2016, \mndoi [\apj] {10.3847/0004-637X/818/2/204},
  \href {http://adsabs.harvard.edu/abs/2016ApJ...818..204V} {818, 204}

\bibitem[\protect\citeauthoryear{{van Weeren} et~al.,}{{van Weeren}
  et~al.}{2017}]{2017NatAs...1E...5V}
{van Weeren} R.~J.,  et~al., 2017, \mndoi [Nature Astronomy]
  {10.1038/s41550-016-0005}, \href
  {http://adsabs.harvard.edu/abs/2017NatAs...1E...5V} {1, 0005}

\makeatother
\end{thebibliography}

\section{Affiliations}
\label{sec_affil}

\parindent 0pt

\setlength{\parskip}{1em}

A. Weltman: Department of Mathematics \& Applied Mathematics, University of Cape Town, 7701 Rondebosch, Cape Town, South Africa

P. Bull: Department of Astronomy, University of California  Berkeley, Berkeley, CA 94720, USA

S. Camera: Dipartimento di Fisica, Universit\`a degli Studi di Torino,
Via P. Giuria 1, 10125 Torino, Italy;
INFN -- Istituto Nazionale di Fisica Nucleare, Sezione di Torino,
Via P. Giuria 1, 10125 Torino, Italy;
INAF -- Istituto Nazionale di Astrofisica, Osservatorio Astrofisico
di Torino, Strada Osservatorio 20, 10025 Pino Torinese, Italy

K. Kelley: International Centre for Radio Astronomy Research (ICRAR), University of Western Australia, Ken and Julie Michael Building, 7 Fairway, Crawley, WA 6009, Australia

H. Padmanabhan: ETH Zurich, Wolfgang-Pauli-Strasse 27, CH 8093 Zurich, Switzerland;
Canadian Institute for Theoretical Astrophysics, University of Toronto, 
60 St George St, Toronto, ON M5S 3H8, Canada

J. Pritchard: Department of Physics, Imperial College London, Prince Consort Road, London SW7 2AZ, UK

A. Raccanelli: Institut de Ci\`encies del Cosmos (ICCUB), Universitat de Barcelona (IEEC-UB), Mart\'{i} Franqu\`es 1, E08028 Barcelona, Spain

S. Riemer-S\o{}rensen: Institute of Theoretical Astrophysics, University of Oslo, P.O. Box 1029 Blindern, N-0315 Oslo, Norway

L. Shao: Kavli Institute for Astronomy and Astrophysics, Peking University, Beijing 100871, China

S. Andrianomena: 
South African Radio Astronomy Observatory (SARAO), The Park, Park Road, Cape Town 7405, South Africa; Department of Physics \& Astronomy, University of the Western Cape, Cape Town 7535, South Africa

E. Athanassoula: Aix Marseille Univ, CNRS, CNES, LAM, Marseille, France

D. Bacon: Institute of Cosmology \& Gravitation, University of Portsmouth, Portsmouth PO1 3FX, United Kingdom

R. Barkana: Raymond and Beverly Sackler School of Physics and Astronomy, Tel Aviv University, Tel Aviv 69978, Israel

G. Bertone: GRAPPA, Institute of Physics, University of Amsterdam, Science Park 904, 1098 XH Amsterdam, Netherlands

C. B{\oe}hm: School of Physics, The University of Sydney, NSW 2006, Australia

C. Bonvin: D\'epartement de Physique Th\'eorique and Center for Astroparticle Physics,
Universit\'e de Gen\`eve, 1211 Gen\`eve 4, Switzerland

A. Bosma: Aix Marseille Univ, CNRS, CNES, LAM, Marseille, France

M.~Br\"uggen: University of Hamburg, Gojenbergsweg 112, 21029 Hamburg, Germany

C.~Burigana: INAF, Istituto di Radioastronomia, Via Piero Gobetti 101,
I-40129 Bologna, Italy; 
Dipartimento di Fisica e Scienze della Terra, Universit\`a di Ferrara, Via
Giuseppe Saragat 1, I-44122 Ferrara, Italy; 
Istituto Nazionale di Fisica Nucleare, Sezione di Bologna,
Via Irnerio 46, I-40126 Bologna, Italy

F. Calore: GRAPPA, Institute of Physics, University of Amsterdam, Science Park 904, 1098 XH Amsterdam, Netherlands;
LAPTh, CNRS, 9 Chemin de Bellevue, BP-110, Annecy-le-Vieux, 74941, Annecy Cedex, France

J. A. R. Cembranos: Departamento de F\'isica Te\'orica I and UPARCOS, Universidad Complutense de Madrid, E-28040 Madrid, Spain

C. Clarkson: School of Physics \& Astronomy, Queen Mary University of London, London E1 4NS, UK;
Department of Mathematics \& Applied Mathematics, University of Cape Town, 7701 Rondebosch, Cape Town, South Africa;
Department of Physics \& Astronomy, University of the Western Cape, Cape Town 7535, South Africa

R.~M.~T.~Connors: Cahill Center for Astronomy and Astrophysics, California Institute of Technology, Pasadena, CA 91125, USA

\'A. de la Cruz-Dombriz: Cosmology and Gravity Group and Mathematics and Applied Mathematics Department, University of Cape Town, 7701 Rondebosch, South Africa

P.~K.~S.~Dunsby: University of Cape Town, 7701 Rondebosch, Cape Town, South Africa;
South African Astronomical Observatory, Observatory 7925, Cape Town, South Africa

J. Fonseca: Dipartimento di Fisica e Astronomia ``G. Galilei'', Università degli Studi di Padova, Via Marzolo 8, 35131 Padova, Italy

N. Fornengo: Dipartimento di Fisica, Universit\`a degli Studi di Torino, Via P. Giuria 1, 10125 Torino, Italy;
INFN, Istituto Nazionale di Fisica Nucleare, Sezione di Torino, Via P. Giuria 1, 10125 Torino, Italy

D. Gaggero: GRAPPA, Institute of Physics, University of Amsterdam, Science Park 904, 1098 XH Amsterdam, Netherlands

I. Harrison: Jodrell Bank Centre for Astrophysics, The University of Manchester, Manchester M13 9PL, UK

J. Larena: Department of Mathematics \& Applied Mathematics, University of Cape Town, 7701 Rondebosch, Cape Town, South Africa

Y.-Z. Ma: School of Chemistry and Physics, University of KwaZulu-Natal,
Westville Campus, Private Bag X54001, Durban, 4000, South Africa; 
NAOC-UKZN Computational Astrophysics Centre (NUCAC),
University of KwaZulu-Natal, Durban, 4000, South Africa;
Purple Mountain Observatory, Chinese Academy of Sciences, Nanjing 210008, China

R. Maartens: Department of Physics \& Astronomy, University of the Western Cape, Cape Town 7535, South Africa;
Institute of Cosmology \& Gravitation, University of Portsmouth, Portsmouth PO1 3FX, United Kingdom

M. M\'endez-Isla: University of Cape Town, 7701 Rondebosch, Cape Town, South Africa

S.~D.~Mohanty: Department of Physics and Astronomy, The University of Texas Rio Grande Valley,
One West University Blvd, Brownsville, TX 78520, USA

S. Murray: International Centre for Radio Astronomy Research (ICRAR), Curtin University,  Bentley, WA 6102, Australia

D. Parkinson: School of Mathematics \& Physics, University of Queensland, St Lucia, QLD 4072, Australia;
Korea Astronomy and Space Science Institute, Daejeon 34055, Korea

A. Pourtsidou: School of Physics and Astronomy, Queen Mary University of London, London E1 4NS, UK; Institute of Cosmology and Gravitation, University of Portsmouth, PO1 3FX, UK 

P. J. Quinn: International Centre for Radio Astronomy Research (ICRAR), University of Western Australia, Ken and Julie Michael Building, 7 Fairway, Crawley, WA 6009, Australia

M. Regis: Dipartimento di Fisica, Universit\`a degli Studi di Torino, Via P. Giuria 1, 10125 Torino, Italy;
INFN, Istituto Nazionale di Fisica Nucleare, Sezione di Torino, Via P. Giuria 1, 10125 Torino, Italy

P. Saha: Department of Physics, University of Zurich, Winterthurerstrasse
190, 8057 Zurich, Switzerland;
Institute for Computational Science, University of Zurich,
Winterthurerstrasse 190, 8057 Zurich, Switzerland

M. Sahl\'en: Department of Physics and Astronomy, Uppsala University, SE-751
20, Uppsala, Sweden

M. Sakellariadou: Theoretical Particle Physics \& Cosmology Group, Department of
Physics, King's College London, University of London, Strand, London WC2R 2LS,
UK

J. Silk: Institut d'Astrophysique, UMR 7095 CNRS, Universit\'e Pierre et Marie Curie, 98bis Blvd Arago, 75014 Paris, France;
AIM-Paris-Saclay, CEA/DSM/IRFU, CNRS, Univ Paris 7, F-91191, Gif-sur-Yvette, France;
Department of Physics and Astronomy, The John Hopkins University, Homewood Campus, Baltimore MD 21218, USA;
Beecroft Institute of Particle Astrophysics and Cosmology, Department of Physics, University of Oxford, Oxford OX1 3RH, UK

T.~Trombetti:
Dipartimento di Fisica e Scienze della Terra, Universit\`a di
Ferrara, Via Giuseppe Saragat 1, I-44122 Ferrara, Italy;
Istituto Nazionale di Fisica Nucleare, Sezione di Ferrara,
Via Giuseppe Saragat 1, I-44122 Ferrara, Italy;
Istituto Nazionale di Astrofisica, Istituto di Radioastronomia, Via Piero Gobetti 101, I-40129 Bologna, Italy

F. Vazza: Dipartimento di Fisica e Astronomia, Universit\`a' di Bologna, Via Gobetti 93/2, 40122, Italy; Istituto Nazionale di Astrofisica, Istituto di Radioastronomia, Via Piero Gobetti 101, I-40129 Bologna, Italy;
University of Hamburg, Gojenbergsweg 112, 21029 Hamburg, Germany

T. Venumadhav: Institute for Advanced Study, 1 Einstein Drive Princeton, NJ 08540, USA

F. Vidotto: University of the Basque Country UPV/EHU, Departamento de F\'isica Te\'orica, Barrio Sarriena s/n, 48940 Leioa, Bizkaia, Spain

F. Villaescusa-Navarro: Center for Computational Astrophysics, Flatiron Institute, 162 5th Avenue, New York, NY, 10010, USA

Y. Wang: School of Physics, Huazhong University of Science and Technology,
  1037 Luoyu Road, Wuhan, Hubei Province 430074, China

C. Weniger: GRAPPA, Institute of Physics, University of Amsterdam, Science Park 904, 1098 XH Amsterdam, Netherlands

L.~Wolz: School of Physics, University of Melbourne, Parkville, 3010, Victoria, Australia

F. Zhang: School of Physics and Electronic Engineering, Guangzhou University, 510006 Guangzhou, China

B. M. Gaensler: Dunlap Institute for Astronomy and Astrophysics, 50 St. George Street, University of Toronto, ON M5S 3H4, Canada

\end{document}